TECHNISCHE UNIVERSITÄT WIEN
Vienna University of Technology

# Dissertation

# How General Is Holography?

*Flat Space and Higher-Spin Holography in 2+1 Dimensions*

ausgeführt zum Zwecke der Erlangung des akademischen Grades eines
Doktors der technischen Wissenschaften

unter der Anleitung von

*Ass.-Prof. Priv.-Doz. Dr.* Daniel Grumiller
Institut für Theoretische Physik (E136)
TU Wien

und mitbetreut durch

*Prof. Dr.* Radoslav Rashkov
Institut für Theoretische Physik (E136)
TU Wien

eingereicht an der Technischen Universität Wien
Fakultät für Physik

von

*Dipl.-Ing.* Max Riegler
Matrikelnummer: 0725216
Strozzigasse 27/7
1080 Wien

Diese Dissertation haben begutachtet:

Ass.-Prof. Priv.-Doz. Dr. Daniel Grumiller | Prof. Dr. Alejandra Castro Anich

Wien, 6.3.2016

Dipl.-Ing. Max Riegler

**Max Riegler**
*How General Is Holography?*
*Flat Space and Higher-Spin Holography in 2+1 Dimensions*, 6.3.2016
Gutachter/Gutachterin: Daniel Grumiller und Alejandra Castro Anich
Betreuer: Daniel Grumiller und Radoslav Rashkov

**TU Wien**
Institut für Theoretische Physik
Wiedner Hauptstrasse 8-10/E136
A-1040, Wien

# Kurzfassung

*"Wie allgemein ist das holographische Prinzip?"*

Dieser Frage werde ich mich in meiner Dissertation widmen. Sie ist von so fundamentaler Natur, dass es sinnvoll erscheint, sich diesem Problem in einer Umgebung zu stellen, die möglichst simpel, aber dennoch interessant und komplex genug ist, um generelle Aussagen tätigen zu können. Da Gravitation in 2+1 Dimensionen diesen Anforderungen entspricht, werde ich mich auf Holographie konzentrieren, deren Gravitationstheorien 2+1 Dimensionen und deren Quantenfeldtheorien 1+1 Dimensionen aufweisen. Die zwei wichtigsten Gründe hierfür sind, dass (i) Gravitation in 2+1 Dimensionen technisch sehr effizient beschrieben werden kann, und (ii) dass die dualen Quantenfeldtheorien unendlich viele Symmetrien haben und somit einen sehr hohen Grad an Kontrolle erlauben. Dies ermöglicht es, neuartige holographische Korrespondenzen exakt zu überprüfen.

Von speziellem Interesse, was die allgemeine Gültigkeit des holographischen Prinzips betrifft, sind so genannte Höhere-Spin-Gravitationstheorien, welche die übliche lokale Koordinateninvarianz mit weiteren verallgemeinerten Symmetrien erweitern.

Zunächst beschäftige ich mich mit Höherer-Spin-Holographie, die auf Raumzeiten basiert, welche nicht asymptotisch Anti-de-Sitter sind. Von einer gegebenen Gravitationstheorie ausgehend, bestimme ich in weiterer Folge die dazugehörigen asymptotischen Symmetrien der dualen Quantenfeldtheorien und unitäre Repräsentationen dieser Symmetriealgebren. Weiters beschreibe ich eine Möglichkeit, im Rahmen dieser "Nicht-Anti-de-Sitter-Holographie" eine duale Quantenfeldtheorie zu erhalten, welche eine beliebig große (aber nicht unendliche) Anzahl von Quantenzuständen zulässt.

Der zweite Teil dieser Dissertation beschäftigt sich mit Holographie für asymptotisch flache Raumzeiten. Zuerst zeige ich, wie man verschiedene Ergebnisse, wie zum Beispiel eine (Höhere-Spin) Cardy-Formel für flache Raumzeiten, welche die Anzahl der Quantenzustände einer konformen Feldtheorie bei einer bestimmten Temperatur angibt, oder die asymptotischen Symmetrien von asymptotisch flachen Raumzeiten als Limes einer verschwindenden kosmologischen Konstante der bekannten Anti-de-Sitter-Ergebnisse erhalten kann.

Weiters setze ich mich mit unitären Repräsentationen der asymptotischen Symmetriealgebren von asymptotisch flachen Raumzeiten auseinander. Dies führt unter bestimmten Annahmen zu einem NO-GO-Theorem, welches nicht gleichzeitig flache Raumzeiten, Höhere-Spin-Symmetrien und Unitarität erlaubt. Ebenso wird eine Möglichkeit, dieses NO-GO-Theorem zu umgehen, explizit behandelt.

Überdies werde ich zeigen, wie man asymptotisch flache Raumzeiten inklusive (Höhere-Spin) chemischer Potentiale holographisch konsistent beschreiben kann. Ebenso werde ich die dazugehörige thermale Entropie von bestimmten kosmologischen asymptotisch flachen Raumzeiten bestimmen.

Den Schluss meiner Dissertation bildet eine explizite Überprüfung des holographischen Prinzips für asymptotisch flache Raumzeiten. Ich präsentiere eine Methode, die eine bestimmte Form einer Wilson-Schleife darstellt, die eine holographische Bestimmung der Verschränkungsentropie von Feldtheorien erlaubt, von denen angenommen wird, dual zu asymptotisch flachen Raumzeiten zu sein. Ich erweitere die Methode überdies, um auch erfolgreich Höhere-Spin-Symmetrie miteinzubeziehen und die thermale Entropie der zugehörigen dualen Feldtheorien bestimmen zu können.

# Abstract

*"How general is the holographic principle?"*

This is the question I will explore in this thesis. As this question is very fundamental, one is well advised to try and tackle the problem in an environment which is as simple as possible but still interesting and complex enough to allow for a general interpretation of the results. Since gravity in 2+1 dimensions satisfies those requirements I will focus on holography involving 2+1 dimensional spacetimes and 1+1 dimensional quantum field theories. The two most important reasons for this are: (i) gravity in 2+1 dimensions can be described very efficiently on a technical level. (ii) The dual quantum field theories have infinitely many symmetries and thus allow for a very high degree of control. This allows one to explicitly and exactly check new holographic correspondences.

Of very special interest regarding the generality of the holographic principle are so-called higher-spin gravity theories which extend the usual local invariance under coordinate changes by a more general set of symmetries.

In this thesis I will first focus on higher-spin holography which is based on spacetimes that do not asymptote to Anti-de Sitter spacetimes. Starting from a given higher-spin theory I will determine the corresponding asymptotic symmetries of the corresponding dual quantum field theories and their unitary representations. Furthermore, using "non-Anti-de Sitter holography" I will describe a dual quantum field theory, which allows for an arbitrary (albeit not infinitely) large number of quantum microstates.

The second part of this thesis is concerned with holography for asymptotic flat spacetimes. First I will show how to obtain various results, like an analogue of a (higher-spin) Cardy formula which counts the number of microstates of a conformal field theory at a given temperature, or the asymptotic symmetries of asymptotically flat spacetimes, as a limit of vanishing cosmological constant from the known Anti-de Sitter results.

Furthermore, I will explore unitary representations of the asymptotic symmetry algebras of asymptotically flat spacetimes, which under certain assumptions, will result in a NO-GO theorem that forbids having flat space, higher-spins and unitarity at the same time. In addition I will elaborate on a specific example that allows to circumvent this NO-GO theorem.

I will also show how to consistently describe asymptotically flat spacetimes with additional (higher-spin) chemical potentials in a holographic setup and how to determine the corresponding thermal entropy of certain cosmological asymptotically flat spacetimes.

The finale of this thesis will be an explicit check of the holographic principle for asymptotically flat spacetimes. I will present a method, using a special version of a Wilson-line, which allows one to determine the entanglement entropy of field theories which are assumed to be dual to asymptotically flat spacetimes in a holographic manner. I will also extend this method in order to be able to also successfully include higher-spin symmetries and determine the thermal entropy of the corresponding dual field theories.

# Acknowledgements

無我夢中 (mugamuchū) is a Japanese expression whose literal meaning can be roughly translated as "without myself, in the middle of a dream". A little bit less literally taken it usually refers to the act of being absorbed in something so much that one completely forgets about other things. If I had to describe myself with one word, I think 無我夢中 would be the most appropriate one. Once something has caught my interest I get completely absorbed in it. Out of the things that caught my interest so far, there are two which deeply influenced me up until today: physics and rock climbing. Luckily both have a lot in common and there are many things rock climbing has taught me that can be applied to the way I approach fundamental physics research and vice versa. One such thing is that it is good to have partners whom you can rely on. In climbing as well as in physics it is possible to accomplish great things being on one's own. Reliable partners, however, make the whole story a bit safer and much more likely to succeed. Thus, I want to thank all the people and institutions who gave me the opportunity to do this research on holography.

First I want to thank my family whose continuous support throughout all the years made it possible for me to study physics and eventually write this thesis. I want to especially thank my mother **Ulrike Riegler** whose continuous support has always been, and will always be invaluable to me. I would not be the physicist I am today without her! Special thanks go also to my aunt **Karin Riegler** who, like my mother, always supported me in whatever I wanted to do. Also my grandparents **Angela** and **Alfred Riegler** as well as my uncle **Günther Riegler** have my gratitude. Last but by no means least I want to thank my fiancé **Nagisa Nakamura** (中村凪沙) who was always there for me during the last two years of my PhD and supported me in everything I did. 宝物、今まで本当にありがとうございました。私は貴方と出会うまで、ほかの人をこんなに好きになると一度も想像できませんでした。これからも宜しくお願いしますね。

Of course I would also not have been to able to write this thesis without my advisor **Daniel Grumiller** whose advice and continuous support was invaluable. Ever since we first met, almost seven years ago, you have guided me towards the path of doing fundamental theoretical physics research. You supported me when I went to Zürich as an ERASMUS student during my masters degree you gave me the opportunity to go to Korea for almost half a year for a research visit with Soo-Jong Rey's group and you supported me when I wanted to go to Tadashi Takayanagi's group in Japan in order to learn more about holographic entanglement entropy. You gave me a lot of freedom in pursuing my interests and always had faith in my work. Thank you very much! Special thanks also to my co-advisor **Radoslav Rashkov** for his continuous support.

I would like to thank the **Austrian Academy of Sciences** for awarding me a *DOC* grant over 70.000 € which financed my PhD studies for almost two years and furthermore allowed me to pursue my research in Vienna. I feel very honored and am very grateful to have received this grant.

Likewise I want to thank the **Doktoratskolleg Particles and Interactions – DKPI** for accepting me as a PhD student. I am very grateful for all the benefits I had via the retreats, summer schools and special travel opportunities. I also want to thank all the people who made it possible to have such a Doktoratskolleg in Vienna.

The fincancial support of the **Austrian Science Fund (FWF)** has played a major role in the early stages of this thesis and also for projects outside of the scope of this thesis. Thus, I would like to thank the FWF for funding the projects **P 27182-N27**, **I 1030-N27** and the **START** project **Y435-N16**.

I want to thank **Soo-Jong Rey** for inviting me to a research visit with him and his group at Seoul National University in Korea for almost half a year at the beginning of my PhD. I had a wonderful time in Korea (also special thanks to **Subong Yu** and **Jiwon Kim**) and learned a lot.

I want to thank **Tadashi Takayanagi** for giving me the opportunity for a research visit with his group at Kyoto University in Japan for four months in order to learn more about holographic entanglement entropy. I really enjoyed being in Kyoto and am very grateful for this wonderful opportunity.

A PhD thesis by itself is of course not a journey one embarks on alone. I want to thank all my colleagues and collaborators and the people I met during my PhD for enlightening discussions. **Hamid Afshar**, **Martin Ammon**, **Arjun Bagchi**, **Rudranil Basu**, **Frederic Brünner**, **Andrea Campoleoni**, **Christian Ecker**, **Mirah Gary**, **Hernan Gonzalez**, **Sung Ho**, **Alexander Jahn**, **Masahiro Nozaki**, **Blagoje Oblak**, **Stefan Prohazka**, **Jan Rosseel** and **Jakob Salzer**. Also special thanks to **Alejandra Castro Anich** for valuable input and being one of my referees for this thesis.

I also want to take this opportunity to thank those of my teachers back in primary and high school who had great influence on me. I want to thank **Friederike Grabner**, **Roswitha Lachmayr**, **Helmut Linhart**, **Elke Renner** and **Harald Straßl**. Thank you for being extraordinary teachers!

Last but by no means least I want to thank everyone who made my time as a PhD, no matter where on this planet, an unforgettable part of my life. First I would like to thank **Jörg Doppler** for introducing me to the world of bouldering and subsequently also to rock climbing almost six years ago. I also want to thank the Austrian climbing crew, **Wladimir Agranat**, **Sandra Gombotz**, **Linda Lam**, **Andreas Neumann**, **Matthias Sablatschan**, **Lukas Schmutzer**, **Edda Marie Rainer**, **Wolfgang Riedler**, **Micha Seiwerth**, **Mario Walther**, the Korean climbing crew, **David Holmes**, **Robin Kimmerling**, **Tammy Lee**, **Dong-Il Ryu** and the Japanese climbing crew, **Michinori Ano** (阿野道徳), **Taimei Dogura** (土倉大明), **Motoko Katsu**, **Hiroaki Oda** (小田博明), **Tom Rush**, **Wataru Yamada** (山田航) and **Mana Yamashita** (山下真奈).

Also special thanks to my friends **Christoph Fuß**, **Julian Rossmann** and **Sebastian Singer** whom I have known for years and no matter how much time has passed, we still have the same connection that we already shared during our time in high school.

# Contents

# List of Figures

# Introduction

# 0

> „DON'T PANIC
>
> **– Douglas Adams**
> The Hitchhiker's Guide to the Galaxy

The holographic principle proposes a duality between a (d+1)-dimensional theory of (quantum) gravity and a d-dimensional quantum field theory located at the boundary of the gravity theory and has been a very active and successful field of research in modern theoretical physics during the last twenty years.
The reason for this huge success is that the holographic principle allows one to access regimes of gravitational and quantum field theories alike, which would otherwise be impossible to access and thus providing an intimate link between geometry and quantum (field) theory. Depending on the observable and the theory one can potentially freely switch between a gravitational description or a quantum field theory description. It is exactly this property, which allows to describe a theory in d+1 dimensions in terms of a d-dimensional theory that gave the holographic principle its name.

Usually when speaking about a hologram one thinks about a two dimensional screen on which not only the intensity but also the phase information of a given object is imprinted on. Thus when looking at the screen our brain takes that information stored on the two dimensional surface and assembles it in such a way that we perceive it as a three dimensional object because that is the most useful way for our mind to perceive that information. From a physical standpoint however it does not matter in which way we describe the object. We can describe the object as the three dimensional thing as we perceive it, or we could equivalently describe it purely in terms of the intensity and phase information stored on the two dimensional holographic screen. Both descriptions are equivalent, and which one we use depends on what we want do describe. For some applications the three dimensional description might be better, while for others the two dimensional one might be more suitable.

**AdS/CFT:** Undoubtedly the most famous explicit realization of the holographic principle is the Anti-de Sitter/Conformal Field Theory (AdS/CFT) correspondence, which was discovered by Juan Maldacena in 1997 [1] relating a theory of quantum gravity with negative cosmological constant with a quantum field theory which is invariant under conformal i.e. angle but not length preserving transformations[1]. One of the main reasons why this particular example has generated such a high interest is that the AdS/CFT duality is a strong-weak duality [2, 3]. If the quantum field theory is strongly coupled, then this regime of the theory is described by a weakly coupled gravity theory – and vice versa. Thus, one of the main applications of the AdS/CFT duality and suitable generalizations thereof up until now has been to study quantum field theories at strong coupling such as quantum chromodynamics (QCD) [4] using a weakly coupled gravity theory where calculations are comparatively easy. This impressive application of the holographic principle to strongly coupled quantum field theories was actually not the original incentive for studying the holographic principle. Originally it was, and still mainly is, intended to study quantum gravity, i.e. gravity at strong coupling. And indeed one of the first great successes of holography and AdS/CFT in particular was a clear path towards a possible resolution of the black hole information paradox.

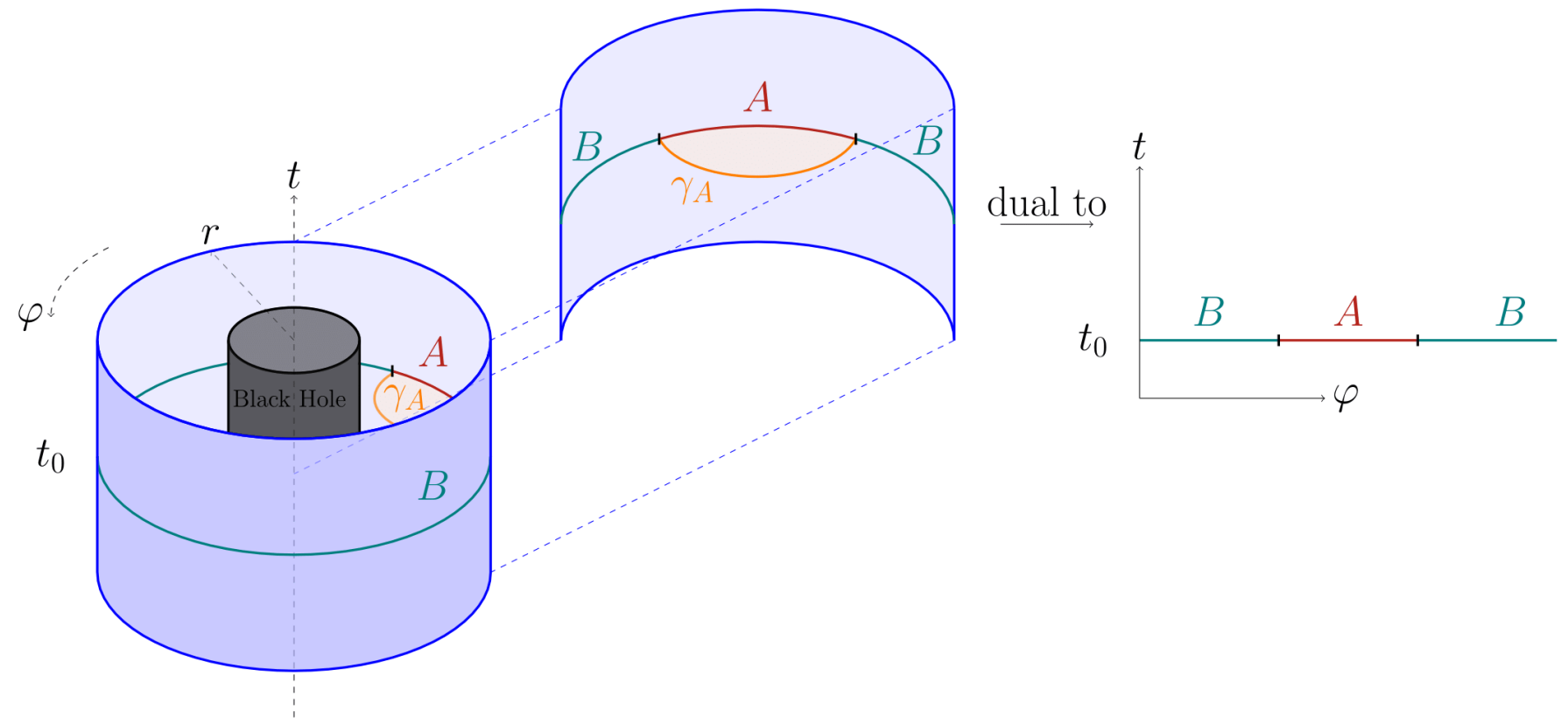

**Fig. 0.1.:** This figure shows a particularly nice example of how the holographic correspondence relates bulk geometry and a quantum field theory observable. The spacetime represents three dimensional Anti-de Sitter space, which is a space of constant negative curvature whereas the dual boundary theory is a two dimensional conformal field theory which at $t = t_0$ can be divided into two subsystems $A$ and its complement $B$. The quantity of interest is the entanglement entropy of the subsystem $A$, which is a measure for the amount of entanglement between $A$ and $B$. One can either determine the entanglement entropy using specific quantum field theory methods, or holographically, by simply calculating the length of the geodesic $\gamma_A$ attached at the boundary in the way it is shown in the figure. Both calculations yield the exact same result and are thus equivalent methods to determine entanglement entropy of the system $A$ with system $B$.

**Black Holes as Origins of Holography:** Black holes as exact solutions of the Einstein equations were originally thought of having no entropy until Bekenstein proposed the following Gedankenexperiment. Assume a cup filled with a hot gas that has a given nonzero entropy and throw this cup into a black hole. If a black hole in general did not have any entropy, then this would clearly violate the second law of thermodynamics as one would have successfully decreased the entropy of the total system by simply throwing that cup into a black hole. This led Bekenstein to think of black holes as objects having entropy. Actually to be more precise, he thought of black holes as objects having the *maximum* amount of entropy possible for a given region of spacetime. Just imagine a hot relativistic gas confined to a spherical region with a given, fixed radius. One can increase the entropy of this system by adding more and more energy up until there is so much energy in that region that everything collapses into a black hole, which now carries all the entropy of the system prior to its collapse. Bekenstein then found that the upper bound of the entropy after this collapse, surprisingly, is proportional to the area of the black hole horizon rather than its volume. This was a rather puzzling feature as entropy usually encodes the density of states of a given system and thus one would naively expect the entropy to scale with the volume of the system rather than the area of its boundary. To add more confusion, Hawking and Bekenstein showed in 1975 [5, 6] that if black holes carry entropy they also have to slowly radiate away energy, which subsequently led to a paradox called "the black hole information loss paradox".
Assume that one would throw some pure quantum state e.g. represented by a wave function into a black hole. After its absorption what would be radiated away from the black hole after some time has passed would not be a pure state anymore, but some thermal state e.g. represented by a density matrix. This process of transforming a pure state into a thermal state is something that is clearly at odds with (unitary) quantum mechanics as there would be information lost about the infalling pure state, which of course is problematic. One possible way to resolve this issue has been formulated by 't Hooft [7]. He noticed that there is a way how incoming particles can influence outgoing, i.e. radiated ones, by deforming the event horizon via their own gravitational field. Thus an undeformed horizon would radiate a different spectrum than a deformed one. Hence the information of the infalling particle would not be lost, but rather is imprinted on the area of the horizon, whose deformations again completely determine the outgoing radiation. Independent from 't Hooft, Susskind found even stronger evidence that indeed such a holographic description can resolve the information paradox by relating long highly excited string states with black holes [8], thus providing an explicit connection to string theory whose 2d world sheet description is basically a holographic description. Historically this was the first example of a new insight on effects on the border between quantum theory and gravity provided by the holographic principle.

[1]To be more precise, the duality relates a type IIB string theory on $AdS_5 \times S^5$ with a $\mathcal{N} = 4$ supersymmetric Yang-Mills theory in four dimensions.

**Holography Without String Theory?** Since the holographic principle first emerged in string theory, it is a natural question to ask how much string theory is hidden in the holographic principle? Or is holography a general principle which is independent of string theory and can thus be applied to basically any spacetime and quantum field theory? In order to answer this question one has to try and check whether or not one can apply and/or generalize similar techniques and arguments which are known from the AdS/CFT correspondence to other setups and establish new holographic correspondences. This is usually most efficiently done in a setup that is as simple and at the same time still as interesting as possible. On the gravity side the first measure to take when trying to simplify things is reducing the number of dimensions as much as possible without losing all the interesting features which one wishes to study. There are two things which should be possible to describe if one wishes to learn more about quantum gravity in general: gravitons and black holes. The lowest number of dimensions where it is possible to have both[2] is $d = 3$.

**Gravity in Three Dimensions:** Not only is gravity in three dimensions a lot easier to handle than in higher dimensions, as it allows a reformulation in terms of a Chern-Simons gauge theory [9], but also one can actually quantize Einstein-Hilbert gravity, as there are no local degrees of freedom present [10]. Furthermore, Brown and Henneaux showed in 1986 [11] that the asymptotic symmetries of $AdS_3$ are given by two copies of the Virasoro algebra, the symmetry algebra of two dimensional conformal field theories. Thus, more than ten years before the great breakthrough of the holographic principle based on the seminal paper [1] of Juan Maldacena, Brown and Henneaux had already found a holographic correspondence which does not (explicitly) rely on string theory. Since conformal field theories in 2d are extremely well studied and on top of that allow for a very high degree of control, the $AdS_3/CFT_2$ correspondence provided a very rich testing ground for the holographic principle, see e.g. for selected references [12–15].

**Higher-Spin Symmetries:** One particularly intensively studied aspect of AdS holography in three dimensions is related to so called higher-spin symmetries inspired by the seminal work by Klebanov and Polyakov [16–18]. From a bulk perspective these symmetries can be considered as generalizations of local coordinate transformations, although their precise geometric meaning like a higher-spin analogue of a Riemann tensor for example is not very well understood as of yet. On the field theory side, however, these symmetries have a more intuitive interpretation as fields with spin[3] $s > 2$, which is also the source for their name. The simplifying magic of gravity in three dimensions also applies to the study of higher-spin symmetries as one can consistently work with a finite tower of massless higher-spin fields [19] in contrast to higher dimensions where the study of higher-spin symmetries is much more involved. See e.g. [20] for a review on higher-spin theories in general. Now why is it of such great interest to study these higher-spin symmetries? There are a couple of reasons for that. Maybe the most important one is that massive higher-spin excitations already appear quite naturally in (super)string theories. Usually quantum field theories are non-renormalizable if there are massive particles with spin $s \geq 1$ present, unless the mass was acquired through some kind of spontaneous symmetry breaking. Thus, it could be possible that string theory with its massive higher-spin excitations is actually just a broken phase of some more general gauge theory with additional, unbroken higher-spin symmetries and the corresponding massless higher-spin excitations [21].

[2]Having either gravitons or black holes or both depends on the specific three dimensional gravity theory in question. In $d = 2$ it is possible to have black holes in some theories, but impossible to have gravitons.

[3]To be more precise, higher-spin fields have in general a conformal weight $h > 1$ which is related to the spin $s$ as $s = h + 1$.

Constructing such higher-spin theories in flat space in dimensions $\mathrm{d} > 3$ is not an easy task to perform as there is a very famous theorem originally formulated by Coleman and Mandula [22] and generalized by Pelc and Horwitz [23] that rules out the possibility of having interacting higher-spin fields. However, in $\mathrm{d} = 3$, one can circumvent this NO-GO theorem even in flat spacetimes, while in $\mathrm{d} > 3$ the introduction of a cosmological constant is enough to circumvent it [24–27]. Thus, most of the research in higher-spin holography up until recently was focused on AdS [28–31] and holographic aspects thereof [32], see [IV, 33–35] for selected reviews. For possible future applications to condensed matter physics or checking the generality of the holographic principle, however, it is of interest to try and formulate holographic correspondences which are not asymptotically AdS.

**Holography Beyond AdS:** At least in three dimensions, higher-spin theories have turned out to be a very fertile ground to study spacetimes, which do not asymptote to AdS [36] such as Lobachevsky [I, II, 37], null warped AdS and their generalizations Schrödinger [38–41], Lifshitz spacetimes [42–44] and de Sitter holography [45] in a very efficient way. Aside from being a valuable testing ground for the generality of the holographic principle, spacetimes which do not asymptote to AdS also play an important role as gravity duals for non-relativistic CFTs, which are a common occurrence in e.g. condensed matter physics and thus may be able to provide new insight in these strongly interacting systems. The aforementioned Schrödinger spacetimes for example can be used as a holographic dual to describe cold atoms [38, 39]. Since advanced applications for such strongly interacting systems require a very good understanding of the underlying holographic correspondence it is of vital importance to first understand the generality of the holographic principle better. In addition higher-spin holography, in contrast to the usual AdS/CFT correspondence, is a weak/weak correspondence [46, 47]. Thus, it is much more interesting for explicit checks of the holographic principle because calculations are often feasible on both sides of the correspondence and allow direct comparison of physical observables obtained in a holographic way.

There is another very prominent example of a holographic correspondence which does not involve AdS spacetimes and I have not mentioned yet, but which will have an important role in this thesis: flat space holography.

Trying to establish a holographic correspondence for asymptotically flat spacetimes is of interest for various reasons. First and foremost, if holography is indeed a fundamental principle of nature, then it must also work for flat space. In addition, flat space is a very good approximation for most purposes in physics and thus establishing a notion of flat space holography and a corresponding holographic dictionary can have a huge potential impact on many areas of research in physics. The first steps toward a flat space holographic correspondence have been taken by Polchinski, Susskind and Giddings [48–50]. In three bulk dimensions Barnich and Compère pioneered the field by finding the analogue of the Brown-Henneuax boundary conditions for flat space [51]. Some further key developments were the BMS/CFT or BMS/GCA correspondences[4] [55, 56], the flat space chiral gravity proposal [57], the counting of flat space cosmology[5] microstates [61, 62], the existence of phase transitions between flat space cosmologies and hot flat space [63]. For a selection of papers on various topics related to flat space holography please refer to [III, V, VI, VII, VIII, IX, 57, 61–76]. Despite these recent advances, there are still a lot of things to learn and to understand regarding a holographic principle in flat space and as of yet one can roughly sum up the two general approaches to flat space holography like this:

**Flat Space as a Limit of Vanishing Cosmological Constant:** Simply take appropriate limits of vanishing cosmological constant from the known AdS results in order to get the corresponding flat space result, see e.g. [V, 64, 65, 77]. This was and is still a very popular approach to flat space holography as there is an abundance of AdS results available. The tricky and highly non-trivial part of this is *how* to take the limit. In addition, this limit is not reliable for every holographic observable. Thus, one has to be careful when to trust the results obtained this way.

**Work Directly in Flat Space:** This is the conceptually and technically more challenging route to take. But at the same time it is also the most trustworthy one since no limit is involved. What makes this approach challenging is that many things are in close analogy to techniques and phenomena already encountered in AdS holography, but at the same time different enough that one cannot simply apply the same techniques used in the AdS case. What makes AdS so special, and is also the reason why it is so well studied, is that whenever a problem seems to arise, the non-zero cosmological appears to save the day[6]. Since for flat space the cosmological constant is zero, one is often faced with new conceptual challenges how to deal with those problems. These are normally circumvented by a non-zero cosmological constant.

[4]BMS stands for Bondi–van der Burg–Metzner–Sachs [52, 53], the asymptotic symmetry algebra of flat spacetimes at null infinity, and GCA for Galilean conformal algebra [54].

[5]Flat space cosmologies [58, 59] are the flat space analogues of the Bañados–Teitelboim–Zanelli (BTZ) black hole [12, 60] solutions in $AdS_3$.

[6]One example would be the one-loop partition function in flat space which has to be regularized by hand in order to obtain a finite result [IX, 78].

# Thesis Structure

This thesis consists of five parts, which I will summarize in the following. I will also state at the beginning of each part on which of my previous publications the respective part is based on.

### Part I

This part serves as a basic introduction to the general ideas and the formalism which underly this thesis. It contains lightning reviews of gravity in three dimensions and its formulation as a Chern-Simons theory. Furthermore, the canonical analysis of Chern-Simons theories will be discussed as well as the connection between asymptotic symmetries and quantum field theories in light of the holographic principle, higher-spin gravity in Anti-de Sitter spacetimes and İnönü–Wigner contractions. Last, but not least, I will present an overview of the connection between entropy and quantum entanglement and its relation to holography.

### Part II

*Based on* [I, II]

In this part I will examine certain aspects of non-Anti-de Sitter holography with a special focus on unitary representations of the asymptotic symmetry algebras, which are the basic symmetries that underly the dual field theories in this holographic correspondence. I will first focus on specific non-Anti-de Sitter higher-spin gravity theories, which will serve as a playground to explore general features of this holographic correspondence. Based on these examples I will then present a special and new class of symmetries which allow for arbitrary (albeit not infinitely) large Virasoro central charge while maintaining unitarity at the same time.

### Part III

*Based on* [III, V, VII]

Here I will present the first part of my work on gaining a better understanding of a holographic correspondence in asymptotically flat spacetimes. In this part I will focus on the gravity side of this new correspondence. I will show how one can employ suitable flat space limits from Anti-de Sitter results in order to gain the corresponding flat space results and study unitarity of the dual (higher-spin) field theories. Furthermore, I will extend previous considerations of flat space (higher-spin) holography and show how to consistently add (higher-spin) chemical potentials to flat space and determine the corresponding thermal entropies of the dual field theory holographically.

**Part IV**

*Based on* [VI, VIII]

In this part, which I consider as the highlight of my thesis, I will perform explicit checks of the holographic correspondence in flat space with a focus on holographic entanglement entropy. I will begin by introducing the basics of Galilean conformal field theories and show how one can use these field theories in order to determine entanglement entropy of the field theory, which is dual to flat space. Following up on this I will briefly mention selected explicit checks of the holographic correspondence. The last part of this chapter will be concerned with a holographic description of (higher-spin) entanglement entropy in flat space using a special kind of Wilson line.

**Part V**

This part will conclude my thesis. I will summarize my results and give a conclusion on how general the holographic principle is, based on the results of my research as well as an outlook on possible follow up projects.

**Conventions**

Unless otherwise stated I will use units where $c = \hbar = k_B = 1$ in this thesis.

# List of Publications

# Part I

## General Ideas and Formalism

This part is dedicated to introducing the main ideas and the formalism underlying my thesis. I will explain why it is interesting and beneficial to work with three dimensional gravity theories for the purpose of getting a better understanding of the holographic principle in general. Following this I will explain the basics of the Chern-Simons formalism which will be heavily used throughout this thesis. Furthermore, the importance of asymptotic symmetries and their relation to the dual (quantum) field theories will be explained. Since a big part of my thesis revolves around holography which does not involve AdS spacetimes, I will also review the most important aspects of AdS (higher-spin) holography. In addition, I will introduce the concept of an İnönü–Wigner contraction, which will play an important role in determining the flat space equivalents of the asymptotic symmetry algebras encountered in AdS as well as non-AdS holography. Finally, I will elaborate on the connection between entanglement, entropy and holography.

# Gravity in Three Dimensions

1

> *Das Gehirn ist der wichtigste Muskel beim Klettern. (Your brain is the most important muscle when climbing.)*
>
> **– Wolfgang Güllich**
> German climber

General Relativity in three dimensions is very special in many regards and, as already mentioned in the introduction, there are a lot of reasons why it is beneficial to study gravity in this setup, especially if one is interested in general features of holography.

First and foremost, gravity in three dimensions is technically much simpler than in four or higher dimensions. For example the Riemann tensor $R_{abcd}$ can be expressed in terms of the Ricci tensor $R_{ab}$, the Ricci scalar $R$ and the metric $g_{ab}$ as

$$R_{abcd} = g_{ac}R_{bd} + g_{bd}R_{ac} - g_{ad}R_{bc} - g_{bc}R_{ad} - \frac{1}{2}R(g_{ac}g_{bd} - g_{ad}g_{bc}). \tag{1.1}$$

Now taking also into account Einstein's equations

$$R_{\mu\nu} + \left(\Lambda - \frac{R}{2}\right) g_{\mu\nu} = 8\pi G_N T_{\mu\nu}, \tag{1.2}$$

where $G_N$ is Newton's constant in three dimensions, $\Lambda$ is the cosmological constant and $T_{\mu\nu}$ the energy-momentum tensor which encodes the local energy-momentum distribution. This implies that the curvature of spacetime in three dimensions is completely determined in terms of the local energy-momentum distribution and the value of the cosmological constant. Thus, if there are no matter sources the curvature of spacetime is completely determined by the value of the cosmological constant. This in turn also means that there are no local propagating (bulk-) degrees of freedom i.e. massless gravitons[1].

At first sight this sounds like bad news since a theory with no local propagating degrees of freedom seems to be trivial. Luckily, both local and global effects play an important role in (three dimensional) gravity so that the theory is physically non-trivial. It is also noteworthy that Einstein gravity in three dimensions is a topological theory.

[1]This is true for Einstein-Hilbert gravity in three dimensions. One could, however, also consider other gravity theories in three dimensions which allow for (typically massive) gravitons.

Probably the most famous example illustrating this feature is the BTZ black hole solution found by Bañados, Teitelboim and Zanelli [12]. This black hole solution is locally AdS, but at the boundary of the AdS spacetime is characterized by canonical charges which differ from the usual AdS vacuum. In addition the BTZ black hole has a horizon, singularity and exhibits an ergoregion in general.

In [11] Brown and Henneaux presented boundary conditions for three dimensional gravity, whose corresponding canonical charges generate two copies of the Virasoro algebra. This ultimately lead to the (holographic) conjecture that AdS in three dimensions can equivalently be described by a two-dimensional conformal field theory located at the boundary of AdS [15].

Since gravity in three dimensions is a purely topological theory one might expect that this theory can also be formulated in a way that makes its topological character explicit i.e. a Chern-Simons formulation. I will review Chern-Simons formulations and its properties in Chapter 2. Before doing so it will be instructive to explain how one has to formulate gravity in three dimensions in order to be able to rewrite the Einstein-Hilbert action

$$I_{\text{EH}} = \frac{1}{16\pi G_N} \int_{\mathcal{M}} \mathrm{d}^3x \sqrt{-g}\,(R - 2\Lambda)\,, \tag{1.3}$$

where $g \equiv \det g_{\mu\nu}$, as a Chern-Simons action.

The action (1.3) takes as the fundamental dynamic field the symmetric tensor $g_{\mu\nu}$ which acts as a symmetric bilinear form on the tangent space of the manifold $\mathcal{M}$. Writing the metric in a given basis thus does not necessarily mean that this basis is orthonormal at each given point of spacetime. For many purposes it is, however, advantageous to have a notion of a local orthonormal laboratory frame i.e. a family of ideal observers embedded in a given spacetime. Such a family of ideal observers can be introduced in General Relativity via frame fields $e^a = e^a{}_\mu\,\mathrm{d}x^\mu$, which are often also called *vielbein*. This frame field is a function of the spacetime coordinates $x^\mu$ and carries spacetime indices, which will be denoted by Greek letters $\mu, \nu, \ldots$ and internal local Lorentz indices denoted by Latin letters $a, b, \ldots$. The frame fields $e^a$ and the metric $g_{\mu\nu}$ are related by

$$g_{\mu\nu} = e^a{}_\mu e^b{}_\nu \eta_{ab}, \tag{1.4}$$

where $\eta_{ab}$ is the 2+1 dimensional Minkowski metric with signature $(-,+,+)$. In this formulation local Lorentz indices can be raised and lowered using the Minkowski metric $\eta_{ab}$, while spacetime indices are raised and lowered using the spacetime metric $g_{\mu\nu}$.

The big advantage of using a formulation in terms of frame fields is that one now can very easily promote objects from a flat, Lorentz invariant setting to a description in a coordinate invariant and curved background[2]. Take for example some object $V^a$ which transforms under local Lorentz transformations $\Lambda\left(x^{\mu}\right)^{a}{}_{b}$ like the components of a vector,

$$\tilde{V}^{a}=\Lambda\left(x^{\mu}\right)^{a}{}_{b}V^{b}. \tag{1.5}$$

[2] One example would be a formulation of the Dirac equation in curved backgrounds.

Then one can easily describe this object in a curved background using the frame field[3] as

$$V^{\mu}=e_{a}{}^{\mu}V^{a}. \tag{1.6}$$

Local Lorentz invariance of the frame fields also means that there should be a gauge field associated to that local Lorentz invariance. This gauge field is the spin connection $\omega^{ab}=\omega^{ab}{}_{\mu}\,\mathrm{d}x^{\mu}$ with $\omega^{ab}{}_{\mu}=-\omega^{ba}{}_{\mu}$ which also allows one to define a covariant derivative acting on generalized tensors i.e. tensors which have both spacetime and Lorentz indices as

$$\mathcal{D}_{\mu}V^{a}{}_{\nu}=\partial_{\mu}V^{a}{}_{\nu}+\omega^{a}{}_{b\mu}V^{b}{}_{\nu}-\Gamma^{\sigma}{}_{\nu\mu}V^{a}{}_{\sigma}, \tag{1.7}$$

where $\Gamma^{\sigma}{}_{\nu\mu}$ denotes the affine connection associated to the metric $g_{\mu\nu}$

$$\Gamma^{\sigma}{}_{\nu\mu}=\frac{1}{2}g^{\sigma\delta}\left(\partial_{\nu}g_{\delta\mu}+\partial_{\mu}g_{\nu\delta}-\partial_{\delta}g_{\nu\mu}\right). \tag{1.8}$$

One particular convenient feature in three dimensions is that one can (Hodge) dualize the spin connection in such a way that it has the same index structure as the vielbein. In terms of Lorentz indices this can be achieved by using the 3d Levi-Civita symbol in order to obtain

$$\omega^{a}=\frac{1}{2}\epsilon^{abc}\omega_{bc}\quad\Leftrightarrow\quad\omega_{ab}=-\epsilon_{abc}\omega^{c}, \tag{1.9}$$

where $\epsilon^{012}=1$. It is exactly this dualization of the spin connection which makes it possible to combine the vielbein and the spin connection into a single gauge field as I will review later in Chapter 2.

Using the dualized spin connection one can write the associated curvature two-form $R^{a}$ as

$$R^{a}=\mathrm{d}\omega^{a}+\frac{1}{2}\epsilon^{a}{}_{bc}\omega^{b}\wedge\omega^{c}, \tag{1.10}$$

and consequently the Einstein-Hilbert-Palatini action (1.3) in terms of these new (first order) variables as

$$I_{EHP}=\frac{1}{8\pi G_{N}}\int_{\mathcal{M}}\left[e_{a}\wedge R^{a}-\frac{\Lambda}{6}\epsilon_{abc}e^{a}\wedge e^{b}\wedge e^{c}\right]. \tag{1.11}$$

The equations of motion of the second order action (1.3) which are obtained by varying the action with respect to the metric $g_{\mu\nu}$ are given by the Einstein equations (1.2). Since in the frame-like formalism one has two independent fields $e^{a}$ and $\omega^{a}$ one has to vary (1.11) with respect to both of these fields and subsequently also obtains two equations which encode curvature and torsion respectively as

$$R^a = \mathrm{d}\omega^a + \frac{1}{2}\epsilon^a{}_{bc}\omega^b \wedge \omega^c = \frac{\Lambda}{2}\epsilon^a{}_{bc}e^b \wedge e^c, \tag{1.12a}$$

$$T^a = \mathrm{d}e^a + \epsilon^a{}_{bc}\omega^b \wedge e^c = 0. \tag{1.12b}$$

[3] To be more precise this is the inverse of the frame field $e^{a}{}_{\mu}$ defined by $e^{a}{}_{\mu}e_{a}{}^{\nu}=\delta^{\nu}_{\mu}$.

This basic knowledge of frame fields, spin connections and how to use those two fields to cast the second order Einstein-Hilbert action (1.3) into a first order form (1.11) is already sufficient to be able to move on to the the next chapter, in which I will describe how to rewrite the Einstein-Hilbert-Palatini action (1.11) as a Chern-Simons action.

# Chern-Simons Formulation of Gravity

# 2

> 蛙の子は蛙です。
> *(Like parent, like child.)*
>
> **– Japanese proverb**

As described in the previous chapter, instead of using a second order formalism, where the fundamental field of the theory is the metric $g_{\mu\nu}$, it can for some purposes be more convenient to use a first order formalism where the fundamental fields of the theory are the vielbein $e$ and the spin connection $\omega$. In three dimensions one finds that the dreibein and dualized spin connection have the same index structure in their Lorentz indices. Thus, one can combine these two quantities into a single gauge field

$$\mathcal{A} \equiv e^a P_a + \omega^a J_a, \tag{2.1}$$

where the generators $P_a$ and $J_a$ generate the following Lie algebra

$$[P_a, P_b] = -\Lambda \epsilon_{abc} J^c, \quad [J_a, J_b] = \epsilon_{abc} J^c, \quad [J_a, P_b] = \epsilon_{abc} P^c. \tag{2.2}$$

- For $\Lambda > 0$, i.e. de Sitter spacetimes this gauge algebra is $\mathfrak{so}(3,1)$.
- For $\Lambda = 0$, i.e. flat spacetimes this gauge algebra is $\mathfrak{isl}(2,\mathbb{R}) \sim \mathfrak{sl}(2,\mathbb{R}) \oplus_s \mathbb{R}^3$.
- For $\Lambda < 0$, i.e. Anti-de Sitter spacetimes this gauge algebra is $\mathfrak{so}(2,2) \sim \mathfrak{sl}(2,\mathbb{R}) \oplus \mathfrak{sl}(2,\mathbb{R})$.

Witten [10] showed in 1988 that the Chern-Simons action [9]

$$S_{\text{CS}}[\mathcal{A}] = \frac{k}{4\pi} \int_{\mathcal{M}} \left\langle \mathcal{A} \wedge \mathrm{d}\mathcal{A} + \frac{2}{3} \mathcal{A} \wedge \mathcal{A} \wedge \mathcal{A} \right\rangle, \tag{2.3}$$

defined on a three dimensional manifold $\mathcal{M} = \Sigma \times \mathbb{R}$, with the invariant bilinear form

$$\langle J_a, P_b \rangle = \eta_{ab}, \quad \langle J_a, J_b \rangle = \langle P_a, P_b \rangle = 0, \tag{2.4}$$

is indeed equivalent (up to boundary terms) to the Einstein-Hilbert-Palatini action in the first order formalism for positive, negative and zero cosmological constant

(1.11), provided one identifies the Chern-Simons level $k$ with Newton's constant $G_N$ in three dimensions as

$$k = \frac{1}{4G_N}. \tag{2.5}$$

**Anti-de Sitter Spacetimes:** One particular convenient feature of spacetimes with negative cosmological constant $\Lambda \equiv -\frac{1}{\ell^2} < 0$ where $\ell$ is called the AdS radius, is that in a Chern-Simons formulation the underlying gauge symmetry $\mathfrak{so}(2,2)$ is a direct sum of two copies of $\mathfrak{sl}(2,\mathbb{R})$. This split can be made explicit by introducing the generators

$$J_a^{\pm} = \frac{1}{2}\left(J_a \pm \ell P_a\right). \tag{2.6}$$

These new generators satisfy

$$\left[J_a^+, J_b^-\right] = 0, \qquad \left[J_a^{\pm}, J_b^{\pm}\right] = \epsilon_{abc} J^{c\pm}. \tag{2.7}$$

One can explicitly realize this split via

$$J_a^+ = \begin{pmatrix} T^a & 0 \\ 0 & 0 \end{pmatrix}, \quad J_a^- = \begin{pmatrix} 0 & 0 \\ 0 & \bar{T}^a \end{pmatrix}, \tag{2.8}$$

where both $T^a$ and $\bar{T}^a$ satisfy an $\mathfrak{sl}(2,\mathbb{R})$ algebra. From (2.4) one can immediately see that

$$\langle T_a, T_b \rangle = \frac{\ell}{2}\eta_{ab}, \quad \left\langle \bar{T}_a, \bar{T}_b \right\rangle = -\frac{\ell}{2}\eta_{ab}. \tag{2.9}$$

The gauge field $\mathcal{A}$ can now be written as

$$\mathcal{A} = \begin{pmatrix} \left(\omega^a + \frac{1}{\ell}e^a\right) T_a & 0 \\ 0 & \left(\omega^a - \frac{1}{\ell}e^a\right) \bar{T}_a \end{pmatrix} \equiv \begin{pmatrix} A^a T_a & 0 \\ 0 & \bar{A}^a \bar{T}_a \end{pmatrix}. \tag{2.10}$$

Thus, after implementing this explicit split of $\mathfrak{so}(2,2)$ into a direct sum of two copies of $\mathfrak{sl}(2,\mathbb{R})$, the Chern-Simons action (2.3) also splits into two contributions

$$S_{\text{EH}}^{\text{AdS}}[A,\bar{A}] = S_{\text{CS}}[A] + S_{\text{CS}}[\bar{A}], \tag{2.11}$$

where the invariant bilinear forms appearing in the Chern-Simons action are given by (2.9). Since both $T^a$ and $\bar{T}^a$ satisfy an $\mathfrak{sl}(2,\mathbb{R})$ algebra it is usually practical to not distinguish between the two generators, i.e. setting $T^a = \bar{T}^a$. This in turn also means that the invariant bilinear form in both sectors will be the same. From (2.9), however, we know that the invariant bilinear form in both sectors should have opposite sign. This is not a real problem since this relative minus sign can be easily introduced by hand by not taking the sum, but rather the difference of the two Chern-Simons actions

$$S_{\text{EH}}^{\text{AdS}} = S_{\text{CS}}[A] - S_{\text{CS}}[\bar{A}]. \tag{2.12}$$

As the factor of $\ell$ in (2.9) only yields an overall factor of $\ell$ to the action (2.12) one can also absorb this factor simply in the Chern-Simons level as

$$k = \frac{\ell}{4G_N}. \tag{2.13}$$

This form of the Chern-Simons connection (2.12) is usually the one discussed in the literature on AdS holography in $2+1$ dimensions. The big advantage of this split into an unbarred and a barred part in the case of AdS holography is that usually one only has to explicitly calculate things for one of the two sectors, as the other sector works in complete analogy, up to possible overall minus signs.
Up to this point I have only presented the basics of the Chern-Simons formulation of gravity in $2+1$ dimensions but did not go into detail as to *why* exactly this formulation is so convenient and powerful for the purpose of studying the holographic principle. Thus, I will spend the remainder of this chapter explaining the benefits of using the Chern-Simons formulation.

## Gravity as a Gauge Theory

Maybe the biggest advantage of this formalism using Chern-Simons gauge fields is that this allows one to use all the techniques and machinery which is familiar from ordinary gauge theories. One can for example use finite gauge transformations of the form

$$\mathcal{A} \to g^{-1}\left(\tilde{\mathcal{A}} + \mathrm{d}\right) g, \tag{2.14}$$

where $g$ is some element of the group $G$ which is generated by some Lie algebra $\mathfrak{g}$ and $\mathcal{A} \in \mathfrak{g}$ to bring the gauge field $\mathcal{A}$ into a form which is convenient for the given task at hand. I will use this gauge freedom at various points in this thesis. One can use for example a special gauge which is very convenient in the asymptotic analysis of AdS and non-AdS spacetimes whereas another gauge will be more convenient when making the transition from AdS to flat space. Since the gauge transformations (2.14) are finite in contrast to infinitesimal gauge transformations generated by a gauge parameter $\xi$ as

$$\delta_\xi \mathcal{A} = \mathrm{d}\xi + [\mathcal{A}, \xi], \tag{2.15}$$

one has to be careful which finite gauge transformations actually leave the Chern-Simons action (2.3) invariant. In general a finite gauge transformation (2.14) changes the Chern-Simons action (2.3) as $S_{\mathrm{CS}}[\mathcal{A}] \to S_{\mathrm{CS}}[\tilde{\mathcal{A}}] + \delta S_{\mathrm{CS}}[\tilde{\mathcal{A}}]$ with [79]

$$\delta S_{\mathrm{CS}}[\tilde{\mathcal{A}}] = -\frac{k}{12\pi}\int_{\mathcal{M}} \left\langle g^{-1}\mathrm{d}g \wedge g^{-1}\mathrm{d}g \wedge g^{-1}\mathrm{d}g \right\rangle - \frac{k}{4\pi}\int_{\partial\mathcal{M}} \left\langle \mathrm{d}g g^{-1} \wedge \tilde{\mathcal{A}} \right\rangle. \tag{2.16}$$

This term vanishes for infinitesimal gauge transformations (2.15) with gauge parameters $\xi \in \mathfrak{g}$ which are continuously connected to the identity $g \sim \mathbb{1} + \xi$ and for finite gauge transformations which approach $g \to \mathbb{1}$ sufficiently fast when approaching the boundary, but not for general finite gauge transformations. This means that there are finite gauge transformations of the form (2.14) which can change the state of the system and thus map between physically distinct setups. Now considering the variation of (2.3) with respect to the gauge field $\mathcal{A}$ one obtains the equations of motion of the Chern-Simons action (2.3) as

$$F = \mathrm{d}\mathcal{A} + \mathcal{A} \wedge \mathcal{A} = 0, \tag{2.17}$$

which means that on-shell the Chern-Simons connection has to be locally flat. Remembering that the connection $\mathcal{A}$ can also be expressed in terms of a vielbein and spin connection as in (2.1), then requiring a flat connection $\mathcal{A}$ is equivalent to the equations (1.12), which encode curvature and torsion. This is another check that the Chern-Simons action indeed correctly describes gravity in $2+1$ dimensions.

To require that the connection is locally flat also means that $\mathcal{A} = 0$ is always a (trivial) solution of the equations of motion. Keeping in mind that finite gauge transformations in general can change the physical state, this in turn also means that for some holographic applications it can be beneficial to first start with the trivial configuration $\mathcal{A} = 0$ and then use a finite gauge transformation (2.14) in order to obtain the desired result of a non-trivial configuration. This is a technique which will be extensively used for example in Chapter 15.

Before continuing onwards to higher-spin symmetries I will briefly elaborate on an important point of three dimensional gravity, namely how diffeomorphisms appear in this gauge theoretic formulation. First consider the infinitesimal gauge transformation (2.15) but now with a special gauge parameter of the form $\xi = \zeta^\nu \mathcal{A}_\nu$. After using the Leibniz rule one obtains

$$\delta_{(\zeta^\nu \mathcal{A}_\nu)} \mathcal{A}_\mu = \partial_\mu \zeta^\nu \mathcal{A}_\nu + \zeta^\nu \partial_\mu \mathcal{A}_\nu + \zeta^\nu [\mathcal{A}_\mu, \mathcal{A}_\nu]. \tag{2.18}$$

Now adding $\zeta^\nu \left(\partial_\nu \mathcal{A}_\mu - \partial_\nu \mathcal{A}_\mu\right)$ to the right hand side of this equation does not really change anything. However, it allows one to rewrite (2.18) in a more suggestive form as

$$\delta_{(\zeta^\nu \mathcal{A}_\nu)} \mathcal{A}_\mu = \mathcal{L}_\zeta \mathcal{A}_\mu + \zeta^\nu F_{\mu\nu}, \tag{2.19}$$

where $\mathcal{L}_\zeta \mathcal{A}_\mu$ is the Lie derivative of the gauge field $\mathcal{A}_\mu$ given by

$$\mathcal{L}_\zeta \mathcal{A}_\mu = \zeta^\nu \partial_\nu \mathcal{A}_\mu + \mathcal{A}_\nu \partial_\mu \zeta^\nu. \tag{2.20}$$

Thus, one can see that diffeomorphisms in three dimensional gravity are on-shell (i.e. for $F = 0$) equivalent to infinitesimal gauge transformations with gauge parameter $\xi = \zeta^\nu \mathcal{A}_\nu$.

## Straightforward Extension to Higher-Spins

Being able to treat gravity as a gauge theory also has the advantage that one has a very efficient tool at hand when describing gravity theories which exhibit more symmetries than just diffeomorphism and local Lorentz invariance. As already mentioned in the introduction one interesting class of theories which exhibit such symmetries are higher-spin gravity theories. Of course one can also describe those theories in $2+1$ dimensions using a second order formalism (see. e.g. [31]), but the extension from ordinary, i.e. spin-2 gravity, to higher-spin gravity can be performed straightforwardly in the Chern-Simons formalism.

Take as an example again AdS, where the gauge algebra is given by two copies of $\mathfrak{sl}(2,\mathbb{R})$. Now going from e.g. spin-2 gravity to spin-3 one simply has to replace the gauge algebra $\mathfrak{sl}(2,\mathbb{R}) \to \mathfrak{sl}(3,\mathbb{R})$ and can perform (almost) the exact same calculations as in the spin-2 case. There is but one subtlety involved in this procedure, which is how to embed usual spin-2 gravity into the higher-spin setting. Thinking in abstract algebraic terms then $\mathfrak{sl}(2,\mathbb{R})$ represents the usual spin-2 gravity setting with diffeomorphism and local Lorentz invariance. Thus, the question on how to embed gravity in the higher-spin context is reduced to the question on how to embed $\mathfrak{sl}(2,\mathbb{R}) \hookrightarrow \mathfrak{sl}(N,\mathbb{R})$. Depending on the embedding one also obtains qualitatively different higher-spin theories. In most of the literature on higher-spin gravity in $2+1$ dimensions whenever the term spin-$N$ gravity is used authors usually refer to the principal embedding of $\mathfrak{sl}(2,\mathbb{R}) \hookrightarrow \mathfrak{sl}(N,\mathbb{R})$ whose spectrum contains fields of spin-$s = 2, 3, \ldots, N$.

In a metric formulation the geometric interpretation of these higher-spin symmetries is often not very clear. Things like Riemannian curvature, the notion of geodesics or even black hole event horizons are not gauge invariant quantities anymore after introducing higher-spin symmetries. The Chern-Simons formulation, however, has a precise notion of gauge invariance which also extends to higher-spin symmetries. Things like regularity or extremality of black hole horizons can be encoded in the holonomies of the connection $\mathcal{A}$ (see e.g. [80]), geodesics and their proposed higher-spin extensions can be related to Wilson lines (see e.g. [81, 82]) as shown in Chapter 15.

# 3 Asymptotic Symmetries and Dual Field Theory

*You can't take the sky from me.*

**– Joss Whedon**
Firefly

In this chapter I will summarize how to perform a canonical analysis[1] of Chern-Simons theories in general. This will ultimately enable one to determine the asymptotic symmetries of a given theory.
First I want to be a little bit more specific about the general setup in which I am going to perform the canonical analysis. I will assume that the manifold $\mathcal{M}$ has the topology of a cylinder $\mathcal{M} = \Sigma \times \mathbb{R}$ and can be parametrized via coordinates $x^\mu = (t, \rho, \varphi)$, $\mu = 0, 1, 2$. In addition, I assume that $\Sigma$ has the topology of a disk and is parameterized by $\varphi$ and $\rho$, where $\rho = \rho_0$ corresponds to the boundary of that disk and $\varphi \sim \varphi + 2\pi$. See Figure 3.1 for a visualization.

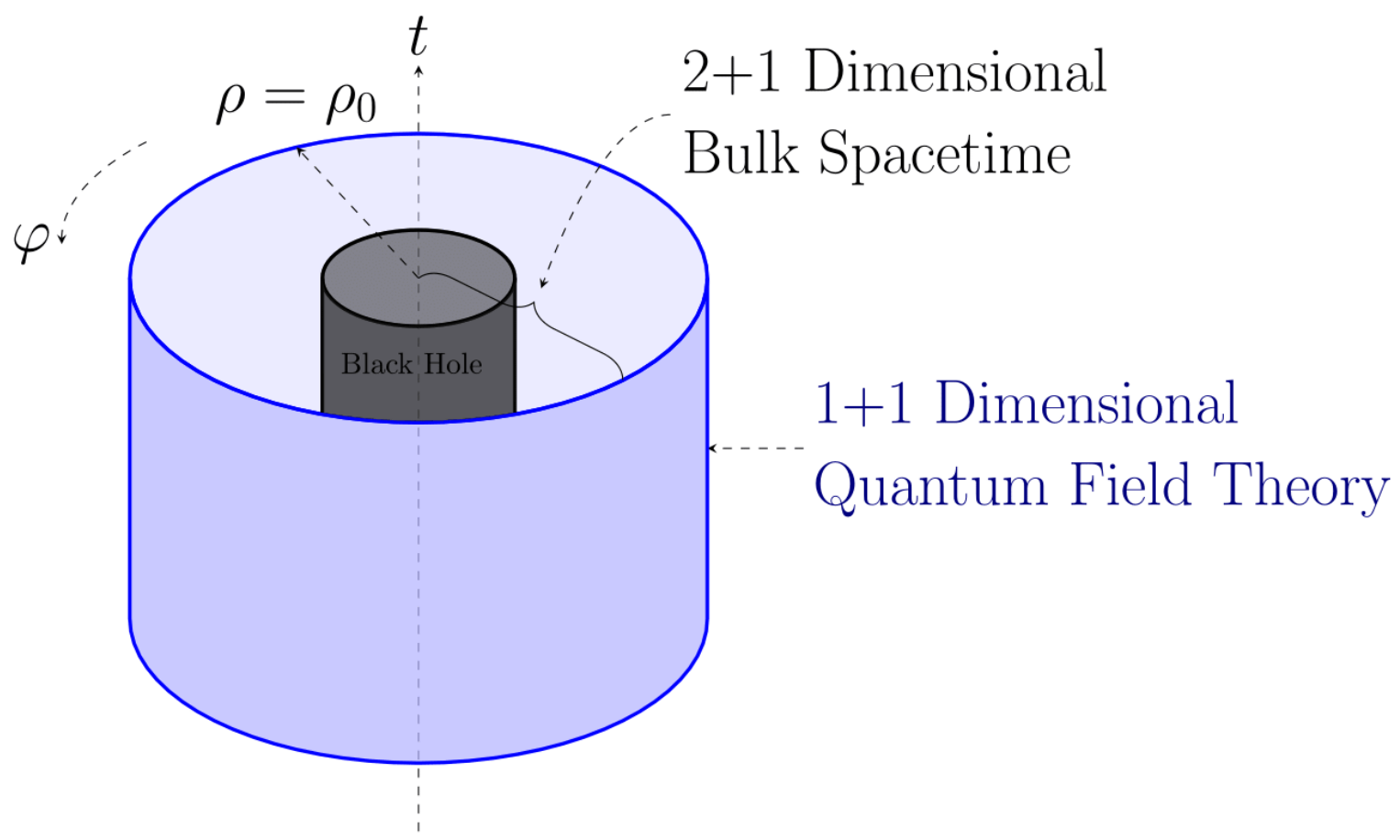

**Fig. 3.1.:** Pictorial representation of the manifold $\mathcal{M}$.

The Chern-Simons gauge field $\mathcal{A}$ is a Lie algebra valued 1-form that can be written as

$$\mathcal{A} = \mathcal{A}^a{}_\mu \, \mathrm{d}x^\mu T_a, \tag{3.1}$$

[1]Please refer to e.g. [79, 83] for details on how to quantize gauge systems in a Hamiltonian formulation.

with $T_a$ being a basis of some Lie algebra $\mathfrak{g}$ with commutation relations

$$[T_a, T_b] = f^c{}_{ab} T_c. \tag{3.2}$$

In addition, I will assume that the Lie algebra admits a non-degenerate invariant bilinear form which will be denoted by

$$h_{ab} = \langle T_a, T_b \rangle \,. \tag{3.3}$$

Lie algebra indices $(a, b, \ldots)$ are raised and lowered with $h_{ab}$ and spacetime indices $(\mu, \nu, \ldots)$ with the spacetime metric $g_{\mu\nu}$.
One can write (2.3) explicitly in components as

$$S_{\mathsf{CS}}[\mathcal{A}] = \frac{k}{4\pi} \int_{\mathcal{M}} \mathrm{d}^3x \epsilon^{\mu\nu\lambda} h_{ab} \left( \mathcal{A}^a{}_\mu \partial_\nu \mathcal{A}^b{}_\lambda + \frac{1}{3} f^a{}_{cd} \mathcal{A}^c{}_\mu \mathcal{A}^d{}_\nu \mathcal{A}^b{}_\lambda \right). \tag{3.4}$$

In order to proceed with the canonical analysis it is convenient to use a 2+1 decomposition [84] of the action (2.3), which explicitly separates the manifold $\mathcal{M}$ into the disk $\Sigma$ and the real line $\mathbb{R}$, i.e. an explicit split into time $t$ and the coordinates which parametrize the disc $x^{\bar{\mu}} \equiv \mathbf{x} = (\rho, \varphi)$, $\bar{\mu} = 1, 2$. This has the advantage that ultimately one only has to deal with integrals over the two-dimensional (spatial) disk $\Sigma$ and its boundary manifold $\partial\Sigma$. The decomposition of (3.4) is given by

$$S_{\mathsf{CS}}[\mathcal{A}] = \frac{k}{4\pi} \int_{\mathbb{R}} \mathrm{d}t \int_{\Sigma} \mathrm{d}^2x \epsilon^{\bar{\mu}\bar{\nu}} h_{ab} \left( \dot{\mathcal{A}}^a{}_{\bar{\mu}} \mathcal{A}^b{}_{\bar{\nu}} + \mathcal{A}^a{}_0 F^b{}_{\bar{\mu}\bar{\nu}} + \partial_{\bar{\nu}} \left( \mathcal{A}^a{}_{\bar{\mu}} \mathcal{A}^b{}_0 \right) \right), \tag{3.5}$$

with $F^a{}_{\bar{\mu}\bar{\nu}} = \partial_{\bar{\mu}} \mathcal{A}^a{}_{\bar{\nu}} - \partial_{\bar{\nu}} \mathcal{A}^a{}_{\bar{\mu}} + f^a{}_{bc} \mathcal{A}^b{}_{\bar{\mu}} \mathcal{A}^c{}_{\bar{\nu}}$ and $\epsilon^{\bar{\mu}\bar{\nu}} \equiv \epsilon^{t\bar{\mu}\bar{\nu}}$. Since there are no $\dot{\mathcal{A}}^a{}_0$ terms appearing in (3.5) one can see $\mathcal{A}^a{}_0$ as a Lagrange multiplier, which in turn also means that on-shell the Chern-Simons gauge field has to be gauge flat, i.e. $F^a{}_{\mu\nu} = 0$. Thus, the dynamical fields of the Chern-Simons action are $\mathcal{A}^a{}_{\bar{\mu}}$.
The starting point of the canonical analysis is a Hamiltonian density. Hence one first has to determine the canonical momenta $\pi_a{}^\mu$ corresponding to the canonical variables $\mathcal{A}^a{}_\mu$ from the Lagrangian density[2] $\mathcal{L}$ via $\pi_a{}^\mu \equiv \frac{\partial \mathcal{L}}{\partial \dot{\mathcal{A}}^a{}_\mu}$.
The canonical momenta are in general not independent quantities. In case of the Chern-Simons action (3.5) one obtains the following relations[3]

$$\phi_a{}^0 := \pi_a{}^0 \approx 0 \qquad \phi_a{}^{\bar{\mu}} := \pi_a{}^{\bar{\mu}} - \frac{k}{4\pi} \epsilon^{\bar{\mu}\bar{\nu}} h_{ab} \mathcal{A}^b{}_{\bar{\nu}} \approx 0, \tag{3.6}$$

which are called primary constraints to emphasize that the equation of motions are not used to obtain these relations. Having identified the canonical variables and

[2]Where by Lagrangian density I mean $S_{\mathsf{CS}} = \int_{\mathcal{M}} \mathrm{d}x^3 \mathcal{L}$.
[3]From this point onward one has to distinguish between weak ($\approx$) and strong ($=$) equalities. Two functions in the phase space, $f$ and $g$, are weakly equal $f \approx g$ if restricted to a constraint surface, but not throughout the whole phase space. If $f$ and $g$ are equal independently of the constraints being satisfied $f = g$, they are called strongly equal.

their corresponding canonical momenta one can also define their Poisson brackets as

$$\{\mathcal{A}^a{}_\mu(\mathbf{x}), \pi_b{}^\nu(\mathbf{y})\} = \delta^a{}_b \delta_\mu{}^\nu \delta^2(\mathbf{x} - \mathbf{y}). \tag{3.7}$$

The next step in this analysis is to calculate the canonical Hamiltonian density using a Legendre transformation

$$\mathcal{H} = \pi_a{}^\mu \dot{\mathcal{A}}^a{}_\mu - \mathcal{L} = -\frac{k}{4\pi} \epsilon^{\bar{\mu}\bar{\nu}} h_{ab} \left( \mathcal{A}^a{}_0 F^b{}_{\bar{\mu}\bar{\nu}} + \partial_{\bar{\nu}} \left( \mathcal{A}^a{}_{\bar{\mu}} \mathcal{A}^b{}_0 \right) \right). \tag{3.8}$$

As the system under consideration is a constrained Hamiltonian system one also has to include the constraints (3.6) in a Hamlitonian description of the system. This can be done by adding the constraints to (3.8) and thus obtaining a total Hamiltonian $\mathcal{H}_T$ using some arbitrary multipliers $u^a{}_\mu$ as

$$\mathcal{H}_T = \mathcal{H} + u^a{}_\mu \phi_a{}^\mu. \tag{3.9}$$

Then, by construction, the equations of motion obtained from $\mathcal{H}_T$ via $\dot{F} \approx \{F, \mathcal{H}_T\}$ for some function $F$, which is function of the canonical variables, are the same as the ones obtained from the Lagrangian description.
Since the primary constraints should be conserved after a time evolution, which is generated by the total Hamiltonian $\mathcal{H}_T$, one has to require

$$\dot{\phi}_a{}^\mu = \{\phi_a{}^\mu, \mathcal{H}_T\} \approx 0, \tag{3.10}$$

which leads to the following secondary constraints

$$\mathcal{K}_a \equiv -\frac{k}{4\pi} \epsilon^{\bar{\mu}\bar{\nu}} h_{ab} F^b{}_{\bar{\mu}\bar{\nu}} \approx 0 \tag{3.11a}$$

$$D_{\bar{\mu}} \mathcal{A}^a{}_0 - u^a{}_{\bar{\mu}} \approx 0, \tag{3.11b}$$

where $D_{\bar{\mu}} X^a = \partial_{\bar{\mu}} X^a + f^a{}_{bc} \mathcal{A}^b{}_{\bar{\mu}} X^c$ is the gauge covariant derivative.
The Lagrange multipliers $u^a{}_{\bar{\mu}}$ can be determined via the Hamilton equations of motion as

$$\dot{\mathcal{A}}^a{}_{\bar{\mu}} = \frac{\partial \mathcal{H}_T}{\partial \pi_a{}^{\bar{\mu}}} = u^a{}_{\bar{\mu}}. \tag{3.12}$$

This allows one to rewrite (3.11b) and in addition yields the following weak equality

$$D_{\bar{\mu}} \mathcal{A}^a{}_0 - u^a{}_{\bar{\mu}} = D_{\bar{\mu}} \mathcal{A}^a{}_0 - \partial_0 \mathcal{A}^a{}_{\bar{\mu}} = F^a{}_{\bar{\mu}0} \approx 0. \tag{3.13}$$

The total Hamiltonian can now be written in the following form

$$\mathcal{H}_T = \mathcal{A}^a{}_0 \bar{\mathcal{K}}_a + u^a{}_0 \phi_a{}^0 + \partial_{\bar{\mu}} (\mathcal{A}^a{}_0 \pi_a{}^{\bar{\mu}}), \tag{3.14}$$

with

$$\bar{\mathcal{K}}_a = \mathcal{K}_a - D_{\bar{\mu}} \phi_a{}^{\bar{\mu}}. \tag{3.15}$$

The canonical commutation relations (3.7) can now be used to determine the following Poisson bracket algebra of constraints

$$\{\phi_a{}^{\bar{\mu}}(\mathbf{x}), \phi_b{}^{\bar{\nu}}(\mathbf{y})\} = -\frac{k}{2\pi}\epsilon^{\bar{\mu}\bar{\nu}} h_{ab}\delta^2(\mathbf{x}-\mathbf{y}), \tag{3.16a}$$

$$\{\phi_a{}^{\bar{\mu}}(\mathbf{x}), \bar{\mathcal{K}}_b(\mathbf{y})\} = -f_{ab}{}^c\phi_c{}^{\bar{\mu}}\delta^2(\mathbf{x}-\mathbf{y}), \tag{3.16b}$$

$$\{\bar{\mathcal{K}}_a(\mathbf{x}), \bar{\mathcal{K}}_b(\mathbf{y})\} = -f_{ab}{}^c\bar{\mathcal{K}}_c\delta^2(\mathbf{x}-\mathbf{y}), \tag{3.16c}$$

which are the only non-vanishing Poisson brackets of the constraints $\phi_a{}^{\mu}$ and $\bar{\mathcal{K}}_a$. This allows one to directly determine which of the constraints are first class and which are second class. First class constraints are constraints whose Poisson brackets vanish weakly with every other constraint. If that is not the case, then the constraint is called second class. This is a crucial distinction of constraints since first class constraints generate gauge transformations whereas second class constraints are used to restrict the phase space and thus promote the Poisson brackets to Dirac brackets, which can be used to consistently quantize the system.
Since only the Poisson brackets of $\phi_a{}^0$ and $\bar{\mathcal{K}}_a$ vanish weakly with all other constraints these are first class constraints. On the other hand $\phi_a{}^{\bar{\mu}}$ are second class constraints since they have non-weakly vanishing Poisson brackets with other constraints. Thus, one can use the second class constraints $\phi_a{}^{\bar{\mu}}$ in order to promote the Poisson brackets to Dirac brackets by eliminating physically irrelevant degrees of freedom. For the case at hand setting the second class constraints $\phi_a{}^{\bar{\mu}}$ strongly to zero $\phi_a{}^{\bar{\mu}} = 0$ one obtains the following Dirac brackets for the dynamical variables

$$\{\mathcal{A}^a{}_{\bar{\mu}}(\mathbf{x}), \mathcal{A}^b{}_{\bar{\nu}}(\mathbf{y})\}_{\text{D.B}} = \frac{2\pi}{k}h^{ab}\epsilon_{\bar{\mu}\bar{\nu}}\delta^2(\mathbf{x}-\mathbf{y}), \tag{3.17}$$

where $\epsilon_{\bar{\mu}\bar{\nu}}$ is obtained using $\epsilon^{\bar{\mu}\bar{\alpha}}\epsilon_{\bar{\alpha}\bar{\nu}} = \delta^{\bar{\mu}}{}_{\bar{\nu}}$.
Having determined all constraints one can also check that the number of local degrees of freedom of the physical system described by the Chern-Simons action (2.3) is indeed equal to zero. The degrees of freedom of a constrained Hamiltonian system are characterized by the dimension of the phase space $N$, the number of first class constraints $M$ and the number of second class constraints $S$ as

$$\text{№ of local physical D.O.F} = \frac{1}{2}(N - 2M - S)\,. \tag{3.18}$$

The phase space for a Chern-Simons theory in three dimensions is determined by the gauge fields $\mathcal{A}^a{}_\mu$. Denoting the dimension of the Lie algebra $\mathfrak{g}$ as $\mathfrak{D}$, then the dimension of the phase space is $N = 6\mathfrak{D}$. Accordingly, the number of first class constraints is $M = 2\mathfrak{D}$ and the number of second class constraints is $S = 2\mathfrak{D}$. Combining $N, M$ and $S$ as in (3.18) one finds that indeed the number of local physical degrees of freedom is zero as expected.

## Constructing the Gauge Generator

It was mentioned in the previous section that the first class constraints ${\phi_a}^0$ and $\bar{\mathcal{K}}_a$ generate gauge transformations. In this section I will explicitly show how to construct the canonical charges which generate those gauge transformations by using Castellani's algorithm [85]. In general one can construct such a gauge generator by

$$\mathcal{G} = \varepsilon(t)\mathcal{G}_0 + \dot{\varepsilon}(t)\mathcal{G}_1, \tag{3.19}$$

with $\dot{\varepsilon}(t) \equiv \frac{d\varepsilon(t)}{dt}$ and where $\varepsilon(t)$ is an arbitrary function of $t$. The constraints $\mathcal{G}_0$ and $\mathcal{G}_1$ have to fulfill the following relations

$$\mathcal{G}_1 = C_{\text{PFC}}, \tag{3.20a}$$

$$\mathcal{G}_0 + \{\mathcal{G}_1, \mathcal{H}_T\} = C_{\text{PFC}}, \tag{3.20b}$$

$$\{\mathcal{G}_0, \mathcal{H}_T\} = C_{\text{PFC}}, \tag{3.20c}$$

where $C_{\text{PFC}}$ denotes a primary first class constraint. These relations are fulfilled for $\mathcal{G}_0 = \bar{\mathcal{K}}_a$ and $\mathcal{G}_1 = {\phi_a}^0 = {\pi_a}^0$.

For the following considerations it will prove to be convenient to work with a smeared generator, which can be obtained by integrating over the spatial surface $\Sigma$ as

$$\mathcal{G}[\varepsilon] = \int_\Sigma \mathrm{d}^2x \left(D_0\varepsilon^a {\pi_a}^0 + \varepsilon^a \bar{\mathcal{K}}_a\right). \tag{3.21}$$

One can show by a straightforward but tedious calculation that this smeared generator generates the following gauge transformations via $\delta_\varepsilon \bullet = \{\bullet, \mathcal{G}[\varepsilon]\}$

$$\delta_\varepsilon {\mathcal{A}^a}_0 = D_0\varepsilon^a, \tag{3.22a}$$

$$\delta_\varepsilon {\mathcal{A}^a}_{\bar{\mu}} = D_{\bar{\mu}}\varepsilon^a, \tag{3.22b}$$

$$\delta_\varepsilon {\pi_a}^0 = -{f_{ab}}^c \varepsilon^b {\pi_c}^0, \tag{3.22c}$$

$$\delta_\varepsilon {\pi_a}^{\bar{\mu}} = \frac{k}{4\pi}\epsilon^{\bar{\mu}\bar{\nu}} h_{ab}\partial_{\bar{\nu}}\varepsilon^b - {f_{ab}}^c\varepsilon^b {\pi_c}^{\bar{\mu}}, \tag{3.22d}$$

$$\delta_\varepsilon {\phi_a}^{\bar{\mu}} = -{f_{ab}}^c\varepsilon^b {\phi_c}^{\bar{\mu}}. \tag{3.22e}$$

The generator $\mathcal{G}$ that has been constructed so far is only a preliminary result. The reason for this is that I am considering a Chern-Simons theory with a boundary which renders the generator $\mathcal{G}$ non-functionally differentiable.

In order to make this statement more precise I will first perform the full variation of the generator for a field independent gauge parameter $\varepsilon^a$

$$\begin{aligned}\delta\mathcal{G}[\varepsilon] = &\int_\Sigma \mathrm{d}^2x(\delta(D_0\varepsilon^a{\pi_a}^0) + \varepsilon^a\delta\bar{K}_a) = \\ = &\int_\Sigma \mathrm{d}^2x \Big( {f^a}_{bc}\varepsilon^c {\pi_a}^\mu \delta {\mathcal{A}^b}_\mu + D_\mu\varepsilon^a\delta{\pi_a}^\mu + \frac{k}{4\pi}\epsilon^{\bar{\mu}\bar{\nu}} h_{ab}\partial_{\bar{\mu}}\varepsilon^a\delta{\mathcal{A}^b}_{\bar{\nu}} - \\ &\partial_{\bar{\mu}}\left(\frac{k}{4\pi}\epsilon^{\bar{\mu}\bar{\nu}}h_{ab}\varepsilon^a\delta{\mathcal{A}^b}_{\bar{\nu}} + \varepsilon^a\delta{\pi_a}^{\bar{\mu}}\right)\Big).\end{aligned} \tag{3.23}$$

The first three terms are regular bulk terms and thus do not spoil functional differentiability. The last term on the other hand is a boundary term that spoils functional differentiability. In order to fix this one has to add a suitable boundary term to the gauge generator in such a way that the variation of this additional boundary term cancels exactly the boundary term in (3.23) i.e.

$$\delta\bar{\mathcal{G}}[\varepsilon] = \delta\mathcal{G}[\varepsilon] + \delta\mathcal{Q}[\varepsilon], \tag{3.24}$$

with

$$\delta\mathcal{Q}[\varepsilon] = \int_\Sigma \mathrm{d}^2x\, \partial_{\bar{\mu}}\left(\frac{k}{4\pi}\epsilon^{\bar{\mu}\bar{\nu}}h_{ab}\varepsilon^a\delta\mathcal{A}^b{}_{\bar{\nu}} + \varepsilon^a\delta\pi_a{}^{\bar{\mu}}\right). \tag{3.25}$$

This expression for the variation of the canonical boundary charge $\delta\mathcal{Q}[\varepsilon]$ can be further simplified by first setting the second class constraints $\phi_a{}^{\bar{\mu}} \approx 0$ strongly equal to zero and thus going to the reduced phase space. One can then use in addition Stoke's theorem[4], which simplifies the variation of the boundary charge even further to

$$\delta\mathcal{Q}[\varepsilon] = \frac{k}{2\pi}\int \mathrm{d}\varphi\, \langle\varepsilon\,\delta\mathcal{A}_\varphi\rangle = \frac{k}{2\pi}\int \mathrm{d}\varphi\, h_{ab}\,\varepsilon^a\delta\mathcal{A}^b{}_\varphi. \tag{3.26}$$

Whether this expression is functionally integrable or not depends on the specific form of the gauge parameter $\varepsilon^a$ and thus differs from theory to theory. It is, however, important to note that the general expression for the variation of the canonical boundary charge (3.26) is independent of the theory under consideration.
One simple example where (3.26) is functionally integrable is when the gauge parameter is field independent. One then obtains the following canonical boundary charge

$$\mathcal{Q}[\varepsilon] = \frac{k}{2\pi}\int \mathrm{d}\varphi\, \langle\varepsilon\,\mathcal{A}_\varphi\rangle = \frac{k}{2\pi}\int \mathrm{d}\varphi\, h_{ab}\varepsilon^a\mathcal{A}^b{}_\varphi. \tag{3.27}$$

### Asymptotic Symmetry Algebra

As a final step one now has to determine the Dirac brackets of the canonical boundary charge $\mathcal{Q}$ with itself in order to determine the asymptotic symmetry algebra generated by those charges[5], which read in general

$$\{\mathcal{Q}[\varepsilon], \mathcal{Q}[\lambda]\} = \mathcal{Q}[\sigma(\varepsilon,\lambda)] + Z[\varepsilon,\lambda], \tag{3.28}$$

where $Z[\varepsilon,\lambda]$ denotes possible central terms and $\sigma(\varepsilon,\lambda)$ is a composite gauge parameter consisting of $\varepsilon$ and $\lambda$. At this point it is important to note that the appearance of those central terms is exactly the mechanism which is responsible for having non-trivial physics at the boundary, as these central terms reduce some of the first class constraints to second class constraints. Therefore, these transformations are not proper gauge transformations anymore but rather correspond to global symmetry transformations which change the state of the physical system.
This asymptotic symmetry algebra can of course be determined by brute force evaluation of the Dirac brackets and the basic relations (3.17). As this is usually a rather tedious calculation I will use the following shortcut when determining asymptotic symmetry algebras in my thesis.
Given two functions $\mathcal{V}$, $\mathcal{W}$ and a canonical boundary charge $\mathcal{Q}[\varepsilon] = \int \mathrm{d}\varphi\, \varepsilon(\mathbf{x})\mathcal{V}(\mathbf{x})$ one can use the fact that this charge generates infinitesimal gauge transformations via $\delta_\varepsilon \bullet = \{\bullet, \mathcal{Q}[\varepsilon]\}$. Thus, knowing for example how $\mathcal{W}$ transforms under an infinitesimal gauge transformation with gauge parameter $\varepsilon$ one can determine the Dirac bracket[6] $\{\mathcal{V}(\varphi), \mathcal{W}(\bar{\varphi})\}$ using

$$\delta_\varepsilon \mathcal{W}(\mathbf{y}) = -\{\mathcal{Q}[\varepsilon], \mathcal{W}(\mathbf{y})\} = -\int \mathrm{d}\varphi\, \varepsilon(\mathbf{x})\{\mathcal{V}(\mathbf{x}), \mathcal{W}(\mathbf{y})\}. \tag{3.29}$$

[4] I assume here that the boundary of the surface $\Sigma$ is parametrized by $\varphi$.
[5] To be more precise the starting point is again the Poisson bracket algebra of the improved canonical charges $\bar{\mathcal{G}}$ which reduces to the Dirac bracket algebra of the canonical boundary charges after setting the second class constraints strongly to zero.

As an example let me consider the function $\mathcal{L}(\varphi)$ which transforms under gauge transformations with gauge parameter $\epsilon(\varphi)$ as

$$\delta_\epsilon \mathcal{L} = \frac{k}{4\pi}\epsilon''' + \left(2\epsilon'\mathcal{L} + \epsilon\mathcal{L}'\right), \tag{3.30}$$

where a prime denotes derivative with respect to $\varphi$. Furthermore assume that the canonical boundary charge is given by

$$\mathcal{Q}[\epsilon] = \int \mathrm{d}\varphi\, \epsilon\, \mathcal{L}. \tag{3.31}$$

The Dirac bracket $\{\mathcal{L}(\varphi), \mathcal{L}(\bar{\varphi})\}$ then is calculated using

$$\delta_\epsilon \mathcal{L}(\bar{\varphi}) = -\{\mathcal{Q}(\epsilon), \mathcal{L}(\bar{\varphi})\} = -\int \mathrm{d}\varphi\, \epsilon(\varphi)\{\mathcal{L}(\varphi), \mathcal{L}(\bar{\varphi})\}. \tag{3.32}$$

Equation (3.32) can be satisfied for

$$\{\mathcal{L}(\varphi), \mathcal{L}(\bar{\varphi})\} = \frac{k}{4\pi}\delta'''(\varphi - \bar{\varphi}) + \left[2\mathcal{L}(\bar{\varphi})\delta'(\varphi - \bar{\varphi}) - \mathcal{L}'(\bar{\varphi})\delta(\varphi - \bar{\varphi})\right], \tag{3.33}$$

with $\delta'(\varphi - \bar{\varphi}) = \partial_\varphi \delta(\varphi - \bar{\varphi})$. This can also be written in terms of $\delta_\epsilon \mathcal{L}$ as

$$\{\mathcal{L}(\varphi), \mathcal{L}(\bar{\varphi})\} = -\delta_\epsilon \mathcal{L}(\bar{\varphi})\Big|_{\partial^n_{\bar{\varphi}}\epsilon(\bar{\varphi}) = (-1)^n \partial^n_{\bar{\varphi}}\delta(\varphi - \bar{\varphi})}. \tag{3.34}$$

The expression (3.34) can be seen as a convenient shortcut that allows one to determine the Dirac bracket algebra of the asymptotic symmetries directly from the transformation behavior of the state dependent fields under infinitesimal gauge transformations.

[6]Please note that for the sake of compactness I will from now on omit the additional subscript that I used previously to distinguish Poisson and Dirac brackets i.e. from now on $\{\cdot,\cdot\} \equiv \{\cdot,\cdot\}_{\text{D.B}}$ whenever I am using the term Dirac bracket.

## Example: $\mathfrak{sl}(2,\mathbb{R}) \oplus \mathfrak{sl}(2,\mathbb{R})$ Chern-Simons Theory

I want to close this chapter by reviewing asymptotically $\text{AdS}_3$ boundary conditions [11] for a $\mathfrak{sl}(2,\mathbb{R}) \oplus \mathfrak{sl}(2,\mathbb{R})$ Chern-Simons theory and how to determine the asymptotic symmetry algebra using the methods previously described in this chapter [86]. In [87] it was shown that the metric

$$\mathrm{d}s^2 = \ell^2 \left[ \mathrm{d}\rho^2 - \frac{2\pi}{k}\left(\mathcal{L}(\mathrm{d}x^+)^2 + \bar{\mathcal{L}}(\mathrm{d}x^-)^2\right) - \left(e^{2\rho} + \frac{4\pi^2}{k^2}\mathcal{L}\bar{\mathcal{L}}e^{-2\rho}\right)\mathrm{d}x^+\,\mathrm{d}x^- \right], \tag{3.35}$$

is a solution of Einstein's equations in 3d for any functions $\mathcal{L} \equiv \mathcal{L}(x^+)$, $\bar{\mathcal{L}} \equiv \bar{\mathcal{L}}(x^-)$, where $x^\pm = \frac{t}{\ell} \pm \varphi$, $\ell$ is the AdS radius and $k$ is given by (2.13).
For constant $\mathcal{L}$ and $\bar{\mathcal{L}}$ one obtains the BTZ black hole [12] with mass $M$ and angular momentum $J$ via the identification

$$\mathcal{L}_{\text{BTZ}} = -\frac{1}{4\pi}(M\ell - J), \tag{3.36a}$$

$$\bar{\mathcal{L}}_{\text{BTZ}} = -\frac{1}{4\pi}(M\ell + J). \tag{3.36b}$$

Global $\text{AdS}_3$ is obtained for $J = 0$ and $M\ell = -\frac{k}{2}$ which corresponds to

$$\mathcal{L}_{\text{AdS}} = \bar{\mathcal{L}}_{\text{AdS}} = \frac{k}{8\pi}, \tag{3.37}$$

while Poincaré patch AdS is obtained for $M = J = 0$ as well as an additional decompactification of the boundary coordinate $\varphi$.
Solutions with different (and in general non-constant) $\mathcal{L}$ and $\bar{\mathcal{L}}$ can be related by global symmetry transformations which correspond to finite transformations of the form (2.14) that change the canonical boundary charges. I again want to emphasize the fact that this is purely due to the presence of a boundary in the theory, making it necessary to introduce the boundary canonical charge (3.27). This makes it possible that some of the first class constraints can be reduced to second class constraints. This in turn changes some of the gauge symmetries in the bulk to global symmetries at the boundary.
In order to describe the metric (3.35) in a Chern-Simons formulation using two Chern-Simons gauge fields $A$ and $\bar{A}$ I will first choose the following basis of $\mathfrak{sl}(2,\mathbb{R})$ generators

$$[L_n, L_m] = (n-m)L_{n+m}, \tag{3.38}$$

with $n, m = \pm 1, 0$ whose invariant bilinear form in the fundamental representation is given by (A.2).
In order to simplify the canonical analysis one can also use some of the gauge freedom provided by Chern-Simons theories to fix the radial dependence as

$$A = b^{-1}\left[a(x^+, x^-) + \mathrm{d}\right] b, \tag{3.39a}$$

$$\bar{A} = b\left[\bar{a}(x^+, x^-) + \mathrm{d}\right] b^{-1}, \tag{3.39b}$$

with $b = b(\rho) = e^{\rho L_0}$. One can then verify using $g_{\mu\nu} = \frac{\ell^2}{2}\left\langle A_\mu - \bar{A}_\mu, A_\nu - \bar{A}_\nu\right\rangle$ that the connections

$$a = \left(L_1 + \frac{2\pi}{k}\mathcal{L}(x^+)L_{-1}\right)\mathrm{d}x^+, \tag{3.40a}$$

$$\bar{a} = -\left(L_{-1} + \frac{2\pi}{k}\bar{\mathcal{L}}(x^-)L_1\right)\mathrm{d}x^-, \tag{3.40b}$$

correctly reproduce the line element (3.35).
The next step consists of determining the gauge transformations $\varepsilon = b^{-1}\epsilon^{(0)}(x^+)b$, $\bar{\varepsilon} = b\bar{\epsilon}^{(0)}(x^-)b^{-1}$ which preserve the structure of (3.40). Those transformations are given by

$$\epsilon^{(0)}(x^+) = \epsilon(x^+)L_1 - \epsilon'(x^+)L_0 + \left[\frac{\epsilon''(x^+)}{2} + \frac{2\pi}{k}\mathcal{L}(x^+)\epsilon(x^+)\right]L_{-1}, \tag{3.41a}$$

$$\bar{\epsilon}^{(0)}(x^-) = \left[\frac{\bar{\epsilon}''(x^-)}{2} + \frac{2\pi}{k}\bar{\mathcal{L}}(x^-)\bar{\epsilon}(x^-)\right]L_1 - \bar{\epsilon}'(x^-)L_0 + \bar{\epsilon}(x^-)L_{-1}, \tag{3.41b}$$

where a prime denotes a derivative with respect to the argument of the function it is acting on i.e. $f'(x^\pm) = \partial_{x^\pm}f(x^\pm)$.
One can now also determine how the functions $\mathcal{L}$ and $\bar{\mathcal{L}}$ transform under those gauge transformations. This transformation behavior is given by

$$\delta_\epsilon\mathcal{L} = \epsilon\mathcal{L}' + 2\mathcal{L}\epsilon' + \frac{k}{4\pi}\epsilon''', \qquad \delta_{\bar{\epsilon}}\bar{\mathcal{L}} = -\left(\bar{\epsilon}\bar{\mathcal{L}}' + 2\bar{\mathcal{L}}\bar{\epsilon}' + \frac{k}{4\pi}\bar{\epsilon}'''\right). \tag{3.42}$$

The corresponding variations of the canonical boundary charges can be integrated and read

$$\mathcal{Q}[\varepsilon] = \int \mathrm{d}\varphi\,\epsilon\,\mathcal{L}, \quad \bar{\mathcal{Q}}[\bar{\varepsilon}] = -\int \mathrm{d}\varphi\,\bar{\epsilon}\,\bar{\mathcal{L}}. \tag{3.43}$$

One can now use (3.34) in order to determine the Dirac bracket algebra[7] of these canonical boundary charges straightforwardly as

$$\{\mathcal{L}(\varphi), \mathcal{L}(\bar{\varphi})\} = \frac{k}{4\pi}\delta'''(\varphi - \bar{\varphi}) + \left[2\mathcal{L}(\bar{\varphi})\delta'(\varphi - \bar{\varphi}) - \mathcal{L}'(\bar{\varphi})\delta(\varphi - \bar{\varphi})\right], \tag{3.44a}$$

$$\{\bar{\mathcal{L}}(\varphi), \bar{\mathcal{L}}(\bar{\varphi})\} = \frac{k}{4\pi}\delta'''(\varphi - \bar{\varphi}) + \left[2\bar{\mathcal{L}}(\bar{\varphi})\delta'(\varphi - \bar{\varphi}) - \bar{\mathcal{L}}'(\bar{\varphi})\delta(\varphi - \bar{\varphi})\right]. \tag{3.44b}$$

For many purposes it is already sufficient to have the asymptotic symmetries available in the form of Dirac brackets as in (3.44). It is, however, often also useful to go one step further and represent the algebra (3.44) in terms of its Fourier modes. In order to cast the algebra into a possibly more familiar form one first has to suitable decompose the functions $\mathcal{L}(\varphi)$ and $\bar{\mathcal{L}}(\varphi)$ in terms of Fourier modes and quantize the system, i.e. replacing $i\{\cdot,\cdot\} \to [\cdot,\cdot]$. After doing so one obtains the following asymptotic symmetry algebra given by

$$[L_n, L_m] = (n-m)L_{n+m} + \frac{c}{12}n(n^2-1)\delta_{n+m,0}, \tag{3.45a}$$

$$[\bar{L}_n, \bar{L}_m] = (n-m)\bar{L}_{n+m} + \frac{\bar{c}}{12}n(n^2-1)\delta_{n+m,0}, \tag{3.45b}$$

where

$$c = \bar{c} = 6k = \frac{3\ell}{2G_N}. \tag{3.46}$$

[7] Dirac brackets of the form $\{A(t \pm \varphi), B(\bar{t} \pm \bar{\varphi})\}$ are evaluated at $t = \bar{t}$, where $A$ and $B$ are some arbitrary functions of their repective argument. Thus, one can also equivalently write $\{A(t \pm \varphi), B(\bar{t} \pm \bar{\varphi})\}\big|_{t=\bar{t}} = \{A(\varphi), B(\bar{\varphi})\}$.

This is the famous result Brown and Henneaux obtained in [11] which showed that there are boundary conditions for $AdS_3$ that can be chosen in such a way that the asymptotic symmetries are given by two copies of the Virasoro algebra as in (3.45), which in turn gave rise to the idea that the holographic dual of $AdS_3$ is a two dimensional conformal field theory.

# Anti-de Sitter Higher-Spin Gravity 4

*Allwissend bin ich nicht; doch viel ist mir bewusst.*
*(Omniscent I am not; yet much is known to me.)*

**– Johann Wolfgang Goethe**
Faust I

Having reviewed how to describe $AdS_3$ Einstein-Hilbert gravity using the Chern-Simons formalism in Chapter 3, I now want to review how one can extend the Chern-Simons formalism in order to describe higher-spin gravity theories.

Since a Chern-Simons gauge theory with gauge algebra $\mathfrak{sl}(2,\mathbb{R}) \oplus \mathfrak{sl}(2,\mathbb{R})$ describes asymptotically $AdS_3$ spin-2 gravity, it is natural to promote the gauge algebra to $\mathfrak{sl}(N,\mathbb{R}) \oplus \mathfrak{sl}(N,\mathbb{R})$[1] in order to describe gravity theories which are still asymptotically $AdS_3$ but have additional higher-spin symmetries. And indeed in [88] it was shown that for $N \geq 3$ such a Chern-Simons theory describes the nonlinear interactions of gravity coupled to a finite tower of massless integer spin-$s \leq N$ fields.

From a holographic perspective one point of interest are again the asymptotic symmetries of these higher-spin gravity theories and thus the procedure outlined in Chapter 3, but this time for general $\mathfrak{sl}(N,\mathbb{R}) \oplus \mathfrak{sl}(N,\mathbb{R})$. This analysis has been performed in [28, 29] and showed that the asymptotic symmetries for the boundary conditions considered in those papers amount to two copies of $\mathcal{W}_N$-algebras [89] which can be considered as (nonlinear) higher-spin extensions of the Virasoro algebra.

Denoting again by $L_n$ the generators of the $\mathfrak{sl}(2,\mathbb{R})$ subalgebra contained in $\mathfrak{sl}(N,\mathbb{R})$ and $\langle \cdot,\cdot \rangle$ the invariant bilinear form of the fundamental representation of $\mathfrak{sl}(N,\mathbb{R})$ then matching the Chern-Simons level $k$ with the normalization of the Einstein-Hilbert action requires

$$k = \frac{\ell}{8G_N \langle L_0, L_0 \rangle}. \tag{4.1}$$

In the same way the corresponding Virasoro central charge of the dual CFT is given by

$$c = 12k \langle L_0, L_0 \rangle . \tag{4.2}$$

[1] As already mentioned in Chapter 2 the spectrum of the higher-spin gravity theory depends on the specific embedding of $\mathfrak{sl}(2,\mathbb{R}) \hookrightarrow \mathfrak{sl}(N,\mathbb{R})$. Everything about higher-spin gravity theories that I will review in this chapter has been done for the principal embedding of $\mathfrak{sl}(2,\mathbb{R}) \hookrightarrow \mathfrak{sl}(N,\mathbb{R})$. Thus it is implicitly understood that everything described in this chapter relates to the principal embedding without explicitly mentioning the specific embedding.

Also the metric is now obtained from the gauge fields $A$ and $\bar{A}$ as [33, 90]

$$g_{\mu\nu} = \frac{\ell^2}{2\langle L_0, L_0\rangle} \left\langle A_\mu - \bar{A}_\mu, A_\nu - \bar{A}_\nu \right\rangle . \tag{4.3}$$

In order to illustrate all this it is instructive to take the first step away from the spin-2 example and look at spin-3 gravity described by a principally embedded $\mathfrak{sl}(3,\mathbb{R}) \oplus \mathfrak{sl}(3,\mathbb{R})$ Chern-Simons theory.

## Spin-3 Anti-de Sitter Gravity

The principal embedding of $\mathfrak{sl}(2,\mathbb{R}) \hookrightarrow \mathfrak{sl}(3,\mathbb{R})$ contains the $\mathfrak{sl}(2,\mathbb{R})$ spin-2 triplet $L_n$ ($n = \pm 1, 0$), the spin-3 quintet $W_m$ ($m = \pm 2, \pm 1, 0$) and can be represented in terms of the following commutation relations

$$[L_n, L_m] = (n-m)L_{n+m}, \tag{4.4a}$$

$$[L_n, W_m] = (2n-m)W_{n+m}, \tag{4.4b}$$

$$[W_n, W_m] = \sigma(n-m)(2m^2 + 2n^2 - nm - 8)L_{n+m}, \tag{4.4c}$$

where $\sigma$ is an arbitrary constant which can be changed by rescaling $W_n$. I will set $\sigma = -\frac{1}{3}$ from now on. The invariant bilinear form on this algebra will be given by the trace which is taken in the fundamental representation of the algebra.
Using the same gauge and notation as in (3.39) one can write down the spin-3 extension of (3.40) as [29]

$$a = \left(L_1 + \frac{2\pi}{k}\mathcal{L}(x^+)L_{-1} - \frac{\pi}{2k}\mathcal{W}(x^+)W_{-2}\right) \mathrm{d}x^+, \tag{4.5a}$$

$$\bar{a} = -\left(L_{-1} + \frac{2\pi}{k}\bar{\mathcal{L}}(x^-)L_1 - \frac{\pi}{2k}\bar{\mathcal{W}}(x^-)W_2\right) \mathrm{d}x^-. \tag{4.5b}$$

From now on I will only display all expressions of the $A$-sector, since all resulting expressions involving the $\bar{A}$ sector can be obtained in the exact same way as in the unbarred sector.
Parametrizing the gauge transformations which are compatible with (4.5) as

$$b^{-1}\varepsilon b, \quad \varepsilon = \sum_{n=-1}^{1} \epsilon^n(x^+)L_n + \sum_{n=-2}^{2} \chi^n(x^+)W_n, \tag{4.6}$$

where

$$\epsilon^1 \equiv \epsilon, \quad \epsilon^2 = -\epsilon', \quad \epsilon^{-1} = \frac{\epsilon''}{2} + \frac{2\pi}{k}\epsilon\mathcal{L} + \frac{4\pi}{k}\chi\mathcal{W}, \tag{4.7a}$$

$$\chi^2 \equiv \chi, \quad \chi^1 = -\chi', \quad \chi^0 = \frac{\chi''}{2} + \frac{4\pi}{k}\mathcal{L}, \tag{4.7b}$$

$$\chi^{-1} = -\frac{\chi'''}{6} - \frac{10\pi}{3k}\chi'\mathcal{L} - \frac{4\pi}{3k}\chi\mathcal{L}', \tag{4.7c}$$

$$\chi^{-2} = \frac{\chi''''}{24} + \frac{4\pi}{3k}\chi''\mathcal{L} + \frac{7\pi}{6k}\chi'\mathcal{L}' + \frac{\pi}{3k}\chi\mathcal{L}'' + \frac{4\pi^2}{k^2}\chi\mathcal{L}^2 - \frac{\pi}{2k}\epsilon\mathcal{W}, \tag{4.7d}$$

one finds that the fields $\mathcal{L}$ and $\mathcal{W}$ transform under these gauge transformations as

$$\delta_\epsilon \mathcal{L} = \epsilon\mathcal{L}' + 2\epsilon'\mathcal{L} + \frac{k}{4\pi}\epsilon''', \tag{4.8a}$$

$$\delta_\epsilon \mathcal{W} = \epsilon\mathcal{W}' + 3\epsilon'\mathcal{W}, \tag{4.8b}$$

$$\delta_\chi \mathcal{L} = 2\chi\mathcal{W}' + 3\chi'\mathcal{W}, \tag{4.8c}$$

$$\delta_\chi \mathcal{W} = -\frac{1}{3}\left[2\chi\mathcal{L}''' + 9\chi'\mathcal{L}'' + 15\chi''\mathcal{L}' + \frac{k}{4\pi}\chi''''' + \frac{64\pi}{k}\left(\chi\mathcal{L}\mathcal{L}' + \chi'\mathcal{L}^2\right)\right]. \tag{4.8d}$$

As in the spin-2 case one can now determine the canonical boundary charges

$$\mathcal{Q}[\varepsilon] = \int \mathrm{d}\varphi \left[\epsilon(\varphi)\mathcal{L}(\varphi) + \chi(\varphi)\mathcal{W}(\varphi)\right], \tag{4.9}$$

and their corresponding Dirack bracket algebra

$$\{\mathcal{L}(\varphi), \mathcal{L}(\bar\varphi)\} = \frac{k}{4\pi}\delta''' + \left[2\mathcal{L}\delta' - \mathcal{L}'\delta\right], \tag{4.10a}$$

$$\{\mathcal{L}(\varphi), \mathcal{W}(\bar\varphi)\} = 3\delta'\mathcal{W} - 2\delta\mathcal{W}', \tag{4.10b}$$

$$\begin{aligned}\{\mathcal{W}(\varphi), \mathcal{W}(\bar\varphi)\} = \frac{1}{3}\Big[&2\delta\mathcal{L}''' - 9\delta'\mathcal{L}'' + 15\delta''\mathcal{L}' - 10\delta'''\mathcal{L} - \frac{k}{4\pi}\delta''''' \\ &+ \frac{64\pi}{k}\left(\delta\mathcal{L}\mathcal{L}' - \delta'\mathcal{L}^2\right)\Big],\end{aligned} \tag{4.10c}$$

where all functions on the right hand side are functions of $\bar\varphi$, $\delta$ denotes $\delta(\varphi - \bar\varphi)$ and a prime denotes derivative with respect to $\bar\varphi$. Choosing appropriate Fourier mode expansions for the fields $\mathcal{L}$, $\mathcal{W}$ and replacing again $i\{\cdot,\cdot\} \to [\cdot,\cdot]$ one obtains the semiclassical version of the $\mathcal{W}_3$ algebra [89, 91, 92]

$$[L_n, L_m] = (n-m)L_{n+m} + \frac{c}{12}n(n^2-1)\delta_{n+m,0}, \tag{4.11a}$$

$$[L_n, W_m] = (2n-m)W_{n+m}, \tag{4.11b}$$

$$\begin{aligned}[W_n, W_m] = -\frac{1}{3}\Big[&(n-m)(2n^2+2m^2-nm-8)L_{n+m} + \frac{96}{c}(n-m)\Lambda_{n+m} \\ &+ \frac{c}{12}n(n^2-4)(n^2-1)\delta_{n+m,0}\Big],\end{aligned} \tag{4.11c}$$

where the Virasoro central charge is again given by $c = 6k$ and

$$\Lambda_{n+m} = \sum_{p\in\mathbb{Z}} L_{n-p}L_p. \tag{4.12}$$

The barred sector yields exactly the same algebra and, thus the (semiclassical) asymptotic symmetries are given by two copies of the $\mathcal{W}_3$ algebra.

The $\mathcal{W}_3$ algebra is the simplest example of a more general class of nonlinear $\mathcal{W}_N$-algebras which describe the asymptotic symmetries of spin-$N$ gravity theories. As such the $\mathcal{W}_3$ algebra already exhibits all the important features which make $\mathcal{W}_N$-algebras special and interesting to study.

For $N \geq 3$ all $\mathcal{W}_N$-algebras are infinite-dimensional, nonlinear and centrally extended algebras which always contain the Virasoro algebra as a subalgebra.
After one has determined the asymptotic symmetries the first thing to do is usually to look for unitary representations of the given algebra and then determine a possible field theory dual. In the usual spin-2 gravity case one can directly proceed from the given Virasoro algebras since the Virasoro algebra is consistent with both small and large central charges. In the case of $\mathcal{W}_N$-algebras one has to be a little bit more careful because of the nonlinear terms such as $\Lambda_n$ in (4.12) in the $\mathcal{W}_3$-case.

## Quantum Asymptotic Symmetries

In [28] it has been pointed out for the first time that the canonical analysis performed previously in general is only valid for large values of the Virasoro central charge $c$, i.e. the asymptotic symmetry algebra obtained is not the full quantum algebra but rather the semiclassical one.
The reason for this is basically the appearance of the nonlinear terms in the $\mathcal{W}_N$ algebras. I will illustrate this using the nonlinear term (4.12) that appeared already in the semiclassical $\mathcal{W}_3$-algebra (4.11).
In a proper quantum field theory which has global symmetries given by (4.4) and local symmetries given by the $\mathcal{W}_3$-algebra (4.11) one has to be careful whenever products of operators appear. Assuming that the theory has a vacuum state $|0\rangle$ which is invariant under the global symmetries and interpreting the Fourier modes of (4.11) as operators, then these operators have to satisfy

$$L_n|0\rangle = 0, \quad W_m|0\rangle = 0 \qquad \forall\, n \geq -1, \text{ and } n \geq -2. \tag{4.13}$$

Thus, having introduced such a vacuum state one can also define a consistent notion of normal ordering of operators in a way that the expectation value of such normal ordered products vanishes.
For the vacuum as defined in (4.13) and the operator $\Lambda_n$ this means that a normal ordered version of $\Lambda_n$ is given by

$$:\!\Lambda\!:_n = \sum_{p\geq -1} L_{n-p}L_p + \sum_{p<-1} L_p L_{n-p}. \tag{4.14}$$

This is a very natural choice of vacuum and thus also the choice of normal ordering is quite natural. However, considering normal ordering and assuming that the algebra (4.11) still describes the correct local symmetries yields a contradiction. This contradiction manifests itself in the form of the Jacobi identities, which are no longer satisfied for the algebra (4.11) and the normal ordered $:\!\Lambda\!:_n$. This is a serious problem as the validity of the Jacobi identities is a fundamental property of the commutator. This means that the algebra (4.11) has to be modified in such a way that it is again compatible with the Jacobi identities.

One straightforward way to do this is assuming that the quantum $\mathcal{W}_3$-algebra is a deformed version of (4.11) and make an ansatz with new arbitrary structure constants. Then demanding compatibility with the Jacobi identities completely fixes those new structure constants and directly yields the correct quantum $\mathcal{W}_3$-algebra which is valid for all values of the Virasoro central charge $c$ and is compatible with the normal ordered nonlinear operators appearing in the algebra.
For the $\mathcal{W}_3$-algebra this procedure yields

$$[L_n, L_m] =(n-m)L_{n+m} + \frac{c}{12}n(n^2-1)\delta_{n+m,0}, \tag{4.15a}$$

$$[L_n, W_m] =(2n-m)W_{n+m}, \tag{4.15b}$$

$$\begin{aligned}[W_n, W_m] =& -\frac{1}{3}\left[(n-m)g(n,m)L_{n+m} + \frac{96}{c+\frac{22}{5}}(n-m) :\Lambda:_{n+m}\right. \\ &\left.+\frac{c}{12}n(n^2-4)(n^2-1)\delta_{n+m,0}\right],\end{aligned} \tag{4.15c}$$

with

$$g(n,m) = \left(2n^2+2m^2-nm-8-\frac{144}{5c+22}\left((n+m)+3\right)\left((n+m)+2\right)\right). \tag{4.16}$$

One can also absorb the additional terms which appear in $g(n,m)$ and which are proportional to $\frac{1}{c}$ in the normal ordering definition of $:\Lambda:_n$ as

$$:\tilde{\Lambda}:_n= \sum_{p\geq -1} L_{n-p}L_p + \sum_{p<-1} L_pL_{n-p} - \frac{3}{10}(n+3)(n+2)L_n. \tag{4.17}$$

This definition has also the advantage that $:\tilde{\Lambda}:_n$ is a conformal quasi primary in contrast to $:\Lambda:_n$, which can be easily checked using the formulas provided in (C.4). Using $:\tilde{\Lambda}:_n$ then yields a possibly more familiar expression for the quantum $\mathcal{W}_3$-algebra, which is usually used in the literature [89, 91, 92]

$$[L_n, L_m] =(n-m)L_{n+m} + \frac{c}{12}n(n^2-1)\delta_{n+m,0}, \tag{4.18a}$$

$$[L_n, W_m] =(2n-m)W_{n+m}, \tag{4.18b}$$

$$\begin{aligned}[W_n, W_m] =& -\frac{1}{3}\left[(n-m)(2n^2+2m^2-nm-8)L_{n+m} + \frac{96}{c+\frac{22}{5}}(n-m) :\tilde{\Lambda}:_{n+m}\right. \\ &\left.+\frac{c}{12}n(n^2-4)(n^2-1)\delta_{n+m,0}\right].\end{aligned} \tag{4.18c}$$

This is the final form of the asymptotic symmetry algebra of AdS spin-3 gravity. One can proceed in exactly the same manner for AdS spin-$N$ gravity and thus determine the asymptotic symmetries as quantum $\mathcal{W}_N$-algebras.
The quantum asymptotic symmetry algebras determined in this way then are the basic symmetries which have to be considered when looking for quantum field theory duals. As such, it is of interest to look for unitary representations of those algebras.

Unitarity in this sense means that the module that consists of the states which are built from repeated application of the operators and which generate the asymptotic quantum symmetry algebra on the vacuum does not contain any states which have negative norm. This analysis usually boils down to determining the eigenvalues of the Gramian matrix ,whose basis contains all the states for a given excitation level[2] of the vacuum.

This analysis usually restricts the possible values of the Virasoro central charge contained in the $\mathcal{W}_N$-algebras. For instance considering as the simplest example the Virasoro algebra with central charge $c$ accompanied by a highest-weight vector state $|h\rangle$, where $L_0|h\rangle = h|0\rangle$ and $L_n|h\rangle = 0$, $\forall\ n > 0$ one can only have unitary irreducible representations of this algebra if either $c \geq 1$ and $h \geq 0$ or $c$ takes one of the values [93, 94]

$$c = 1 - \frac{6}{m(m+1)}, \qquad \forall\ m = 2, 3, 4, \ldots, \tag{4.19}$$

and at the same time $h$ has the values

$$h = h_{r,s}(c) = \frac{((m+1)r - ms)^2 - 1}{4m(m+1)}, \tag{4.20}$$

for $r = 1, 2, 3, \ldots, m-1$ and $s = 1, 2, 3, \ldots, r$. In the case $m = \frac{q}{p-q}$, $0 < r < q$, $0 < s < p$ for $p$ and $q$ coprime integers and $r$ and $s$ integers the corresponding representations have their own name and are called minimal models [95] which are related to various condensed matter systems such as the Ising model.

For $\mathcal{W}_N$-algebras one can find analogous restrictions on the central charges which lead to the $\mathcal{W}_N$ minimal models [32].

[2]The term *level* in this context refers to the $L_0$ eigenvalue of a state created by acting on the vacuum with a given number of the generators of the quantum symmetry algebra.

# İnönü–Wigner Contractions

# 5

> *Some allies are more dangerous than enemies.*
>
> **– George R. R. Martin**
> A Song of Ice and Fire

In 1953 Erdal İnönü and Eugene Paul Wigner discussed in [96] the possibility of group contractions in order to obtain certain groups via a limiting procedure from other groups. This concept of a group contraction can be motivated by the following example. When we are walking on the surface of the earth then for all practical purposes we can assume that we are moving on a two-dimensional plane. Thus, we are performing Euclidean transformations on this plane, i.e. Euclidean translations and rotations around an axis perpendicular to this plane. This is of course only part of the truth. What we are actually doing are rotations around the center of the earth albeit with a very large radius. This motivated İnönü andl Wigner originally to think about the Euclidean group in two dimensions as a limit of the group of rotations in three dimensions. One of their central results in this paper is the following theorem:

**Theorem 1.** *Every Lie group can be contracted with respect to any of its continuous subgroups and only with respect to these.*
The subgroup with respect to which the contraction is undertaken will be called S. *The contracted infinitesimal elements form an abelian invariant subgroup of the contracted group. The subgroup S with respect to which the contraction was undertaken is isomorphic with the factor group of this invariant subgroup. Conversely, the existence of an abelian invariant subgroup and the possibility to choose from each of its cosets an element so that these form a subgroup S, is a necessary condition for the possibility to obtain the group from another group by contraction.*

In practice this usually amounts to a (singular) limiting procedure involving the generators of the Lie algebra $\mathfrak{g}$ which generate the group $G$ to be contracted. Putting the example mentioned in the beginning in a more formal form, then let me denote the special orthogonal group in three dimensions as $\mathrm{SO}(3)$. The generators of the corresponding Lie algebra $\mathfrak{so}(3)$ can be denoted as $X_a$ with $a = 1, 2, 3$ with commutation relations

$$[X_1, X_2] = X_3, \quad [X_3, X_1] = X_2, \quad [X_2, X_3] = X_1. \tag{5.1}$$

Performing a change of variables as

$$Y_1 = \epsilon X_1, \quad Y_2 = \epsilon X_2, \quad Y_3 = X_3, \tag{5.2}$$

yields

$$[Y_1, Y_2] = \epsilon^2 Y_3, \quad [Y_3, Y_1] = Y_2, \quad [Y_2, Y_3] = Y_1. \tag{5.3}$$

Now after performing the limit $\epsilon \to 0$ the first commutator vanishes and the resulting algebra is the one of the Euclidean algebra on the plane $\mathfrak{iso}(2)$. This is one specific example of an İnönü–Wigner contraction, and also the one I mentioned in the beginning, but not the only one possible. As stated in theorem 1 every Lie group can be contracted with respect to any of its continuous subgroups under suitable circumstances. Another famous example would be the contraction of the Poincaré group to the Galilei group as the speed of light tends to infinity. Thus, İnönü-Wigner contractions can serve as a precise tool to make certain physical limits manifest on the level of the underlying symmetries. I will demonstrate this with the following example of contracting relativistic conformal symmetries, which will be an important concept for Part III of this thesis.

## Contracting Relativistic Conformal Symmetries in 2d

Relativistic conformal symmetries play an important role as the asymptotic symmetries associated to AdS spacetimes. For the special case of $\text{AdS}_3$ these symmetries are even more special as the corresponding symmetry algebra of the 2-dimensional boundary is infinite-dimensional and consists of two copies of the Virasoro algebra $\mathfrak{vir}$ with commutation relations

$$[\mathcal{L}_n, \mathcal{L}_m] = (n-m)\mathcal{L}_{n+m} + \frac{c}{12} n(n^2-1)\delta_{n+m,0}, \tag{5.4a}$$

$$[\bar{\mathcal{L}}_n, \bar{\mathcal{L}}_m] = (n-m)\bar{\mathcal{L}}_{n+m} + \frac{\bar{c}}{12} n(n^2-1)\delta_{n+m,0}, \tag{5.4b}$$

$$[\mathcal{L}_n, \bar{\mathcal{L}}_m] = 0, \tag{5.4c}$$

where $c$ and $\bar{c}$ denote the respective central charges. Since this algebra describes the symmetries of a relativistic conformal field theory in 2d, there should also be corresponding limiting expressions which describe conformal symmetries at very high speeds close to the speed of light (ultrarelativistic limit) and at speeds which are much smaller than the speed of light (nonrelativistic limit). For the following discussion I will focus on 2d, as this will be the case which is most relevant for my thesis.

### Nonrelativistic Limit

In order to get a better understanding of how the nonrelativistic limit translates into an appropriate contraction of the relativistic conformal algebra it is instructive first to look at the non-centrally extended version of the Virasoro algebra, i.e. the Witt algebra, instead of looking at the centrally extended version (5.4).
The vector fields defined on a plane which generate (two copies of) the Witt algebra

$$[\mathcal{L}_n, \mathcal{L}_m] =(n-m)\mathcal{L}_{n+m}, \tag{5.5a}$$

$$[\bar{\mathcal{L}}_n, \bar{\mathcal{L}}_m] =(n-m)\bar{\mathcal{L}}_{n+m}, \tag{5.5b}$$

$$[\mathcal{L}_n, \bar{\mathcal{L}}_m] =0, \tag{5.5c}$$

in two dimensions are given by

$$\mathcal{L}_n = -z^{n+1}\partial_z, \quad \bar{\mathcal{L}}_n = -\bar{z}^{n+1}\partial_{\bar{z}}. \tag{5.6}$$

In terms of space and time coordinates, $z = t + x$, $\bar{z} = t - x$ the differentials are then given by $\partial_z = \frac{1}{2}(\partial_t + \partial_x)$ and $\partial_{\bar{z}} = \frac{1}{2}(\partial_t - \partial_x)$.
The nonrelativistic limit corresponds to taking the velocities to zero, i.e.

$$t \to t, \quad x \to \epsilon x, \tag{5.7}$$

for $\epsilon \to 0$. One can now take the following linear combinations [77, 97]

$$L_n := \mathcal{L}_n + \bar{\mathcal{L}}_n, \quad M_n := -\epsilon\left(\mathcal{L}_n - \bar{\mathcal{L}}_n\right), \tag{5.8}$$

whose expressions in terms of vector fields in the limit $\epsilon \to 0$ read

$$\begin{aligned} L_n =& - t^{n+1}\partial_t - (n+1)t^n x\partial_x + \mathcal{O}(\epsilon^2), \\ M_n =& t^{n+1}\partial_x + \mathcal{O}(\epsilon^2). \end{aligned} \tag{5.9}$$

These vector field generate the Galilean Conformal algebra (GCA) in 2 dimensions whose commutation relations are given by

$$[L_n, L_m] =(n-m)L_{n+m}, \tag{5.10a}$$

$$[L_n, M_m] =(n-m)M_{n+m}, \tag{5.10b}$$

$$[M_n, M_m] =0. \tag{5.10c}$$

It is important to understand at this point that the nonrelativistic limit precisely corresponds to the linear combinations (5.8). With this knowledge one can also straightforwardly take the nonrelativistic limit of two copies of the Virasoro algebra (5.4) using (5.8). This in turn yields the centrally extended GCA, which reads

$$[L_n, L_m] =(n-m)L_{n+m} + \frac{c_L}{12}n(n^2-1)\delta_{n+m,0}, \tag{5.11a}$$

$$[L_n, M_m] =(n-m)M_{n+m} + \frac{c_M}{12}n(n^2-1)\delta_{n+m,0}, \tag{5.11b}$$

$$[M_n, M_m] =0, \tag{5.11c}$$

with

$$c_L = c + \bar{c}, \quad c_M = -\epsilon\left(c - \bar{c}\right). \tag{5.12}$$

It is also worth noting at this point that he contraction (5.8) corresponding to the nonrelativistic limit can also be applied to higher-spin extensions of the Virasoro algebra, i.e. various $\mathcal{W}$-algebras [III].

### Ultrarelativistic Limit

Previously I explored the nonrelativistic limit of relativistic conformal symmetries which corresponds to vanishing velocities. One can of course also take a look at the other extreme limit, i.e. the one where the velocities approach the speed of light, the ultrarelativistic limit. Again starting from the Witt algebra (5.5) one can realize this limit by the following scaling

$$t \to \epsilon t, \quad x \to x. \tag{5.13}$$

In contrast to the nonrelativistic limit I will now not use a CFT defined on the plane, but rather use a CFT on the cylinder. One can then consider the following vector fields

$$\mathcal{L}_n = ie^{in\omega}\partial_\omega, \quad \bar{\mathcal{L}}_n = ie^{in\bar{\omega}}, \tag{5.14}$$

where $z = e^{i\omega}$ and $\omega = t + x$, $\bar{\omega} = t - x$. Now taking the linear combinations [65]

$$L_n := \mathcal{L}_n - \bar{\mathcal{L}}_{-n}, \quad M_n := \epsilon \left(\mathcal{L}_n + \bar{\mathcal{L}}_{-n}\right), \tag{5.15}$$

one obtains

$$\mathcal{L}_n - \bar{\mathcal{L}}_{-n} = ie^{inx}\left[i\sin(nt)\partial_t + \cos(nt)\partial_x\right], \tag{5.16a}$$

$$\mathcal{L}_n + \bar{\mathcal{L}}_{-n} = ie^{inx}\left[\cos(nt)\partial_t + i\sin(nt)\partial_x\right]. \tag{5.16b}$$

In the limit $\epsilon \to 0$ the vector fields $L_n$ and $M_n$ are well defined and read

$$L_n = ie^{inx}\left(\partial_x + int\partial_t\right), \quad M_n = ie^{inx}\partial_t. \tag{5.17}$$

One can check that these two new vector fields generate the same algebra that also appeared in the nonrelativistic limit i.e. (5.10). This is a special feature of two dimensions where one can simply exchange the meaning of time and space. Thus, the nonrelativistic and the ultrarelativistic limit[1] of the relativistic conformal algebra in two dimensions lead to two isomorphic algebras. This feature is often called the BMS/GCA correspondence [65], since the algebra (5.10) which is obtained after taking the ultrarelativistic limit is the $\mathfrak{bms}_3$ algebra.
As in the nonrelativistic case one can extend the contraction (5.15) to the contraction

[1]The ultrarelativistic limit of the relativistic conformal algebra is usually called conformal Carrol algebra, see e.g. [98–100].

of two Virasoro algebras, which again yields the algebra (5.11) but with different central charges

$$c_L = c - \bar{c}, \quad c_M = \epsilon\,(c + \bar{c})\,. \tag{5.18}$$

These two examples show that İnönü–Wigner contractions can be a very convenient and efficient tool to employ certain physical limits on an algebraic level. It should be noted that there are also instances of $\mathcal{W}$-algebras which allow for İnönü–Wigner contractions and thus give rise to a whole new class of nonlinear contracted algebras as I will show in Chapter 10.

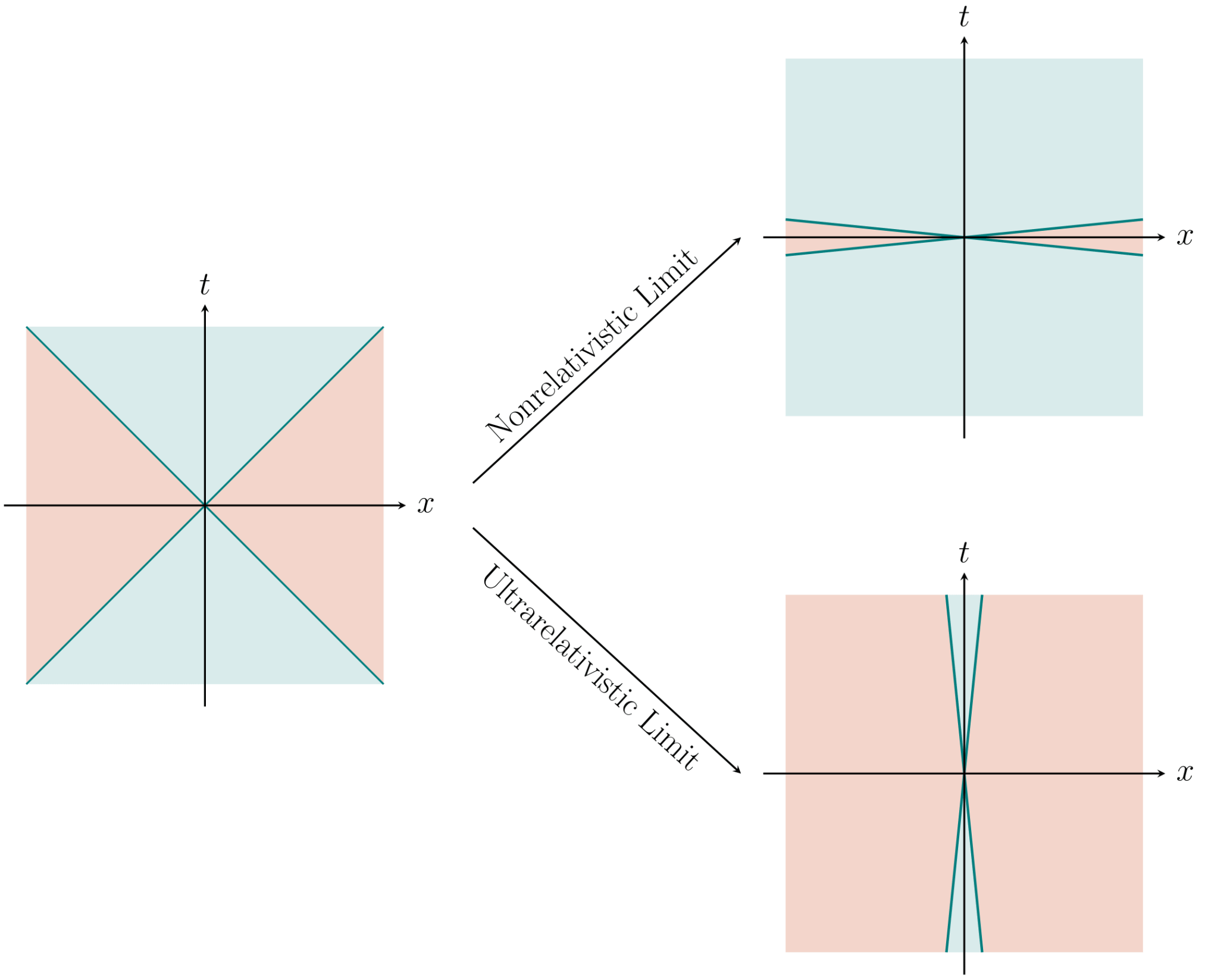

**Fig. 5.1.:** Degenerate light cones in the ultra- and nonrelativistic limits. In the nonrelativistic limit, where the speed of light approaches infinity (i.e. $v \approx 0$), the light cone flattens out and everything is causally connected, whereas in the ultrarelativistic limit (i.e. $v \approx c$) nothing is causally connected anymore.

By looking at Figure 5.1 one can also see that there is a conceptually important consequence after performing the İnönü–Wigner contraction of the *relativistic* conformal algebra. The resulting symmetry algebras cannot describe a Lorentz invariant field theory anymore since the light cone structure is degenerate in either of the two limits. Therefore one must not forget the physical consequences when applying such a convenient and efficient tool like an İnönü–Wigner contraction.

# 6 Entanglement, Entropy and Holography

> 一度あったことは忘れないもんさ... 思い出せないだけで。
> *(You don't forget things you've experienced ... It's just you can't recall them.)*
>
> **– 宮崎駿 (Hayao Miyazaki)**
> 千と千尋の神隠し (Spirited Away)

Originally coined "Verschränkung" by Schrödinger [101], entanglement is one of the most intriguing, fascinating and unique properties of quantum mechanics. Whenever a pair or a group of particles cannot be described in terms of independent quantum states but rather in terms of a single quantum state for the whole system, one usually refers to the particles as entangled. This can either happen because the states have been specifically prepared to be in an entangled state, or the interactions between the particles are of such a form that the system is entangled because of the interactions.
Quantum entanglement is a very active field of research and has triggered many new fields of research such as quantum information and communication, see e.g. [102]. For many purposes, theoretical and experimental alike it is of interest to quantify the amount of entanglement of a quantum system. One particularly useful way of quantifying entanglement from a theoretical perspective is given by the entropy of the entangled quantum system, which is the main topic of this chapter.

## Basics of Quantum Entanglement

In order to understand the basics of entanglement entropy I will first consider a quantum system with a large amount of degrees of freedom such as spin chains which is described in terms of a density matrix $\rho$ and a total Hilbert space $\mathcal{H}$. Now one can divide the total system into two subsystems, system $A$ and its complement which I denote by $B$. This amounts to writing the total Hilbert space $\mathcal{H}$ as a product of two Hilbert spaces, i.e. $\mathcal{H} = \mathcal{H}_A \otimes \mathcal{H}_B$ corresponding to the two subsystems $A$ and $B$. The subsystem $A$ can then be described by the reduced density matrix $\rho_A$ as

$$\rho_A = \mathrm{Tr}_B\, \rho = \sum_i \langle \psi_B^i | \rho | \psi_B^i \rangle, \tag{6.1}$$

where the trace is taken only over $\mathcal{H}_B = \{|\psi_B^1\rangle, |\psi_B^2\rangle, \ldots\}$.
One can now define the entanglement entropy of the subsystem $A$ as the von Neumann entropy associated to the reduced density matrix $\rho_A$

$$S_A = -\text{Tr}_A\,[\rho_A \ln \rho_A]\,. \tag{6.2}$$

The definition (6.2) has some very nice properties which make it a good entanglement measure.

- $S_A(\rho_A)$ is zero iff $\rho_A$ represents a pure state since the density matrix is idempotent for pure states $\rho_A^2 = \rho_A$.
- $S_A(\rho_A)$ has a maximum which is $\ln(\dim \mathcal{H})$, for a maximally mixed state.
- $S_A(\rho_A)$ is invariant under a unitary change of basis $S_A(\rho_A) = S_A(U\rho_A U^\dagger)$.

Thus $S_A(\rho_A)$ gives a basis invariant measure of how much a system fails to be able to be factored as a product of states.
For a pure product state one has $S_A = 0$ and for a maximally entangled state $S_A = \ln(\dim \mathcal{H}_A)$. In the case of bipartite systems at zero temperature which are divided into a part $A$ and its complement $B$ the entanglement entropy also satisfies $S_A = S_B$. Let me illustrate this using a simple example of a bipartite system consisting of two states each as $\mathcal{H}_A = \{|\uparrow\rangle_A, |\downarrow\rangle_A\}$ and $\mathcal{H}_B = \{|\uparrow\rangle_B, |\downarrow\rangle_B\}$.

**Pure Product State:** Consider the following pure state

$$|\Psi\rangle = \frac{1}{2}\left(|\uparrow\rangle_A + |\downarrow\rangle_A\right) \otimes \left(|\uparrow\rangle_B + |\downarrow\rangle_B\right), \tag{6.3}$$

whose reduced density matrix reads

$$\rho_A = \text{Tr}_B |\Psi\rangle\langle\Psi| = \frac{1}{2}\left(|\uparrow\rangle_A + |\downarrow\rangle_A\right)\left(|\uparrow\rangle_A + |\downarrow\rangle_A\right). \tag{6.4}$$

This density matrix has eigenvalues 1 and 0 and is thus, as expected, idempotent. Hence one finds for the entanglement entropy

$$S_A = -\text{Tr}_A\,[\rho_A \ln \rho_A] = -\text{Tr}_A\left[\rho_A^2 \ln \rho_A^2\right] = -\text{Tr}_A\,[2\rho_A \ln \rho_A] = 0. \tag{6.5}$$

**Maximally Entangled State:** Consider now the following maximally entangled state

$$|\Psi\rangle = \frac{1}{\sqrt{2}}\left(|\uparrow\rangle_A \otimes |\downarrow\rangle_B + |\downarrow\rangle_A \otimes |\uparrow\rangle_B\right), \tag{6.6}$$

with reduced density matrix

$$\rho_A = \text{Tr}_B |\Psi\rangle\langle\Psi| = \frac{1}{2}\left(|\uparrow\rangle_A\langle\uparrow|_A + |\downarrow\rangle_A\langle\downarrow|_A\right), \tag{6.7}$$

which has eigenvalues $\frac{1}{2}$ and $\frac{1}{2}$. This yields the following value for the entanglement entropy

$$S_A = -\text{Tr}\begin{pmatrix} \frac{1}{2} & 0 \\ 0 & \frac{1}{2} \end{pmatrix}\begin{pmatrix} \log(\frac{1}{2}) & 0 \\ 0 & \log(\frac{1}{2}), \end{pmatrix} = \ln 2 \tag{6.8}$$

which equals, also as expected, the natural logarithm of the dimension of the Hilbert space $\mathcal{H}_A$. One can also straightforwardly check that indeed $S_A = S_B$ for this system.

**(Strong) Subadditivity:** Another important property that entanglement entropy satisfies is called (strong) subadditivity. A bipartite system with parts $A$ and $B$ is called subadditive if it satisfies the following inequality

$$S_A + S_B \geq S_{A\cup B} + S_{A\cap B}. \tag{6.9}$$

A tripartite system with parts $A$, $B$ and $C$ is called strong subadditive if it satisfies

$$S_{A\cup B\cup C} + S_B \leq S_{A\cup B} + S_{B\cup C}, \tag{6.10}$$

or equivalently [103]

$$S_A + S_C \leq S_{A\cup B} + S_{B\cup C}. \tag{6.11}$$

These strong subadditivity inequalities for quantum systems have been proven in [104]. I will also review later how these inequalities can be derived in a simple way using a holographic approach.

## Entanglement Entropy in Quantum Field Theories

Consider a relativistic quantum field theory on a lattice with lattice spacing $a$ in d+1 dimensions with d$\geq 2$. At a given fixed time $t = t_0$ one can again divide the total system into a subsystem $A$ and its complement $B$, where $\partial A$ is the boundary of the subsystem. In [105, 106] it was then shown that the entanglement entropy of such a system diverges in the continuum limit $a \to 0$ as

$$S_A \sim \frac{\text{Area}(\partial A)}{a^{d+1}} + \ldots \tag{6.12}$$

This behavior is called *area law* and is very similar to the behavior of the Bekenstein-Hawking entropy of a black hole, which also scales with the area of the black hole horizon. This similarity will play an important role for motivating a holographic description of entanglement entropy but first I want to elaborate on one key technique which is used for calculating entanglement entropy in quantum field theories, the *replica trick*.

Since the replica trick is by now a well established technique for calculating entan-glement entropy there exists also excellent literature explaining the trick and its application to entanglement entropy in detail e.g. [107] (for arbitrary dimensions and finite temperature) or [108] (for 2d conformal field theories) whose structure I will also follow in this section. For the sake of simplicity and also relevance for this thesis I will stick to two dimensions and zero temperature in what follows.

Assume that the 2d system is defined on a (Euclidean) plane with coordinates $(x, t_E)$ and can be described in terms of an Euclidean action $S_E[\phi(x, t_E)]$. The entangling region $A$ will be an interval in $x$ direction at $t_E = 0$.

Calculating entanglement entropy for such a 2d quantum field theory using (6.2) is often not very efficient since calculating the logarithm of the density matrix can be potentially very involved. The replica trick allows one to circumvent this problem by effectively reducing the problem of calculating entanglement entropy to determining the partition function of an $n$-sheeted Riemann surface. The starting point of this trick are the so-called Tsallis[1] entropies which are given by

$$S_n^{\text{Tsallis}} = \frac{\text{Tr}_A\, \rho_A^n - 1}{1-n}. \tag{6.13}$$

The entanglement entropy can be obtained from the Tsallis entropies in the limit[2] $n \to 1$ which can also be written as[3]

$$S_A = \lim_{n\to 1} S_n^{\text{Tsallis}} = -\partial_n \text{Tr}_A\, \rho_A^n\big|_{n=1} = -\partial_n \ln\left(\text{Tr}_A\, \rho_A^n\right)\big|_{n=1}, \tag{6.14}$$

which at this point shifted the problem of calculating entanglement entropy from calculating a logarithm to calculating powers of the density matrix, which is computationally easier to perform. Now in order to make contact with my statement about $n$-sheeted Riemann surfaces in the beginning one has to calculate $\text{Tr}_A\, \rho_A^n$ and perform some path integral gymnastics.

As in usual quantum mechanics the density matrix of the whole system is given by the ground state wave $\Psi$ function and its hermitian conjugate $\Psi^\dagger$ as $\rho = \Psi\Psi^\dagger$. In a path integral representation $\Psi[\phi(x,0)]$ is given by path integrating from $t_E = \infty$ up to $t_E = 0$ and the hermitian conjugate $\Psi^\dagger[\phi'(x,0)]$ by integrating from $t_E = 0$ to $t_E = \infty$ i.e.

$$\Psi[\phi(x,0)] = \int_{t_E=-\infty}^{\phi(x,0)} \mathcal{D}\phi\, e^{-S_E[\phi]}, \quad \Psi^\dagger[\phi'(x,0)] = \int_{\phi'(x,0)}^{t_E=\infty} \mathcal{D}\phi\, e^{-S_E[\phi]}. \tag{6.15}$$

See Fig. 6.1 for a pictorial interpretation of this path integration. The reduced density matrix $\rho_A$ can now be obtained by combining $\Psi[\phi(x,0)]$ and $\Psi^\dagger[\phi'(x,0)]$ into

[1] The Tsallis entropies are related to the maybe better known Rényi entropies $S_n^{\text{Rényi}}$ as $S_n^{\text{Tsallis}} = \frac{1}{1-n}\left[e^{(1-n)S_n^{\text{Rényi}}} - 1\right]$ [109].

[2] $\text{Tr}_A\, \rho_A^n$ is analytic for $\text{Re}(n)>1$. Thus the limit $n \to 1^+$ is well defined.

[3] Assuming $\text{Tr}_A\, \rho_A = 1$.

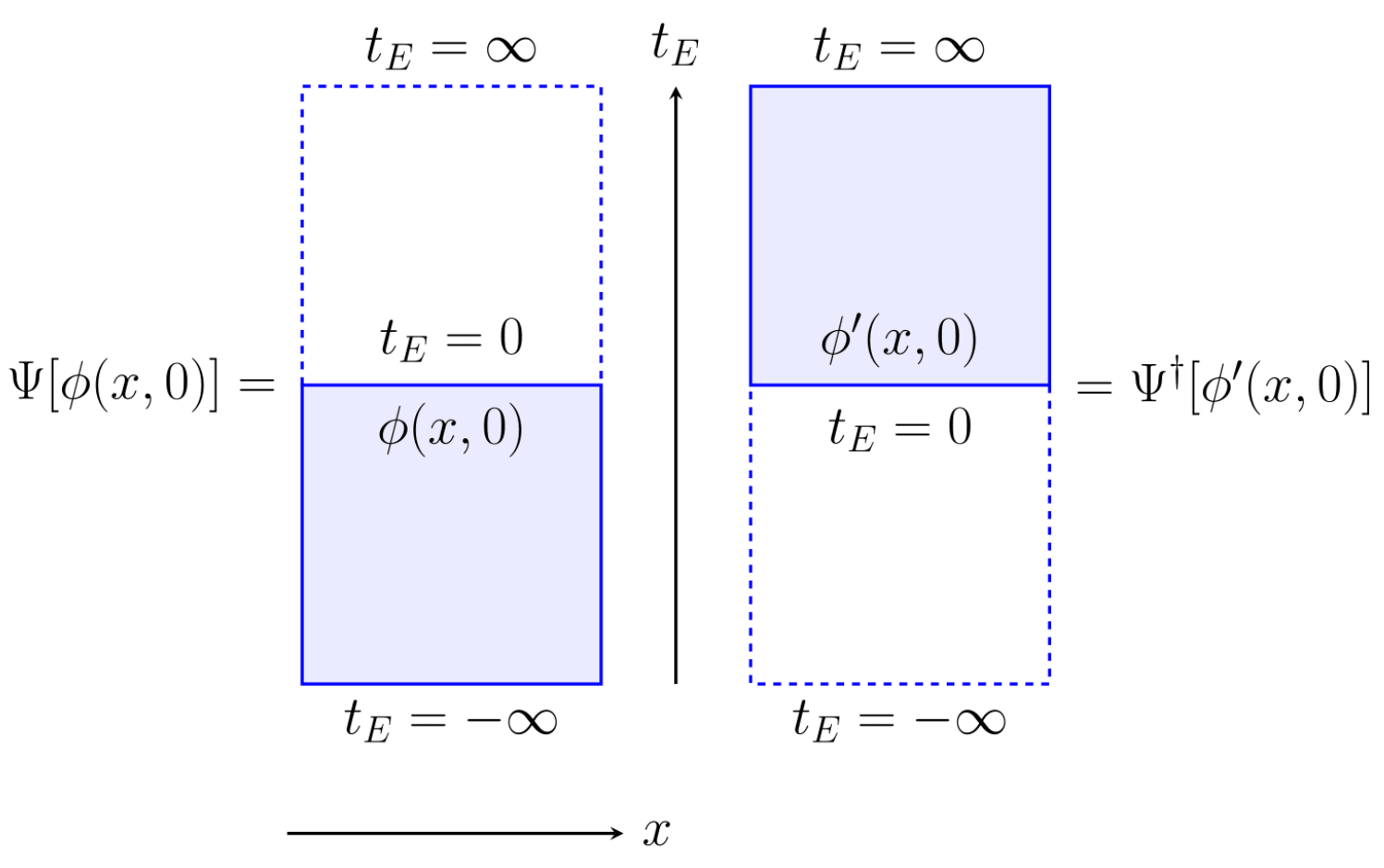

**Fig. 6.1.:** Pictorial interpretation of the ground state written as a path integral.

$[\rho_A]_{\phi^+\phi^-}$, such that $\forall x \in B \to \phi(x,0) = \phi'(x,0)$ and $\forall x \in A \begin{cases} \phi(x,0) \equiv \phi^- \\ \phi'(x,0) \equiv \phi^+ \end{cases}$. In therms of a path integral this is equivalent to gluing together the two patches in Fig. 6.1 along the region $B$ as shown in Fig. 6.2 leaving an open cut along region $A$ with boundaries $\phi^\pm$.

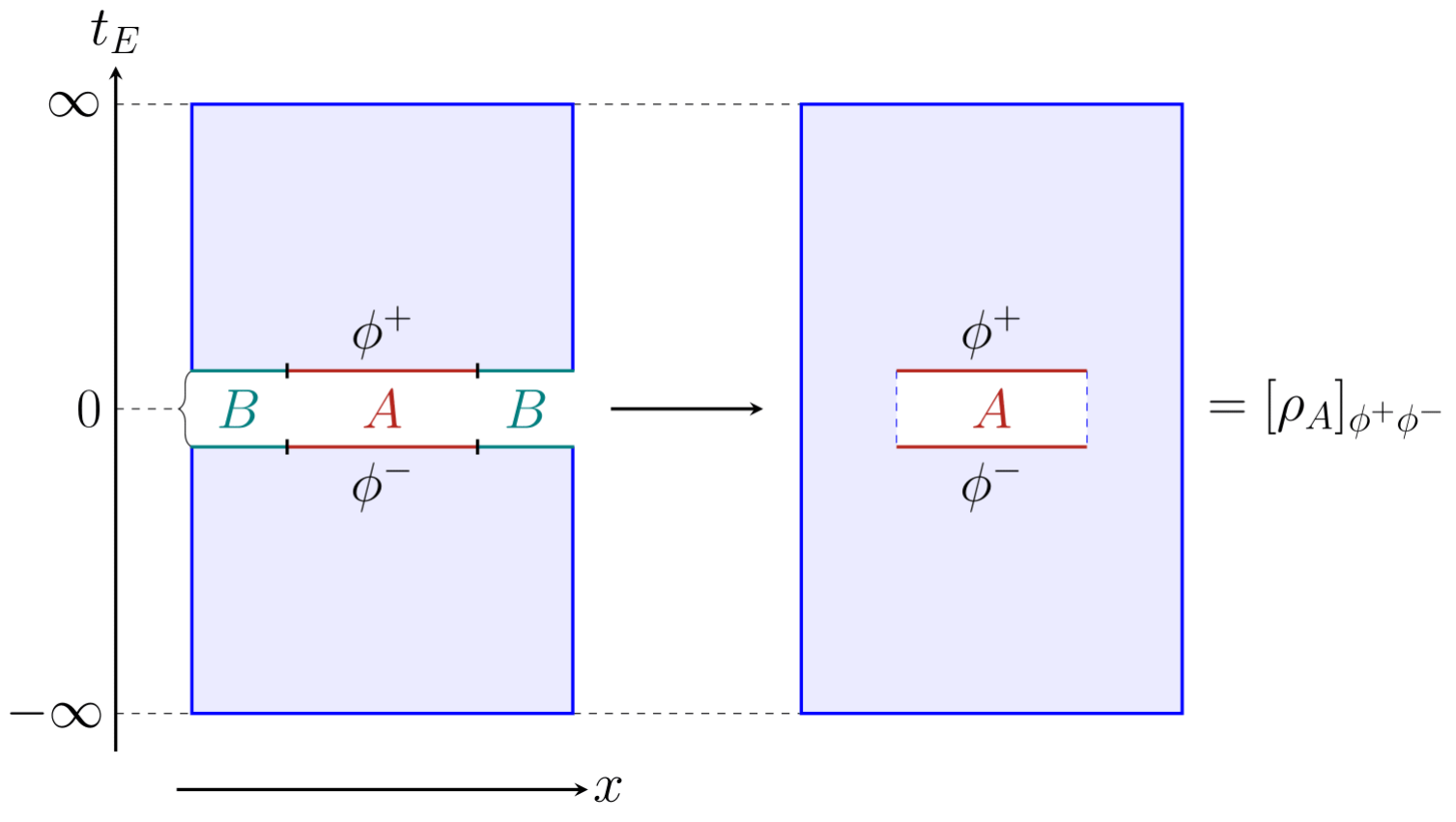

**Fig. 6.2.:** Path integrating out the degrees of freedom of the system $B$ from $A$ amounts to gluing together the two sheets as depicted, leaving a plane with a cut along $A$.

Then in order to determine $\text{Tr}_A\, \rho_A^n$ one makes $n$ replicas $[\rho]_{\phi_1^+\phi_1^-}[\rho]_{\phi_2^+\phi_2^-}\dots[\rho]_{\phi_n^+\phi_n^-}$ of such a reduced density matrix and takes the trace successively alongside those copies. Again, this can be done in a very elegant way in the path integral formalism where this operation amounts to gluing together the $\phi_j^\pm$'s as $\phi_j^- = \phi_{j+1}^+$, where $\phi_n^- = \phi_1^+$ and then integrating over $\phi_j^+$, see Fig. 6.3 for another pictorial interpretation. This procedure effectively then corresponds to a partition function $Z_n$ calculated on an $n$-sheeted Riemann surface $\mathcal{R}_n$.

$$\text{Tr}_A\, \rho_A^n = (Z_1)^{-n} \int_{(x,t_E)\in\mathcal{R}_n} \mathcal{D}\phi e^{-S_E[\phi]} \equiv \frac{Z_n}{(Z_1)^n}. \tag{6.16}$$

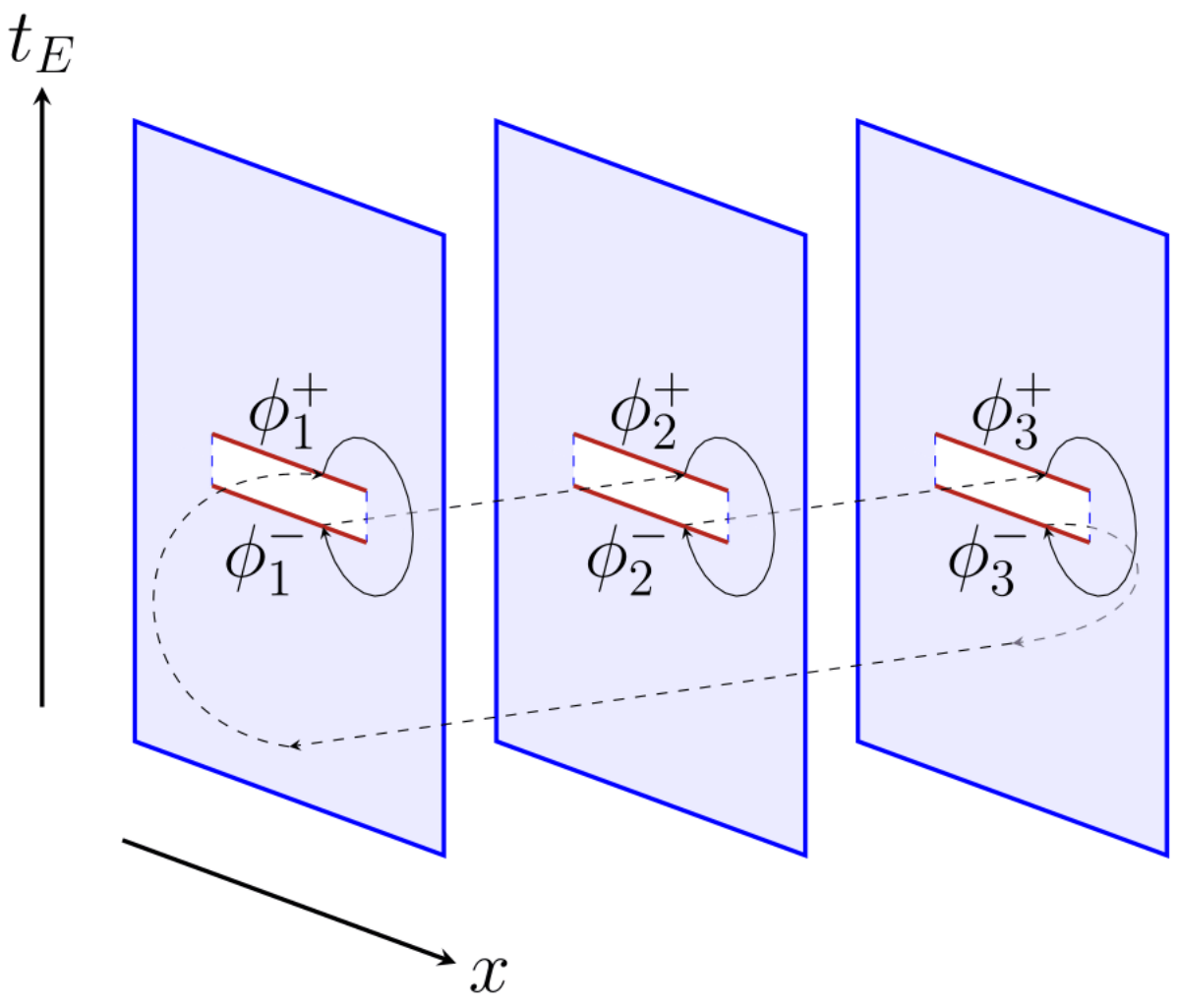

**Fig. 6.3.:** Determining $\text{Tr}_A\, \rho_A^n$ using a path integral, where I set $n = 3$ for illustrative purposes.

With this knowledge one can rewrite (6.14) as

$$S_A = -(\partial_n - 1) \ln Z_n \big|_{n=1}. \tag{6.17}$$

Hence all one needs to know in order to determine entanglement entropy is the partition function $Z_n$ on the given $n$-sheeted Riemann surface.

### Example: 2d Conformal Field Theory

In a 2d conformal field theory with central charge $c$ calculating entanglement entropy is particularly easy to perform because of the infinite number of symmetries in a conformally invariant system. I will review this for three different examples of conformal field theories.

**Infinitely Long System at Zero Temperature:** Here I will consider a conformal field theory which corresponds to a quantum system which extends from $-\infty < x < \infty$ at zero temperature. First I will change to complex coordinates $w = x + it_E$ and map the $n$-sheeted Riemann surface $\mathcal{R}_n$ to the complex plane $\mathbb{C}$ via a conformal transformation as

$$z = \left(\frac{w-u}{w-v}\right)^{\frac{1}{n}}, \tag{6.18}$$

where $|u - v| = l$ are the boundary points of the entangling interval $A$.
Using the transformation properties of the energy-momentum tensor one can relate the expressions of the energy momentum tensor on $\mathcal{R}_n$ with the ones on the plane as

$$T(w) = \left(\frac{\mathrm{d}z}{\mathrm{d}w}\right)^2 T(z) + \frac{c}{12}\{z, w\}, \tag{6.19}$$

where $\{\cdot,\cdot\}$ denotes the Schwarzian derivative $\{z,w\} = \frac{z'''}{z'} - \frac{3}{2}\left(\frac{z''}{z'}\right)^2$. Since $\langle T(z)\rangle_{\mathbb{C}} = 0$ by translational and rotational invariance one finds for the expectation value of the energy-momentum tensor on $\mathcal{R}_n$

$$\langle T(w)\rangle_{\mathcal{R}_n} = \frac{c}{12}\{z,w\} = \frac{c}{24}\left(1-\frac{1}{n^2}\right)\frac{(v-u)^2}{(w-u)^2(w-v)^2}. \tag{6.20}$$

In [107] it has then been shown that the conformal ward identities imply

$$\langle T(w)\rangle_{\mathcal{R}_n} = \frac{\langle T(w)\Phi_n(u)\Phi_{-n}(v)\rangle_{\mathbb{C}}}{\langle \Phi_n(u)\Phi_{-n}(v)\rangle_{\mathbb{C}}}, \tag{6.21}$$

where $\Phi_n(u)$ and $\Phi_{-n}(v)$ are two primary operators with scaling dimensions[4] $\Delta_n = \bar{\Delta}_n = \frac{c}{24}\left(1-\frac{1}{n^2}\right)$. The key observation made in [107] then was that $\mathrm{Tr}_A\,\rho_A^n$ behaves like the $n^{\text{th}}$ power of the two-point function of a primary operator $\Phi_n$ with scaling dimensions $\Delta_n = \bar{\Delta}_n$ as specified before. Thus one has $\mathrm{Tr}_A\,\rho_A^n \propto \langle \Phi_n(u)\Phi_{-n}(v)\rangle_{\mathbb{C}}^n$. Normalizing $\langle \Phi_n(u)\Phi_{-n}(v)\rangle_{\mathbb{C}}^n = |v-u|^{-2n(\Delta_n+\bar{\Delta}_n)}$, and introducing an infinitesimal parameter $a$ (e.g. the lattice spacing) in order to render the resulting expression dimensionless one obtains

$$\mathrm{Tr}_A\,\rho_A^n \propto \left(\frac{|v-u|}{a}\right)^{-\frac{c}{6}\left(n-\frac{1}{n}\right)}. \tag{6.22}$$

Using (6.14) and $|u-v| = l$ one immediately obtains the famous expression of the entanglement entropy of an infinitely long 2d conformally invariant system at zero temperature [107, 110]

$$S_A = \frac{c}{3}\ln\frac{l}{a}. \tag{6.23}$$

**Other Geometries:** The fact that $\mathrm{Tr}_A\,\rho_A^n$ transforms under a general coordinate transformation as a product of $n$ two-point function of primary operators with specific scaling dimensions also means that it is particularly easy to compute that quantity in other geometries, simply by using conformal transformations $z \to \tilde{z} = w(z)$ and [95]

$$\langle \Phi(z_1,\bar{z}_1)\Phi(z_2,\bar{z}_2)\ldots\rangle = \prod_j |w'(z_j)|^{2\Delta_n}\,\langle \Phi(w_1,\bar{w}_1)\Phi(w_2,\bar{w}_2)\ldots\rangle\,. \tag{6.24}$$

Assuming that the $x$-direction is compactified to a circle of circumference $L$ one can use the conformal map $w = \tan\left(\frac{\pi w'}{L}\right)$ to map from $w \in \mathcal{R}_n$ to an $n$-sheeted cylinder with complex coordinates $w'$. This yields the following entanglement entropy [107, 110]

$$S_A = \frac{c}{3}\ln\left[\frac{L}{\pi a}\sin\left(\frac{\pi l}{L}\right)\right]. \tag{6.25}$$

[4] These kind of primary operators are also often called *twist operators* in the literature.

In the same way one can immediately obtain the entanglement entropy for a system on the circle in a thermal mixed state at finite inverse temperature $\beta$. This corresponds to the conformal map $w = e^{\frac{2\pi}{\beta}w'}$ which maps each sheet of $\mathcal{R}_n$ onto an infinitely long cylinder with circumference $\beta$. This yields then the following entropy [107]

$$S_A = \frac{c}{3}\ln\left[\frac{\beta}{\pi a}\sinh\left(\frac{\pi l}{\beta}\right)\right]. \tag{6.26}$$

## Holographic Entanglement Entropy

Even though the replica trick is a very convenient method to determine entanglement entropy it nevertheless gets increasingly difficult to actually perform the necessary calculations as the complexity and the dimensions of the system increase. Thus one might ask if there is an even more efficient method of computing entanglement entropy and, quite remarkably, there is one coming from holography.

In [111] Ryu and Takayanagi originally proposed that one can compute the entanglement entropy of an entangling region $A$ in a d-dimensional conformal field theory holographically using the following area law

$$S_A = \frac{\text{Area}(\gamma_A)}{4G_N}, \tag{6.27}$$

where $\gamma_A$ is a $(d-1)$-dimensional static, minimal surface in $\text{AdS}_{\text{d}+1}$ whose boundary is given by $\partial A$.

This prescription by now is known as the Ryu-Takayanagi proposal and although it has not been proven in full generality as of yet[5], there is strong evidence that the proposal is indeed correct. For a selection of references elaborating on the connection between gravity an entanglement see [105, 106, 108, 115–127].

Again the case of d$=$ 2 is special since in that case one can derive the area law explicitly using the holographic correspondence. Since $\text{Tr}_A\,\rho_A^n$ is proportional to $n$ products of two point functions of primary operators as specified in the previous section the starting point for a holographic prescription is a two point function $\langle\Phi_n(u)\Phi_{-n}(v)\rangle$. Such a two point function can be determined holographically for $\Delta_n \gg 1$ as

$$\langle\Phi_n(u)\Phi_{-n}(v)\rangle \sim e^{-\frac{2n\Delta_n L_{\gamma_A}}{\ell}}, \tag{6.28}$$

where $L_{\gamma_A}$ is the length of a geodesic that is attached at the points $u$ and $v$ at the AdS boundary[6] and $\ell$ is the AdS radius. See also Fig. 6.4 .

[5] There are recent developments towards a proof, see e.g. [112–114].

[6] The geodesic is actually not attached directly to the AdS boundary, since the geodesic length diverges at the boundary, but rather is attached to some cutoff surface which is placed very close to the boundary. This surface has the same purpose as the lattice spacing $a$ introduced previously. It introduces a cutoff and renders the resulting entanglement entropy finite.

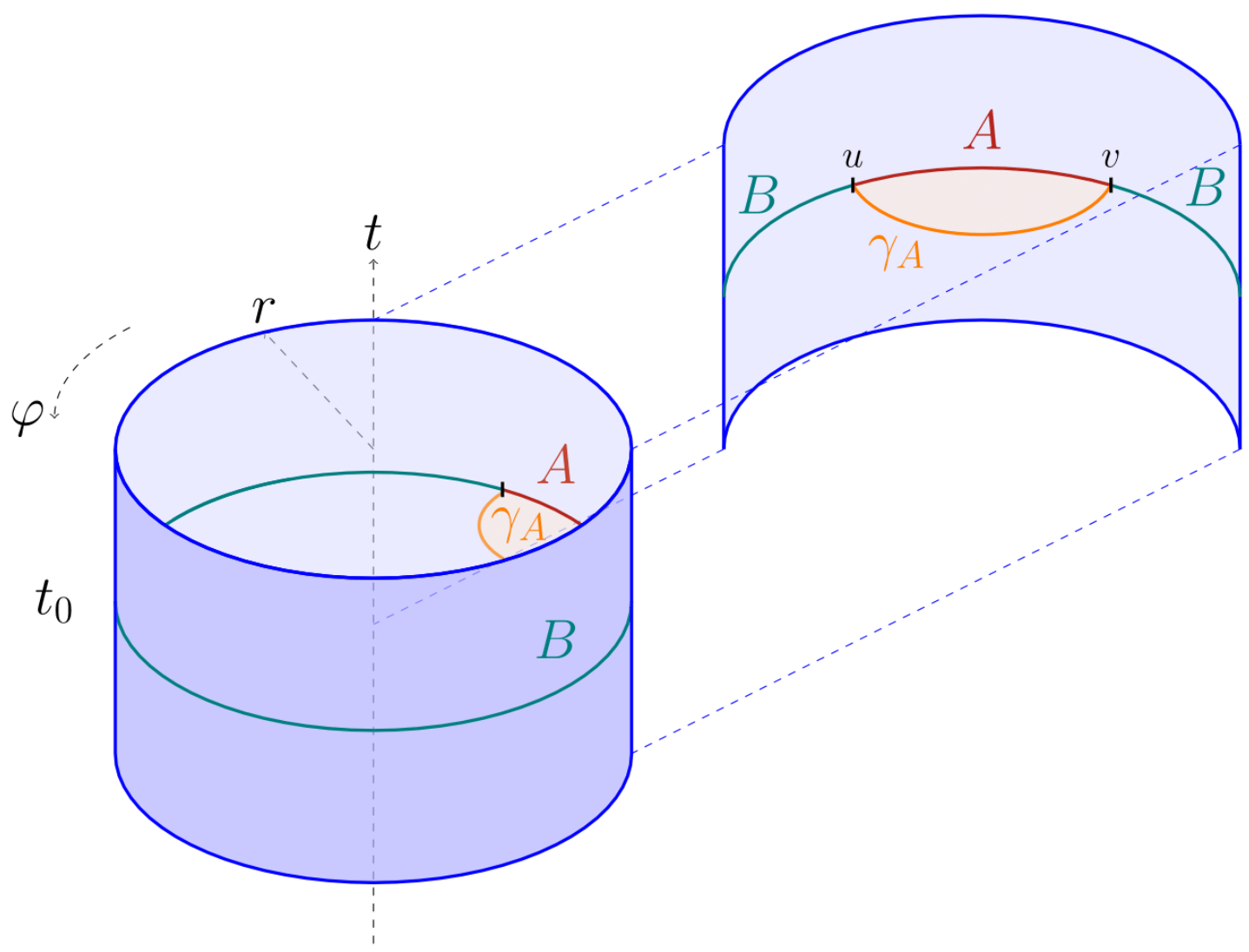

**Fig. 6.4.:** Spacelike geodesic $\gamma_A$ attached to the boundary of $\mathrm{AdS}_3$.

Now using (6.14) one finds

$$S_A = 2\left(\partial_n(n\Delta_n)\big|_{n=1}\right)\frac{L_{\gamma_A}}{\ell} = \frac{L_{\gamma_A}}{4G_N}, \tag{6.29}$$

which is exactly (6.27) for $\mathrm{d} = 2$.

The beauty of the Ryu-Takayanagi prescription is that it gives a connection between pure geometric bulk information and the entanglement entropy even without knowing explicitly what the corresponding dual field theory actually is. In addition, one has the advantage that static minimal surfaces in AdS are also more feasible to compute than performing replica tricks on the dual field theory side. This allows for even more sophisticated calculations of entanglement entropy using holography.

### Example: Geodesics in $\mathrm{AdS}_3$

In order to get a little bit of intuition for this holographic description of entanglement entropy I will present a couple of explicit examples of a holographic derivation of the entanglement entropy of (6.23), (6.25) and (6.26).

I will start first with $\mathrm{AdS}_3$ in Poincaré coordinates

$$\mathrm{d}s^2 = \frac{\ell^2}{z^2}\left(\mathrm{d}z^2 - \mathrm{d}t^2 + \mathrm{d}x^2\right), \tag{6.30}$$

where $0 \leq z < \infty$, $-\infty < x, t < \infty$ and the boundary is located at $z = 0$. I will place the cutoff surface at $z = a$ and consider an entangling interval $A$ at constant $t = t_0$ and $-\frac{l}{2} \leq x \leq \frac{l}{2}$. One can then parametrize the geodesic attached to this surface as

$$(x, z) = \frac{l}{2}(\cos s, \sin s), \qquad \epsilon \leq s \leq \pi - \epsilon, \tag{6.31}$$

where the infinitesimal parameter $\epsilon$ is related to the cutoff as $a = \frac{l\epsilon}{2}$.

The length of this geodesic is given by

$$L_{\gamma_A} = 2\ell \int_{\epsilon}^{\frac{\pi}{2}} \frac{\mathrm{d}s}{\sin s} = -2\ell \ln \frac{\epsilon}{2} = 2\ell \ln \frac{l}{a}, \tag{6.32}$$

which after applying (6.27) yields exactly (6.23).
Global $\mathrm{AdS}_3$ is given by the line element

$$\mathrm{d}s^2 = \ell^2 \left( -\cosh^2 \rho \, \mathrm{d}t^2 + \mathrm{d}\rho^2 + \sinh^2 \rho \, \mathrm{d}\theta^2 \right), \tag{6.33}$$

where $0 < \rho < \infty$, $-\infty < t < \infty$, $\theta \sim \theta + 2\pi$ and the boundary is located at $\rho \to \infty$. As in the Poincaré case one has to choose a cutoff surface to attach the geodesic to. This surface will be located at $\rho = \rho_0$ and the entangling interval with length $l$ is given by $t = t_0$ and $0 \leq \theta \leq \frac{2\pi l}{L}$ with $L$ being the circumference of the boundary cycle. After some gymnastics involving the geodesic equation in these coordinates and for these boundary conditions one can write the length of the geodesic compactly as

$$\cosh\left(\frac{L_{\gamma_A}}{\ell}\right) = 1 + 2\sinh^2 \rho_0 \sin^2\left(\frac{\pi l}{L}\right). \tag{6.34}$$

Assuming that the cutoff surface is very close to the $\mathrm{AdS}_3$ boundary $\rho_0 \to \infty$ one obtains the following expression using (6.27)

$$S_A \sim \frac{\ell}{4G_N} \ln\left[e^{2\rho_0} \sin^2\left(\frac{\pi l}{L}\right)\right] = \frac{c}{3} \ln\left[e^{\rho_0} \sin\left(\frac{\pi l}{L}\right)\right], \tag{6.35}$$

which is exactly (6.25) upon identifying the respective cutoffs as $e^{-\rho_0} = \frac{\pi a}{L}$.
As another example I use the Euclidean BTZ black hole in the following coordinates [12]

$$\mathrm{d}s^2 = (r^2 - r_+^2)\,\mathrm{d}\tau^2 + \frac{\ell^2}{r^2 - r_+^2}\,\mathrm{d}r^2 + r^2\,\mathrm{d}\varphi, \tag{6.36}$$

where the Euclidean time is compactified as $\tau \sim \tau + \frac{2\pi\ell}{r_+}$ and $\varphi \sim \varphi + 2\pi$. The Euclidean BTZ black hole horizon is located at $r_+$ and the boundary at $r \to \infty$. We will choose the same subsystem $A$ as in the global AdS case.
In order to perform the calculation of the geodesic length it is convenient to remember that the Euclidean BTZ at inverse temperature $\beta^{-1}$ is equivalent to thermal $\mathrm{AdS}_3$ with temperature $\beta$. This can be explicitly seen using the following coordinate transformation

$$r = r_+ \cosh \rho, \qquad \tau = \frac{\ell}{r_+}\theta, \qquad \varphi = \frac{\ell}{r_+}t, \tag{6.37}$$

which yields exactly (6.33) with $t \to it$. Thus, one can determine the geodesic length along the same lines as before and simply replace $\sinh\rho \to \cosh\rho$ and $\sin t \to \sinh t$. This yields the following compact expressions for the geodesic length

$$\cosh\left(\frac{L_{\gamma_A}}{\ell}\right) = 1 + 2\cosh^2 \rho_0 \sinh^2\left(\frac{\pi l}{\beta}\right). \tag{6.38}$$

Again using (6.27) and $e^{-\rho_0} = \frac{\pi a}{\beta}$ one obtains (6.26) for large $\rho_0$.

## (Strong) Subadditivity Revisited

I will close this chapter by reviewing how the holographic description of entanglement entropy can provide a very elegant way to intuitively understand and prove (strong) subadditivity [128, 129]. I start first with subadditivity (6.9) of a bipartite system[7], which is depicted in Fig. 6.5.

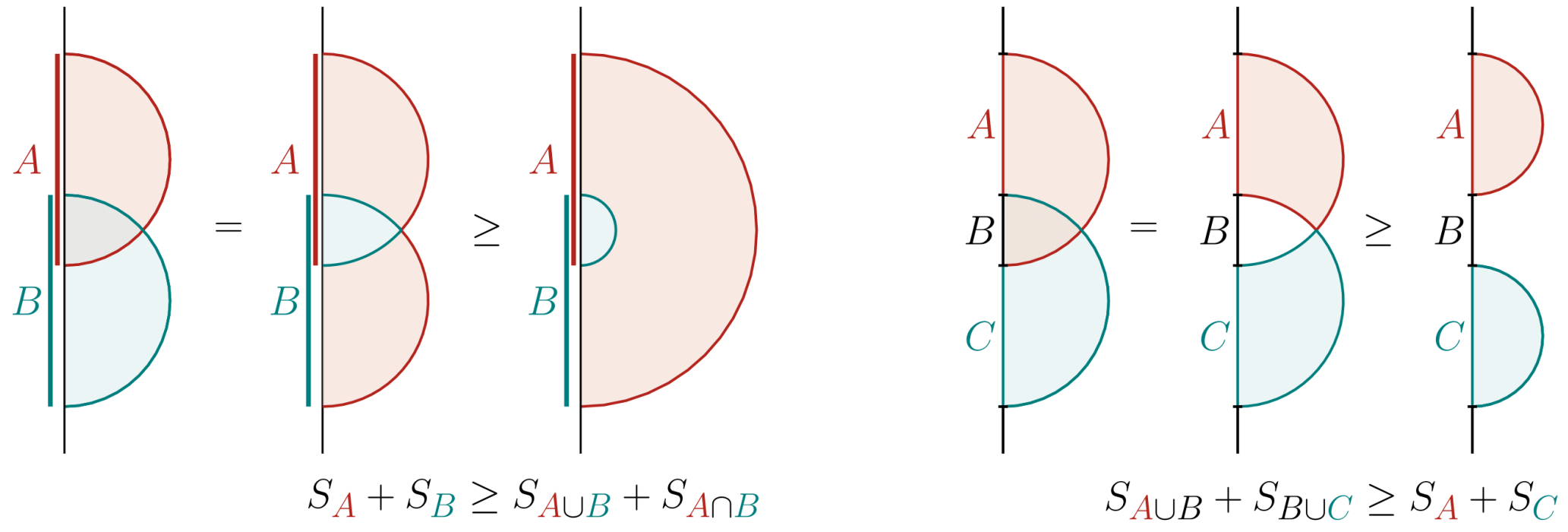

(a) (b)

**Fig. 6.5.:** (a) The leftmost part of the figure is the holographic entanglement entropy of $S_A + S_B$. The middle part merely represents a different colorization of some parts of the curves, whereas the rightmost part depicts the holograpahic entanglement entropy of $S_{A\cup B} + S_{A\cap B}$. (b) Holographic entanglement entropy for a tripartite system with subsystems $A$, $B$ and $C$. The leftmost part of the figure depicts the geodesics whose length corresponds to $S_{A\cup B} + S_{B\cup C}$. The middle part shows the same setup, just with a different choice of coloring. The rightmost part then shows the geodesics whose length corresponds to $S_A + S_C$.

The sum of the entanglement entropy of region $A$ and $B$ is given by the two colored half circles respectively which are assumed to be geodesics and have thus minimal length. One can now perform a simple change of color for some parts of the two geodesics, as depicted, to obtain two curves that now enclose the union and intersection of $A$ and $B$, respectively. This is exactly the right hand side of the subadditivity inequality (6.9). The curves one started out with are geodesics, i.e. curves which minimize the length between two given points. Therefore the resulting curves after this recoloring must be longer or equal in length of the curves which one would attach to the boundary in order to determine the entanglement entropy of the union and intersection of $A$ and $B$.

The holographic proof of strong subadditivity (6.11) for a tripartite system works in the exact same way and is also depicted in Fig. 6.5

[7]For illustrative purposes I will consider a 2d system so that the corresponding dual surface in the bulk is given by a geodesic.

# Part II

## Non-AdS Higher-Spin Holography

In this part of my thesis I explore certain aspects of non-AdS holography. I will begin this part with a short introduction on the general formalism used in non-AdS holography, which extends the formalism already known from AdS holography. I will then focus on a specific case of non-AdS holography, namely spin-4 Lobachevsky holography, and determine the asymptotic symmetries for various theories. In concluding, I will check for unitary representations of the corresponding asymptotic quantum symmetry algebras and provide an explicit example of a topological theory of gravity in three dimensions whose dual field theory allows for an arbitrary (albeit not infinitely) large Virasoro central charge without violating unitarity.

# 7 General Formalism

*Viele Steine, (Many stones,)*
*müde Beine, (tired legs,)*
*Aussicht keine, (scenic view; none.)*
*Heinrich Heine.*

**– Heinrich Heine**
Gipfelbuch des Brockens

Non-AdS holography in $2+1$ dimensions is in many regards very similar to AdS holography. There are, however, certain subtleties which one has to be aware of when leaving the comfort zone of the well studied AdS examples. The main purpose of this chapter is to describe a general algorithm for non-AdS higher-spin holography, point out the subtle differences to AdS higher-spin holography and to motivate why it is interesting to study holography in $2+1$ dimensions for spacetimes which are not AdS.
Many applications of the holographic principle require a generalization of the famous AdS/CFT correspondence to a more general gauge/gravity correspondence. Some examples are: null warped AdS, and their generalization Schrödinger spacetimes that feature an arbitrary scaling exponent and are used as holographic duals of non-relativistic CFTs that describe cold atoms [38–40], Lifshitz spacetimes, which are the gravity duals of Lifshitz-like fixed points [42], and the AdS/log CFT correspondence [130, 131].
This chapter is organized as follows: I will first present a general algorithm which can be used to determine the (quantum) asymptotic symmetries of non-AdS spacetimes. I will then use an explicit example, namely spin-3 Lobachevsky holography, in order to demonstrate how this algorithm works in detail.

## 7.1 A General Algorithm for Non-AdS Holography

The general algorithm for non-AdS holography which I will review in this section originally goes back to the work of Brown and Henneaux [11] and has been first presented in [I, 37]. This algorithm is a generalization of the procedure outlined in Chapter 3 and can roughly be summarized by the following steps, which are also summarized in form of a flowchart shown in Figure 7.1.

**Identify Bulk Theory and Variational Principle:** The first step in this algorithm consists of identifying the bulk theory[1] one wants to describe and then proposing a suitable generalized variational principle which is consistent with the theory under consideration. For non-AdS higher-spin theories such a generalized variational principle has been first described in [36], which I will review briefly.

Varying the bulk Chern-Simon action (2.3) one obtains on the one hand the equations of motion and on the other hand boundary conditions on the connection[2] $A$

$$\int_{\partial\mathcal{M}} \langle A \wedge \delta A \rangle = 0, \tag{7.1}$$

for some manifold $\mathcal{M} = \Sigma \times \mathbb{R}$, which I assume to have the topology of a cylinder and which can be described in terms of a radial coordinate $\rho$ and boundary coordinates $\varphi, t$ as shown in Figure 3.1. Choosing light-cone coordinates $x^{\pm} = \frac{t}{\ell} \pm \varphi$ this expression can be written as

$$\int_{\partial\mathcal{M}} \langle A_+ \wedge \delta A_- - A_- \wedge \delta A_+ \rangle = 0. \tag{7.2}$$

These boundary conditions can be satisfied for example by setting one part of the gauge fields, e.g. $A_- = 0$, at the boundary. This is, however, a rather strict boundary condition which would not allow for general non-AdS backgrounds. Thus, in order to be able to describe non-AdS backgrounds one has to relax these boundary conditions most of the time[3]. This can be achieved by adding a boundary term $B[A]$ to the Chern-Simons action. This boundary term can be written as

$$B[A] = \frac{k}{4\pi} \int_{\partial\mathcal{M}} \langle A_+ A_- \rangle\,. \tag{7.3}$$

Adding this boundary term changes the boundary conditions (7.2) to

$$\int_{\partial\mathcal{M}} \langle A_+ \wedge \delta A_- \rangle = 0. \tag{7.4}$$

One can now also choose to set $\delta A_-|_{\partial\mathcal{M}} = 0$ instead of $A_- = 0$ and similarly for $\bar{A}$. Choosing either $\delta A_-|_{\partial\mathcal{M}} = 0$ or $A_-|_{\partial\mathcal{M}} = 0$ depends on the specific theory one is looking at but for most of the non-AdS backgrounds it turns out that one has to use the more relaxed boundary condition $\delta A_-|_{\partial\mathcal{M}} = 0$.

[1] This usually boils down to choosing an appropriate embedding of $\mathfrak{sl}(2,\mathbb{R}) \hookrightarrow \mathfrak{sl}(N,\mathbb{R})$ and then fix the Chern-Simons connections $A$ and $\bar{A}$ in such a way that they correctly reproduce the desired gravitational background.

[2] Of course one also obtains similar conditions for $\bar{A}$, which I will not explicitly write down in this chapter as all formulas for $\bar{A}$ can be obtained from the formulas provided for $A$ by simply exchanging $A \leftrightarrow \bar{A}$.

[3] One example where one would not have to relax these boundary conditions is Einstein gravity with vanishing cosmological constant described by an $\mathfrak{isl}(2,\mathbb{R})$ valued connection.

**Impose Suitable Boundary Conditions:** After having chosen the bulk theory and set up a consistent variational principle the next step in the holographic analysis is choosing appropriate boundary conditions for the Chern-Simons connections $A$ and $\bar{A}$. This is the most crucial step in the whole analysis as the boundary conditions essentially determine the physical content of the dual field theory at the boundary. Gauging away the radial dependence of the gauge fields as in (3.39)

$$A_\mu = b^{-1}\left(\mathfrak{a}_\mu + a_\mu^{(0)} + a_\mu^{(1)}\right) b, \quad \bar{A}_\mu = b^{-1}\left(\bar{\mathfrak{a}}_\mu + \bar{a}_\mu^{(0)} + \bar{a}_\mu^{(1)}\right) b, \quad b = e^{\rho L_0}, \tag{7.5}$$

one can then identify the following three contributions to the Chern-Simons connections.

- $\mathfrak{a}_\mu$ and $\bar{\mathfrak{a}}_\mu$ denote the (fixed) background which was chosen in the previous step.
- $a_\mu^{(0)}$ and $\bar{a}_\mu^{(0)}$ correspond to state dependent leading contributions in addition to the background that contain all the physical information about the field degrees of freedom at the boundary.
- $a_\mu^{(1)}$ and $\bar{a}_\mu^{(1)}$ are subleading contributions.

Choosing suitable boundary conditions in this context thus means choosing $a_\mu^{(0)}$ ($\bar{a}_\mu^{(0)}$) and $a_\mu^{(1)}$ ($\bar{a}_\mu^{(1)}$) in such a way that there exist gauge transformations which preserve these boundary conditions i.e.

$$\delta_\varepsilon A_\mu = \mathcal{O}\left(b^{-1} a_\mu^{(0)} b\right) + \mathcal{O}\left(b^{-1} a_\mu^{(1)} b\right), \tag{7.6}$$

for some gauge parameter $\varepsilon$ which can also be written as

$$\varepsilon = b^{-1}\left(\epsilon^{(0)} + \epsilon^{(1)}\right) b, \tag{7.7}$$

and similarly for the barred quantities. The transformations $\epsilon^{(0)}$ usually belong to the asymptotic symmetry algebra while $\epsilon^{(1)}$ are trivial gauge transformations.

**Perform Canonical Analysis and Check Consistency of Boundary Conditions:** Once the boundary conditions and the gauge transformations which preserve these boundary conditions have been fixed one has to determine the canonical boundary charges. This is a standard procedure which is described in great detail for example in [79, 83] and in a bit less detail also in e.g. [37]. This procedure eventually leads to the variation of the canonical boundary charge[4]

$$\delta \mathcal{Q}[\varepsilon] = \frac{k}{2\pi}\int_{\partial\Sigma} \left\langle \epsilon^{(0)} \delta a_\varphi^{(0)}\, \mathrm{d}\varphi \right\rangle, \tag{7.8}$$

[4] See also e.g. (3.26).

where $\varphi$ parametrizes the cycle of the boundary cylinder. A similar expression holds again for the barred quantities. Of course one also has to check whether or not the boundary conditions chosen at the beginning of the algorithm are actually physically admissible. That is, the variation of the canonical boundary charge is finite, conserved in time and integrable in field space. If all these conditions are met then one can proceed with determining the (semiclassical) asymptotic symmetry algebra. Otherwise one has to start over again and choose a new set of boundary conditions and repeat the algorithm up to this point until a finite, conserved and integrable canonical boundary charge is obtained.

**Determine Semiclassical Asymptotic Symmetry Algebra:** This step consists in working out the Dirac brackets between the canonical generators $\mathcal{G}$ which directly yields the semiclassical asymptotic symmetry algebra. There is a well known trick which can be used to simplify calculations at this point. Let us assume that one has two charges with Dirac bracket $\{\mathcal{G}[\varepsilon_1], \mathcal{G}[\varepsilon_2]\}$. Then one can exploit the fact that these brackets generate a gauge transformation as $\{\mathcal{G}[\varepsilon_1], \mathcal{G}[\varepsilon_2]\} = \delta_{\epsilon_2}\mathcal{G}$, and read of the Dirac brackets by evaluating $\delta_{\epsilon_2}\mathcal{G}$. This relation for the canonical gauge generators is on-shell equivalent to a corresponding relation only involving the canonical boundary charges

$$\{\mathcal{Q}[\varepsilon_1], \mathcal{Q}[\varepsilon_2]\} = \delta_{\epsilon_2}\mathcal{Q}, \tag{7.9}$$

which in most cases is comparatively easy and straightforward to calculate. This directly leads to the semiclassical asymptotic symmetry algebra including all possible semiclassical central extensions.

**Determine the Quantum Asymptotic Symmetry Algebra:** This part of the algorithm first appeared in [28]. One insight of this paper was that the asymptotic symmetry algebra derived in the previous steps is only valid for large values of the central charges. When taking into account small values of the central charge and in addition introducing normal ordering it can happen that the asymptotic symmetry algebra violates the Jacobi identities. The simplest way to fix this is to assume suitable deformations of the asymptotic symmetry algebra for finite values of the central charge and normal ordered expressions and then determine the exact form of the deformations by demanding compatibility with the Jacobi identities, see e.g. [132]. In practice the most efficient approach is to allow all possible deformations of the algebra that are consistent with the specific operator content and then check the Jacobi identities. If a deformation is not allowed, then the Jacobi identities will automatically require this deformation to be absent yielding as a final result the correct quantum asymptotic symmetry algebra.

**Look for Unitary Representations of the Quantum Asymptotic Symmetry Algebra:** This point works in principle in exact the same way as in a conformal field theory.

First one defines a suitable highest-weight state and which modes of the operator algebra annihilate this highest-weight state and which modes generate new states. Then one requires that the resulting module does not contain any states with negative norm. This usually restricts the central charges appearing in the quantum asymptotic symmetry algebra and thus also in turn the Chern-Simons level $k$ to certain fixed values. These values can be continuous or discrete, infinitely, finitely many, or even none at all depending on the specific theory in question.

**Identify the Dual Field Theory:** With the results from all the previous steps one can then finally proceed in trying to identify or put possible restrictions on a quantum field theory which realizes all these quantum asymptotic symmetries in a unitary way. Once this dual field theory is identified or conjectured as a dual theory one can perform further checks of the holographic conjecture like calculating partition functions or determining correlation functions on the gravity side.

## 7.2 Example: Spin-3 Lobachevsky Holography

Perhaps the simplest example of non-AdS holography is non-principally embedded spin-3 Lobachevsky holography first described in [I, 37]. I will use this example to briefly demonstrate how the algorithm presented earlier works in practice. I will, however, only recapitulate the main points as this analysis has already been elaborated fully in my Master's thesis [37].
Spin-3 Lobachevsky holography can be described by two Lie algebra valued Chern-Simons connections $A$ and $\bar{A}$ that take values in the non-principal embedding $\mathfrak{sl}(2,\mathbb{R}) \hookrightarrow \mathfrak{sl}(3,\mathbb{R})$, which results in an algebra with non-vanishing commutation relations of the form

$$[L_n,\, L_m] =(n-m)L_{m+n}, \tag{7.10a}$$

$$[S,\, G^\pm_n] = \pm\, G^\pm_n, \tag{7.10b}$$

$$[L_n,\, G^\pm_m] =(\frac{n}{2}-m)\, G^\pm_{n+m}, \tag{7.10c}$$

$$[G^+_n,\, G^-_m] =L_{m+n} - \frac{3}{2}(n-m)S, \tag{7.10d}$$

where $L_n$ $(n = \pm 1, 0)$ is the $\mathfrak{sl}(2,\mathbb{R})$ spin-2 gravity triplet, $G^\pm_n$ $(n = \pm\frac{1}{2})$ a bosonic spin-$\frac{3}{2}$ doublet and $S$ the spin-1 singlet.
The Lobachevsky background is fixed by

$$\mathfrak{a}_\rho = L_0, \qquad \mathfrak{a}_\varphi = -\frac{1}{4}L_1, \qquad \mathfrak{a}_t = 0, \tag{7.11a}$$

$$\bar{\mathfrak{a}}_\rho = -L_0, \qquad \bar{\mathfrak{a}}_\varphi = -L_{-1}, \qquad \bar{\mathfrak{a}}_t = \sqrt{3}S, \tag{7.11b}$$

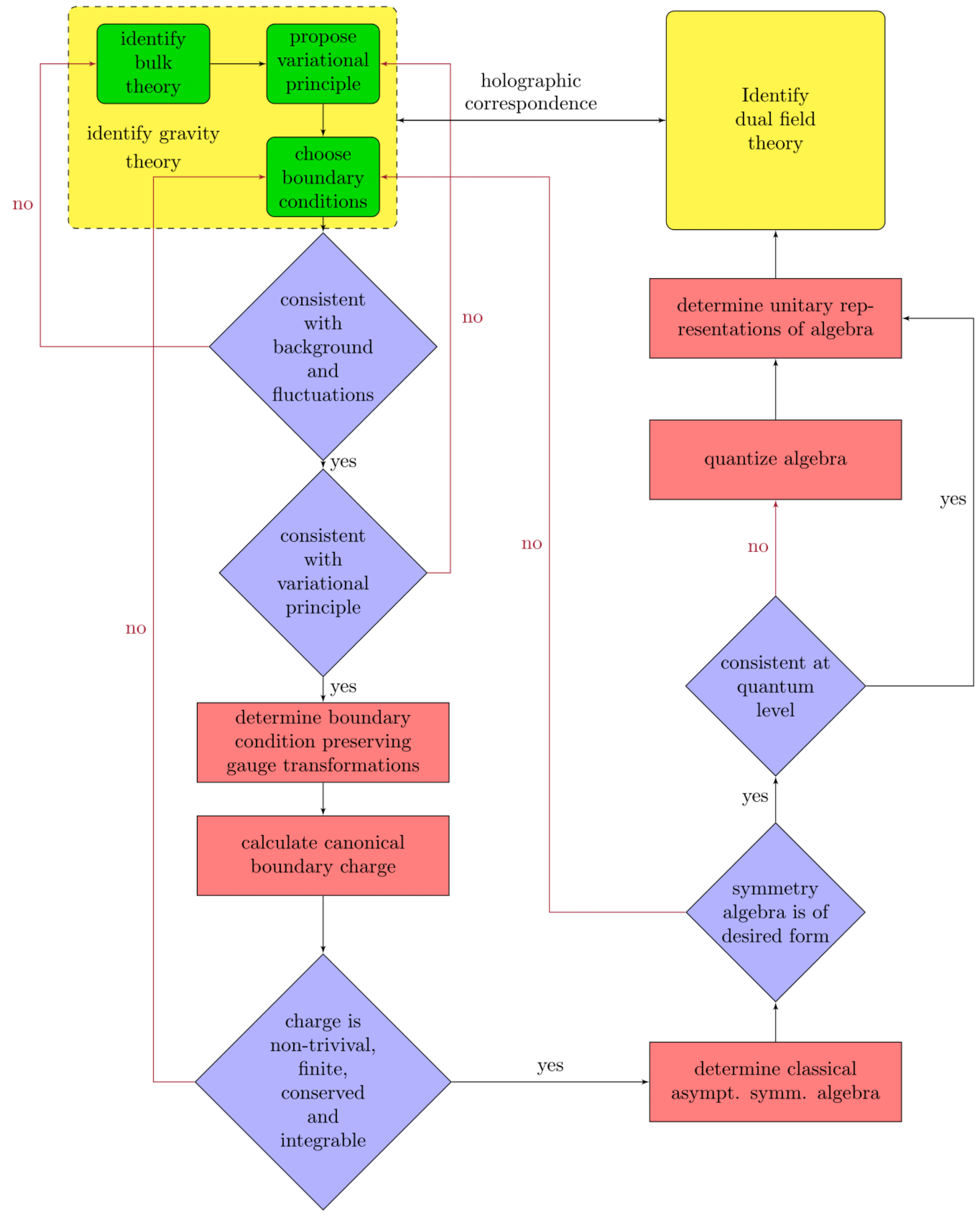

**Fig. 7.1.:** Flowchart depicting the procedure of analyzing higher-spin gravity theories.

where I used again the gauge (7.5). The state dependent fluctuations are fixed to

$$a_\rho^{(0)} = 0, \qquad a_\varphi^{(0)} = \frac{2\pi}{k}\left(\mathcal{L}L_{-1} + \mathcal{G}^{\pm}G^{\pm}_{-\frac{1}{2}} + \mathcal{J}S\right), \qquad a_t^{(0)} = 0, \tag{7.12a}$$

$$\bar{a}_\rho^{(0)} = 0, \qquad \bar{a}_\varphi^{(0)} = \frac{2\pi}{k}\bar{\mathcal{J}}S, \qquad \bar{a}_t^{(0)} = 0, \tag{7.12b}$$

where the functions $\mathcal{L}, \mathcal{G}^{\pm}$ and $\mathcal{J}$ all only depend on the angular coordinate $\varphi$ parametrizing the cycle of the boundary cylinder. The trivial, subleading fluctuations are simply $a_\mu^{(0)} = \bar{a}_\mu^{(1)} = \mathcal{O}\left(e^{-2\rho}\right)$.

Determining the gauge transformations that preserve these boundary conditions, the asymptotic symmetry algebra and then finally the quantum asymptotic symmetry algebra leads to asymptotic symmetries consisting of one copy of the Polyakov–Bershadsky $\mathcal{W}_3^{(2)}$-algebra [133, 134] and a $\hat{\mathfrak{u}}(1)$ current algebra.
Defining $\hat{k} = -k - 3/2$ and denoting normal ordering by ::, the $\mathcal{W}_3^{(2)}$-algebra is given by

$$[L_n, L_m] = (n-m)L_{m+n} + \frac{c}{12}\, n(n^2-1)\, \delta_{n+m,\,0}, \tag{7.13a}$$

$$[L_n, J_m] = -\, mJ_{n+m}, \tag{7.13b}$$

$$[L_n, G^{\pm}_m] = (\frac{n}{2} - m)\, G^{\pm}_{n+m}, \tag{7.13c}$$

$$[J_n, J_m] = \kappa\, n\, \delta_{n+m,\,0}, \tag{7.13d}$$

$$[J_n, G^{\pm}_m] = \pm\, G^{\pm}_{m+n}, \tag{7.13e}$$

$$\begin{aligned}[G^{+}_n, G^{-}_m] =& \frac{\lambda}{2}\,(n^2 - \frac{1}{4})\, \delta_{n+m,\,0} - (\hat{k}+3)L_{m+n} \\ &+ \frac{3}{2}(\hat{k}+1)(n-m)J_{m+n} + 3\sum_{p\in\mathbb{Z}} :J_{m+n-p}J_p:,\end{aligned} \tag{7.13f}$$

with the $\hat{\mathfrak{u}}(1)$ level

$$\kappa = \frac{2\hat{k}+3}{3}, \tag{7.14}$$

the Virasoro central charge

$$c = 25 - \frac{24}{\hat{k}+3} - 6(\hat{k}+3), \tag{7.15}$$

and the central term in the $G^{\pm}$ commutator

$$\lambda = (\hat{k}+1)(2\hat{k}+3)\,. \tag{7.16}$$

One constraint for unitary highest weight representations of this algebra requires non-negativity of $c$ which fixes the level $\hat{k}$ to lie in the interval $-\frac{1}{3} \geq \hat{k} \geq -\frac{3}{2}$. Another obstruction to unitarity comes from the $G^{\pm}$ sector. It turns out that the two level $\frac{3}{2}$ descendants of the vacuum, $G^{\pm}_{-3/2}|0\rangle$ lead to a Gram matrix proportional to $\lambda$ with positive and negative eigenvalue. Therefore, generic values of the level $\hat{k}$ lead to positive and negative norm states, which renders the theory non-unitary. The only exception arises if $\lambda$ vanishes, in which case the $G^{\pm}$ descendants become null states. Thus, the number of possible values of $\hat{k}$ compatible with unitarity is reduced to two values, $\hat{k} = -1$ and $\hat{k} = -3/2$. The latter leads to a trivial ($c = 0$) theory, the former to a rather simple one ($c = 1$), which is a theory of a single free boson.
The unitarity analysis above applies to AdS holography in the non-principal embedding as well, where the asymptotic symmetry algebra consists of two copies of the Polyakov–Bershadsky $\mathcal{W}_3^{(2)}$-algebra (see [30, 135]).

# Non-AdS Higher-Spin Gravity in $2+1$ Dimensions

# 8

*Gibt es dazu auch eine geometrische Interpretation? (Is there a geometrical interpretation of the problem?)*

**– Peter C. Aichelburg**
Austrian Theoretical Physicist

Lobachevsky holography in $2+1$ dimensions has been first explored in [I, 37]. Lobachevsky spacetimes are particularly interesting to study using the Chern-Simons formalism because such a description in terms of a Chern-Simons gauge field necessarily forces one to go beyond the well known spin-2 AdS gauge algebra $\mathfrak{sl}(2,\mathbb{R}) \oplus \mathfrak{sl}(2,\mathbb{R})$. To be more precise, the Chern-Simons gauge connection corresponding to the asymptotic line element

$$\mathrm{d}s^2 = \mathrm{d}t^2 + \mathrm{d}\rho^2 - \sigma e^{2\rho}\,\mathrm{d}\varphi^2, \tag{8.1}$$

where depending on the sign of $\sigma$ this metric is asymptotically $\mathrm{AdS}_2 \times \mathbb{R}$ ($\sigma = 1$) or $\mathbb{H}_2 \times \mathbb{R}$ ($\sigma = -1$), with $\mathbb{H}_2$ being the two dimensional Lobachevsky plane, necessarily needs a singlet $S$ with $\mathrm{Tr}\,(S^2) \neq 0$ in order to properly describe the asymptotic background (8.1). This setup cannot be realized simply by a Chern-Simons connection taking values in $\mathfrak{sl}(2,\mathbb{R}) \oplus \mathfrak{sl}(2,\mathbb{R})$ as there is no singlet present in this gauge algebra. The simplest case to consider where one has a singlet in combination with an $\mathfrak{sl}(2,\mathbb{R})$ triplet present is given by the non-principal embedding of $\mathfrak{sl}(2,\mathbb{R}) \hookrightarrow \mathfrak{sl}(3,\mathbb{R})$ as described in [I, 37].

In this chapter I will focus on extensions of the work in [I, 37] by first considering spin-4 Lobachevsky holography and finishing with a generalization of the 3-1 embedding of $\mathfrak{sl}(2,\mathbb{R}) \hookrightarrow \mathfrak{sl}(4,\mathbb{R})$ to an $N-1$ embedding of $\mathfrak{sl}(2,\mathbb{R}) \hookrightarrow \mathfrak{sl}(N,\mathbb{R})$.

## 8.1 $\mathfrak{sl}(4,\mathbb{R})$ Lobachevsky Holography

In order to get a better understanding of the constraints on unitarity of quantum field theories whose holographic gravity duals are described by non-principally embedded Chern-Simons connections it is instructive to go beyond the non-principal embedding of $\mathfrak{sl}(2,\mathbb{R}) \hookrightarrow \mathfrak{sl}(3,\mathbb{R})$. This will be the primary goal of this section. I will take all possible non-principal embeddings of $\mathfrak{sl}(2,\mathbb{R}) \hookrightarrow \mathfrak{sl}(4,\mathbb{R})$ and find appropriate boundary conditions for the Chern-Simons gauge connection so that the corresponding metric is again (8.1). I will then perform a canonical analysis, determine the boundary condition preserving gauge transformations, the corresponding boundary charges and their asymptotic symmetry algebra. Since this procedure will be the same for all embeddings treated in this thesis I will use the 2-1-1 embedding as an explicit example to show all the necessary calculations explicitly. The details of the canonical analyses for the other embeddings can be found in Appendix B.

### 8.1.1 2-1-1 Embedding of $\mathfrak{sl}(2,\mathbb{R}) \hookrightarrow \mathfrak{sl}(4,\mathbb{R})$

The 2-1-1 embedding of $\mathfrak{sl}(2,\mathbb{R}) \hookrightarrow \mathfrak{sl}(4,\mathbb{R})$ has three $\mathfrak{sl}(2,\mathbb{R})$ spin-2 generators $L_n$ $(n=\pm1,0)$, four spin-$\frac{3}{2}$ doublets $G_n^{a|b}$ $(a,b=\pm1; n=\pm\frac{1}{2})$, one spin-1 triplet of singlets $J^a$ $(a=0,\pm1)$ and another spin-1 singlet $S$. They obey the following non-vanishing algebraic relations

$$\begin{aligned}
[L_n,L_m] &=(n-m)L_{n+m}, && \text{(8.2a)}\\
[L_n,G_m^{a|b}] &=(\frac{n}{2}-m)G_{n+m}^{a|b}, && \text{(8.2b)}\\
[J^a,G_m^{b|c}] &=\frac{(a-b)}{2}G_m^{2a+b|c}, && \text{(8.2c)}\\
[J^a,J^b] &=(a-b)J^{a+b}, && \text{(8.2d)}\\
[S,G_m^{a|b}] &=2bG_m^{a|b}, && \text{(8.2e)}\\
[G_n^{a|b},G_m^{c|-b}] &=\delta_{a+c,0}\left(L_{n+m}+(m-n)\left(J^0+\frac{b}{2}S\right)\right)+(m-n)J^a\delta_{a,c}. && \text{(8.2f)}
\end{aligned}$$

This basis for the generators is a convenient choice because, as in the $\mathfrak{sl}(3,\mathbb{R})$ case, one can characterize the doublets by their $S$ and $J^0$ charge, respectively. As in the $\mathfrak{sl}(3,\mathbb{R})$ case I will denote the connections $A$, $\bar{A}$ and the boundary condition preserving gauge transformations $\varepsilon$, $\bar{\varepsilon}$ as

$$\begin{aligned}
A_\mu &=b^{-1}(a_\mu^{(0)}+a_\mu^{(1)}), & \bar{A}_\mu &=b(\bar{a}_\mu^{(0)}+\bar{a}_\mu^{(1)})b^{-1},\\
\varepsilon &=b^{-1}(\epsilon^{(0)}+\epsilon^{(1)})b, & \bar{\varepsilon} &=b(\bar{\epsilon}^{(0)}+\bar{\epsilon}^{(1)})b^{-1},
\end{aligned} \tag{8.3}$$

with

$$b=e^{\rho L_0}. \tag{8.4}$$

The background fluctuations of the metric (8.1) I will assume to be

$$g_{\mu\nu}=\begin{pmatrix} 1+\mathcal{O}(e^{-2\rho}) & \mathcal{O}(e^{-2\rho}) & \mathcal{O}(1)\\ \cdot & 1+\mathcal{O}(e^{-2\rho}) & \mathcal{O}(1)\\ \cdot & \cdot & -\sigma e^{2\rho}+\mathcal{O}(1)\end{pmatrix}_{\mu\nu}, \tag{8.5}$$

then the corresponding connection $A$ has to obey the following boundary conditions

$$a_\rho^{(0)} = L_0, \qquad a_t^{(0)} = 0, \qquad a_\mu^{(1)} = \mathcal{O}(e^{-2\rho}), \tag{8.6a}$$

$$a_\varphi^{(0)} = \sigma L_1 - \frac{2\pi}{k}\left(\mathcal{L}L_{-1} + \sum_{a,b=\pm} a\mathcal{G}^{a|b}G^{a|b}_{-\frac{1}{2}} - \sum_{a=-1}^{1}\frac{2}{(1-3a^2)}\mathcal{J}^a J^a - \frac{\mathcal{S}}{4}S\right), \tag{8.6b}$$

where the state dependent functions $\mathcal{L}, \mathcal{G}^{a|b}, \mathcal{J}^a$ and $\mathcal{S}$ all depend only on the angular coordinate $\varphi$.
For the $\bar{A}$-sector the connection has to obey asymptotically

$$\bar{a}_\rho^{(0)} = -L_0, \qquad \bar{a}_t^{(0)} = \frac{1}{\sqrt{2}}S, \qquad \bar{a}_\mu^{(1)} = \mathcal{O}(e^{-2\rho}), \tag{8.7a}$$

$$\bar{a}_\varphi^{(0)} = -L_{-1} + \frac{2\pi}{k}\left(\sum_{a=-1}^{1}\frac{2}{(1-3a^2)}\bar{\mathcal{J}}^a J^a - \frac{\bar{\mathcal{S}}}{4}S\right), \tag{8.7b}$$

where again the state dependent functions $\bar{\mathcal{J}}^a$ and $\bar{\mathcal{S}}$ all depend only on the angular coordinate $\varphi$.

#### Canonical analysis for the $A$-sector:

In the following I will describe how to perform a canonical analysis for the $A$ part of the Chern-Simons connection, which will result in the semiclassical asymptotic symmetry algebra of the $A$-sector.
The gauge transformations preserving the boundary conditions (8.6) are given by

$$\begin{aligned}\epsilon^{(0)} =& \epsilon^1 L_1 + \epsilon^2 L_0 + \epsilon^3 L_{-1} + \epsilon^4 G^{+|+}_{\frac{1}{2}} + \epsilon^5 G^{+|+}_{-\frac{1}{2}} + \epsilon^6 G^{+|-}_{\frac{1}{2}} + \epsilon^7 G^{+|-}_{-\frac{1}{2}} + \epsilon^8 G^{-|+}_{\frac{1}{2}} \\ &+ \epsilon^9 G^{-|+}_{-\frac{1}{2}} + \epsilon^{10} G^{-|-}_{\frac{1}{2}} + \epsilon^{11} G^{-|-}_{-\frac{1}{2}} + \epsilon^{12} J^+ + \epsilon^{13} J^0 + \epsilon^{14} J^- + \epsilon^{15} S,\end{aligned} \tag{8.8}$$

with

$$\epsilon^1 = \epsilon, \qquad \epsilon^2 = -\sigma\epsilon', \qquad \epsilon^3 = \frac{1}{2}\epsilon'' + \frac{\pi\sigma}{k}\left(-2\mathcal{L}\epsilon - \sum_{a,b=\pm}\mathcal{G}^{a|b}\epsilon^{-a|-b}_{\frac{1}{2}}\right), \tag{8.9a}$$

$$\epsilon^4 = \epsilon^{+|+}_{\frac{1}{2}}, \quad \epsilon^5 = -\sigma\epsilon^{+|+\prime}_{\frac{1}{2}} - \frac{\pi\sigma}{k}\left(2\mathcal{G}^{+|+}\epsilon - 2\mathcal{J}^+\epsilon^{-|+}_{\frac{1}{2}} + \left(\mathcal{S} - 2\mathcal{J}^0\right)\epsilon^{+|+}_{\frac{1}{2}}\right), \tag{8.9b}$$

$$\epsilon^6 = \epsilon^{+|-}_{\frac{1}{2}}, \quad \epsilon^7 = -\sigma\epsilon^{+|-\prime}_{\frac{1}{2}} - \frac{\pi\sigma}{k}\left(2\mathcal{G}^{+|-}\epsilon - 2\mathcal{J}^+\epsilon^{-|-}_{\frac{1}{2}} - \left(\mathcal{S} + 2\mathcal{J}^0\right)\epsilon^{+|-}_{\frac{1}{2}}\right), \tag{8.9c}$$

$$\epsilon^8 = \epsilon^{-|+}_{\frac{1}{2}}, \quad \epsilon^9 = -\sigma\epsilon^{-|+\prime}_{\frac{1}{2}} + \frac{\pi\sigma}{k}\left(2\mathcal{G}^{-|+}\epsilon - 2\mathcal{J}_0^-\epsilon^{+|+}_{\frac{1}{2}} - \left(\mathcal{S} + 2\mathcal{J}^0\right)\epsilon^{-|+}_{\frac{1}{2}}\right), \tag{8.9d}$$

$$\epsilon^{10} = \epsilon^{-|-}_{\frac{1}{2}}, \quad \epsilon^{11} = -\sigma\epsilon^{-|-\prime}_{\frac{1}{2}} + \frac{\pi\sigma}{k}\left(2\mathcal{G}^{-|-}\epsilon - 2\mathcal{J}^-\epsilon^{+|-}_{\frac{1}{2}} + \left(\mathcal{S} - 2\mathcal{J}^0\right)\epsilon^{-|-}_{\frac{1}{2}}\right), \tag{8.9e}$$

$$\epsilon^{12} = \epsilon_0^+, \qquad \epsilon^{13} = \epsilon_0^0, \qquad \epsilon^{14} = \epsilon_0^-, \qquad \epsilon^{15} = \epsilon_0, \tag{8.9f}$$

and

$$\epsilon^{(1)} = \mathcal{O}(e^{-2\rho}). \tag{8.10}$$

As before I suppressed the explicit dependence of the gauge parameters $\epsilon$, $\epsilon^{a|b}_{\frac{1}{2}}$, $\epsilon^a_0$ and $\epsilon_0$ on the angular coordinate $\varphi$ for the sake of readability. In that sense it is also understood that a prime denotes differentiation with respect to $\varphi$. In order to simplify expressions I also used that $\sigma^2 = 1$.

These boundary condition preserving gauge transformations imply the following non-trivial transformation behavior of the state dependent functions under the asymptotic symmetries

$$\delta_\epsilon \mathcal{L} = \sigma(2\mathcal{L}\epsilon' + \mathcal{L}'\epsilon) - \frac{k}{4\pi}\epsilon''', \tag{8.11a}$$

$$\begin{aligned}\delta_{\epsilon^{a|b}_{\frac{1}{2}}} \mathcal{L} =& \frac{\sigma}{2}\left(\mathcal{G}^{-a|-b\prime}\epsilon^{a|b}_{\frac{1}{2}} + 3\mathcal{G}^{-a|-b}\epsilon^{a|b\prime}_{\frac{1}{2}}\right) \\ &+ \frac{\pi\sigma}{k}\left(\mathcal{G}^{-a|-b}\left(b\mathcal{S} - 2a\mathcal{J}^0\right) + 2a\mathcal{G}^{a|-b}\mathcal{J}^{-a}\right)\epsilon^{a|b}_{\frac{1}{2}},\end{aligned} \tag{8.11b}$$

$$\begin{aligned}\delta_\epsilon \mathcal{G}^{a|b} =& \sigma\left(\mathcal{G}^{a|b\prime}\epsilon + \frac{3}{2}\mathcal{G}^{a|b}\epsilon'\right) \\ &+ \frac{\pi\sigma}{k}\left(\mathcal{G}^{a|b}\left(b\mathcal{S} - 2a\mathcal{J}^0_0\right) + 2a\mathcal{G}^{-a|b}\mathcal{J}^a_0\right)\epsilon,\end{aligned} \tag{8.11c}$$

$$\begin{aligned}\delta_{\epsilon^{a|b}_{\frac{1}{2}}} \mathcal{G}^{c|d} =& \delta_{b,d}\left(\delta_{a+c,0}\sigma\left(-\mathcal{J}^{c\prime}\epsilon^{a|b}_{\frac{1}{2}} - 2\mathcal{J}^c\epsilon^{a|b\prime}_{\frac{1}{2}} - \frac{2\pi}{k}d\mathcal{S}\mathcal{J}^c\epsilon^{a|b}_{\frac{1}{2}}\right)\right. \\ &\delta_{a,c}\left(\frac{\sigma}{2}\left(\left(-2c\sigma\mathcal{L} + cd\mathcal{S}' - 2\mathcal{J}^{0\prime}\right)\epsilon^{a|b}_{\frac{1}{2}} + 2\left(cd\mathcal{S} - 2\mathcal{J}^0\right)\epsilon^{a|b\prime}_{\frac{1}{2}}\right)\right. \\ &\left.\left.+\frac{\pi\sigma}{2k}\left(c\left(\mathcal{S} - 2cd\mathcal{J}^0\right)^2 - 4c\mathcal{J}^-\mathcal{J}^+\right)\epsilon^{a|b}_{\frac{1}{2}} + \frac{k\sigma}{2\pi}c\epsilon^{a|b\prime\prime}_{\frac{1}{2}}\right)\right),\end{aligned} \tag{8.11d}$$

$$\delta_{\epsilon^a_0} \mathcal{G}^{b|c} = \frac{(a+b)}{2}\mathcal{G}^{b-2a|c}, \tag{8.11e}$$

$$\delta_{\epsilon_0} \mathcal{G}^{a|b} = -2b\mathcal{G}^{a|b}\epsilon_0, \tag{8.11f}$$

$$\delta_{\epsilon^{a|b}_{\frac{1}{2}}} \mathcal{J}^c = \frac{(a+c)}{2}\mathcal{G}^{2c-a|-b}\epsilon^{a|b}_{\frac{1}{2}}, \tag{8.11g}$$

$$\delta_{\epsilon^a_0} \mathcal{J}^b = (a+b)\mathcal{J}^{b-a}\epsilon^a_0 + \frac{k}{4\pi}(1-3a^2)\epsilon^{a\prime}_0\delta_{a,b}, \tag{8.11h}$$

$$\delta_{\epsilon^{a|b}_{\frac{1}{2}}} \mathcal{S} = -2b\mathcal{G}^{-a|-b}\epsilon^{a|b}_{\frac{1}{2}}, \tag{8.11i}$$

$$\delta_{\epsilon_0} \mathcal{S} = \frac{2k}{\pi}\epsilon_0{}', \tag{8.11j}$$

where again a prime denotes derivative with respect to the argument of the function which in this case would be $\varphi$. The canonical boundary charge for the boundary conditions (8.6) and the corresponding gauge transformations which preserve these boundary conditions is given by

$$\mathcal{Q}[\varepsilon] = \int \mathrm{d}\varphi \left(\mathcal{L}\epsilon + \sum_{a,b=\pm}\mathcal{G}^{a|b}\epsilon^{-a|-b}_{\frac{1}{2}} + \sum_{a=-1}^{1}\mathcal{J}^a\epsilon^{-a}_0 + \mathcal{S}\epsilon_0\right). \tag{8.12}$$

The non-vanishing Dirac brackets which are obtained from (8.11) using the canonical boundary charge (8.12) are given by

$$
\begin{aligned}
\{\mathcal{L}(\varphi), \mathcal{L}(\bar{\varphi})\} =& \sigma\left(2\mathcal{L}\delta' - \mathcal{L}'\delta\right) - \frac{k}{4\pi}\delta''', && \text{(8.13a)}\\
\{\mathcal{L}(\varphi), \mathcal{G}^{a|b}(\bar{\varphi})\} =& \sigma\left(\frac{3}{2}\mathcal{G}^{a|b}\delta' - \mathcal{G}^{a|b\prime}\delta\right) \\
& + \frac{\pi\sigma}{k}\left(\mathcal{G}^{a|b}\left(2a\mathcal{J}^0 - b\mathcal{S}\right) - 2a\mathcal{G}^{-a|b}\mathcal{J}^a\right)\delta, && \text{(8.13b)}\\
\{\mathcal{J}^a(\varphi), \mathcal{J}^b(\bar{\varphi})\} =& (a-b)\mathcal{J}^{a+b}\delta + \frac{k}{4\pi}\left(1-3a^2\right)\delta_{a+b,0}\delta', && \text{(8.13c)}\\
\{\mathcal{J}^a(\varphi), \mathcal{G}^{b|c}(\bar{\varphi})\} =& \frac{a-b}{2}\mathcal{G}^{2a+b|c}\delta, && \text{(8.13d)}\\
\{\mathcal{S}(\varphi), \mathcal{G}^{a|b}(\bar{\varphi})\} =& 2b\mathcal{G}^{a|b}\delta, && \text{(8.13e)}\\
\{\mathcal{S}(\varphi), \mathcal{S}(\bar{\varphi})\} =& \frac{2k}{\pi}\delta', && \text{(8.13f)}\\
\{\mathcal{G}^{a|b}(\varphi), \mathcal{G}^{c|d}(\bar{\varphi})\} =& \delta_{b+d,0}\left(\delta_{a,c}\sigma\left(\mathcal{J}^{c\prime}\delta - 2\mathcal{J}^c\delta' + \frac{2\pi}{k}d\mathcal{S}\mathcal{J}^c\delta\right)\right. \\
& + \delta_{a+c,0}\frac{\sigma}{2}\left(\left(2c\sigma\mathcal{L} - cd\mathcal{S}' + 2\mathcal{J}^{0\prime}\right)\delta + 2\left(cd\mathcal{S} - 2\mathcal{J}^0\right)\delta'\right. \\
& \left.\left. + \frac{\pi}{k}c\left(4\mathcal{J}^-\mathcal{J}^+ - \left(\mathcal{S} - 2cd\mathcal{J}^0\right)^2\right)\delta - \frac{k}{\pi}c\delta''\right)\right), && \text{(8.13g)}
\end{aligned}
$$

where all state dependent functions appearing on the r.h.s depend on $\bar{\varphi}$ and a $\delta$ without indices denotes the dirac $\delta$-distribution as $\delta\left(\varphi - \bar{\varphi}\right)$ and $\delta' \equiv \partial_{\bar{\varphi}}\left(\varphi - \bar{\varphi}\right)$. In order to bring this algebra into a form in which all fields are proper Virasoro primaries one can implement the following shift

$$
\mathcal{L} \to \hat{\mathcal{L}} = \mathcal{L} + \frac{\pi\sigma}{4k}\mathcal{S}\mathcal{S} + \frac{2\pi\sigma}{k}\left(\mathcal{J}^0\mathcal{J}^0 - \mathcal{J}^+\mathcal{J}^-\right). \tag{8.14}
$$

This yields the following algebra

$$
\begin{aligned}
\{\hat{\mathcal{L}}(\varphi), \hat{\mathcal{L}}(\bar{\varphi})\} =& \sigma\left(2\hat{\mathcal{L}}\delta' - \mathcal{L}'\delta\right) - \frac{k}{4\pi}\delta''', && \text{(8.15a)}\\
\{\hat{\mathcal{L}}(\varphi), \mathcal{J}^a(\bar{\varphi})\} =& \sigma\mathcal{J}^a(\varphi)\delta', && \text{(8.15b)}\\
\{\hat{\mathcal{L}}(\varphi), \mathcal{S}(\bar{\varphi})\} =& \sigma\mathcal{S}(\varphi)\delta', && \text{(8.15c)}\\
\{\hat{\mathcal{L}}(\varphi), \mathcal{G}^{a|b}(\bar{\varphi})\} =& \sigma\left(\frac{3}{2}\mathcal{G}^{a|b}\delta'(\varphi - \bar{\varphi}) - \mathcal{G}^{a|b\prime}\delta\right), && \text{(8.15d)}\\
\{\mathcal{J}^a(\varphi), \mathcal{J}^b(\bar{\varphi})\} =& (a-b)\mathcal{J}^{a+b}\delta + \frac{k}{4\pi}\left(1-3a^2\right)\delta_{a+b,0}\delta', && \text{(8.15e)}\\
\{\mathcal{J}^a(\varphi), \mathcal{G}^{b|c}(\bar{\varphi})\} =& \frac{a-b}{2}\mathcal{G}^{2a+b|c}\delta, && \text{(8.15f)}\\
\{\mathcal{S}(\varphi), \mathcal{G}^{a|b}(\bar{\varphi})\} =& 2b\mathcal{G}^{a|b}\delta, && \text{(8.15g)}\\
\{\mathcal{S}(\varphi), \mathcal{S}(\bar{\varphi})\} =& \frac{2k}{\pi}\delta', && \text{(8.15h)}
\end{aligned}
$$

$$\{\mathcal{G}^{a|b}(\varphi),\mathcal{G}^{c|d}(\bar\varphi)\} =\delta_{b+d,0}\left(\delta_{a,c}\sigma\left(\left(\mathcal{J}^{c\prime}+\frac{2\pi}{k}d\mathcal{S}\mathcal{J}^c\right)\delta-2\mathcal{J}^c\delta'\right)\right.$$
$$+\delta_{a+c,0}\frac{\sigma}{2}\left(\left(2c\sigma\hat{\mathcal{L}}-cd\mathcal{S}'+2\mathcal{J}^{0\prime}\right.\right.$$
$$\left.\left.+\frac{\pi}{k}c\left(8\mathcal{J}^-\mathcal{J}^+-\frac{3}{2}\mathcal{S}\mathcal{S}+4cd\mathcal{S}\mathcal{J}^0-8\mathcal{S}^0\mathcal{J}^0\right)\right)\delta\right.$$
$$\left.\left.+2\left(cd\mathcal{S}-2\mathcal{J}^0\right)\delta'-\frac{k}{\pi}c\delta''\right)\right). \tag{8.15i}$$

Using the mode expansions

$$\hat{\mathcal{L}}(\varphi)=\frac{\sigma}{2\pi}\sum_{n\in\mathbb{Z}}\hat{L}_n e^{-in\varphi},\qquad \mathcal{J}^a(\varphi)=\frac{i}{2\pi}\sum_{n\in\mathbb{Z}}J^a_n e^{-in\varphi}, \tag{8.16a}$$
$$\mathcal{S}(\varphi)=\frac{i}{2\pi}\sum_{n\in\mathbb{Z}}S_n e^{-in\varphi},\qquad \mathcal{G}^{a|b}(\varphi)=\frac{(i\sigma)^{\frac{1+b}{2}}}{2\pi}\sum_{n\in\mathbb{Z}}G^{a|b}_n e^{-in\varphi}, \tag{8.16b}$$
$$\delta(\varphi-\bar\varphi)=\frac{1}{2\pi}\sum_{n\in\mathbb{Z}}e^{-in(\varphi-\bar\varphi)}, \tag{8.16c}$$

and replacing $i\{.,.\}\to[.,.]$ one readily obtains the (semiclassical) algebra

$$[\hat{L}_n,\hat{L}_m] =(n-m)\hat{L}_{n+m}+\frac{\mathfrak{c}}{12}n(n^2-1)\delta_{m+n,0}, \tag{8.17a}$$
$$[\hat{L}_n,J^a_m] =-mJ^a_{n+m}, \tag{8.17b}$$
$$[\hat{L}_n,S_m] =-mS_{n+m}, \tag{8.17c}$$
$$[\hat{L}_n,G^{a|b}_m] =\left(\frac{n}{2}-m\right)G^{a|b}_{n+m}, \tag{8.17d}$$
$$[J^a_n,J^b_m] =(a-b)J^{a+b}_{m+n}-\frac{k}{2}\left(1-3a^2\right)n\delta_{a+b,0}\delta_{n+m,0}, \tag{8.17e}$$
$$[J^a_n,G^{b|c}_m] =\frac{a-b}{2}G^{2a+b|c}_{n+m}, \tag{8.17f}$$
$$[S_n,G^{a|b}_m] =2bG^{a|b}_{n+m}, \tag{8.17g}$$
$$[S_n,S_m] =-4kn\delta_{n+m}, \tag{8.17h}$$
$$[G^{a|b}_n,G^{c|d}_m] =\delta_{b+d,0}\left(\delta_{a,c}\left((m-n)J^c_{m+n}-\frac{d}{2k}\{SJ\}^c\right)+\delta_{a+c,0}\left(c\hat{L}_{n+m}\right.\right.$$
$$-\frac{cd}{2}(m-n)S_{m+n}+(m-n)J^0_{m+n}+ck\left(n^2-\frac{1}{4}\right)\delta_{m+n,0}$$
$$\left.\left.+\frac{c}{4k}\left(-4\{JJ\}^{+|-}+\frac{3}{2}SS-2cd\{SJ\}^0+8JJ^{0|0}\right)\right)\right). \tag{8.17i}$$

with $\mathfrak{c}=6k$.

In order to write the nonlinear $\mathcal{W}$-algebras treated in this thesis as compact as possible I also use the following shorthand notations

$$AB^{a|b}_n=\sum_{p\in\mathbb{Z}}A^a_{n-p}B^b_p, \tag{8.18a}$$
$$\Omega AB^{a|b}_n=\sum_{p\in\mathbb{Z}}pA^a_{n-p}B^b_p, \tag{8.18b}$$

$$\Omega^2 AB_n^{a|b} = \sum_{p\in\mathbb{Z}} p^2 A^a_{n-p} B^b_p, \tag{8.18c}$$

$$ABC_n^{a|b|c} = \sum_{p,q\in\mathbb{Z}} A^a_{n-p-q} B^b_p C^c_q, \tag{8.18d}$$

$$ABCD_n^{a|b|c|d} = \sum_{p,q,r\in\mathbb{Z}} A^a_{n-p-q-r} B^b_p C^c_q D^d_r, \tag{8.18e}$$

$$AB_n^{\{a|b\}} = AB_n^{a|b} + AB_n^{b|a}, \tag{8.18f}$$

$$\begin{aligned} ABC_n^{\{a|b|c\}} =& ABC_n^{a|b|c} + ABC_n^{c|a|b} + ABC_n^{b|c|a} \\ &+ ABC_n^{a|c|b} + ABC_n^{c|b|a} + ABC_n^{b|a|c}, \end{aligned} \tag{8.18g}$$

$$\{AB\}_n^{a|b} = AB_n^{a|b} + BA_n^{a|b}, \tag{8.18h}$$

$$\{\Omega AB\}_n^{a|b} = \Omega AB_n^{a|b} + \Omega BA_n^{a|b}, \tag{8.18i}$$

$$\begin{aligned} \{ABC\}_n^{a|b|c} =& ABC_n^{a|b|c} + ACB_n^{a|b|c} + BAC_n^{a|b|c} \\ &+ BCA_n^{a|b|c} + CAB_n^{a|b|c} + CBA_n^{a|b|c}, \end{aligned} \tag{8.18j}$$

$$AB_n^{[a|b]} = AB_n^{a|b} - AB_n^{b|a}, \tag{8.18k}$$

where it is understood that if the nonlinear terms have no additional "color" index one has to simply replace $A^a \to A$ in the corresponding expressions.

**Canonical analysis for the $\bar{A}$-sector:**

For the $\bar{A}$-sector the gauge transformations preserving (8.7) are given by

$$\bar{\epsilon}^{(0)} = \bar{\epsilon}_0^+ J^+ + \bar{\epsilon}_0^0 J^0 + \bar{\epsilon}_0^- J^- + \bar{\epsilon}_0(\varphi) S. \tag{8.19}$$

The state dependent functions $\bar{\mathcal{J}}^a$ and $\mathcal{S}$ transform under these gauge transformations non-trivially as

$$\delta_{\bar{\epsilon}_0^a} \mathcal{J}^b = (a+b)\mathcal{J}^{b-a}\bar{\epsilon}_0^a - \frac{k}{4\pi}(1-3a^2)\bar{\epsilon}_0^{a\prime}\delta_{a,b}, \tag{8.20a}$$

$$\delta_{\bar{\epsilon}_0} \mathcal{S} = -\frac{k}{2\pi}\bar{\epsilon}_0{}'. \tag{8.20b}$$

The canonical boundary charge is given by

$$\bar{\mathcal{Q}}[\bar{\varepsilon}] = \int \mathrm{d}\varphi \left( \bar{\mathcal{S}} + \sum_{a=-1}^{1} \bar{\mathcal{J}}^{-a}\bar{\epsilon}_0^a \right). \tag{8.21}$$

This canonical boundary charge obeys the following non-vanishing Dirac brackets

$$\{\bar{\mathcal{J}}^a(\varphi), \bar{\mathcal{J}}^b(\bar{\varphi})\} = (a-b)\bar{\mathcal{J}}^{a+b}\delta - \frac{k}{4\pi}(1-3a^2)\delta'\delta_{a+b,0}, \tag{8.22a}$$

$$\{\bar{\mathcal{S}}(\varphi), \bar{\mathcal{S}}(\bar{\varphi})\} = -\frac{k}{2\pi}\delta'. \tag{8.22b}$$

Using the mode expansions

$$\bar{S}(\varphi) = \frac{1}{2\pi}\sum_{n\in\mathbb{Z}} \bar{S}_n e^{-in\varphi}, \qquad \bar{\mathcal{J}}^0(\varphi) = \frac{i}{2\pi}\sum_{n\in\mathbb{Z}} \bar{J}^0_n e^{-in\varphi}, \tag{8.23a}$$

$$\bar{\mathcal{J}}^{\pm}_0(\varphi) = \frac{\pm 1}{2\pi}\sum_{n\in\mathbb{Z}} \bar{J}^{\pm}_n e^{-in\varphi}, \quad \delta(\varphi-\bar{\varphi}) = \frac{1}{2\pi}\sum_{n\in\mathbb{Z}} e^{-in(\varphi-\bar{\varphi})}, \tag{8.23b}$$

one obtains the following algebra of Fourier modes

$$[\bar{J}_n, \bar{J}_m] = -kn\delta_{n+m,0}, \tag{8.24a}$$

$$[\bar{J}^a_n, \bar{J}^b_m] = (a-b)\bar{J}^{a+b}_{n+m} - \frac{k}{2}n(1-3a^2)\delta_{n+m,0}\delta_{a+b,0}. \tag{8.24b}$$

$$\tag{8.24c}$$

In order to bring this algebra into a more familiar form one can make the redefinition

$$\bar{U}^0_n := i\bar{J}^0_n, \quad \bar{U}^+_n := \frac{1}{2}\left(\bar{J}^+_n + \bar{J}^-_n\right), \quad \bar{U}^-_n := \frac{1}{2i}\left(\bar{J}^+_n - \bar{J}^-_n\right), \tag{8.25}$$

which leads to

$$[\bar{J}_n, \bar{J}_m] = -kn\delta_{n+m,0}, \tag{8.26a}$$

$$[\bar{U}^a_n, \bar{U}^b_m] = \epsilon^{ab}{}_c\bar{U}^c_{n+m} - \frac{k}{2}n\delta_{n+m,0}\delta_{a,b}, \tag{8.26b}$$

with $\epsilon^{+-}{}_0 = 1$. This algebra is consistent with the Jacobi identities, thus it is also valid for finite $k$. Hence one obtains an affine $\hat{\mathfrak{u}}(1)\oplus\hat{\mathfrak{su}}(2)$ as the asymptotic symmetry algebra in the $\bar{A}$-sector.

### 8.1.2 2-2 Embedding of $\mathfrak{sl}(2,\mathbb{R}) \hookrightarrow \mathfrak{sl}(4,\mathbb{R})$

The $2-2$ embedding of $\mathfrak{sl}(2,\mathbb{R})$ into $\mathfrak{sl}(4,\mathbb{R})$ contains one $\mathfrak{sl}(2,\mathbb{R})$ spin-2 triplet $L_n$ ($n=\pm1,0$), three more spin-2 triplets $T^a_n$ ($a=\pm1,0; n=\pm1,0$) and three spin-1 singlets $S^a$ $a=\pm1,0$ with non-vanishing commutation relations

$$[L_n, L_m] = (n-m)L_{n+m}, \tag{8.27a}$$

$$[L_n, T^a_m] = (n-m)T^a_{n+m}, \tag{8.27b}$$

$$[S^a, S^b] = (a-b)S^{a+b}, \tag{8.27c}$$

$$[S^a, T^b_m] = (a-b)(1+a(1-a+2ab))T^{a+b}_m, \tag{8.27d}$$

$$[T^0_n, T^0_m] = \frac{1}{4}(n-m)L_{n+m}, \tag{8.27e}$$

$$[T^{\pm}_n, T^0_m] = -\frac{1}{4}(-1+n^2-nm+m^2)S^{\pm}, \tag{8.27f}$$

$$[T^-_n, T^+_m] = \frac{1}{2}(n-m)L_{n+m} - \frac{1}{2}(-1+n^2-nm+m^2)S^0. \tag{8.27g}$$

In the following I will use the same notation which has already been used in the previous section for the gauge fields $A, \bar{A}$ and the boundary condition preserving gauge transformations $\varepsilon, \bar{\varepsilon}$ as given in (8.3).
In order to reduce clutter in the formulas I will also from now on drop all trivial fluctuations which do not influence the asymptotic symmetries i.e. $a_\mu^{(1)}$ and $\bar{a}_\mu^{(1)}$. More details can be found in Appendix B.1. The relevant connection which obeys (8.5) is then given by

$$a_\rho^{(0)} = L_0, \qquad a_t^{(0)} = 0, \tag{8.28a}$$

$$a_\varphi^{(0)} = \sigma L_1 - \frac{2\pi}{k}\left(\frac{\mathcal{L}}{2}L_{-1} + \sum_{a=-1}^{1}\left[(2-a^2)\mathcal{T}^a T^a_{-1} - (1-\frac{3a^2}{2})\mathcal{S}^a S^a\right]\right), \tag{8.28b}$$

where the state dependent functions $\mathcal{L}, \mathcal{T}^a$ and $\mathcal{S}^a$ only depend on the angular coordinate $\varphi$.
The connection $\bar{A}$ on the other hand has to abide the following asymptotic behavior

$$\bar{a}_\rho^{(0)} = -L_0, \qquad \bar{a}_t^{(0)} = 2S^0, \tag{8.29a}$$

$$\bar{a}_\varphi^{(0)} = -L_{-1} + \frac{2\pi}{k}\bar{\mathcal{S}}S^0. \tag{8.29b}$$

Performing the canonical analysis in the same way as in the preceding section whose details can again be found in Appendix B.1 then leads to the following semiclassical asymptotic symmetry algebra which I will denote $\mathcal{W}_4^{(2-1-1)}$ in the $A$-sector

$$[L_n, L_m] =(n-m)L_{n+m} + \frac{c}{12}n(n^2-1)\delta_{n+m,0}, \tag{8.30a}$$

$$[L_n, T^a_m] =(n-m)T^a_{n+m}, \tag{8.30b}$$

$$[L_n, S^a_m] = -mS^a_{n+m}, \tag{8.30c}$$

$$[S^a_n, S^b_m] =(a-b)S^{a+b}_{n+m} - \frac{c}{12}(1-3a^2)n\delta_{a+b,0}\delta_{n+m,0}, \tag{8.30d}$$

$$[S^a_n, T^b_m] =f(a,b)T^{a+b}_{n+m}, \tag{8.30e}$$

$$[T^\pm_n, T^\pm_m] = -\frac{9}{2c}(n-m)SS^{\pm|\pm}_{n+m}, \tag{8.30f}$$

$$[T^0_n, T^0_m] =(n-m)\left(\frac{1}{4}L_{n+m} + \frac{3}{2c}SS^{0|0}_{n+m} - \frac{3}{c}SS^{\{+|-\}}_{n+m}\right) + \frac{c}{48}n(n^2-1)\delta_{n+m,0}, \tag{8.30g}$$

$$\begin{aligned}[T^\pm_n, T^0_m] =& -\frac{1}{4}g(n,m)S^\pm_{n+m} - \frac{3}{c}\{LS\}^\pm_{n+m} + \frac{24}{c^2}\left(SSS^{\{\pm|\pm|\mp\}}_{n+m} - SSS^{\{0|0|\pm\}}_{n+m}\right)\\ &\mp\frac{3}{2c}\left((n-2m)SS^{0|\pm}_{n+m} + (2n-m)SS^{\pm|0}_{n+m} + \Omega SS^{[0|\pm]}_{n+m}\right),\end{aligned} \tag{8.30h}$$

$$\begin{aligned}[T^-_n, T^+_m] =&(n-m)\left(\frac{1}{2}L_{n+m} + \frac{12}{c}SS^{0|0}_{n+m}\right) - \frac{1}{2}g(n,m)S^0_{n+m} + \frac{c}{24}n(n^2-1)\delta_{n+m,0}\\ &-\frac{3}{2c}\left((3n-2m)SS^{+|-}_{n+m} + (2n-3m)SS^{-|+}_{n+m} + \Omega SS^{[-|+]}_{n+m}\right)\\ &-\frac{6}{c}\{LS\}^0_{n+m} + \frac{24}{c^2}SSS^{\{-|+|0\}}_{n+m} - \frac{144}{c^2}SSS^{0|0|0}_{n+m},\end{aligned} \tag{8.30i}$$

with

$$f(a,b) = (a-b)(1+a(1-a+2ab)), \tag{8.31a}$$
$$g(n,m) = (-1+n^2-nm+m^2), \tag{8.31b}$$

and the the same shorthand notation as in (8.18). The central charge $c$ in terms of the Chern-Simons level $k$ is given by $c = 12k$.
For the barred sector one obtains an affine $\hat{\mathfrak{u}}(1)$ current algebra of the form

$$[\bar{S}_n, \bar{S}_m] = kn\delta_{n+m,0}. \tag{8.32}$$

### 8.1.3 3-1 Embedding of $\mathfrak{sl}(2,\mathbb{R}) \hookrightarrow \mathfrak{sl}(4,\mathbb{R})$

The 3-1 embedding of $\mathfrak{sl}(2,\mathbb{R}) \hookrightarrow \mathfrak{sl}(4,\mathbb{R})$ contains an $\mathfrak{sl}(2,\mathbb{R})$ spin-2 triplet $L_n$ $(n = \pm, 1)$, two more spin-2 triplets $T_n^{\pm}$ $(n = \pm 1, 0)$ one spin-1 singlet $S$ and a spin-3 quintet $W_n$ $(n = \pm 2, \pm 1, 0)$ whose commutation relations are given by

$$[L_n, L_m] = (n-m)L_{n+m}, \tag{8.33a}$$
$$[L_n, J_n] = -mJ_{n+m}, \tag{8.33b}$$
$$[L_n, T_m^{\pm}] = (n-m)T_{n+m}^{\pm}, \tag{8.33c}$$
$$[L_n, W_m] = (2n-m)W_{n+m}, \tag{8.33d}$$
$$[J_n, T_m^{\pm}] = \pm T_{n+m}^{\pm}, \tag{8.33e}$$
$$[T_n^{\pm}, T_m^{\mp}] = -\frac{1}{4}(n-m)L_{n+m} \mp W_{n+m} \pm \frac{2}{3}(-1+n^2-nm+m^2)J_{n+m}, \tag{8.33f}$$
$$[W_n, T_m^{\pm}] = \frac{1}{12}(\pm 4 \mp n^2 \pm 3mn \mp 6m^2)T_{n+m}^{\pm}, \tag{8.33g}$$
$$[W_n, W_m] = \frac{1}{48}(m-n)(2n^2-nm+2m^2-8)L_{n+m}. \tag{8.33h}$$

This embedding is somehow special in the sense that it is the first non-principal embedding whose highest spin actually exceeds $s = 2$ and it is quite analogue to the spin-3 non-principal embedding. It is also worth noting that one part of the asymptotic symmetry algebra of this embedding and the spin-3 non-principal embedding ($\mathcal{W}_4^{(2)}$ and $\mathcal{W}_3^{(2)}$) actually belong to a very large class of $\mathcal{W}$-algebras called Feigin-Semikhatov algebras which are denoted by $\mathcal{W}_N^{(2)}$. As these algebras will play a prominent role for unitary holographic models which can exhibit an arbitrary (but not infinitely) large central charge it is very instructive to study the first couple of representatives of this family of $\mathcal{W}$-algebras in detail in order to see which features will play a crucial role in discussing unitarity later on.
As in the preceding section one can write down a connection $A$ in this embedding whose state dependent functions only depend on the angular coordinate $\varphi$ and which satisfies (8.5) as

$$a_\rho^{(0)} = L_0, \qquad a_t^{(0)} = 0, \tag{8.34a}$$

$$a_\varphi^{(0)} = \sigma L_1 + \frac{2\pi}{k}\left(-\frac{\mathcal{L}}{4}L_{-1} + \mathcal{T}^+T_{-1}^+ + \mathcal{T}^-T_{-1}^- + \mathcal{W}W_{-2} + \frac{4}{3}\mathcal{S}S\right). \tag{8.34b}$$

The connection $\bar{A}$ on the other hand has to satisfy the following boundary conditions

$$\bar{a}_\rho^{(0)} = -L_0, \qquad \bar{a}_\varphi^{(0)} = \frac{8\pi}{3k}\bar{\mathcal{S}}S, \qquad \bar{a}_t^{(0)} = \frac{4\sqrt{2}}{\sqrt{3}}S. \tag{8.35}$$

The canonical analysis performed in Appendix B.2 then yields

$$[L_n, L_m] = (n-m)L_{n+m} + \frac{c}{12}n(n^2-1)\delta_{n+m,0}, \tag{8.36a}$$

$$[L_n, J_n] = -mJ_{n+m}, \tag{8.36b}$$

$$[L_n, \hat{T}_m^\pm] = (n-m)\hat{T}_{n+m}^\pm, \tag{8.36c}$$

$$[L_n, \hat{W}_m] = (2n-m)\hat{W}_{n+m}, \tag{8.36d}$$

$$[J_n, J_m] = -\frac{3k}{4}n\delta_{n+m}, \tag{8.36e}$$

$$[J_n, \hat{T}_m^\pm] = \pm\hat{T}_{n+m}^\pm, \tag{8.36f}$$

$$\begin{aligned}[\hat{T}_n^\pm, \hat{T}_m^\mp] = &-\frac{1}{4}(n-m)L_{n+m} \mp \hat{W}_{n+m} \pm \frac{2}{3}(-1+n^2-nm+m^2)J_{n+m} \\ &-\frac{k}{2}n(n^2-1)\delta_{n+m,0} - \frac{1}{4k}(n-m)JJ_{n+m} \pm \frac{1}{3k}\{LJ\}_{n+m} \\ &\pm \frac{44}{27k^2}JJJ_{n+m},\end{aligned} \tag{8.36g}$$

$$\begin{aligned}[\hat{W}_n, \hat{T}_m^\pm] = &\frac{1}{12}(\pm 4 \mp n^2 \pm 3mn \mp 6m^2)\hat{T}_{n+m}^\pm \mp \frac{1}{6k}\{L\hat{T}^\pm\}_{n+m} \\ &\mp \frac{4}{9k^2}\{JJ\hat{T}^\pm\}_{n+m} - \left(\frac{1}{2k}n - \frac{1}{3k}m\right)J\hat{T}_{n+m}^\pm + \frac{1}{3k}\Omega J\hat{T}_{n+m}^\pm \\ &-\left(\frac{1}{6k}n - \frac{2}{3k}m\right)\hat{T}^\pm J_{n+m} - \frac{1}{3k}\Omega\hat{T}^\pm J_{n+m},\end{aligned} \tag{8.36h}$$

$$\begin{aligned}[\hat{W}_n, \hat{W}_m] = &\frac{1}{48}(m-n)(2n^2-nm+2m^2-8)L_{n+m} - \frac{k}{24}(n^2-4)(n^2-1)n\delta_{n+m,0} \\ &+\frac{1}{72k}(m-n)(2n^2-nm+2m^2-8)JJ_{n+m} - \frac{1}{12k}(n-m)LL_{n+m} \\ &-\frac{1}{2k}(n-m)\{\hat{T}^+\hat{T}^-\}_{n+m} - \frac{1}{27k^2}(n-m)\{LJJ\}_{n+m} \\ &-\frac{1}{27k^3}JJJJ_{n+m},\end{aligned} \tag{8.36i}$$

with $c = 24k$, as the asymptotic symmetry algebra for the $A$-sector which I will denote as $\mathcal{W}_4^{(2)}$.

The barred sector is again not quite as elaborate as the unbarred sector and has an affine $\hat{\mathfrak{u}}(1)$ governing its asymptotic symmetries

$$[\bar{S}_n, \bar{S}_m] = -\frac{3k}{4}n\delta_{n+m,0}. \tag{8.37}$$

## 8.2 $\mathcal{W}_N^{(2)}$ Lobachevsky Boundary Conditions

As a concluding part of this chapter and also to provide a smoother transition to Chapter 9, which will be mainly focused on unitarity, I present in this section a generalized version of the boundary conditions displayed in (8.6) and (8.34), which first appeared in [II].

$$a_t = 0, \qquad \bar{a}_t = \sqrt{\frac{2}{\text{Tr}\,[S^2]}}\,S, \qquad a_\rho = L_0, \qquad \bar{a}_\rho = -L_0, \tag{8.38a}$$

$$a_\varphi = \frac{1}{4\text{Tr}[L_1 L_{-1}]}\,L_1 + \frac{2\pi}{k}\,(\mathcal{L}L_{-1} + \mathcal{T}^{\pm}T^{\pm}_{-\frac{N}{2}+1} + \sum_{\ell=3}^{N-1}\mathcal{W}_\ell W^{\ell}_{-\ell+1} + \mathcal{S}S) \tag{8.38b}$$

$$\bar{a}_\varphi = -L_{-1} + \frac{2\pi}{k}\,\bar{\mathcal{S}}S \tag{8.38c}$$

The generators $S$, $L_n$, $T_n^{\pm}$ and $W_n^{\ell}$ refer to the spin-1 singlet, gravity spin-2 triplet, the two spin-$\frac{N}{2}$ $(N-1)$-plets and the spin-$\ell$ generators, respectively. This connection corresponds to the asymptotic Lobachevsky line-element

$$\mathrm{d}s^2 = 2\text{Tr}\,[L_0]^2\,\mathrm{d}\rho^2 + \mathrm{d}t^2 + \left(\tfrac{1}{4}\,e^{2\rho} + \mathcal{O}(1)\right)\,\mathrm{d}\varphi^2 + \mathcal{O}(1)\,\mathrm{d}t\,\mathrm{d}\varphi\,. \tag{8.39}$$

Following the algorithm outlined in [I] with the boundary conditions above leads to the $\mathcal{W}_N^{(2)}$ algebra discussed in section 9.2 times a $\hat{\mathfrak{u}}(1)$ current algebra with level $\kappa$ (9.45) as quantum asymptotic symmetry algebra. The current algebra part within the $\mathcal{W}_N^{(2)}$ algebra is given by

$$[J_n,\, J_m] = -\text{Tr}\left[S^2\right]k\delta_{n+m,\,0}\,. \tag{8.40}$$

Comparing with (9.38a) one can relate the level $\hat{\mathfrak{u}}(1)$ level $\kappa$ with the Chern–Simons level $k$.

$$\kappa = -\text{Tr}\left[S^2\right]k. \tag{8.41}$$

In order to compute Tr $\left[S^2\right]$ I use the following definition of the Killing form, $K_{ab} = f^d{}_{ac}\,f^c{}_{bd}$, where $f^d{}_{ac}$ are the structure constants, and normalize it in such a way that the $\mathfrak{sl}(2,\mathbb{R})$ part is given by

$$\text{Tr}\,[L_a L_b] = \begin{pmatrix} 0 & 0 & -1 \\ 0 & \frac{1}{2} & 0 \\ -1 & 0 & 0 \end{pmatrix}_{ab}\,. \tag{8.42}$$

The boundary conditions (8.38) can also be straightforwardly modified for asymptotically $\text{AdS}_3$ backgrounds by simply replacing the boundary conditions for $\bar{A}$ by an expression which is similar to the one in the $A$-sector. This then leads to two copies of the $\mathcal{W}_N^{(2)}$-algebra [136] as an asymptotic symmetry algebra, which in turn also means that all results in the following chapter can also be applied to AdS holography.

# 9 Unitarity in Non-AdS Higher-Spin Gravity

> *Sine ira et studio.*
> *(Without anger and partiality.)*
>
> **– Publius Cornelius Tacitus**
> Annals 1.1

One crucial property a consistent family of quantum gravity models should have is a corresponding QFT dual that allows for both small and large central charges of the underlying quantum symmetry algebras. This has a very simple physical reason as one should be able to discuss both the semi-classical limit (large central charge) and the quantum limit (central charge of $\mathcal{O}(1)$). If a model allowed only large central charges, then one would miss subtleties related to quantum corrections. If, on the other hand, only small central charges were possible, then the theory would not allow any holographic interpretations of things such as the huge entropy of black holes, which would require a massive amount of states and thus also a large central charge.

Having both large and small central charges at one's disposal is not the only requirement one would expect from such a family of quantum gravity models. Another often discussed aspect is whether or not there are unitary representations of the corresponding field theory duals. While one cannot completely rule out that there are satisfactory quantum gravity models which could actually be non- unitary [130, 131, 137], it is usually very hard to study such models on the gravity side.

So far most of the models studied were not able to incorporate both of these requirements. Therefore it is highly motivating to search for models which allow for both unitary representations and finite and arbitrary large central charge. Higher-spin gravity in $2+1$ dimensions can provide such a family of models, which I will show in this chapter. To be more precise it will turn out that a specific class of non-principal embeddings of $\mathfrak{sl}(2,\mathbb{R}) \hookrightarrow \mathfrak{sl}(N,\mathbb{R})$ will lead to a QFT which allows both unitary representations and arbitrary small and large (but not infinite) values of the central charge.

Before going into more detail I want to emphasize an important property of non-principal embeddings which actually already puts a constraint on the central charge.

All non-principal embeddings contain a singlet factor which turns into a Kac-Moody algebra when looking at the asymptotic symmetries

$$[J_n,\, J_m] = \kappa\, n\, \delta_{n+m,0} + \dots \tag{9.1}$$

where the ellipsis refers to possible non-abelian terms. Now considering a Verma module built only from descendants using (9.1) one finds that in order to have no states with negative norm, i.e. a unitary representation, the level $\kappa$ has to be non-negative. On the other hand, all embeddings have by construction an $\mathfrak{sl}(2,\mathbb{R})$ factor that translates to a Virasoro algebra as part of the asymptotic symmetry algebra,

$$[L_n,\, L_m] = (n-m)\, L_{n+m} + \frac{c}{12}\, n(n^2-1)\, \delta_{n+m,0}\,. \tag{9.2}$$

Using the same reasoning as above one finds that also the central charge $c$ has to be non-negative. In [138] it was shown that $\kappa$ and $c$ are not independent quantities, and that for large values of the central charge the signs of $c$ and $\kappa$ are always opposite to each other

$$\text{sign}(c) = -\text{sign}(\kappa), \qquad \text{if } |c| \to \infty. \tag{9.3}$$

This argument shows already that for non-principal embeddings one cannot make the central charge infinitely large without violating unitarity. The argument does not, however, rule out the possibility that the central charge can be very large and the corresponding representations are unitary at the same time. As long as the central charge is not infinite one does not necessarily violate unitarity. The main point of this chapter will be to show that there indeed exists such a non-principal embedding corresponding to a $\mathcal{W}_N^{(2)}$ algebra where one can tune the central charge to be very large and retain unitarity at the same time. The main result of this chapter will be that the maximum value of the central charge $c$ of this non-principal embedding is bounded by $N$ which one can assume to be arbitrarily large but finite. This bound for odd $N$ is given by

$$c \leq \frac{N}{4} - \frac{1}{8} - \mathcal{O}(1/N). \tag{9.4}$$

For even $N$ it has been shown in [139] that the Virasoro central charge in unitary representations of $\mathcal{W}_N^{(2)}$ is allowed to take the values

$$c = (1-\beta)\frac{1 - 2N + \beta N \frac{N-2}{N-1}}{1 + \frac{\beta}{N-1}} \sim (\beta-1)(2-\beta)N, \tag{9.5}$$

where $\beta$ is some parameter which depends on $N$ and a level $k$ (which is not to be confused with the Chern-Simons level) whose allowed range for large $N$ is $[1,2]$. Thus, for arbitrarily large $N$ the central charge can also become arbitrarily large as long as the bounds (9.4) and (9.5) are obeyed.

In this chapter I will first analyze in detail the possible values of the central charge which allow for unitary representations of the asymptotic symmetry algebras derived in the previous chapter. I will then continue in analyzing unitary representations of $\mathcal{W}_N^{(2)}$-algebras, also known as Feigin-Semikhatov algebras, which will eventually lead to (9.4) and (9.5).

## 9.1 Unitarity in Spin-4 Lobachevsky Holography

Spin-4 gravity is the easiest higher-spin model where several different non-principal embeddings exist (see [36, 140] for explicit results). I have already shown in section 8.1 how one can implement all of the three possible non-principal embeddings of $\mathfrak{sl}(2,\mathbb{R}) \hookrightarrow \mathfrak{sl}(4,\mathbb{R})$ that are of relevance in the context of Lobachevsky holography and determined the corresponding semiclassical symmetry algebras. In this section I will first present the quantum versions of these asymptotic symmetry algebras and then continue in analyzing possible unitary representations thereof. Since the asymptotic symmetry algebras, which are related to the connection $\bar{A}$ for all these embeddings, are either affine $\hat{\mathfrak{u}}(1)$ or $\hat{\mathfrak{u}}(1) \oplus \hat{\mathfrak{su}}(2)$ and their central terms all have the same sign as their counterparts in the $A$-sector I will only focus on finding unitary representations of the $\mathcal{W}$-algebras appearing in the $A$-sector.

### 9.1.1 2-1-1 Embedding

Since the asymptotic analysis presented in Section 8.1.1 is a semiclassical one, the resulting asymptotic symmetry algebra is also semiclassical, i.e. it is only valid for large central charges. As I am also interested in finite values of the central charges, the first thing to do is to find the corresponding quantum algebra which is also valid for small values of the central charges. One way to determine this algebra is to first define a vacuum state and fix the action of the generators of the corresponding symmetry algebra on this vacuum. One is dealing with nonlinear algebras in all the cases treated in this thesis, and therefore it is also important to think about a proper normal ordering description so that the vacuum expectation values of these nonlinear operators vanishes. After having introduced this normal ordering one has to check that the resulting algebra still satisfies the Jacobi identities

$$[A_n,[B_m,C_l]]+[C_l,[A_n,B_m]]+[B_m,[C_l,A_n]]=0, \tag{9.6}$$

for some generators $A_n, B_m, C_l$. This requirement of being consistent with the Jacobi identities is usually sufficient to completely fix all the structure constants of the quantum algebra. Since determining the quantum algebra via this procedure involves rather lengthy calculations, I decided to put the details of these computations for all the embeddings of $\mathfrak{sl}(2,\mathbb{R}) \hookrightarrow \mathfrak{sl}(4,\mathbb{R})$ in Appendix C.

The first important ingredient for everything that follows is the definition of the vacuum state which will be used. I will define the vacuum to be the state with the highest amount of symmetry i.e.

$$A_n^s|0\rangle = 0, \quad \forall n \geq 1-s, \tag{9.7}$$

where $A_n^s$ denotes any of the Fourier modes with spin $s$ generating the asymptotic symmetry algebra in question. Another way to phrase this is that this is a vacuum which is invariant under the global part of the asymptotic symmetry algebra. For the 2-1-1 embedding this would be (8.2).
States can then be generated from that vacuum as

$$A_n^s|0\rangle \neq 0, \quad \forall n < 1-s. \tag{9.8}$$

One can now define a normal ordering description as

$$:AB:_n = \sum_{p\geq -b+1} A_{n-p}B_p + \sum_{p<-b+1} B_p A_{n-p}, \tag{9.9}$$

where $b$ is the spin of the operator $B$. Consequently, repeated application of this normal ordering procedure yields the normal ordered expressions for trilinear and quadrilinear terms with an example for trilinear terms shown in (C.2).
After having properly defined a vacuum state and subsequently introduced a normal ordering prescription one can now proceed in determining how this prescription changes the asymptotic symmetry algebras in question as I show in detail in Appendix C. For the $\mathcal{W}_4^{(2-1-1)}$ algebra this leads to the following quantum algebra

$$[\hat{L}_n, \hat{L}_m] =(n-m)\hat{L}_{n+m} + \frac{\mathfrak{c}}{12}n(n^2-1)\delta_{m+n,0}, \tag{9.10a}$$

$$[\hat{L}_n, J_m^a] =-mJ_{n+m}^a, \tag{9.10b}$$

$$[\hat{L}_n, S_m] =-mS_{n+m}, \tag{9.10c}$$

$$[\hat{L}_n, G_m^{a|b}] = \left(\frac{n}{2}-m\right)G_{n+m}^{a|b}, \tag{9.10d}$$

$$[J_n^a, J_m^b] =(a-b)J_{m+n}^{a+b} + \kappa_2\left(1-3a^2\right)n\delta_{a+b,0}\delta_{n+m,0}, \tag{9.10e}$$

$$[J_n^a, G_m^{b|c}] =\frac{a-b}{2}G_{n+m}^{2a+b|c}, \tag{9.10f}$$

$$[S_n, G_m^{a|b}] =2bG_{n+m}^{a|b}, \tag{9.10g}$$

$$[S_n, S_m] =\kappa n\delta_{n+m}, \tag{9.10h}$$

$$\begin{aligned}[G_n^{a|b}, G_m^{c|d}] =&\delta_{b+d,0}\left(\delta_{a,c}\left(-k(m-n)J_{n+m}^c + \frac{d}{2}\{:SJ:\}_{n+m}^c\right)\right.\\ &+\delta_{a+c,0}\left(-c(k-2)\hat{L}_{n+m} + \frac{cd}{2}(k+1)(m-n)S_{m+n}\right.\\ &+c\left(:JJ:_{n+m}^{\{+|-\}} - \frac{3}{8}:SS:_{n+m} + \frac{cd}{2}\{:SJ:\}_{n+m}^0 - 2:JJ:_{n+m}^{0|0}\right)\\ &\left.\left.-k(m-n)J_{m+n}^0 - c\lambda_{N-1}(N,k)\left(n^2-\frac{1}{4}\right)\delta_{m+n,0}\right)\right),\end{aligned} \tag{9.10i}$$

with the Virasoro central charge

$$\mathfrak{c} = \frac{3(k+2)(2k+1)}{k-2}, \tag{9.11}$$

$\hat{\mathfrak{u}}(1)$ level

$$\kappa = -4k, \tag{9.12}$$

$\hat{\mathfrak{su}}(2)$ central extension

$$\kappa_2 = -\frac{k+1}{2}, \tag{9.13}$$

and the central term appearing in the $G_n^{a|b}$ commutator

$$\lambda = k(k+1). \tag{9.14}$$

I will look for unitary representations of this algebra by taking a closer look at the norm of states which can be produced by repeated application of the generators (9.10) on the vacuum. In order to do this one also has to define a sensible way of the action of hermitian conjugation on the generators. For the generators $L_n, S_n, G_n^{a|b}$ and $J_n^0$ this can be defined as

$$L_n^\dagger := L_{-n}; \quad S_n^\dagger := S_{-n}, \quad \left(G_n^{a|b}\right)^\dagger := G_{-n}^{-a|-b}, \quad \left(J_n^0\right)^\dagger := J_{-n}^0. \tag{9.15}$$

The hermitian conjugate of the operators $J_n^\pm$ is a little bit trickier. This can can be seen by considering the following ansatz

$$\left(J_n^\pm\right)^\dagger = \alpha J_{-n}^\mp. \tag{9.16}$$

Looking at

$$\left([J_n^\pm, J_m^\mp]\right)^\dagger = -\alpha^2 [J_{-n}^\mp, J_{-m}^\pm] = \pm J_{-(n+m)}^0 + (k+1) n \delta_{n+m,0}, \tag{9.17}$$

one finds that $\alpha^2 = 1$ in order to satisfy (9.10e). Next looking at

$$\left([J_n^\pm, G_m^{\mp|a}]\right)^\dagger = \alpha [J_{-n}^\mp, G_{-m}^{\pm|-a}] = \left(\pm G_{n+m}^{\pm|a}\right)^\dagger = \pm G_{-n+m}^{\mp|-a}, \tag{9.18}$$

one even finds that $\alpha = -1$ is necessary in order for the definition of hermitian conjugation in (9.15) to be compatible with the algebraic relations in (9.10).
The first level where one can create states from the vacuum is at $n = \frac{3}{2}$. At this level there are two states generated by $G_{-\frac{3}{2}}^{\pm|\pm}$ whose Gramian matrix is given by

$$K^{(\frac{3}{2})} = 2\lambda_{N-1}(N,k) \left( \begin{array}{c|cc} & G_{-\frac{3}{2}}^{+|+} & G_{-\frac{3}{2}}^{-|-} \\ \hline \left(G_{-\frac{3}{2}}^{+|+}\right)^\dagger & 1 & 0 \\ \left(G_{-\frac{3}{2}}^{-|-}\right)^\dagger & 0 & -1 \end{array} \right). \tag{9.19}$$

Since these states have a norm with opposite sign the only way to have a unitary representation, i.e. a representation without negative norm states, is for $\lambda = 0$ which fixes[1] $k = -1$. For $k = -1$ the $\hat{\mathfrak{su}}(2)$ central extension $\kappa_2$ vanishes as well, which corresponds to a simple theory of a single $\hat{\mathfrak{u}}(1)$ current algebra with Virasoro central charge $\mathfrak{c} = 1$. Thus, the 2-1-1 embedding and its associated asymptotic symmetry algebra is not a good candidate for a QFT that has both unitary representations and a central charge which can take values much larger than $\mathcal{O}(1)$.

## 9.1.2 2-2 Embedding

Since the $\mathcal{W}_4^{(2-1-1)}$ algebra treated in section 9.1.1 did not prove to be a good candidate for having both unitary representations and a central charge larger than $\mathcal{O}(1)$, I now want to take a look at the 2-2 embedding and its corresponding asymptotic $\mathcal{W}_4^{(2-2)}$-algebra. Starting from (8.30) and following the procedure described in Appendix C.3.2 commutators of the quantum $\mathcal{W}_4^{(2-2)}$-algebra which will be relevant for the following discussion on unitary representations are given by

$$[L_n, L_m] =(n-m)L_{n+m} + \frac{c}{12}n(n^2-1)\delta_{n+m,0}, \tag{9.20a}$$

$$[S_n^a, S_m^b] =(a-b)S_{n+m}^{a+b} + \kappa(1-3a^2)n\delta_{a+b,0}\delta_{n+m,0}, \tag{9.20b}$$

$$[T_n^a, T_m^b] =c_T n(n^2-1)\delta_{n+m,0}\delta_{a+b,0} + \ldots, \tag{9.20c}$$

$$\tag{9.20d}$$

with

$$\kappa = \frac{1}{24}\left(7 - c - \sqrt{c^2 - 110c + 145}\right), \tag{9.21}$$

the $T_n^a$ central term

$$c_T = -\frac{(2\kappa-1)(3\kappa+2)}{24(\kappa+2)}, \tag{9.22}$$

and the Virasoro central charge

$$c = 12k. \tag{9.23}$$

The ellipsis in the last commutator refers to terms which are known and are explicitly shown in (C.15) but irrelevant when looking for unitary representations. I will use the same definition for hermitian conjugation of the operators generating $\mathcal{W}_4^{(2-2)}$ as in section 9.1.1, i.e.

$$L_n^\dagger := L_{-n}; \quad \left(S_n^0\right)^\dagger := S_{-n}^0, \quad (T_n^a)^\dagger := T_{-n}^{-a}, \quad (S_n^\pm)^\dagger := -S_{-n}^\mp. \tag{9.24}$$

In order to check for unitary representations I will now again check the Gramian matrices up to level 3 for the absence of negative eigenvalues.

[1]The value $k = 0$ is a trivial solution since for this case the Chern-Simons action simply is zero.

### Level 1

At level 1 there are three states present

$$S^0_{-1}|0\rangle, \quad S^+_{-1}|0\rangle, \quad \text{and} \quad S^-_{-1}|0\rangle, \tag{9.25}$$

and the Gramian is given by

$$K^{(1)} = \kappa \left( \begin{array}{c|ccc} & S^+_{-1} & S^0_{-1} & S^-_{-1} \\ \hline \left(S^+_{-1}\right)^\dagger & 2 & 0 & 0 \\ \left(S^0_{-1}\right)^\dagger & 0 & 1 & 0 \\ \left(S^-_{-1}\right)^\dagger & 0 & 0 & 2 \end{array} \right). \tag{9.26}$$

Thus at level 1 one has to require $\kappa \geq 0$, in order for negative norm states to be absent.

### Level 2

At level 2 there are 13 states present

$$L_{-2}|0\rangle, \quad T^a_{-2}|0\rangle, \quad S^a_{-2}|0\rangle, \quad S^a_{-1}S^+_{-1}|0\rangle, \quad \left(S^0_{-1}\right)^2|0\rangle, \quad S^-_{-1}S^0_{-1}|0\rangle, \quad \left(S^-_{-1}\right)^2|0\rangle, \tag{9.27}$$

with $a = \pm, 0$. The corresponding Gramian matrix can be determined straightforwardly. I will, however, refrain from showing it explicitly at this stage because the resulting matrix is simply too large to be displayed properly. Looking at the eigenvalues of the Gramian, at least on first sight, only two values of $\kappa$ allow for unitary representations

$$\kappa_1 = 0 \quad \text{and} \quad \kappa_2 = \frac{1}{2}. \tag{9.28}$$

- $\kappa_1 = 0$: For this value of $\kappa$, $c$ and $c_T$ would take the following values

  $$c = c_T = 1. \tag{9.29}$$

  Since some of the structure constants of (C.15) are proportional to $\frac{1}{\kappa}$ the whole algebra is ill defined for that value of $\kappa$, which is why this solution will be discarded.

- $\kappa_2 = \frac{1}{2}$: For this value of $\kappa$, $c$ and $c_T$ take the following values

  $$c = 1 \quad \text{and} \quad c_T = 0. \tag{9.30}$$

  Thus, in this case the only states at level 2 which are not null are $L_{-2}$, $S^a_{-2}$ and $(S^0_{-1})^2$.

#### Level 3

Since the number of states grows very fast with increasing level it is more efficient to only work with the states which are not null after the analysis of level 1 and 2. Thus for $\kappa = \frac{1}{2}$ a total of 17 states are present

$$L_{-3}|0\rangle, \quad S^a_{-3}|0\rangle, \quad S^a_{-1}S^b_{-2}|0\rangle, \quad S^a_{-1}\left(S^0_{-1}\right)^2|0\rangle, \quad S^+_{-1}S^0_{-1}S^-_{-1}|0\rangle, \tag{9.31}$$

with $a, b = \pm, 0$. The Gramian has again too many entries to be displayed in this thesis. Calculating the eigenvalues of that matrix shows that for $\kappa_2 = \frac{1}{2}$ the Gramian is positive semidefinite and thus allows for a unitary representation.
Hence looking at the first three levels one can already infer that there is only one value of $\kappa$ that allows for unitary representations, namely $\kappa = \frac{1}{2}$, which corresponds to $c = 1$. The information obtained up until this point, i.e. up until level 3, suggests that all states at higher levels have non-negative norm as well for $\kappa = \frac{1}{2}$. Thus, as in the 2-1-1 embedding, the only value of the Virasoro central charge where the $\mathcal{W}_4^{(2-2)}$-algebra allows for unitary representations is given for $c = 1$.

### 9.1.3 3-1 Embedding

Last but not least I will review unitary representations of the $\mathcal{W}_4^{(2)}$-algebra, which made an appearance as part of the asymptotic symmetry algebra of Lobachevsky holography in the 3-1 embedding and have been studied in [139]. Again starting from the semiclassical commutation relations (8.36) and following the steps outlined in Appendix C.3.3 one finds the quantum $\mathcal{W}_4^{(2)}$ algebra as shown in (C.18). However, for the purpose of discussing unitary representations of this algebra a presentation closer to the one in [136] is more convenient. Again only focusing on the terms which are relevant for the discussion of unitary representations the relevant terms of the $\mathcal{W}_4^{(2)}$-algebra are given by

$$[J_n,\, J_n] = \kappa n \delta_{n+m,\,0}, \tag{9.32a}$$

$$[J_n,\, L_m] = n J_{n+m}, \tag{9.32b}$$

$$[J_n,\, T^\pm_m] = \pm\, T^\pm_{n+m}, \tag{9.32c}$$

$$[L_n,\, L_m] = (n-m)L_{n+m} + \frac{c}{12}n(n^2-1)\delta_{n+m,\,0}, \tag{9.32d}$$

$$[L_n,\, T^\pm_m] = \left(n(\tfrac{N}{2}-1) - m\right) T^\pm_{n+m}, \tag{9.32e}$$

$$[T^+_n,\, T^-_m] = \lambda_3(4, k_{\text{FS}})\, f(n)\, \delta_{n+m,\,0} + g(n,m)\lambda_2(4, k_{\text{FS}}) J_{n+m} \ldots, \tag{9.32f}$$

$$[W_n,\, \text{anything}] = \ldots, \tag{9.32g}$$

with $\hat{\mathfrak{u}(1)}$ level

$$\kappa = \frac{3}{4}k_{\text{FS}} + 2, \tag{9.33}$$

the Virasoro central charge

$$c = -\frac{(3k_{\text{FS}}+8)(8k_{\text{FS}}+17)}{k_{\text{FS}}+4}, \tag{9.34}$$

and

$$\lambda_n(N, k_{\text{FS}}) = \prod_{i=1}^{n}(i(k_{\text{FS}}+N-1)-1). \tag{9.35}$$

The relation between $k_{\text{FS}}$ and the Chern-Simons level $k$ can be found in (C.21). The exact form of the functions $f(n)$ and $g(n,m)$ does not matter for the discussion of unitary representations.

As argued in [139], there are two ways to study this algebra. The first is for $\lambda_2(4, k_{\text{FS}}) = 0$ and the other one is for $\lambda_2(4, k_{\text{FS}}) \neq 0$.

**$\lambda_2(4, k_{\text{FS}}) = 0$:** Following the exact same procedure as in sections 9.1.1 and 9.1.2 one finds two possible solutions for $k_{\text{FS}}$ where negative norm states are absent and $\lambda_2(4, k_{\text{FS}}) = 0$, $k_{\text{FS}} = -\frac{8}{3}$ leading to $\kappa = c = 0$, and $k_{\text{FS}} = -\frac{5}{2}$ leading to $\kappa = \frac{1}{8}$ and $c = 1$. Thus, the number of values for $c$ that allow for unitary representations as compared to section 9.1.1 and 9.1.2 has not increased.

**$\lambda_2(4, k_{\text{FS}}) \neq 0$:** In [139] it has been argued that one can define new generators as

$$X_n = \frac{i(T_n^+ + T_n^-)}{2\sqrt{\lambda_2(4, k_{\text{FS}})}}, \quad Y_n = \frac{(T_n^+ - T_n^-)}{2\sqrt{\lambda_2(4, k_{\text{FS}})}}, \tag{9.36}$$

in addition to $\tilde{J}_n = iJ_n$ (and $Z_n = iW_n$). Using these generators the authors found that for

$$\frac{4}{3} \leq k_{\text{FS}} + 4 \leq \frac{15}{8}, \tag{9.37}$$

negative norm states are absent, thus yielding a unitary representation of $\mathcal{W}_4^{(2)}$.

It is encouraging that there are additional solutions as compared to the spin 3 case and it is likely that this trend continues for analogues of $\mathcal{W}_4^{(2)}$ which contain spins larger than $s = 3$. I will show in the next section that this trend indeed continues for $\mathcal{W}_N^{(2)}$ and allows one to find unitary models for arbitrarily large values of the central charge, provided one tunes $N$ to take sufficient large values.

## 9.2 Unitarity in $\mathcal{W}_N^{(2)}$ Gravity

As mentioned in the beginning of this chapter I want to find a family of (quantum) gravity models that are unitary and allow for $\mathcal{O}(1)$ as well as arbitrarily large values of the central charge. In section 9.1 I showed that one promising candidate for such a family of models could be given the next-to-principal embedding of $\mathfrak{sl}(2,\mathbb{R})$ into $\mathfrak{sl}(N,\mathbb{R})$ (the $(N-1)$-1 embedding) and called the corresponding dual theory $\mathcal{W}_N^{(2)}$ gravity. In this context *next-to-principal* means that the algebra contains all integer spins up to $N-1$, but not spin-$N$, which I will denote by $L_n$ (spin-2) and $W_n^s$ (spin-3...$s$). In addition, there is always a pair of spin-$\frac{N}{2}$ generators, $T_n^{\pm}$ and a singlet $J_n$. The first member of this family of $\mathcal{W}$-algebras which follows this classification is found at $N=3$ and leads to the Polyakov–Bershadsky algebra (7.13). The next-simplest case, $N=4$, leads to the $W_4^{(2)}$ algebra encountered in section 9.1.3.

Since the main question I want to answer is related to unitary representations I will first focus on the field theory side, which is related to the boundary conditions presented in section 8.2. The expected quantum asymptotic symmetry algebra for AdS (Lobachevsky) holography consists of two copies (a $\hat{u}(1)$ current algebra and one copy) of the $\mathcal{W}_N^{(2)}$ algebra, introduced by Feigin and Semikhatov [136]. One way to understand the algebraic structure of the $\mathcal{W}_N^{(2)}$-algebra is as a Drinfeld–Sokolov reduction of a specific non-principal embedding of $\mathfrak{sl}(2,\mathbb{R})$ into $\mathfrak{sl}(N,\mathbb{R})$.

First I will review the most relevant aspects of the $\mathcal{W}_N^{(2)}$-algebra [136], which depends on the parameter $N$ and a level $k$. This should not be confused with the Chern-Simons level, which appeared previously as a parameter in the $\mathcal{W}$-algebras treated in this thesis.

$$[J_n,\, J_m] = \kappa\, n\, \delta_{n+m,\,0}, \tag{9.38a}$$
$$[J_n,\, L_m] = n J_{n+m}, \tag{9.38b}$$
$$[J_n,\, T_m^{\pm}] = \pm\, T_{m+n}^{\pm}, \tag{9.38c}$$
$$[L_n,\, L_m] = (n-m)L_{m+n} + \frac{c}{12}\, n(n^2-1)\, \delta_{n+m,\,0}, \tag{9.38d}$$
$$[L_n,\, T_m^{\pm}] = (n(\tfrac{N}{2}-1)-m)\, T_{n+m}^{\pm}, \tag{9.38e}$$
$$[T_n^{+},\, T_m^{-}] = \lambda_{N-1}(N,k)\, f(n)\, \delta_{n+m,\,0} + g(n,m)\lambda_{N-2}(N,k) J_{n+m} \dots, \tag{9.38f}$$
$$[W_n^s,\, \text{anything}] = \dots. \tag{9.38g}$$

The commutators which involve the generators $J_n$, $L_n$ and $T_n^{\pm}$ all exhibit central terms that will play a crucial role when looking for unitary representations. The commutators involving higher spin generators $W_n^s$, with $s = 3, 4, \dots, (N-1)$, on the other hand do not play a vital role, because their central charge is determined by the Virasoro central charge $c$ and is positive if $c$ is bigger than unity. The $\hat{\mathfrak{u}}(1)$ level is given by

$$\kappa = \frac{N-1}{N}\, k + N - 2, \tag{9.39}$$

the Virasoro central charge by

$$c = -\frac{((N+k)(N-1)-N)((N+k)(N-2)N-N^2+1)}{N+k}, \tag{9.40}$$

and the central term in the $T^{\pm}$ commutator by

$$\lambda_{N-1}(N,k) = \prod_{m=1}^{N-1} \left(m(N+k-1)-1\right), \tag{9.41}$$

where again the exact form of $f(n)$ and $g(n,m)$ does not matter for the discussion of unitary representations.

At this point it is also interesting to note that for each value of the level $k$ there exists a dual value $\tilde{k}$ that leads to the same expression for the central charge (9.40)

$$\tilde{k} = \frac{N+1}{N-2}\frac{1}{N+k} - N. \tag{9.42}$$

This duality is involutive, i.e. $\tilde{\tilde{k}} = k$.

Since there are different constraints on unitarity if $N$ is even or odd, I will also split the following discussion in an even and an odd part.

### 9.2.1 Odd $N$

For the purpose of taking the large (but finite) $N$ limit, it turns out to be useful to parametrize the level $k$ in terms of a constant $\alpha$, defined by

$$k = -N + 1 + \frac{\alpha+1}{N-1}. \tag{9.43}$$

The duality (9.42) acts on the parameter $\alpha$ as follows.

$$\tilde{\alpha} = \frac{N(N(1-\alpha)+2\alpha-1)+1}{(N-2)(N+\alpha)} = 1 - \alpha + \mathcal{O}(1/N). \tag{9.44}$$

For large positive values of $N$, the level $k$ approaches large negative values for $\alpha \sim \mathcal{O}(1)$, but in such a way that the sum of $N+k$ remains close to one. The subleading term proportional to $\alpha$ in the definition (9.43) will play a crucial role in the following discussion of unitarity.

Inserting the parametrization (9.43) into the result for $\kappa$ in (9.39) yields

$$\kappa = \frac{\alpha}{N}. \tag{9.45}$$

From the analyses in section 9.1 it is already clear that the $\hat{\mathfrak{u}(1)}$ level $\kappa$ and the Virasoro central charge $c$ both have to be positive as a minimal requirement for unitary representations of non-principally embedded $\mathcal{W}$-algebras. Thus, non-negativity of $\kappa$ imposes a first restriction on the parameter $\alpha$,

$$0 \le \alpha. \tag{9.46}$$

A second, and even stronger constraint on $\alpha$ originates from the central charge (9.40), which written in terms of $\alpha$ reads

$$c = \alpha(1-\alpha)\,N + \alpha\,(\alpha^2 + \alpha - 1) - \sum_{m=1}^{\infty}(1+\alpha)^2(1-\alpha)\Big(-\frac{\alpha}{N}\Big)^m . \qquad (9.47)$$

Positivity of the central charge restricts $\alpha$ to the following interval

$$0 \leq \alpha \leq \frac{N(N-1)+1}{N(N-2)}. \qquad (9.48)$$

In the large $N$ limit $\alpha$ is then essentially restricted to the interval $[0,\,1]$. The inequalities (9.48) and (9.46) are compatible with each other. Therefore, non-negativity of the $\hat{\mathfrak{u}}(1)$ level is not in contradiction with non-negativity of the central charge. Moreover, the central charge (9.47) scales linearly with $N$, and thus can get arbitrarily large if one allows arbitrarily large spins, see Fig. 9.1 where $c$ is plotted for $N = 101$. Taking only these two constraints into account one has very encouraging results, i.e. an arbitrary large Virasoro central charge which is not at odds with unitarity. This is true at least if one is taking only the Virasoro and affine $\hat{\mathfrak{u}}(1)$ modes into account.
However, there is another even more restrictive constraint for odd $N$, tightly linked with the $T^\pm$ modes. As in the $\mathcal{W}_3^{(2)}$ case studied in [I], the $T_n^\pm$ sector (denoted $G_n^\pm$ in [I]) generically leads to states whose norms have opposite signs. Thus, the only way to retain unitarity is to force these states to be null, i.e. the central term $\lambda_{N-1}(N,k)$ from (9.41) has to vanish. Requiring $\lambda_{N-1}(N,k)$ to vanish establishes a polynomial equation for $\alpha$ of degree $N-1$:

$$\lambda_{N-1}(N,k) = \prod_{m=1}^{N-1}\Big(m\,\frac{\alpha+1}{N-1} - 1\Big) \stackrel{!}{=} 0. \qquad (9.49)$$

Thus, there are $N-1$ solutions for $\alpha$ compatible with vanishing $\lambda$.

$$\lambda_{N-1}(N,k) = 0 \quad \Leftrightarrow \quad \alpha \in \Big\{0,\, \frac{1}{N-2},\, \frac{2}{N-3},\, \dots,\, \frac{N-4}{3},\, \frac{N-3}{2},\, N-2\Big\}. \qquad (9.50)$$

All of these solutions are real and non-negative, but not all of them obey the inequalities (9.48). Selecting those $\alpha$ from (9.50) that obey the inequalities (9.48) leads to $\frac{(N+1)}{2}$ solutions for odd $N$. In conclusion, one obtains the following list of allowed rational values for $\alpha$:

$$\alpha = \Big\{0,\, \frac{1}{N-2},\, \frac{2}{N-3},\, \frac{3}{N-4},\, \dots,\, \frac{N-7}{N+5},\, \frac{N-5}{N+3},\, \frac{N-3}{N+1},\, 1\Big\} \qquad (9.51)$$

For each $\alpha$ in (9.51) the $\hat{\mathfrak{u}}(1)$ level $\kappa$ and the Virasoro central charge $c$ are non-negative, and the $T_n^\pm$ descendants of the vacuum are all null states. Moreover, for non-vanishing $c$ the inequality $c \geq 1$ holds. Since the central terms appearing in the $W^s$ part of the algebra are all proportional to $c - 1$ also these central terms are always non-negative. Thus, one can have unitary representations of the algebra (9.38) for the values of $\alpha$ appearing in (9.51).

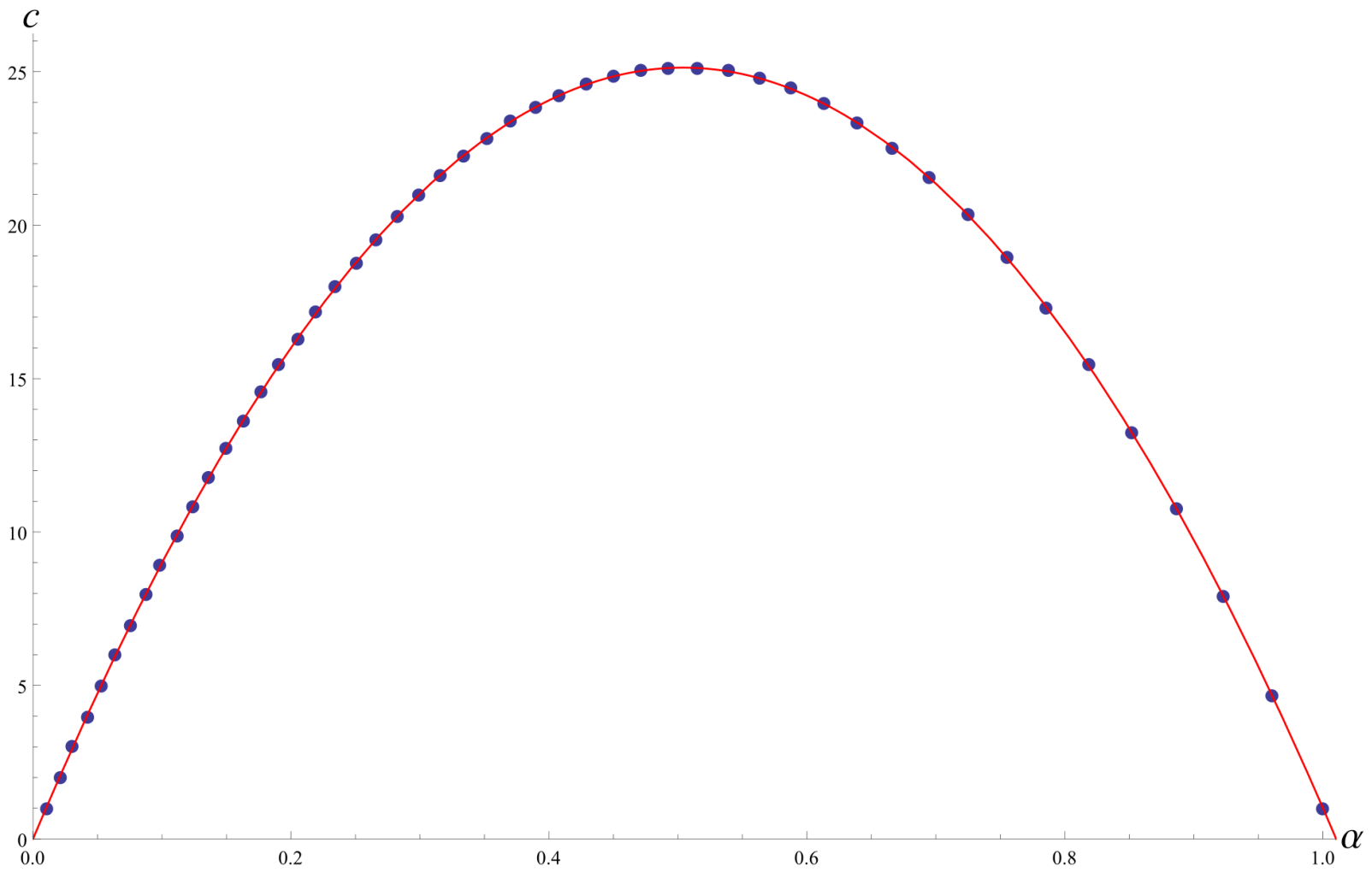

**Fig. 9.1.:** Virasoro central charge $c$ as function of the parameter $\alpha$ for $N = 101$. Red solid curve allowed by positivity. Blue dots allowed by unitarity.

### 9.2.2 Even $N$

The even $N$ case which was first discussed in [139] is an extension of Section 9.1.3 and thus also incorporates the following redefinition of the generators

$$X_n = \frac{i(T_n^+ + T_n^-)}{2\sqrt{\lambda_{N-2}(N, k_{\text{FS}})}}, \quad Y_n = \frac{(T_n^+ - T_n^-)}{2\sqrt{\lambda_{N-2}(N, k_{\text{FS}})}}, \tag{9.52}$$

in addition to $\tilde{J}_n = iJ_n$ (and $Z_n^s = iW_n^s$). This redefinition[2] can be done as long as $\lambda_{N-2}(N, k_{\text{FS}}) \neq 0$.

In order for the algebra (9.38) to allow for unitary representations one then finds the following restriction on $k_{\text{FS}}$

$$\frac{N}{N-1} \leq k_{\text{FS}} + N \leq \frac{N^2 - 1}{N(N-2)}. \tag{9.53}$$

In order to perform a large $N$ limit it is useful to introduce a constant $\beta$ by

$$\beta = (N-1)(k_{\text{FS}} + N - 1). \tag{9.54}$$

Then the condition (9.53) translates to

$$1 \leq \beta \leq \frac{(2N-1)(N-1)}{N(N-2)} = 2 + \frac{N+1}{N(N-2)}, \tag{9.55}$$

[2] For $\lambda_{N-2}(N, k_{\text{FS}}) = 0$ one finds similar restrictions as for odd $N$. The parameter $\alpha$ then takes the values $\alpha = \left\{0, \frac{1}{N-2}, \frac{2}{N-3}, \frac{3}{N-4}, \ldots, \frac{N-8}{N+6}, \frac{N-6}{N+4}, \frac{N-4}{N+2}, 1 - \frac{2}{N}\right\}$.

which in the large $N$ limit restricts $\beta$ to lie in the interval $[1,2]$ and the Virasoro central charge to scale as in (9.5).
Thus, one can see that there is a profound difference between even and odd $N$. Whereas in the odd case only a discrete set of unitary values of the central charge has been found, in the even case there is a continuous interval of possible values[3].

## 9.3 Physical Discussion

I will now provide a physical discussion for odd $N$ in terms of the $\hat{u}(1)$ level $\kappa$ (9.45) and the Virasoro central charge $c$ (9.47) first given in [II]. For a more detailed discussion of the even case, please refer to [139].
Continuing in the list (9.51) the allowed values of $\alpha$ are given by

$$\alpha = \frac{\hat{N}}{N - \hat{N} - 1}, \qquad \hat{N} \in \mathbb{N},\ N \geq 2\hat{N} + 1\,. \tag{9.56}$$

Defining $m := N - 2\hat{N} - 1$, the central charge (9.40) can be rewritten as

$$c(\hat{N},\, m) - 1 = (\hat{N} - 1)\left(1 - \frac{\hat{N}(\hat{N}+1)}{(m+\hat{N})(m+\hat{N}+1)}\right). \tag{9.57}$$

Thus, one obtains for the central charge (9.57) the values of the $W_{\hat{N}}$-minimal models (see e.g. [141]), shifted by 1 from the $\hat{u}(1)$ current algebra. For $\hat{N} = 2$ and arbitrary, but even $m$, half of the values of the Virasoro minimal models are obtained for the bare central charge $c - 1$.
For the following discussion I will divide the spectrum of allowed values of $\alpha$ (9.51) into three regimes, a quantum regime for small values of $\alpha$, a semi-classical regime for generic values of $\alpha$ in the interval $(0,\, 1)$ and a "dual"[4] quantum regime for values of $\alpha$ close to one.

### Quantum Regime

The strong coupling limit on the gravity side corresponds to a very small Chern–Simons level $k$, which implies that one should choose the smallest possible values of $\alpha$. The value $\alpha = 0$ is always possible and trivial, i.e. $\kappa = c = 0$, and the only state in the theory is the vacuum. The next possible value is $\alpha = \frac{1}{N-2}$, leading to a small value for the $\hat{u}(1)$ level, $\kappa \sim \mathcal{O}(\frac{1}{N^2})$, and a central charge, $c = 1$. In this case not only the $T_n^{\pm}$ sector decouples, but also the $W^s$ sector. The dual field theory consists of a free boson, just like in the Polyakov–Bershadsky case [I]. These first examples confirm the general belief that quantum gravity in the ultra-quantum limit might be dual to a very simple field theory. In the quantum regime at large $N$ the central charge takes values close to integers, $c = 0,\, 1,\, 2 - \mathcal{O}(\frac{1}{N^2}),\, \dots$

[3] It is likely that there are more restrictions on unitarity in the even case coming from more sophisticated analyses than solely looking for negative norm states.
[4] I am referring to the duality (9.44) which maps values of $\alpha$ close to zero to values of $\tilde{\alpha}$ close to one. However, this duality does not necessarily map values of $\alpha$ at the beginning of (9.51) to values of $\alpha$ at the end of the lists, but instead can lead to values of $\alpha$ not contained in (9.51).

### Semi-Classical Regime

The semi-classical regime can be approached by making the coupling constant smaller. In this regime the central charge becomes proportional to $N$ and has a quasi-continuous spacing. The region where the central charge is closest to its maximum value corresponds to $m \approx \frac{2}{3}N$ in (9.50) or $\alpha \approx \frac{1}{2}$. Parametrizing the integer $m$ as $m = \frac{2}{3}(N-1+n_0)+n_1$, where $n_1$ is some varying integer with absolute value much smaller than $N$ and $n_0 \in \{0,\, 1,\, 2\}$ is fixed so that $m$ is an integer[5] one obtains

$$\alpha = \frac{1}{2} + \frac{3(3n_1 - 2n_0)}{4N}\,. \tag{9.58}$$

The central charge (9.40) then simplifies to

$$c(n_1) = \frac{N}{4} - \frac{1}{8} - \frac{\delta c}{N} + \mathcal{O}(\tfrac{1}{N^2}), \tag{9.59}$$

with $\delta c = 9(9n_1^2 - 3(4n_0+1)n_1 + 4n_0^2 + 2n_0 - 1)/16$. The maximum of the continuous curve $c$ (see e.g. Fig. 9.1) is never reached, but for large $N$ one can get arbitrarily close to it, provided $n_1$ is a small integer. The result (9.59) proves the bound (9.4). Other semi-classical regimes can be obtained for $m = \frac{p}{q}(N-1+n_0)+n_1$ with some co-prime integers $p$ and $q$ so that $\frac{1}{2} < \frac{p}{q} < 1$, a suitable choice for $n_0 \in \{0,\, 1,\, \dots,\, q-1\}$ and some varying integer $n_1$ with absolute value much smaller than $N$. In those regimes the central charge scales as

$$c(n_1;\, \tfrac{p}{q}) = (2 - \tfrac{q}{p})(\tfrac{q}{p} - 1)\, N + \delta c + \mathcal{O}(\tfrac{1}{N})\,. \tag{9.60}$$

with $\delta c = (2n_1 + 1)\frac{q^3}{p^3} - (3n_1 - 2n_0 + 2)\frac{q^2}{p^2} - 3n_0\frac{q}{p} + 1$.

### Dual Quantum Regime

Further decreasing the coupling constant leads to an interesting phenomenon which is at first sight puzzling. That is, the central charge starts do decrease and approaching small values of $c$ up to $c \sim \mathcal{O}(1)$, which again can be seen in Figure 9.1. Since the central charge is proportional to the Chern-Simons level $k$ and thus also inverse proportional to Newton's constant, this means that one is again in a regime where quantum effects dominate, in contrast to the case of large central charge treated previously where semi-classical effects dominate. Thus quantum in this context means $\mathcal{O}(1)$ values of the Chern-Simons level. Looking back at the duality (9.42) that relates different values of the Chern-Simons coupling $k$, leading to the same value of the central charge, it is not so puzzling anymore that one encounters such a

[5] To be more precise, $n_0 = (2N+1) \bmod 3$.

"dual" quantum regime by increasing the coupling. However, as already mentioned previously in this chapter, the duality (9.44) does not necessarily map an allowed value for $\alpha$ on another entry in (9.51). In addition one has also to distinguish between even and odd values of $N$.
For odd $N$ the allowed values for $\alpha$ are given by

$$\alpha = \frac{N-2m-1}{N+2m-1}, \qquad N \gg m \in \mathbb{N}\,. \tag{9.61}$$

The central charge (9.40) then simplifies to

$$c(m) = 1 + 4m - \mathcal{O}(\tfrac{1}{N}), \qquad N \text{ odd}. \tag{9.62}$$

Thus, the dual quantum regime leads to a central charge with level spacing of four to leading order in a $\frac{1}{N}$ expansion. This implies that in the large $N$ limit only a quarter of the quantum regime levels gets mapped to corresponding levels in the dual quantum regime.

## 9.4 Conclusions

In the beginning of this part I have looked at Lobachevsky holography for all relevant non-principal embeddings of $\mathfrak{sl}(2,\mathbb{R}) \hookrightarrow \mathfrak{sl}(4,\mathbb{R})$ and checked their asymptotic symmetry algebras for unitary representations. I then continued to perform the same analysis for $\mathcal{W}_N^{(2)}$-algebras and was able to show explicitly that these symmetry algebras are unitary for arbitrary large, albeit not infinitely large, Virsasoro central charges, provided one also increases $N$ up to very large values. Thus, unitarity does not necessarily rule out the possibility to have very large Virasoro central charges, which circumvents the NO-GO result of [138].

# Part III

## Flat Space Holography–Gravity Side

In this part of my thesis I will explore flat space holography mainly from the gravity side. I will start by showing how to obtain certain results, like a flat space analogue of the Cardy formula, or various new classes of $\mathcal{W}$-algebras which will be called $\mathcal{FW}$-algebras, using a proper limit from the known Anti-de Sitter results. In addition, unitary representations of the newly found $\mathcal{FW}$-algebras will be determined, which under certain assumptions, results in a NO-GO theorem for non-linear $\mathcal{FW}$-algebras that rules out the possibility of having flat space, higher-spins and unitary representations at the same time. Furthermore I will elaborate on an explicit example which involves a linear $\mathcal{FW}$-algebra that circumvents this NO-GO theorem. I will then present how to implement higher-spin symmetries in the framework of flat space holography. Building up on this I will finish this part by a holographic description of flat space including (higher-spin) chemical potentials.

# Flat Space as a Limit from AdS | 10

> *There is nothing like looking, if you want to find something. You certainly usually find something, if you look, but it is not always quite the something you were after.*
>
> **– J.R.R. Tolkien**
> The Hobbit

Flat space in $2+1$ dimensions is usually described in terms of the Minkowski metric $\eta_{\mu\nu} = \text{diag}\,(-1,1,1)$. For many holographic purposes, however, it will be beneficial to work with a different, more general form of the metric similar to the one in the AdS case (3.35), which also covers more general solutions to the Einstein equations for the case of a vanishing cosmological constant.
Flat space in the so-called BMS gauge can be characterized by the line element [64]

$$\mathrm{d}s^2 = \mathcal{M}\,(\varphi)\,\mathrm{d}u^2 - 2\,\mathrm{d}u\,\mathrm{d}r + 2\mathcal{N}\,(u,\varphi)\,\mathrm{d}u\,\mathrm{d}\varphi + r^2\,\mathrm{d}\varphi^2, \tag{10.1}$$

where $0 \leq r < \infty$ is a radial coordinate, $-\infty < u < \infty$ is a retarded time coordinate and $\varphi \sim \varphi + 2\pi$ is an angular coordinate parametrizing the boundary circle of the manifold which has the topology of a cylinder. The functions $\mathcal{M}$ and $\mathcal{N}$ are, similar to the AdS case, state dependent functions that characterize the field content of the dual boundary theory and which, in addition, have to obey

$$\partial_u \mathcal{M} = 0, \qquad 2\partial_u \mathcal{N} = \partial_\varphi \mathcal{M}. \tag{10.2}$$

These equations are satisfied iff $\mathcal{N}\,(u,\varphi) = \mathcal{L}\,(\varphi) + \frac{u}{2}\mathcal{M}'\,(\varphi)$, for some function $\mathcal{L}\,(\varphi)$ and $\mathcal{M} = \mathcal{M}\,(\varphi)$. Depending on the values of $\mathcal{M}$ and $\mathcal{N}$ one obtains different flat space solutions. For the case of constant $\mathcal{M}$ and $\mathcal{N}$ there are three different types of solutions which are of interest for this thesis.

- For $\mathcal{M} = \mathcal{N} = 0$ and additional decompactification of the coordinate $\varphi \to x$, $-\infty < x < \infty$, this solution is known as the null orbifold [142–145]. This orbifold describes flat space modulo points which can be identified via a null rotation, hence the name null orbifold.

- For $\mathcal{M} = -1$, $\mathcal{N} = 0$ the solution is global flat space written in Eddington-Finkelstein coordinates.

- For generic $\mathcal{M} \geq 0$ and $\mathcal{N} \neq 0$ one obtains a family of cosmological solutions called "flat space cosmologies" (FSC's) [58, 59]. This family of solutions is of similar importance to flat space holography as the BTZ black hole to AdS$_3$ holography. These solutions carry mass $\mathcal{M}$ and angular momentum $\mathcal{N}$ and exhibit a cosmological horizon which also carries entropy and thus is of great interest for flat space holography to study flat space gravity duals with non-zero temperature.

In [64] it was shown that the asymptotic symmetries, which leave the space of allowed metrics (10.1) invariant is given by the centrally extended $\mathfrak{bms}_3$ algebra

$$[L_n, L_m] =(n-m)L_{n+m}, \tag{10.3a}$$

$$[L_n, M_m] =(n-m)M_{n+m} + \frac{c_M}{12}n(n^2-1)\delta_{n+m,0}, \tag{10.3b}$$

$$[M_n, M_m] =0, \tag{10.3c}$$

with $c_M = \frac{3}{G_N}$. Since the $\mathfrak{bms}_3$ algebra is the asymptotic symmetry algebra of flat space at null infinity[1], one can already see that my statement about flat space holography in the beginning was not quite complete. As pointed out in [146], the appropriate boundary from a conformal point of view, for establishing a flat space holographic correspondence is null infinity. So to be more precise, the metric (10.1) actually describes flat space at null infinity.

I mentioned in Chapter 5 that one can also obtain the $\mathfrak{bms}_3$ algebra (10.3) as a limiting procedure starting from relativistic conformal symmetries. This can be done either as a nonrelativistic limit or as an ultrarelativistic one. For the case at hand the limit that correctly reproduces (10.3) starting from the asymptotic symmetries of AdS$_3$ is the ultrarelativistic one. One can imagine this ultrarelativistic limit also as an ultrarelativistic boost of the whole AdS$_3$ cylinder as it is schematically depicted in Figure 10.1. This limit then yields the conformal boundary of flat space at null infinity. There is also something very important which I want to mention at this point. Flat space holography, as it is usually understood at the moment, is not a complete holographic description of flat space as of yet. This is related to the fact that one is describing either only future $\mathscr{I}^+$ or past null infinity $\mathscr{I}^-$, whereas a complete holographic description should also be able to describe the other half of null infinity (and also spacelike infinity $i^0$). For the purpose of this thesis, however, this is not an issue, as flat space holography itself is a rather new concept and thus one has to first get a good grasp on how to set up holography at $\mathscr{I}^\pm$ anyways before one can even try to tackle flat space holography on $\mathscr{I}^+ \cup \mathscr{I}^-$ including $i^0$. The important thing to keep in mind is that one is actually doing holography on $\mathscr{I}^\pm$ in the current setup.

[1] Either future $\mathscr{I}^+$ or past null infinity $\mathscr{I}^-$.

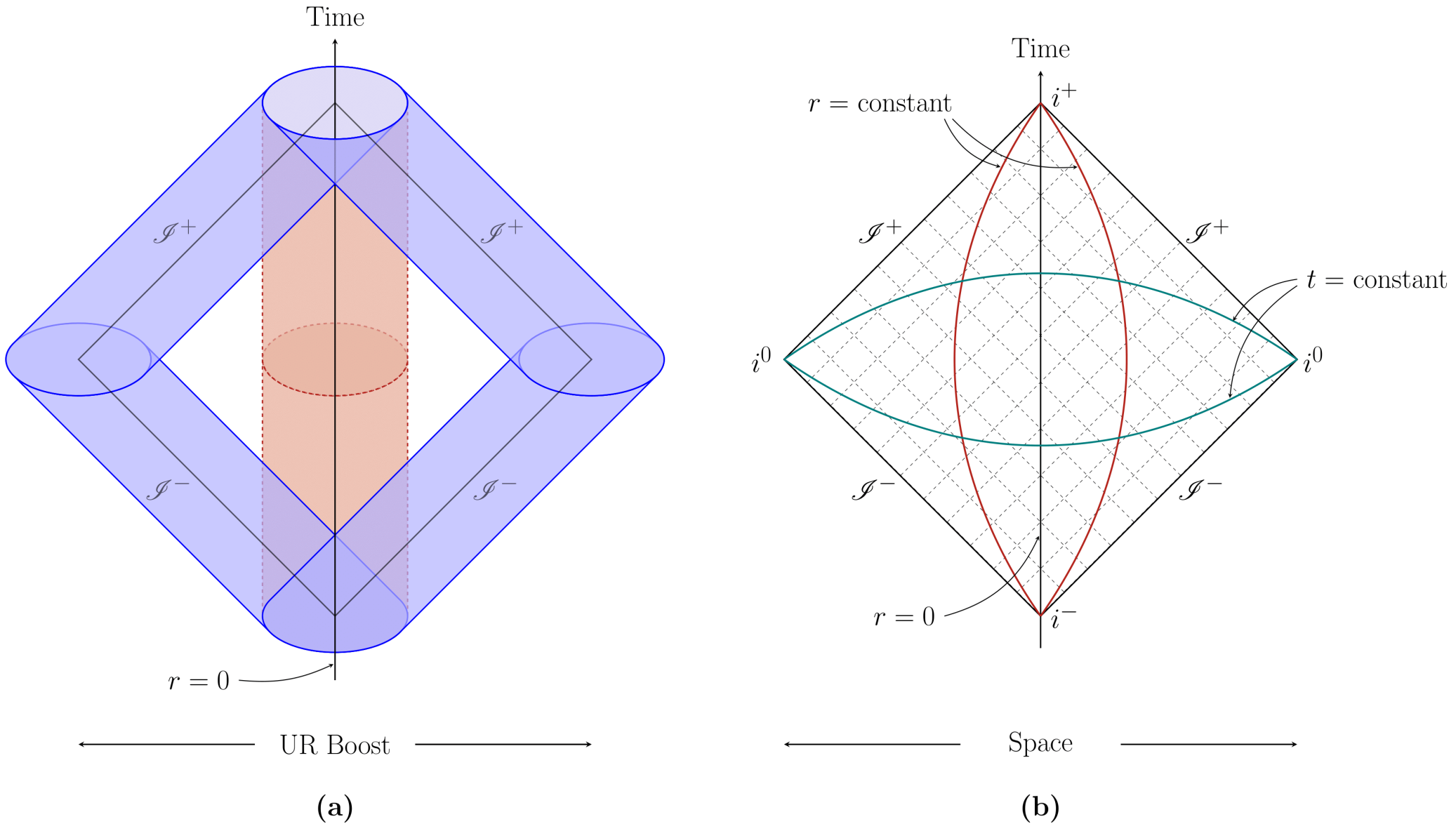

**Fig. 10.1.:** (a) Different ultrarelativistic boosts of global $AdS_3$. Depending on which end of the $AdS_3$ cylinder is boosted one obtains either $\mathscr{I}^+$ or $\mathscr{I}^-$. (b) Penrose–Carter diagram of global Minkowski space. Comparing this Penrose–Carter diagram with (a) one can readily see that flat space holography on $\mathscr{I}^\pm$ only captures a certain part of Minkowski space.

**FSC as a Limit of BTZ:** There are more instances than just the asymptotic symmetries where one can take a meaningful limit from AdS results. One particular example I will describe in the following is the flat space limit of the BTZ black hole [12]. It has been shown in [58] that it is indeed possible to take a flat limit of the BTZ black hole which results in a cosmological solution in flat space. Starting from the BTZ black hole [12]

$$\mathrm{d}s^2 = -\frac{(r^2 - r_+^2)(r^2 - r_-^2)}{r^2\ell^2}\,\mathrm{d}t^2 + \frac{r^2\ell^2}{(r^2 - r_+^2)(r^2 - r_-^2)}\,\mathrm{d}r^2 + r^2\left(\mathrm{d}\varphi - \frac{r_+ r_-}{\ell r^2}\,\mathrm{d}t\right)^2, \tag{10.4}$$

with $r_\pm = \sqrt{2G_N\ell(\ell M + J)} \pm \sqrt{2G_N\ell(\ell M - J)}$ and where $M$ and $J$ are the mass and the angular momentum of the BTZ black hole. Now taking the limit[2] $\ell \to \infty$ the outer horizon scales as $r_+ \to \ell\hat{r}_+$ in this limit, where $\hat{r}_+ = \sqrt{8G_N M}$, and is thus pushed to infinity. The inner horizon survives the limit and takes the value $r_- \to r_0 = \sqrt{\frac{2G_N}{M}}J$. The resulting metric

$$\mathrm{d}s^2 = \hat{r}_+^2\,\mathrm{d}t^2 - \frac{r^2\,\mathrm{d}r^2}{\hat{r}_+^2(r^2 - r_0^2)} + r^2\,\mathrm{d}\varphi^2 - 2\hat{r}_+ r_0\,\mathrm{d}t\,\mathrm{d}\varphi, \tag{10.5}$$

is a FSC with a cosmological horizon at $r_0$.

This limit is sometimes useful to gain a better intuition on the geometrical interpretation of some, at first sight, unusual features of flat space holography. Geometrically this limit can be envisioned as zooming into the region between inner ($r_-$) and outer ($r_+$) BTZ horizon. The former outer BTZ horizon $r_+$ effectively becomes the new asymptotic boundary in flat space, while the inner horizon effectively turns into the FSC horizon. Thus, the main conceptual point to gain from this application of a flat space limit is that whenever one wants to obtain holographic results related to FSCs as a limit from the BTZ case one has to work with *inner* instead of *outer* BTZ black hole mechanics. This in turn also means that it would not be surprising to find a predecessor of the $\mathfrak{bms}_3$ symmetries (10.3) as a near horizon geometry of BTZ black holes, see e.g. [147].
As in the BTZ case one can associate a Hawking temperature $T_H = \frac{\kappa}{2\pi}$ and entropy $S_{\text{FSC}} = \frac{\pi |r_0|}{2G_N}$ to the cosmological horizon of the FSC via surface gravity $\kappa = \frac{\hat{r}_+^2}{r_0}$ and the Bekenstein-Hawking area law, respectively. The charges, mass $\mathcal{M}$ and angular momentum $\mathcal{N}$ as in (10.1), associated with the FSC, then obey a first law of thermodynamics

$$\mathrm{d}\mathcal{M} = -T_H\,\mathrm{d}S_{\text{FSC}} + \Omega_{\text{FSC}}\,\mathrm{d}\mathcal{N}, \tag{10.6}$$

where $\Omega_{\text{FSC}}$ is the angular velocity of the FSC horizon. The weird sign in front of the $T_H\,\mathrm{d}S_{\text{FSC}}$ term is again a reminder of the fact that the first law of FSC arises as a limit from the BTZ *inner* horizon dynamics [148, 149].

[2] Since the AdS radius is related to the cosmological constant $\Lambda$ as $\Lambda = -\frac{1}{\ell^2}$ this limit is equivalent to $\Lambda \to 0$.

**Grassmann Trick:** Once the correct way to obtain the flat space limit has been found the only thing left to do is to actually perform the limit. Since a lot of the work in this thesis is related to the Chern-Simons formulation of gravity and many flat space limits are related to İnönü–Wigner contractions of $\mathfrak{so}(2,2)$ to $\mathfrak{isl}(2,\mathbb{R})$ it is convenient to use the so called Grassmann trick. This trick has been first presented in [70] and is a neat way of writing the limit of diverging AdS radius in a compact way. This can be done by denoting $\epsilon = \frac{1}{\ell}$ and replacing $\ell \to \frac{1}{\epsilon}$ in all AdS expressions. Taking the flat space limit now amounts to treating $\epsilon$ as a Grassmann variable, i.e. $\epsilon^2 = 0$, which is where the name of the trick comes from. This might look like a triviality at first, but there will be several encounters in this thesis where this approach will prove to be useful. Particular examples are the explicit matrix representations of $\mathfrak{isl}(2,\mathbb{R})$ and $\mathfrak{isl}(3,\mathbb{R})$ found in Appendix A.2 where the Grassmann trick allows one to construct matrix representations in a very efficient way and also explicitly realize the invariant bilinear forms on $\mathfrak{isl}(N,\mathbb{R})$ in form of various trace functions.

This chapter is organized as follows. I will first show how to obtain an expression for a Chern-Simons connection describing asymptotically flat spacetimes from the known $\text{AdS}_3$ expressions using the Grassmann trick. I will then proceed with showing how one can obtain various new algebras which are of interest for flat space (higher-spin) holography using an İnönü–Wigner contraction of certain $\text{AdS}_3$ asymptotic symmetry algebras. I will conclude this chapter with an expression for a flat space analogue of the Cardy formula for both spin-2 and spin-3 charged flat space cosmologies which is obtained by employing a suitable limit of vanishing cosmological constant of the known (higher-spin) Cardy formula.

## 10.1 Chern-Simons Formulation of Flat Space

As a first example where the Grassmann trick is able to provide useful insights for flat space holography is given by the transition from the AdS Chern-Simons formulation of gravity to the flat space one. The starting point of this limit is the AdS Chern-Simons formalism (2.12), that is described by the two gauge fields

$$\mathfrak{A} = \omega + \epsilon e, \qquad \bar{\mathfrak{A}} = \omega - \epsilon e, \tag{10.7}$$

which take values in $\mathfrak{sl}(2,\mathbb{R})$ whose commutation relations are given by

$$[\mathfrak{L}_n, \mathfrak{L}_m] = (n-m)\mathfrak{L}_{n+m}, \tag{10.8}$$

for $n, m = \pm 1, 0$. Next, one has to rewrite the connection (10.7) in a form that is convenient to use for the purpose of taking this limit. After the rewriting, the limit to flat space can now be simply taken by using the Grassmann trick and setting $\epsilon^2 = 0$. This yields expressions for the dreibein and spin connection components, which I will denote by $e_{\text{Flat}}$ and $\omega_{\text{Flat}}$, respectively. The dreibein and spin connection can then be combined into the flat $\mathfrak{isl}(2,\mathbb{R})$ valued Chern-Simons gauge field as

$$\mathcal{A} = e^a_{\text{Flat}} M_a + \omega^a_{\text{Flat}} L_a, \tag{10.9}$$

where $L_n$ and $M_n$ are the generators of the $\mathfrak{isl}(2,\mathbb{R})$ algebra $(n, m = \pm 1, 0)$

$$[L_n, L_m] = (n-m)L_{n+m}, \tag{10.10a}$$
$$[L_n, M_m] = (n-m)M_{n+m}, \tag{10.10b}$$
$$[M_n, M_m] = 0. \tag{10.10c}$$

### Spin-2

I will first demonstrate this procedure explicitly for the spin-2 case. The starting point will be a slight generalization of the metric (3.35)

$$\mathrm{d}s^2 = \ell^2 \left[ \mathrm{d}r^2 + \frac{6}{c}\mathcal{L}(\mathrm{d}x^+)^2 + \frac{6}{\bar{c}}\bar{\mathcal{L}}(\mathrm{d}x^-)^2 - \left( e^{2r} + \frac{36}{c\bar{c}}\mathcal{L}\bar{\mathcal{L}}e^{-2r} \right) \mathrm{d}x^+ \,\mathrm{d}x^- \right], \tag{10.11}$$

with $x^{\pm} = \epsilon t \pm \varphi$, and where the functions $\mathcal{L}$ and $\bar{\mathcal{L}}$ again depend on $x^+$ and $x^-$ respectively. In order to perform the flat space limit it will, however, be more convenient to work in the following coordinates

$$\mathrm{d}s^2 = \left(\mathcal{M}\left(u,\varphi\right) - r^2\epsilon^2\right)\mathrm{d}u^2 - 2\,\mathrm{d}u\,\mathrm{d}r + 2\mathcal{N}\left(u,\varphi\right)\mathrm{d}u\,\mathrm{d}\varphi + r^2\,\mathrm{d}\varphi^2, \qquad (10.12)$$

where $u$ is a timelike coordinate, $r$ is again a radius and $\varphi \sim \varphi + 2\pi$ an angular coordinate. The functions $\mathcal{M}$ and $\mathcal{N}$ are related to $\mathcal{L}$ and $\bar{\mathcal{L}}$ as

$$\mathcal{M} = 12\left(\frac{\mathcal{L}}{c} + \frac{\bar{\mathcal{L}}}{\bar{c}}\right), \qquad \mathcal{N} = \frac{6}{\epsilon}\left(\frac{\mathcal{L}}{c} - \frac{\bar{\mathcal{L}}}{\bar{c}}\right), \qquad (10.13)$$

or equivalently [64]

$$\frac{\mathcal{M} + 2\epsilon\mathcal{N}}{4} = \frac{6}{c}\mathcal{L}, \qquad \frac{\mathcal{M} - 2\epsilon\mathcal{N}}{4} = \frac{6}{\bar{c}}\bar{\mathcal{L}}, \qquad (10.14)$$

where $x^{\pm} = \epsilon u \pm \varphi$.
This connection can be rewritten in the so called BMS gauge as

$$\mathfrak{A} = \mathfrak{b}^{-1}\,\mathrm{d}\mathfrak{b} + \mathfrak{b}^{-1}\mathfrak{a}\mathfrak{b}, \qquad \bar{\mathfrak{A}} = \mathfrak{b}\,\mathrm{d}\mathfrak{b}^{-1} + \mathfrak{b}\bar{\mathfrak{a}}\mathfrak{b}^{-1}, \qquad \mathfrak{b} = e^{\frac{\epsilon}{2}rL_{-1}}, \qquad (10.15a)$$

$$\mathfrak{a} = \left(\mathfrak{L}_1 - \frac{(\mathcal{M} + 2\mathcal{N}\epsilon)}{4}\mathfrak{L}_{-1}\right)(\epsilon\,\mathrm{d}u + \mathrm{d}\varphi)\,, \qquad (10.15b)$$

$$\bar{\mathfrak{a}} = -\left(\mathfrak{L}_1 - \frac{(\mathcal{M} - 2\mathcal{N}\epsilon)}{4}\mathfrak{L}_{-1}\right)(\epsilon\,\mathrm{d}u - \mathrm{d}\varphi)\,, \qquad (10.15c)$$

One can easily check that this connection satisfies the equations of motion
$\mathrm{d}\mathfrak{A} + \mathfrak{A} \wedge \mathfrak{A} = 0$ and $\mathrm{d}\bar{\mathfrak{A}} + \bar{\mathfrak{A}} \wedge \bar{\mathfrak{A}} = 0$ provided

$$\partial_u\mathcal{M} = 2\epsilon^2\partial_\varphi\mathcal{N}, \quad 2\partial_u\mathcal{N} = \partial_\varphi\mathcal{M}. \qquad (10.16)$$

The next step is to determine $e_{\text{Flat}}$ and $\omega_{\text{Flat}}$ by setting $\epsilon^2 = 0$ in (10.15b) and (10.15c). This yields

$$\mathfrak{a}\big|_{\epsilon^2=0} = \left(\mathfrak{L}_1 - \frac{\mathcal{M}}{4}\mathfrak{L}_{-1}\right)\epsilon\,\mathrm{d}u + \left(\mathfrak{L}_1 - \frac{(\mathcal{M} + 2\mathcal{N}\epsilon)}{4}\mathfrak{L}_{-1}\right)\mathrm{d}\varphi, \qquad (10.17a)$$

$$\bar{\mathfrak{a}}\big|_{\epsilon^2=0} = -\left(\mathfrak{L}_1 - \frac{\mathcal{M}}{4}\mathfrak{L}_{-1}\right)\epsilon\,\mathrm{d}u + \left(\mathfrak{L}_1 - \frac{(\mathcal{M} - 2\mathcal{N}\epsilon)}{4}\mathfrak{L}_{-1}\right)\mathrm{d}\varphi. \qquad (10.17b)$$

The dreibein and spin connection in this limit can now be directly read off from these expressions using (10.7) as

$$e\big|_{\epsilon^2=0} = \frac{1}{2}\mathfrak{L}_{-1}\,\mathrm{d}r + \left(\mathfrak{L}_1 - \frac{\mathcal{M}}{4}\mathfrak{L}_{-1}\right)\mathrm{d}u + \left(r\mathfrak{L}_0 - \frac{\mathcal{N}}{2}\mathfrak{L}_{-1}\right)\mathrm{d}\varphi, \qquad (10.18a)$$

$$\omega\big|_{\epsilon^2=0} = \left(\mathfrak{L}_1 - \frac{\mathcal{M}}{4}\mathfrak{L}_{-1}\right)\mathrm{d}\varphi. \qquad (10.18b)$$

The beauty of the Grassman trick is now that one can identify the components of the dreibein and spin connection in (10.18) with the ones in (10.9), i.e. the $\mathfrak{isl}(2,\mathbb{R})$ valued connection $\mathcal{A}$. One only has to replace the generators $\mathfrak{L} \to M$ in the expression for the dreibein in (10.18) in order to obtain $e_{\text{Flat}}$ i.e. $e\big|_{\epsilon^2=0} \overset{\mathfrak{L}\to M}{=} e_{\text{Flat}}$ and similarly for the spin connection $\omega_{\text{Flat}}$ one has to replace $\mathfrak{L} \to L$. Thus, one obtains as the flat $\mathfrak{isl}(2,\mathbb{R})$ connection $\mathcal{A}$

$$\begin{aligned}\mathcal{A} =& \frac{1}{2}M_{-1}\,\mathrm{d}r + \left(M_1 - \frac{\mathcal{M}}{4}M_{-1}\right)\mathrm{d}u + \left(rM_0 - \frac{\mathcal{N}}{2}M_{-1}\right)\mathrm{d}\varphi \\ &+ \left(L_1 - \frac{\mathcal{M}}{4}L_{-1}\right)\mathrm{d}\varphi.\end{aligned} \tag{10.19}$$

This connection exactly reproduces (10.1) via

$$\mathrm{d}s^2 = -2\eta_{ab}e^a_{\text{Flat}}e^b_{\text{Flat}}, \tag{10.20}$$

with $\eta_{ab} = \text{antidiag}(1, -\frac{1}{2}, 1)$. Which shows that one can indeed obtain the $\mathfrak{isl}(2,\mathbb{R})$ valued connection $\mathcal{A}$ for flat space by starting with the $\mathfrak{sl}(2,\mathbb{R})$ valued connection (10.15) and treating $\epsilon$ as a Grassman parameter.
As a neat consequence of this limiting procedure it follows that (10.19) can also be written in the following way

$$\mathcal{A} = b^{-1}\,\mathrm{d}b + b^{-1}ab, \quad b = e^{\frac{r}{2}M_{-1}}, \tag{10.21a}$$

$$a = \left(M_1 - \frac{\mathcal{M}}{4}M_{-1}\right)\mathrm{d}u - \frac{\mathcal{N}}{2}M_{-1}\,\mathrm{d}\varphi + \left(L_1 - \frac{\mathcal{M}}{4}L_{-1}\right)\mathrm{d}\varphi, \tag{10.21b}$$

which provides similar simplifications for a holographic description of flat space as the gauge (3.39) in the AdS case.

## 10.2 İnönü–Wigner Contractions and $\mathcal{FW}$-Algebras

In Chapter 5 I gave a brief introduction on İnönü–Wigner contractions and presented as a specific example two different limits of the relativistic conformal algebra in two dimensions (5.4). In this section I show how one can perform such a contraction also for $\mathcal{W}$-algebras and describe the subtleties one has to take care of.
Let me begin by recalling the basic features of non- and ultrarelativistic contractions of two copies of the relativistic conformal algebra (5.4) with generators $\mathcal{L}_n$, $\bar{\mathcal{L}}_n$ and central charges $c$, $\bar{c}$. These generators are related to the generators $L_n$, $M_n$ of the $\mathfrak{bms}_3$ algebra, which is isomorphic to the Galilean conformal algebra in two dimensions $\mathfrak{gca}_2$, as

$$\text{Nonrelativistic contraction:} \quad L_n := \mathcal{L}_n + \bar{\mathcal{L}}_n, \quad M_n := -\epsilon\left(\mathcal{L}_n - \bar{\mathcal{L}}_n\right), \tag{10.22}$$

$$\text{Ultrarelativistic contraction:} \quad L_n := \mathcal{L}_n - \bar{\mathcal{L}}_{-n}, \quad M_n := \epsilon\left(\mathcal{L}_n + \bar{\mathcal{L}}_{-n}\right). \tag{10.23}$$

The $\mathfrak{bms}_3$ algebra, or equivalently the Galilean conformal algebra is obtained in the $\epsilon \to 0$ limit as

$$[L_n, L_m] = (n-m)L_{n+m} + \frac{c_L}{12}n(n^2-1)\delta_{n+m,0}, \tag{10.24a}$$

$$[L_n, M_m] = (n-m)M_{n+m} + \frac{c_M}{12}n(n^2-1)\delta_{n+m,0}, \tag{10.24b}$$

$$[M_n, M_m] = 0, \tag{10.24c}$$

where the central charges $c_L$, $c_M$ depend on the type of contraction:

$$\text{Nonrelativistic contraction:} \quad c_L = c + \bar{c}, \quad c_M = \lim_{\epsilon\to 0} \epsilon(-c+\bar{c}), \tag{10.25}$$

$$\text{Ultrarelativistic contraction:} \quad c_L = c - \bar{c}, \quad c_M = \lim_{\epsilon\to 0} \epsilon(c+\bar{c}). \tag{10.26}$$

Note that in particular the $c_M$ central charge has a dimension of inverse length, because it involves the contraction parameter $\epsilon$. Thus, the exact value of the central charge $c_M$ cannot have any physical meaning since an arbitrary rescaling would alter its value. Therefore only the sign of $c_M$ will be physically relevant.

While from a physical point of view the ultrarelativistic contraction is the relevant one for the purpose of doing flat space holography, from an algebraic point of view the nonrelativistic one is more convenient to apply. The reason for this is that the ultrarelativistic contraction mixes operators with positive and negative mode numbers. This is not a problem, as long as there are no nonlinear operators present. As soon as there are nonlinear operators and a notion of normal ordering present, mixing of operators with positive and negative mode numbers compromises the notion of normal ordering of the algebras before the contraction. Thus, the resulting algebra after an ultrarelativistic contraction will not be compatible anymore with the normal ordering prescription employed before the contraction, i.e. the Jacobi identities will not hold anymore. This is to be expected, since one is tempering with the notion of normal ordering in this particular limit. For all practical purposes, however, this does not pose a problem. One can also start from two copies of semiclassical $\mathcal{W}$-algebras, perform the İnönü–Wigner contraction and then introduce normal ordering and determine the deformations of the resulting quantum algebra by demanding that the Jacobi identities hold.

Since contracting two semiclassical algebras and then quantizing them is a bit inefficient I will employ a simple shortcut. This shortcut relies on the fact that the nonrelativistic contraction does not temper with normal ordering. As such one can simply contract two quantum $\mathcal{W}$-algebras, whose nonlinear terms are normal ordered with respect to a given highest-weight state, and automatically obtain a quantum $\mathcal{FW}$-algebra which inherits the same normal ordering prescription before the contraction. Also assuming normal ordering with respect to some highest-weight state in the ultrarelativistic limit, the resulting quantum $\mathcal{FW}$ algebras can be obtained from the nonrelativistic expressions, provided one replaces the nonrelativistic expressions for $c_L$ and $c_M$ (10.25) with the corresponding ultrarelativistic ones (10.26).

## 10.3 Contracting $\mathcal{W}_3$ and $\mathcal{W}_3^{(2)}$

For illustrative purposes I will show how the nonrelativistic contraction from two copies each of $\mathcal{W}_3$ and $\mathcal{W}_3^{(2)}$ to $\mathcal{FW}_3$ and $\mathcal{FW}_3^{(2)}$ are performed. More examples of contracted $\mathcal{FW}$-algebras can be found in Appendix D.

### 10.3.1 $\mathcal{W}_3 \to \mathcal{FW}_3$

The simplest contraction to perform from given AdS results is spin-3 AdS gravity, whose asymptotic symmetry algebra consists of two copies of the $\mathcal{W}_3$ algebra [28, 29]. The non-trivial commutation relations between the generators of a single copy of $\mathcal{W}_3$ are given by [89, 91, 92]

$$[\mathcal{L}_n,\, \mathcal{L}_m] = (n-m)\,\mathcal{L}_{n+m} + \frac{c}{12}\, n(n^2-1)\,\delta_{n+m,\,0}, \tag{10.27a}$$

$$[\mathcal{L}_n,\, \mathcal{W}_m] = (2n-m)\,\mathcal{W}_{n+m}, \tag{10.27b}$$

$$\begin{aligned}[\mathcal{W}_n,\, \mathcal{W}_m] = {} & (n-m)(2n^2+2m^2-nm-8)\,\mathcal{L}_{n+m} + \frac{96}{c+\frac{22}{5}}\,(n-m)\ :\!\mathcal{L}\mathcal{L}\!:_{n+m} \\ & + \frac{c}{12}\, n(n^2-4)(n^2-1)\,\delta_{n+m,\,0},\end{aligned} \tag{10.27c}$$

with the normal ordering prescription

$$:\!\mathcal{L}\mathcal{L}\!:_n = \sum_{p\geq -1} \mathcal{L}_{n-p}\mathcal{L}_p + \sum_{p<-1} \mathcal{L}_p\mathcal{L}_{n-p} - \frac{3}{10}(n+3)(n+2)\mathcal{L}_n\,. \tag{10.28}$$

The generators of the other copy of $\mathcal{W}_3$ will be denoted with bar on top, $\bar{\mathcal{L}}_n$ and $\bar{\mathcal{W}}_n$. I define the nonrelativistic contraction in analogy to the spin-2 case (10.22)

$$L_n := \mathcal{L}_n + \bar{\mathcal{L}}_n, \qquad M_n := -\epsilon\left(\mathcal{L}_n - \bar{\mathcal{L}}_n\right), \tag{10.29a}$$

$$U_n := \mathcal{W}_n + \bar{\mathcal{W}}_n, \qquad V_n := -\epsilon\left(\mathcal{W}_n - \bar{\mathcal{W}}_n\right). \tag{10.29b}$$

Taking the $\epsilon \to 0$ limit, one obtains the contracted algebra [150]

$$[L_n,\, L_m] = (n-m)L_{n+m} + \frac{c_L}{12}\,(n^3-n)\,\delta_{n+m,\,0}, \tag{10.30a}$$

$$[L_n,\, M_m] = (n-m)M_{n+m} + \frac{c_M}{12}\,(n^3-n)\,\delta_{n+m,\,0}, \tag{10.30b}$$

$$[L_n,\, U_m] = (2n-m)U_{n+m}, \tag{10.30c}$$

$$[L_n,\, V_m] = (2n-m)V_{n+m}, \tag{10.30d}$$

$$[M_n,\, U_m] = (2n-m)V_{n+m}, \tag{10.30e}$$

$$\begin{aligned}[U_n,\, U_m] = {} & (n-m)(2n^2+2m^2-nm-8)L_{n+m} + (n-m)\left(\frac{192}{c_M}\,\Lambda_{n+m}\right. \\ & \left. - \frac{96(c_L+\frac{44}{5})}{c_M^2}\,\Theta_{n+m}\right) + \frac{c_L}{12}\, n(n^2-1)(n^2-4)\,\delta_{n+m,\,0},\end{aligned} \tag{10.30f}$$

$$\begin{aligned}[U_n,\, V_m] = {} & (n-m)(2n^2+2m^2-nm-8)M_{n+m} + \frac{96}{c_M}\,(n-m)\Theta_{n+m} \\ & + \frac{c_M}{12}\, n(n^2-1)(n^2-4)\,\delta_{n+m,\,0},\end{aligned} \tag{10.30g}$$

where

$$\Theta_n = \sum_p M_p\, M_{n-p}, \qquad \Lambda_n = \sum_p :L_p\, M_{n-p}: -\tfrac{3}{10}(n+2)(n+3)M_n, \tag{10.31}$$

and the normal ordering prescription is given by

$$:L_n\, M_m:= L_n\, M_m \text{ if } n < -1, \qquad :L_n\, M_m:= M_m\, L_n \text{ if } n \geq -1\,. \tag{10.32}$$

The central charges are given by (10.25) in terms of the original central charges $c$ and $\bar{c}$, assuming that $c_L$ and $c_M$ can take arbitrary (real) values. Assuming normal order with respect to a highest-weight state as in (10.32) then the ultrarelativistic version of that algebra has the exact same form as (10.30), but with central charges (10.26) instead of (10.25).

### 10.3.2 $\mathcal{W}_3^{(2)} \to \mathcal{FW}_3^{(2)}$

In order to show how the nonrelativistic contraction can be implemented for non-principal embeddings, I will take as an example the Polyakov–Bershadsky algebra, $\mathcal{W}_3^{(2)}$.
The (quantum) $\mathcal{W}_3$ algebra is generated by $\mathcal{L}_n$, $\hat{\mathcal{G}}^\pm_n$ and $\mathcal{J}_n$ whose non-vanishing commutation relations read [133, 134]

$$\begin{aligned}
[\mathcal{L}_n,\, \mathcal{L}_m] &= (n-m)\mathcal{L}_{m+n} + \frac{c}{12}\, n(n^2-1)\, \delta_{n+m,\,0}, && (10.33a)\\
[\mathcal{L}_n,\, \mathcal{J}_m] &= -\, m\mathcal{J}_{n+m}, && (10.33b)\\
[\mathcal{L}_n,\, \hat{\mathcal{G}}^\pm_m] &= (\frac{n}{2} - m)\hat{\mathcal{G}}^\pm_{n+m}, && (10.33c)\\
[\mathcal{J}_n,\, \mathcal{J}_m] &= \frac{2k+3}{3}\, n\, \delta_{n+m,\,0}, && (10.33d)\\
[\mathcal{J}_n,\, \hat{\mathcal{G}}^\pm_m] &= \pm\, \hat{\mathcal{G}}^\pm_{m+n}, && (10.33e)\\
[\hat{\mathcal{G}}^+_n,\, \hat{\mathcal{G}}^-_m] &= -\,(k+3)\mathcal{L}_{m+n} + \frac{3}{2}(k+1)(n-m)\mathcal{J}_{m+n} + 3:\mathcal{J}\mathcal{J}:_{n+m}\\
&\quad + \frac{(k+1)(2k+3)}{2}\,(n^2 - \frac{1}{4})\, \delta_{m+n,\,0}, && (10.33f)
\end{aligned}$$

with the central charge

$$c = 25 - \frac{24}{k+3} - 6(k+3), \tag{10.34}$$

and the normal ordering prescription

$$:\mathcal{J}\mathcal{J}:_n = \sum_{p\geq 0} \mathcal{J}_{n-p}\mathcal{J}_p + \sum_{p<0} \mathcal{J}_p\mathcal{J}_{n-p}\,. \tag{10.35}$$

By replacing $\mathcal{L}_n \to \bar{\mathcal{L}}_n$, $\hat{\mathcal{G}}^\pm_n \to \hat{\bar{\mathcal{G}}}^\pm_n$, $\mathcal{J}_n \to \bar{\mathcal{J}}_n$, $k \to \bar{k}$ and $c \to \bar{c}$ in (10.33) one obtains the commutation relations for the second copy of the $\mathcal{W}_3^{(2)}$ algebra needed for the nonrelativistic contraction. In order to properly contract these two algebras a rescaling of $\hat{\mathcal{G}}_n$ and $\hat{\bar{\mathcal{G}}}_n$ with a suitable factor, e.g. $\sqrt{-k-1}$, is necessary. Otherwise terms of $\mathcal{O}(\frac{1}{\epsilon})$ would spoil the limit $\epsilon \to 0$. I drop the hat for the rescaled generators, $\hat{\mathcal{G}}^{\pm}_n = \sqrt{-k-1}\,\mathcal{G}^{\pm}_n$ and similarly for $\bar{\mathcal{G}}^{\pm}_n$.

The linear combinations that will lead to the nonrelativistic contraction of the Polyakov–Bershadsky algebra are defined analogous to (10.22) and (10.29)

$$L_n := \mathcal{L}_n + \bar{\mathcal{L}}_n, \qquad M_n := -\epsilon\left(\mathcal{L}_n - \bar{\mathcal{L}}_n\right), \tag{10.36a}$$

$$J_n := \mathcal{J}_n + \bar{\mathcal{J}}_n, \qquad K_n := -\epsilon\left(\mathcal{J}_n - \bar{\mathcal{J}}_n\right), \tag{10.36b}$$

$$U^{\pm}_n := \mathcal{G}^{\pm}_n + \bar{\mathcal{G}}^{\pm}_n, \qquad V^{\pm}_n := -\epsilon\left(\mathcal{G}^{\pm}_n - \bar{\mathcal{G}}^{\pm}_n\right). \tag{10.36c}$$

It should also be noted that there are some ambiguities in the normalizations of the generators $G^{\pm}$, $\bar{G}^{\pm}$, $U^{\pm}$ and $V^{\pm}$. I fixed these ambiguities already in a convenient way by choosing the specific rescaling $\hat{\mathcal{G}}^{\pm}_n = \sqrt{-k-1}\,\mathcal{G}^{\pm}_n$ previously, with no loss of generality.

The limit $\epsilon \to 0$ then yields the contracted algebra

$$[L_n,\, L_m] = (n-m)L_{m+n} + \frac{c_L}{12}\,n(n^2-1)\,\delta_{n+m,\,0}, \tag{10.37a}$$

$$[L_n,\, M_m] = (n-m)M_{m+n} + \frac{c_M}{12}\,n(n^2-1)\,\delta_{n+m,0}, \tag{10.37b}$$

$$[L_n,\, J_m] = -\,mJ_{n+m}, \tag{10.37c}$$

$$[L_n,\, K_m] = -\,mK_{n+m}, \tag{10.37d}$$

$$[L_n,\, U^{\pm}_m] = (\frac{n}{2}-m)\,U^{\pm}_{n+m}, \tag{10.37e}$$

$$[L_n,\, V^{\pm}_m] = (\frac{n}{2}-m)\,V^{\pm}_{n+m}, \tag{10.37f}$$

$$[M_n,\, J_m] = -\,mK_{n+m}, \tag{10.37g}$$

$$[M_n,\, U^{\pm}_m] = (\frac{n}{2}-m)\,V^{\pm}_{n+m}, \tag{10.37h}$$

$$[J_n,\, J_m] = \frac{32-c_L}{9}\,n\,\delta_{n+m,\,0}, \tag{10.37i}$$

$$[J_n,\, K_m] = -\,\frac{c_M}{9}\,n\,\delta_{n+m,\,0}, \tag{10.37j}$$

$$[J_n,\, U^{\pm}_m] = \pm\,U^{\pm}_{n+m}, \tag{10.37k}$$

$$[J_n,\, V^{\pm}_m] = \pm\,V^{\pm}_{n+m}, \tag{10.37l}$$

$$[K_n,\, U^{\pm}_m] = \pm\,V^{\pm}_{n+m}, \tag{10.37m}$$

$$\begin{aligned}[U^{+}_n,\, U^{-}_m] = {}& L_{n+m} - \frac{3}{2}(n-m)J_{n+m} - \frac{18\,(c_L-26)}{c_M^2}\,:KK:_{n+m} \\ &+ \frac{18}{c_M}\,:JK:_{n+m} + \frac{(c_L-32)}{6}\,(n^2-\frac{1}{4})\,\delta_{n+m,\,0},\end{aligned} \tag{10.37n}$$

$$\begin{aligned}[U^{\pm}_n,\, V^{\mp}_m] = {}& \pm\,M_{n+m} - \frac{3}{2}(n-m)K_{n+m} \pm \frac{18}{c_M}\,:KK:_{n+m} \\ &\pm \frac{c_M}{6}\,(n^2-\frac{1}{4})\,\delta_{n+m,\,0},\end{aligned} \tag{10.37o}$$

with the central charges (10.25) and the normal ordering prescription

$$:KK:_n \equiv \sum_{p\geq 0} K_{n-p}K_p + \sum_{p<0} K_p K_{n-p}, \tag{10.38a}$$

$$:JK:_n \equiv \sum_{p\geq 0} (K_{n-p}J_p + J_{n-p}K_p) + \sum_{p<0} (J_p K_{n-p} + K_p J_{n-p}) \ . \tag{10.38b}$$

Under the same assumptions as for $\mathcal{FW}_3$, the ultra relativistic version of the algebra (10.37) has the exact same form but with the central charges given by (10.26). I will call the algebra (10.37) from now on $\mathcal{FW}_3^{(2)}$-algebra.
Thus, at least at the level of (quantum) asymptotic symmetries, a flat space limit of the nonlinear $\mathcal{W}$-algebras can be implemented in a rather straightforward way.

## 10.4 A Flat Space (Higher-Spin) Cardy Formula

In this section I will show how to obtain a flat space analogue of the Cardy formula[3], which allows one to holographically compute the thermal entropy of a flat space cosmology. While taking the limit itself is simple, it is an essential and important point to note that one has to take the *inner horizon* limit of the Cardy formula and not the well known result for the outer BTZ horizon. The reason for this is, as explained in the beginning of this chapter, that the limit from BTZ to FSCs can be geometrically interpreted as zooming in between the region between the inner and outer BTZ horizon. In order to make this point explicit I will first show that taking the flat space limit of the (standard) outer horizon Cardy formula does not yield the correct result for the microstate counting of FSCs.
The Cardy formula that determines the entropy of the outer horizon of a BTZ black hole is given by [151]:

$$S_{A_{\text{out}}} = \frac{A_{\text{out}}}{4G} = 2\pi\sqrt{\frac{c\,\mathcal{L}}{6}} + 2\pi\sqrt{\frac{\bar{c}\,\bar{\mathcal{L}}}{6}} = S_{\text{outer}}. \tag{10.39}$$

The left hand side is the Bekenstein–Hawking entropy associated with the outer horizon $A_{\text{out}}$, while the right hand side is the Cardy formula for outer horizons. The central charges $c$ and $\bar{c}$ are given by

$$[\mathfrak{L}_n,\, \mathfrak{L}_m] = (n-m)\,\mathfrak{L}_{n+m} + \frac{c}{12}\,(n^3-n)\,\delta_{n+m,\,0}, \tag{10.40a}$$

$$[\bar{\mathfrak{L}}_n,\, \bar{\mathfrak{L}}_m] = (n-m)\,\bar{\mathfrak{L}}_{n+m} + \frac{\bar{c}}{12}\,(n^3-n)\,\delta_{n+m,\,0}, \tag{10.40b}$$

[3] Originally the Cardy formula provided a way to count the number of states of a CFT at non-zero temperature [151]. However, it also coincides with the thermal entropy of BTZ black holes [13] and thus provides a holographic way to count the microstates of the BTZ.

where $\mathcal{L} \equiv \mathfrak{L}_0$, $\bar{\mathcal{L}} \equiv \bar{\mathfrak{L}}_0$ and their respective conformal weights $\mathfrak{h}, \bar{\mathfrak{h}}$ when acting on a highest weight state $|\mathfrak{h}, \bar{\mathfrak{h}}\rangle$ are related to the mass and angular momentum in the usual way

$$\mathcal{L} = \mathfrak{h} = \frac{1}{2}(M\ell - J), \qquad \bar{\mathcal{L}} = \bar{\mathfrak{h}} = \frac{1}{2}(M\ell + J). \tag{10.41}$$

In order to perform the İnönü–Wigner contraction I make the same identifications as in (10.22) which in turn also means that the eigenvalues, $h_L$, $h_M$, of $L_0$ and $M_0$ when acting on a highest weight state $|h_L, h_M\rangle$ are given by

$$h_L = \mathfrak{h} - \bar{\mathfrak{h}}, \qquad h_M = \frac{1}{\ell}\left(\mathfrak{h} + \bar{\mathfrak{h}}\right). \tag{10.42}$$

In addition, I define the quantities $\mathcal{M}$ and $\mathcal{N}$ in a similar fashion as in (10.14)

$$\mathcal{M} = 12\left(\frac{\mathcal{L}}{c} + \frac{\bar{\mathcal{L}}}{\bar{c}}\right), \qquad \mathcal{N} = 6\ell\left(\frac{\mathcal{L}}{c} - \frac{\bar{\mathcal{L}}}{\bar{c}}\right). \tag{10.43}$$

Expressing $c$, $\bar{c}$, $\mathcal{L}$ and $\bar{\mathcal{L}}$ in terms of $c_L$, $c_M$, $\mathcal{M}$ and $\mathcal{N}$ and inserting into $S_{\text{outer}}$ from (10.39) yields

$$S_{\text{outer}} = 2\pi\sqrt{\frac{c\,\mathcal{L}}{6}} + 2\pi\sqrt{\frac{\bar{c}\,\bar{\mathcal{L}}}{6}} = \frac{\pi}{6}\ell\, c_M\sqrt{\mathcal{M}} + \mathcal{O}(\tfrac{1}{\ell}), \tag{10.44}$$

which is obviously not the correct result as its $\ell \to \infty$ limit diverges with $\ell$ as expected.

As mentioned earlier the cosmological horizon of a FSC is obtained as a limit of the inner BTZ horizon. Thus, one has to consider a modified Cardy formula for the BTZ in order to take the limit. This modified Cardy formula should count the microstates of the inner BTZ horizon in order to be a valid starting point for a flat space contraction. Such a modified Cardy formula is given by [148, 149]:

$$S_{A_{\text{int}}} = \frac{A_{\text{int}}}{4G} = \left|2\pi\sqrt{\frac{c\,\mathcal{L}}{6}} - 2\pi\sqrt{\frac{\bar{c}\,\bar{\mathcal{L}}}{6}}\right| = S_{\text{inner}}. \tag{10.45}$$

The modification in comparison to the (standard) Cardy formula (10.39) consists of a relative minus sign between the right-($\mathcal{L}$) and left-($\bar{\mathcal{L}}$) moving contributions.

In order to perform the İnönü–Wigner contraction I repeat the same steps as before but use now (10.45) instead of (10.39). This yields

$$S_{\text{inner}} = \left|2\pi\sqrt{\frac{c\,\mathcal{L}}{6}} - 2\pi\sqrt{\frac{\bar{c}\,\bar{\mathcal{L}}}{6}}\right| = \frac{\pi}{6}\left|c_L\sqrt{\mathcal{M}} + c_M\frac{\mathcal{N}}{\sqrt{\mathcal{M}}}\right| + \mathcal{O}(\tfrac{1}{\ell}). \tag{10.46}$$

Taking the $\ell \to \infty$ limit gives a prediction for the microscopic entropy of the dual quantum field theory:

$$S_{\text{FSC}} = \frac{\pi}{6}\left|c_L\sqrt{\mathcal{M}} + c_M\frac{\mathcal{N}}{\sqrt{\mathcal{M}}}\right| = \pi\sqrt{\frac{c_M h_M}{6}}\left|\frac{h_L}{h_M} + \frac{c_L}{c_M}\right|. \tag{10.47}$$

This agrees precisely with the results obtained in [61, 62, 71].
Another sanity check is to compare the results obtained this way with the results for Einstein gravity where $c = \bar{c}$ and hence $c_L = 0$. The expression (10.47) then simplifies to

$$S_{\text{FSC}}^{\text{Einstein}} = \frac{\pi}{6}\left| c_M \frac{\mathcal{N}}{\sqrt{\mathcal{M}}} \right| = 2\pi\,|h_L|\sqrt{\frac{c_M}{24\,h_M}}. \tag{10.48}$$

The result (10.48) (after translating conventions for $c$-normalization) agrees perfectly with the results in [61, 62].

**Flat Space Chiral Gravity:** One can also use the contractions used previously to determine the microscopic entropy of flat space chiral gravity (FS$\chi$G), a theory that can be obtained as a limit [57] of Topologically Massive Gravity (TMG) [152]. The corresponding action is given by

$$I_{\text{TMG}} = \frac{1}{16\pi G_N}\int \mathrm{d}^3x\sqrt{-g}\left(R + \frac{1}{\mu}CS[\Gamma]\right), \tag{10.49}$$

where $CS[\Gamma] = \varepsilon^{\lambda\mu\nu}\Gamma^{\rho}{}_{\lambda\sigma}\left(\partial_\mu\Gamma^{\sigma}{}_{\rho\nu} + \frac{2}{3}\Gamma^{\sigma}{}_{\mu\tau}\Gamma^{\tau}{}_{\nu\rho}\right)$ is a gravitational Chern-Simons term and $\mu$ is the corresponding Chern-Simons coupling. Flat space chiral gravity arises in the limit $G_N \to \infty$ while keeping fixed $\mu G_N$ so that $\mu G_N = \frac{c_L}{3}$ remains finite. This is particularly interesting as the central charges of the dual field theory are of a form that allow for unitary highest-weight representations of the $\mathfrak{bms}_3 \sim \mathfrak{gca}_2$ algebra [III], i.e. $c_L \neq 0$, $c_M = 0$, a topic I will also elaborate on in Chapter 11.
In [71] it has been shown that the entropy formula for FSCs in TMG takes exactly the same form as (10.47) but with $c_L = \frac{3}{\mu G_N}$, $c_M = \frac{3}{G_N}$, $h_L = M + \frac{1}{8G_N}$ and $h_M = J + \frac{M}{\mu}$. In this limit it is easy to see that $c_M \to 0$ and $\sqrt{\frac{c_M}{h_M}} = \sqrt{\frac{c_L}{h_L}}$. Thus, the entropy for FSCs in flat space chiral gravity is given by

$$S_{\text{FSC}}^{\text{FS}\chi\text{G}} = 2\pi\sqrt{\frac{c_L h_L}{6}} = S_{\text{CFT}}^{\text{Chiral}}. \tag{10.50}$$

The result (10.50) coincides precisely with what one would expect of one chiral half of a CFT [63]. This fits very nicely with the suggestion that flat space chiral gravity is indeed the chiral half of a CFT [57, 153].

**FSC with Spin-3 Charges:** Even though I have not yet introduced higher-spin symmetries in flat space, it is nevertheless instructive to consider a flat space limit from a rotating BTZ black hole with spin-3 charges $\mathcal{W}$ and $\bar{\mathcal{W}}$ [150]. This will enable one to make a prediction for a Cardy-like formula for FSCs that carry spin-3 charges.
Following [135] one can write a Cardy-like formula for the *outer* horizon entropy of a spin-3 charged BTZ as

$$S_{\text{outer}} = 2\pi\left(\sqrt{\frac{c\mathcal{L}}{6}}\sqrt{1 - \frac{3}{4C}} + \sqrt{\frac{\bar{c}\bar{\mathcal{L}}}{6}}\sqrt{1 - \frac{3}{4\bar{C}}}\right), \tag{10.51}$$

where $C$ and $\bar{C}$ are dimensionless constants defined via

$$\sqrt{\frac{c}{6\mathcal{L}^3}}\frac{\mathcal{W}}{4} = \xi = \frac{C-1}{C^{\frac{3}{2}}}, \qquad \sqrt{\frac{\bar{c}}{6\bar{\mathcal{L}}^3}}\frac{\bar{\mathcal{W}}}{4} = \bar{\xi} = \frac{\bar{C}-1}{\bar{C}^{\frac{3}{2}}}, \tag{10.52}$$

and the $C \to \infty$ ($\bar{C} \to \infty$) limit corresponds to the limit of vanishing spin-3 charges. In order to successfully perform a contraction that yields a Cardy-like formula for a spin-3 charged FSC one has again to determine the *inner* horizon spin-3 BTZ formula. In addition one has to find an expression of $C$ and $\bar{C}$ in terms of flat space analogues of these constants, which I will call $\mathcal{R}$ and $\mathcal{P}$. This is actually a non-trivial problem since $C$ and $\bar{C}$ are related to the canonical charges $(\mathcal{L}, \bar{\mathcal{L}}, \mathcal{W}, \bar{\mathcal{W}})$ in a nonlinear way. In the following I present a method to solve this problem. First I will introduce the flat space analogues of the spin-3 charges $\mathcal{W}$ and $\bar{\mathcal{W}}$ as

$$\mathcal{V} = 12\left(\frac{\mathcal{W}}{c} + \frac{\bar{\mathcal{W}}}{\bar{c}}\right), \qquad \mathcal{Z} = 6\ell\left(\frac{\mathcal{W}}{c} - \frac{\bar{\mathcal{W}}}{\bar{c}}\right). \tag{10.53}$$

Using these relations and replacing $\mathcal{W}$ and $\bar{\mathcal{W}}$ by $\mathcal{V}$ and $\mathcal{Z}$ in (10.52) one can deduce a suitable ansatz for $C$ and $\bar{C}$ in terms of $\mathcal{R}$ and $\mathcal{P}$ by demanding that up to $\mathcal{O}(\frac{1}{\ell^2})$ the l.h.s and the r.h.s of (10.52) have to agree. It turns out that a suitable ansatz for $C$ and $\bar{C}$ is given by

$$C = \mathcal{R} + \frac{2}{\ell}D(\mathcal{R},\mathcal{P},\mathcal{M},\mathcal{N}), \qquad \bar{C} = \mathcal{R} - \frac{2}{\ell}D(\mathcal{R},\mathcal{P},\mathcal{M},\mathcal{N}). \tag{10.54}$$

If one identifies

$$\frac{\mathcal{V}}{2\mathcal{M}^{\frac{3}{2}}} = \frac{\mathcal{R}-1}{\mathcal{R}^{\frac{3}{2}}}, \qquad \frac{\mathcal{Z}}{\mathcal{N}\sqrt{\mathcal{M}}} = \mathcal{P}, \tag{10.55}$$

then $D(\mathcal{R},\mathcal{P},\mathcal{M},\mathcal{N})$ is given by

$$D(\mathcal{R},\mathcal{P},\mathcal{M},\mathcal{N}) = \frac{\mathcal{N}}{\mathcal{M}}\frac{\mathcal{R}\left(\mathcal{R}^{\frac{3}{2}}\mathcal{P} + 3\mathcal{R} - 3\right)}{(\mathcal{R}-3)}. \tag{10.56}$$

Again, as in the spin-2 case, the outer horizon limit does not yield the correct result as can be easily shown by replacing all AdS quantities in (10.51) by their flat space counterparts and taking the $\ell \to \infty$ limit. This yields the following expression

$$S_{\text{outer}} = \frac{\pi}{6}\ell c_M\sqrt{\mathcal{M}}\sqrt{1 - \frac{3}{4\mathcal{R}}} + \mathcal{O}(\tfrac{1}{\ell}), \tag{10.57}$$

which is divergent with $\ell$ and thus, as expected, the outer horizon formula (10.51) is not the correct expression for a contraction to flat space. In close analogy to the spin-2 case I assume that the following formula provides an appropriate microstate counting of the *inner* horizon entropy of a spin-3 charged BTZ

$$S_{\text{inner}} = 2\pi\left|\sqrt{\frac{c\mathcal{L}}{6}}\sqrt{1 - \frac{3}{4C}} - \sqrt{\frac{\bar{c}\bar{\mathcal{L}}}{6}}\sqrt{1 - \frac{3}{4\bar{C}}}\right|, \tag{10.58}$$

whose only difference, as compared to the outer horizon formula, is again a relative minus sign between the two left- and right-moving contributions. In addition the $C \to \infty$ and $\bar{C} \to \infty$ limit yields the correct expression for the spin-2 inner horizon formula (10.45) as it should be. Using (10.58) and performing the same steps as before one obtains the following result after taking the $\ell \to \infty$ limit

$$S_{\text{FSC}}^{\text{Spin-3}} = \frac{\pi}{6}\left| c_L \sqrt{\mathcal{M}}\sqrt{1-\frac{3}{4\mathcal{R}}} + c_M \frac{\mathcal{N}\left(4\mathcal{R} - 6 + 3\mathcal{P}\sqrt{\mathcal{R}}\right)}{4\sqrt{\mathcal{M}}(\mathcal{R}-3)\sqrt{1-\frac{3}{4\mathcal{R}}}}\right|. \tag{10.59}$$

As expected the $\mathcal{R} \to \infty$ limit yields again (10.47). It is also important to note that the part of (10.59) which is proportional to $c_M$ can alternatively also be derived by solving holonomy conditions (as in the AdS case) [VII], which I will also show explicitly in Chapter 11. This shows that my assumption (10.58) is not only plausible but seems indeed to be the correct expression for the inner horizon entropy of a spin-3 charged BTZ black hole as its İnönü–Wigner contraction leads to the correct flat space result.

With this derivation I want to close this chapter on flat space as a limit from AdS and proceed with the next chapter, in which I will study unitarity of highest-weight representations of $\mathcal{FW}$-algebras.

# 11 Unitarity in Flat Space Higher-Spin Gravity

*Curiously enough, the only thing that went through the mind of the bowl of petunias as it fell was Oh no, not again.*

**– Douglas Adams**
The Hitchhiker's Guide to the Galaxy

Having derived some of the $\mathcal{FW}$-algebras, which play a prominent role as the asymptotic symmetries of higher-spin theories in flat space in Chapter 10, using İnönü–Wigner contractions, I will now proceed in determining unitary highest-weight representations of these algebras. In the following I will assume that the $\mathcal{FW}$-algebras derived previously are genuine algebras in the sense that the central charges $c_L$, $c_M$ can take arbitrary values. The unitarity analysis will follow the same logic as in Chapter 9, i.e. first defining a suitable vacuum state, hermitian conjugation of the operators and then determining the Gramian matrix for the first couple of excitation levels. Before proceeding to higher-spins, however, I will first consider the simple case of spin-2 gravity whose asymptotic symmetry algebra is given by the $\mathfrak{bms}_3$ algebra (10.24). This example already yields a heavy restriction on unitarity as I will show shortly.
In order to proceed one has to choose a vacuum. My choice is defined by the highest weight conditions

$$L_n|0\rangle = M_n|0\rangle = 0, \quad \forall n \geq -1\,. \tag{11.1}$$

While the vacuum conditions (11.1) seem pretty natural from a CFT point of view and lead to a Poincaré-invariant vacuum, it is nevertheless also possible that there might be other sensible choices of vacuum. For the purpose of this thesis, however, I will stick to the vacuum definition (11.1) and appropriate higher-spin generalizations thereof. Thus, whenever I am making statements about unitarity of $\mathcal{FW}$-algebras I mean unitarity with respect to the highest-weight conditions (and higher-spin versions thereof) (11.1).
With similar caveats, I define also hermitian conjugation in a standard way,

$$L_n^\dagger := L_{-n}, \qquad M_n^\dagger := M_{-n}\,. \tag{11.2}$$

Having defined the vacuum and hermitian conjugation, the next step is to address the issue of unitarity. Calculating the inner products of all level-2 descendants yields the Gramian matrix

$$K^{(2)} = \left( \begin{array}{c|cc} & L_{-2} & M_{-2} \\ \hline (L_{-2})^\dagger & \frac{c_L}{2} & \frac{c_M}{2} \\ (M_{-2})^\dagger & \frac{c_M}{2} & 0 \end{array} \right). \tag{11.3}$$

which has determinant $\det\left(K^{(2)}\right) = -\frac{c_M^2}{4}$. Thus, if $c_M \neq 0$ then there is always a positive and a negative norm state, regardless of the signs of the central charges. Therefore, as long as $c_M \neq 0$ the algebra (10.24) does not have any unitary representations[1]. Note that this argument also applies to more general algebras that contain the $\mathfrak{bms}_3$ algebra as a subalgebra such as $\mathcal{FW}$-algebras.

For $c_M = 0$ the Gramian matrix (11.3) has vanishing determinant, which means there is at least one null state. Assuming $c_L \neq 0$, there is exactly one null state and one state whose norm depends on the sign of $c_L$. In this case one can mod out all $M_{-n}$ descendants of the vacuum, since they are all null states [57, 77], and one is left with just a single copy of the Virasoro algebra and the corresponding Virasoro descendants of the vacuum, $L_{-n}|0\rangle$. Then standard CFT considerations of unitary representations of the Virasoro algebra apply. In particular this means that the central charge $c_L$ must be positive in order to have unitary representations.

In conclusion, the necessary conditions for unitarity of $\mathfrak{bms}_3$ algebras of the form (10.24) with vacuum (11.1) and hermitian conjugation (11.2) are

$$c_M = 0, \qquad c_L \geq 0\,, \tag{11.4}$$

whereas the resulting unitary theory is only non-trivial for $c_L > 0$.

This chapter is organized as follows. I will first discuss unitarity of two simple examples of $\mathcal{FW}$-algebras, namely $\mathcal{FW}_3$ and $\mathcal{FW}_3^{(2)}$. I will then discuss unitarity of $\mathcal{FW}_4^{(2-1-1)}$ and $\mathcal{FW}_N^{(2)}$ algebras which will yield certain restrictions on the central charge $c_L$. Then, using arguments based on dimensional analysis, I will show that for *nonlinear* $\mathcal{FW}$-algebras, under certain assumptions, it is not possible to have flat space, unitarity and higher-spin excitations at the same time. I will then use this NO-GO result to show the absence of multi-graviton excitations in $\mathcal{FW}_4^{(2-2)}$. I will close this chapter by presenting an example of a YES-GO result using the *linear* $\mathcal{FW}_\infty$ algebra, which shows that for *linear* $\mathcal{FW}$-algebras it is indeed possible to have flat space, unitarity and higher-spins at the same time.

## 11.1 Unitarity of $\mathcal{FW}_3$

Let me now consider $\mathcal{FW}_3$ whose commutation relations are given by (10.30) as the first example of a nonlinear $\mathcal{FW}$-algebra and address unitary representations of this algebra.

[1] Again, provided one sticks to the vacuum (11.1) and hermitian conjugation (11.2).

The vacuum is again defined by the conditions (11.1) and in addition

$$U_n|0\rangle = V_n|0\rangle = 0, \qquad \forall n \geq -2\,. \tag{11.5}$$

Similarly, the generators (11.2) are supplemented by $U_n^\dagger := U_{-n}$, $V_n^\dagger := V_{-n}$. The spin-2 result for level-2 descendants (11.3) still applies, which means that unitarity again requires the necessary conditions (11.4). As discussed in [150], the condition $c_M = 0$ leads to a further contraction of the algebra (10.30). This is due to the appearance of inverse powers of $c_M$ in the commutation relations of flat space higher-spin generators $U_n$ and $V_n$. Singularities in the limit $c_M \to 0$ can be avoided if one rescales the generators

$$U_n \to c_M U_n, \tag{11.6}$$

before taking the limit $c_M \to 0$. The contracted algebra then simplifies considerably and the non-vanishing commutators read

$$[L_n,\, L_m] = (n-m)L_{n+m} + \frac{c_L}{12}\,(n^3-n)\,\delta_{n+m,\,0}, \tag{11.7a}$$

$$[L_n,\, M_m] = (n-m)M_{n+m}, \tag{11.7b}$$

$$[L_n,\, U_m] = (2n-m)U_{n+m}, \tag{11.7c}$$

$$[L_n,\, V_m] = (2n-m)V_{n+m}, \tag{11.7d}$$

$$[U_n,\, U_m] \propto [U_n, V_m] = 96(n-m)\,\Theta_{n+m}. \tag{11.7e}$$

As in the spin-2 case, the remaining non-trivial part of the algebra is a single copy of the Virasoro algebra, if $c_L \neq 0$. In particular, all the descendants of higher-spin generators $U_{-n}$, $V_{-n}$ and of the supertranslations $M_{-n}$ are null states. This means that, at least for the $\mathcal{FW}_3$ algebra, unitarity in flat space higher-spin gravity leads to an elimination of all physical higher-spin states. The only physical states that arise as descendants of the vacuum are the usual Virasoro descendants.

This result does not depend on the specific $\mathcal{FW}_3$ algebra discussed in this example. The exact same conclusions are reached for $\mathcal{FW}_N$ algebras with $N > 3$, which can be constructed by generalizing the contraction (11.5) to all higher-spin generators [154]. The resulting $\mathcal{FW}_N$ algebras that are compatible with unitarity are again similar to (11.7) and all higher-spin states decouple, which only leaves the Virasoro descendants of the vacuum.

Since $\mathcal{FW}_N$ only covers a very specific class of $\mathcal{FW}$-algebras, it is of course also of interest to check whether or not the same restrictions on unitarity and higher-spin states apply to other classes of $\mathcal{FW}$- algebras that arise as Drinfeld-Sokolov reductions of the various non-principally embedded algebras, or the more general $\mathfrak{ihs}[\lambda]$ case[2].

[2] This is a contraction of two copies of $\mathfrak{hs}[\lambda]$. See Appendix D.4 for more details.

## 11.2 Unitarity of Contracted Polyakov-Bershadsky $\mathcal{FW}_3^{(2)}$

Since the contracted Polyakov-Bershadsky algebra $\mathcal{FW}_3^{(2)}$ contains the $\mathfrak{bms}_3$ algebra as a subalgebra, there are no unitary representations of the algebra (10.37) for $c_M \neq 0$, as outlined in the beginning of this chapter. Thus, in order to obtain unitary representations one has to take the limit $c_M \to 0$, which requires appropriate rescalings of the generators as

$$U_n^\pm \to \hat{U}_n^\pm = c_M U_n^\pm, \qquad V_n^\pm \to \hat{V}_n^\pm = c_M V_n^\pm. \tag{11.8}$$

Taking the limit $c_M \to 0$ leads to a further contraction of the $\mathcal{FW}_3^{(2)}$-algebra. The non-vanishing commutators are given by

$$[L_n, L_m] = (n-m)L_{m+n} + \frac{c_L}{12} n(n^2-1)\delta_{n+m,0}, \tag{11.9a}$$
$$[L_n, J_m] = -\,mJ_{n+m}, \tag{11.9b}$$
$$[J_n, J_m] = \frac{32-c_L}{9}\, n\,\delta_{n+m,0}, \tag{11.9c}$$
$$[L_n, M_m] = (n-m)M_{m+n}, \tag{11.9d}$$
$$[L_n, K_m] = [M_n, J_m] = -mK_{n+m}, \tag{11.9e}$$
$$[L_n, \hat{U}_m^\pm] = \left(\frac{n}{2}-m\right)\hat{U}_{n+m}^\pm, \tag{11.9f}$$
$$[L_n, \hat{V}_m^\pm] = [M_n, \hat{U}_m^\pm] = \left(\frac{n}{2}-m\right)\hat{V}_{n+m}^\pm, \tag{11.9g}$$
$$[J_n, \hat{U}_m^\pm] = \pm\,\hat{U}_{n+m}^\pm, \tag{11.9h}$$
$$[J_n, \hat{V}_m^\pm] = [K_n, \hat{U}_m^\pm] = \pm\hat{V}_{n+m}^\pm, \tag{11.9i}$$
$$[\hat{U}_n^+, \hat{U}_m^-] = -\,18\,(c_L-26)\ :KK:_{n+m}\ . \tag{11.9j}$$

I will define the hermitian conjugates of the operators $L_n$, $M_n$, $J_n$, $K_n$, $U_n^\pm$, and $V_n^\pm$ as

$$L_n^\dagger := L_{-n}, \qquad M_n^\dagger := M_{-n}, \qquad J_n^\dagger := J_{-n}, \qquad K_n^\dagger := K_{-n}, \tag{11.10a}$$
$$(U_n^\pm)^\dagger := U_{-n}^\mp, \qquad (V_n^\pm)^\dagger := V_{-n}^\mp, \tag{11.10b}$$

and the vacuum in the usual way

$$L_n|0\rangle = M_n|0\rangle = 0, \qquad \{\forall n \in \mathbb{Z}| n \geq -1\}, \tag{11.11a}$$
$$J_m|0\rangle = K_m|0\rangle = 0, \qquad \{\forall m \in \mathbb{Z}| m \geq 0\}, \tag{11.11b}$$
$$U_p^\pm|0\rangle = V_p^\pm|0\rangle = 0, \qquad \{\forall p \in \mathbb{Z}+\frac{1}{2}| p \geq -\frac{1}{2}\}. \tag{11.11c}$$

The central terms proportional to $c_L$ in the Virasoro algebra (11.9a) and the current algebra (11.9c) must have non-negative signs in order to have a unitary representa-tion of this algebra. This immediately implies lower and upper bounds on the central charge $c_L$

$$0 \leq c_L \leq 32. \tag{11.12}$$

States generated by $M_n$, $K_n$, $U_n^\pm$ and $V_n^\pm$ have zero norm and are orthogonal to all other states. Thus one can mod out these states and extend the previous definition of the vacuum in the following way

$$L_n|0\rangle = J_{n+1}|0\rangle = 0, \qquad \{\forall n \in \mathbb{Z} | n \geq -1\}, \tag{11.13a}$$
$$M_m|0\rangle = K_m|0\rangle = 0, \qquad \forall m \in \mathbb{Z}, \tag{11.13b}$$
$$U_p^\pm|0\rangle = V_p^\pm|0\rangle = 0, \qquad \forall p \in \mathbb{Z} + \frac{1}{2}. \tag{11.13c}$$

The only states that remain in the theory for $0 \leq c_L \leq 32$ are descendants of the vacuum $L_{-n}|0\rangle$ with $n > 1$, $J_{-m}|0\rangle$ with $m > 0$ or combinations thereof. In order to have a well defined basis of states at level $N$, I will use the following ordering of operators

$$J_{-n_1}^{m_1} \dots J_{-n_p}^{m_p} L_{-n_{p+1}}^{m_{p+1}} \dots L_{-n_N}^{m_N} |0\rangle\,, \tag{11.14}$$

with

$$m_i \in \mathbb{N}, \qquad \sum_{i=1}^{N} m_i n_i = N, \qquad n_1 > \dots > n_p, \qquad n_{p+1} > \dots > n_N, \tag{11.15a}$$
$$n_1, \dots, n_p \in \mathbb{N}\backslash\{0\}, \qquad n_{p+1}, \dots, n_N \in \mathbb{N}\backslash\{0,1\}. \tag{11.15b}$$

At level 1 there is only one state generated by $J_{-1}$. The norm of this state is given by $C_J = -\frac{(c_L-32)}{9}$, which is non-negative only if the bound on the central charge (11.12) holds.

At level 2 three states generated by $L_{-2}$, $J_{-2}$ and $J_{-1}^2$ are present. The Gramian matrix $K^{(2)}$ is given by

$$K^{(2)} = \left(\begin{array}{c|ccc} & L_{-2} & J_{-2} & J_{-1}^2 \\ \hline (L_{-2})^\dagger & \frac{c_L}{2} & 0 & C_J \\ (J_{-2})^\dagger & 0 & 2C_J & 0 \\ (J_{-1}^2)^\dagger & C_J & 0 & 2C_J^2 \end{array}\right). \tag{11.16}$$

For $C_J = 0$ the two $J$-descendants are null states. For $c_L = 0$ descendants generated by $L_n$ have zero norm, but are not orthogonal to all other states in the theory, so they are not null states and cannot be modded out. Two of the three eigenvalues of the Gramian (11.16), $\lambda_0 = 2C_J$ and $\lambda_+ = \frac{1}{4}(c_L + 4C_J^2 + \sqrt{(c_L + 4C_J^2)^2 + (1 - c_L)16C_J^2})$ are non-negative in the whole range $0 \leq c_L \leq 32$. The third eigenvalue, however, $\lambda_- = \frac{1}{4}(c_L + 4C_J^2 - \sqrt{(c_L + 4C_J^2)^2 + (1 - c_L)16C_J^2})$, changes its sign at $c_L = 1$ and is positive in the range $1 \leq c_L \leq 32$. At $c_L = 1$ the descendant associated to $L_{-2}$ is proportional to the $J_{-1}^2$ descendant.

$$L_{-2}|0\rangle = \tfrac{9}{62}\, J_{-1}^2|0\rangle. \tag{11.17}$$

I have checked explicitly that the key features discussed above also persist for level 3 and 4 descendants of the vacuum. In particular, non-negativity of the eigenvalues of the Gramian matrix always restricts $c_L$ to values that are larger or equal to 1. I expect the same to hold true for arbitrary levels larger than four.
Thus, the region $0 \leq c_L \leq 1$ is also excluded for unitary representations and the necessary conditions for the central charge to be consistent with unitarity based on this analysis are

$$1 \leq c_L \leq 32\,. \tag{11.18}$$

For $c_L = 1$, the determinant of the Gram matrix vanishes and some states become linearly dependent. In this case only $J_{-n}$ descendants remain in the theory and all $L_{-n}$ descendants depend linearly on them. It is noteworthy that this is also the value where the Polyakov–Bershadsky algebra has its only non-trivial unitary representation [I]. For $c_L = 32$, on the other hand, the states corresponding to the $\hat{\mathfrak{u}}(1)$ part of the algebra become null states and only the Virasoro modes remain.
While it is possible that there are further restrictions coming from unitarity, the results above show already two remarkable features:

- Requiring the absence of negative norm states implies that all higher-spin descendants of the vacuum become null states and drop out of the physical spectrum.
- This analysis also implies a lower and upper bound on the central charge, in this case $1 \leq c_L \leq 32$.

Since one can not conclude that this is a general feature by simply looking at one specific example of a $\mathcal{FW}$-algebra, I will generalize the results of this section to more generic $\mathcal{FW}$-algebras in the following section.

## 11.3 Unitarity of General $\mathcal{FW}$-Algebras

I already argued previously that flat space higher-spin gravity cannot be unitary for $\mathcal{FW}_N$ algebras which correspond to the Drinfeld-Sokolov reduction of principally embedded $\mathfrak{isl}(2,\mathbb{R}) \hookrightarrow \mathfrak{isl}(N,\mathbb{R})$ for arbitrary values of $c_L$ and $c_M$. In order to obtain such a unitary representations one has to perform an additional contraction, sending $c_M \to 0$, where all higher-spin states become null. In addition, the results found in Section 11.2 suggest that this is a general feature of nonlinear $\mathcal{FW}$-algebras. In this section I will show that this feature indeed persists for other types of nonlinear $\mathcal{FW}$-algebras, which is a very strong indication that one cannot have flat space, higher-spins and unitarity at the same time. Furthermore, I will also present an example of a linear $\mathcal{FW}$ algebra, namely $\mathcal{FW}_\infty$, where it is possible to have flat space, higher-spins and unitarity at the same time.

### 11.3.1 Upper Bound on the Central Charge of $\mathcal{FW}_4^{(2-1-1)}$

In this section I will show that the existence for an upper bound on the central charge arises not only in the Polyakov–Bershadsky algebra, but also in the contraction derived from the $\mathcal{W}_4^{(2-1-1)}$-algebra, which I denote $\mathcal{FW}_4^{(2-1-1)}$, in accordance with the conventions chosen previously. Details regarding the contraction and explicit expressions of the commutation relations of $\mathcal{FW}_4^{(2-1-1)}$ can be found in Appendix D.1. The $\mathcal{FW}_4^{(2-1-1)}$-algebra (D.4) is generated by $L_n$, $M_n$, $O_n$, $P_n$, $Q_n^a$, $R_n^a$, $U_n^{a|b}$ and $V_n^{a|b}$, with $a, b = \pm$. As in the $\mathcal{FW}_3^{(2)}$ case one has to require $c_M = 0$ and as such some of the generators have to be rescaled as

$$\hat{U}_n^{\pm|a} = c_M U_n^{\pm|a}, \quad \text{and} \quad \hat{V}_n^{\pm|a} = c_M V_n^{\pm|a}, \tag{11.19}$$

before taking $c_M \to 0$. This is, however not the only subtlety one has to take care of for this algebra. Since there is an affine $\hat{\mathfrak{su}}(2)$ (D.4q) subalgebra contained in this algebra one has to be a bit more careful on how to properly define hermitian conjugation of these operators. Thus, I will make the following ansatz

$$\left(Q_n^0\right)^\dagger = Q_{-n}^0, \qquad (Q_n^\pm)^\dagger = \gamma Q_{-n}^\mp, \tag{11.20a}$$
$$\left(R_n^0\right)^\dagger = R_{-n}^0, \qquad (R_n^\pm)^\dagger = \gamma R_{-n}^\mp, \tag{11.20b}$$
$$\left(\hat{U}_n^{\pm|a}\right)^\dagger = \mu^\pm \nu^a \hat{U}_{-n}^{\mp|-a}, \qquad \left(\hat{V}_n^{\pm|a}\right)^\dagger = \mu^\pm \nu^a \hat{V}_{-n}^{\mp|-a}, \tag{11.20c}$$

where $\gamma$, $\mu^\pm$ and $\nu^\pm$ are some real numbers which will be determined by demanding consistency with the contracted algebra at hand. I will first look at

$$([Q_n^\pm, Q_m^\mp])^\dagger = -\gamma^2 [Q_{-n}^\mp, Q_{-m}^\pm] = \pm 2 Q_{-(n+m)}^0 - \frac{42 - c_L}{6} n \delta_{n+m,0}, \tag{11.21}$$

which suggests that $\gamma^2 = 1$ in order to satisfy (D.4q). Now choosing $\gamma = -1$ the norm of the states $Q_{-n}^0|0\rangle$, $Q_{-n}^\pm|0\rangle$ has the same sign, as desired. What remains is to check whether or not a choice for $\mu^\pm$ and $\nu^\pm$ exists that is compatible with the algebra (D.4). In order to do this I will look at

$$\left(\left[Q_n^\pm, \hat{U}_m^{\mp|a}\right]\right)^\dagger = \mu^\mp \nu^a \left[Q_{-n}^\mp, \hat{U}_{-m}^{\pm|-a}\right] = \pm \left(\hat{U}_{n+m}^{\pm|a}\right)^\dagger = \pm \mu^\pm \nu^a \hat{U}_{-(n+m)}^{\mp|-a}. \tag{11.22}$$

Thus, in order to be consistent with (D.4s) the parameters $\mu^\pm$ must satisfy

$$\frac{\mu^+}{\mu^-} = -1. \tag{11.23}$$

The last thing to be checked is the rescaled $\left[\hat{U}_n^{\pm|\pm}, \hat{U}_m^{\pm|\mp}\right]$ commutator. Proceeding in the same manner as before one finds that

$$\nu^+\nu^- = -1. \tag{11.24}$$

This shows that one can consistently define hermitian conjugation in such a way that the norm of the states generated by operators associated with the affine $\hat{\mathfrak{su}}(2)$ subalgebra contained in (D.4) all have the same sign.
Looking at the Gramian matrix of the first few levels yields constraints on possible values of $c_L$. In contrast to the $\mathcal{FW}_3^{(2)}$ case there is not only a restriction on the lower bound of $c_L$ but also on possible maximal values of $c_L$,

$$29 - \sqrt{661} \approx 3,29 \leq c_L \leq 36 \quad \text{or} \quad c_L = 42. \tag{11.25}$$

In conclusion the $\mathcal{FW}_4^{(2-1-1)}$-algebra is unitary for $c_M = 0$ and $c_L$ obeying (11.25).

## 11.3.2 Contracted Feigin-Semikhatov Algebras $\mathcal{FW}_N^{(2)}$

$\mathcal{FW}_N^{(2)}$-algebras can be obtained as a contraction of two copies of $\mathcal{W}_N^{(2)}$ Feigin-Semikhatov algebras as described in Appendix D.3. Since I am again mostly interested in unitary representations of the algebra (D.14), the relevant algebras for this unitarity analysis will be the ones obtained after the $c_M \to 0$ limit.
The $\mathcal{FW}_N^{(2)}$-algebras (D.14) are generated by $L_n$, $M_n$, $O_n$, $K_n$, $U_n^{\pm}$, $V_n^{\pm}$, $W_n^s$ and $X_n^s$. In order to be able to properly take the $c_M \to 0$ limit one again has to rescale $U_n^{\pm}$, $V_n^{\pm}$, $W_n^s$, $X_n^s$ by appropriate powers of $c_M$ since the commutators in (D.14n)-(D.14q) have nonlinear terms proportional to powers of $\frac{1}{c_M}$. After performing the limit all the central terms in (D.14n)-(D.14q) are eliminated and the only issue one still has to worry about is whether or not terms proportional to powers of $L_{n+m}$ and $O_{n+m}$ can appear on the right hand side of (D.14n)-(D.14q). If that were the case then one could not simply mod out the states since the inner product of theses states with other states would yield a non-zero result. There is, however, a simple argument as to why such terms cannot appear.
The generators $M_n$, $K_n$, $V_n^{\pm}$ and $X_n^s$ acquire an additional dimension of inverse length via the parameter $\epsilon = \frac{1}{\ell}$ with $\ell$ being the AdS radius during the contraction (D.13a). Thus, terms proportional to powers purely consisting of $L_{n+m}$ and $O_{n+m}$ cannot carry inverse powers of $c_M$, which also is dimensionful. The only possibility for the appearance of terms that are powers of $L_{n+m}$ and $O_{n+m}$ and carry inverse powers of $c_M$ is via mixing with $M_{n+m}$, $P_{n+m}$, $V_{n+m}^{\pm}$ and $X_{n+m}^s$. Such terms only appear as cross terms during the contraction of the nonlinear operators and as such are not the ones with the highest inverse power of $c_M$ on the right hand side of (D.14n)-(D.14q). They are gone after a rescaling with the highest power of $c_M$. This in turn means that all the higher-spin states $W_{-n}^2|0\rangle$, $X_{-n}^s|0\rangle$ and descendants of $U^{\pm}_{-n}|0\rangle$, $V^{\pm}_{-n}|0\rangle$ are null states for unitary representations of the $\mathcal{FW}_N^{(2)}$-algebras. Repeating the same analysis as in the preceding section one finds that demanding the absence of negative norm states restricts the range of possible values of $c_L$ that could allow for unitary representations as

$$1 \leq c_L \leq 2(N-1)^2(N+1)\,. \tag{11.26}$$

This means that the value of $c_L$, for which one obtains a chiral CFT grows as $N^3$ for large $N$ in contrast to linear growth in $N$ in the Feigin-Semikhatov case [II, 139].

### 11.3.3 General NO-GO Result

The previously treated examples of $\mathcal{FW}$-algebras all point towards the conclusion that, under the assumptions I have been working with so far, it is not possible to find unitary representations of flat space higher-spin algebras that contain non-trivial higher-spin states. In this section, I will argue, on dimensional grounds, that this conclusion is generic for nonlinear flat space higher-spin algebras that can be obtained via İnönü–Wigner contractions of AdS higher-spin $\mathcal{W}$-algebras.
Suppose one starts from two copies of an inherently nonlinear $\mathcal{W}$-algebra. These algebras contain higher-spin generators that will be denoted by $W_n$ and $\bar{W}_n$, whose commutation relations can be schematically written as

$$[W_n,\, W_m] = \ldots + f(c)\ :AB:_{n+m} + \ldots + \omega(c) \prod_{j=-(s-1)}^{s-1} (n+j)\,\delta_{n+m,\,0}\,,$$
$$[\bar{W}_n,\, \bar{W}_m] = \ldots + f(\bar{c})\ :\bar{A}\bar{B}:_{n+m} + \ldots + \omega(\bar{c}) \prod_{j=-(s-1)}^{s-1} (n+j)\,\delta_{n+m,\,0}\,. \tag{11.27a}$$

These commutators contain nonlinear terms, which are given by infinite sums of products of other generators of the algebra that are denoted by $A_n$, $B_n$ and $\bar{A}_n$, $\bar{B}_n$, respectively[3]. The two copies of the $\mathcal{W}$-algebra are characterized by central charges $c$, $\bar{c}$. These central charges appear in the nonlinear terms and in the central charge terms in (11.27) via functions $f(c)$ and $\omega(c)$. I will keep these functions arbitrary for the sake of generality, apart from the following restriction on $f(c)$

$$\lim_{c\to\infty} f(c) = 0\,, \tag{11.28}$$

which means that the nonlinear terms are subleading contributions in the semi-classical regime of large central charges. The ellipses in (11.27) denote additional linear and other nonlinear terms which can possibly appear on the right-hand side of the commutator whose exact form is not important for this argument.
Starting from these two copies one can obtain a flat space higher-spin algebra via a nonrelativistic contraction, involving a contraction parameter $\epsilon$ of dimension $[\text{length}]^{-1}$. In analogy to the considerations in Chapter 10 I define the new generators of the contracted $\mathcal{FW}$-algebra as

$$U_n := W_n + \bar{W}_n\,, \qquad V_n := -\epsilon\left(W_n - \bar{W}_n\right)\,, \tag{11.29a}$$

$$C_n := A_n + \bar{A}_n\,, \qquad D_n := -\epsilon\left(A_n - \bar{A}_n\right)\,, \tag{11.29b}$$

$$E_n := B_n + \bar{B}_n\,, \qquad F_n := -\epsilon\left(B_n - \bar{B}_n\right)\,. \tag{11.29c}$$

[3] Of course for arbitrary $\mathcal{W}$-algebras the nonlinear terms can contain an arbitrary large number of products of an arbitrary number of generators. For the sake of simplicity and without loss of generality, however, I will only use bilinear terms in order to make my argument clear in what follows.

This means that the central charges $c_L$ and $c_M$ in terms of $c$ and $\bar{c}$ are given by (10.25). From these definitions, one can already infer that the generators $U_n$, $C_n$, $E_n$, as well as the central charge $c_L$ are dimensionless. Similarly, the generators $V_n$, $D_n$, $F_n$ and the central charge $c_M$ have dimension $[\text{length}]^{-1}$.
In order to obtain unitarity under the assumptions I am making, i.e. there is a well defined vacuum which is invariant under the wedge algebra of the $\mathcal{FW}$ algebra, one has to take the limit $c_M \to 0$ for the reasons explained in the beginning of this chapter. The main point of the following NO-GO theorem is that one cannot take the $c_M \to 0$ limit in such a way that non-trivial central terms remain on the r.h.s of the higher-spin commutators, which renders all higher-spin states to be null states that can be modded out from the resulting module.
For the commutator $[U_n, V_m]$, this is immediate on dimensional grounds. This commutator has dimensions of $[\text{length}]^{-1}$, and in order to have the same dimension, the central term appearing on the right hand side necessarily has to be proportional to $c_M$, implying that the $[U_n, V_m]$ commutator will be center-less in the limit $c_M \to 0$. The commutator $[V_n, V_m]$ is zero upon contraction, so the only non-trivial commutator remaining to be examined is $[U_n, U_m]$.
I will first look at the structure of the nonlinear terms of this commutator. Performing the contraction, one obtains

$$\begin{aligned}\lim_{\epsilon\to 0}[U_n,\,U_m] = \ldots + \lim_{\epsilon\to 0}\frac{1}{4\epsilon^2}\,&\big((f(c)+f(\bar{c}))(\epsilon^2\;{:}CE{:}_{n+m} + \;{:}DF{:}_{n+m})\\ &+\frac{(f(\bar{c})-f(c))}{\epsilon}({:}CF{:}_{n+m} + \;{:}DE{:}_{n+m})\big) + \ldots\,.\end{aligned} \tag{11.30}$$

For a generic function $f(c)$, so that (11.28) holds, one finds that

$$f(c)+f(\bar{c}) \sim \mathcal{O}(\epsilon^2) \quad \text{and} \quad f(c)-f(\bar{c}) \sim \mathcal{O}(\epsilon)\,. \tag{11.31}$$

The contraction $\epsilon \to 0$ can thus consistently be performed. The nonlinear terms that generically survive the contraction are the ones of the form ${:}DF{:}_{n+m}$, ${:}CF{:}_{n+m}$ and ${:}DE{:}_{n+m}$, which have dimensions $[\text{length}]^{-2}$, $[\text{length}]^{-1}$ and $[\text{length}]^{-1}$, respectively. Since the commutator $[U_n, U_m]$ is dimensionless, these nonlinear terms have to appear with prefactors that depend on $c_M$, in order to compensate for their dimension. The structure of the $[U_n, U_m]$ commutator is thus schematically given by

$$
\begin{aligned}
[U_n, U_m] = \ldots + \mathcal{O}(\tfrac{1}{c_M^2})\ :DF:_{n+m} + \mathcal{O}(\tfrac{1}{c_M})\,(:CF:_{n+m} + :DE:_{n+m}) \\
+ \tilde{\omega}(c_L) \prod_{j=-(s-1)}^{s-1} (n+j)\, \delta_{n+m,0},
\end{aligned}
\tag{11.32}
$$

where the central term has to be a function of $c_L$ on dimensional grounds.
In order to take the limit $c_M \to 0$, one thus has to rescale[4] $U_n \to \tilde{U}_n := c_M U_n$. The central terms in the $[\tilde{U}_n, \tilde{U}_m]$ commutator, however, do not survive this contraction, showing that all higher spin generators of the flat space algebra lead to null states when acting on the vacuum. Although I have given the argument for the case in which the nonlinear terms are quadratic, it can easily be extended to the case where nonlinear terms of higher order appear. I have thus shown that unitary representations that contain non-trivial higher spin states are not possible for inherently nonlinear flat space higher spin algebras, at least not under the assumptions I have been working with.
Since every NO-GO theorem is only as good as the assumptions it is built upon I want to emphasize that there are at least three possible loopholes that could circumvent this theorem.

- I have chosen a highest-weight state that satisfies the conditions (11.1). It could also very well be that a highest-weight representation is simply not the correct representation to consider. One possible candidate for a suitable unitary representation of $\mathcal{FW}$-algebras could be so-called “rest-frame” states which were proposed in [IX] to explain unitarity of the one-loop partition function in flat space.

- Assuming the highest-weight conditions (11.1) hold, then another possible loophole to the NO-GO theorem could be to use a different prescription when determining the norm of states as in comparison to the one I used in this analysis.

- If both the highest-weight conditions (11.1) and the prescription to determine the norm of states are the same as the one I used previously, then the most obvious loophole remaining would be the nonlinearity of the higher-spin algebra. If the higher-spin algebra under consideration is an inherently linear one, then the whole argument in Subsection 11.3.3 would not be valid anymore.

[4]This rescaling applies for bilinear terms. For higher order terms one would have to use higher powers of $c_M$ as well in order to be able to take the proper $c_M \to 0$ limit.

In Section 11.4 I will show how such a loophole can be used to circumvent the NO-GO theorem presented in this subsection using the linear $\mathcal{FW}_\infty$ algebra.

### 11.3.4 Elimination of Multi-Graviton Excitations in $\mathcal{FW}_4^{(2-2)}$

As a first application of this NO-GO theorem I want to show how this theorem can be used to eliminate multi-graviton excitations by demanding both flat space and unitarity. I will focus on a specific example given by the $\mathcal{FW}_4^{(2-2)}$ algebra. This algebra is a contraction of two copies of $\mathcal{W}_4^{(2-2)}$ algebras and is given by the commutation relations (D.7) generated by the operators $L_n$, $M_n$, $O_n^a$, $P_n^a$, $\hat{U}_m^a$ and $\hat{V}_m^a$ with $a = \pm, 0$.
As in the cases treated before one has to require $c_M = 0$ in order to have a chance at finding unitary representations. Thus, I will make the following rescaling

$$U_n^a \to \hat{U}_n^a = (c_M)^{\frac{3}{2}}\, U_n^a, \tag{11.33a}$$
$$V_n^a \to \hat{V}_n^a = \sqrt{c_M} V_n^a, \tag{11.33b}$$

before taking the limit $c_M \to 0$. After performing the limit $c_M \to 0$ one obtains the following non-vanishing commutation relations.

$$[L_n, L_m] =(n-m)L_{m+n} + \frac{c_L}{12}n(n^2-1)\,\delta_{n+m,0}, \tag{11.34a}$$
$$[L_n, M_m] =(n-m)M_{m+n}, \tag{11.34b}$$
$$[L_n, O_m^a] =-\,mO_{n+m}^a, \tag{11.34c}$$
$$[L_n, P_m^a] =-\,mP_{n+m}^a, \tag{11.34d}$$
$$[L_n, \hat{U}_m^a] =(n-m)\hat{U}_{n+m}^a, \tag{11.34e}$$
$$[L_n, \hat{V}_m^a] =(n-m)\hat{V}_{n+m}^a, \tag{11.34f}$$
$$[M_n, O_m^a] =-\,mP_{n+m}^a, \tag{11.34g}$$
$$[O_n^a, O_m^b] =(a-b)O_{n+m}^{a+b} + \frac{62-c_L}{12}(1-3a^2)n\,\delta_{a+b,0}\delta_{n+m,0}, \tag{11.34h}$$
$$[O_n^a, P_m^b] =(a-b)P_{n+m}^{a+b}, \tag{11.34i}$$
$$[O_n^a, \hat{U}_m^b] =f(a,b)\hat{U}_{n+m}^{a+b}, \tag{11.34j}$$
$$[O_n^a, \hat{V}_m^b] =f(a,b)\hat{V}_{n+m}^{a+b}, \tag{11.34k}$$
$$[\hat{U}_n^\pm, \hat{U}_m^0] =-\,48(c_L-86)\left(:PPP:_{n+m}^{\{\pm|\pm|\mp\}} - :PPP:_{n+m}^{\{0|0|\pm\}}\right), \tag{11.34l}$$
$$[\hat{U}_n^-, \hat{U}_m^+] =-\,48(c_L-86)\left(:PPP:_{n+m}^{\{-|+|0\}} -6:PPP:_{n+m}^{0|0|0}\right), \tag{11.34m}$$
$$[\hat{U}_n^\pm, \hat{V}_m^0] =24\left(:PPP:_{n+m}^{\{\pm|\pm|\mp\}} - :PPP:_{n+m}^{\{0|0|\pm\}}\right), \tag{11.34n}$$
$$[\hat{U}_n^-, \hat{V}_m^+] =24\left(:PPP:_{n+m}^{\{-|+|0\}} -6:PPP:_{n+m}^{0|0|0}\right). \tag{11.34o}$$

Using definitions of the vacuum and hermitian conjugation similar to the ones found in (11.20) and looking at the commutator (11.34h) in analogy to section 11.3.1 one obtains again similar bounds on the central charge $c_L$ consistent with unitarity given by

$$\frac{1}{2}\left(77-\sqrt{5185}\right)\approx 2,49\leq c_L\leq 56 \quad \text{or} \quad c_L=62. \tag{11.35}$$

In addition, one finds that from the contracted algebra (11.34) the only non-trivial vacuum descendants are $L_{-n}|0\rangle$ with $n>1$, $O^a_{-n}|0\rangle$ with $n>0$ or combinations thereof. All other descendants are null states, in particular the ones generated by the other spin-2 generators $U^a_n$ and $V^a_n$. Thus, at least for the present example, unitarity in flat space eliminates multi-graviton states. The generalization of this statement to arbitrary $\mathcal{FW}$-algebras that contain multiple spin-2 states should work along the lines of the NO-GO result in section 11.3.2.

## 11.4 YES-GO: Unitarity of $\mathcal{FW}_\infty$

The previous examples, as well as the general NO-GO argument of 11.3.2, make it clear that, under the assumptions mentioned at the beginning of this chapter, unitarity is not compatible with having non-trivial higher spin states for flat space higher-spin algebras that are inherently nonlinear. The nonlinear character of the algebras is however crucial for the argument. In this section, I will show that focusing on linear higher-spin algebras one can easily evade the NO-GO result of Section 11.3.2.

A straightforward way to obtain linear flat space higher-spin algebras is by performing an İnönü–Wigner contraction of two copies of a linear AdS higher-spin algebra of $\mathcal{W}$-type. $\mathcal{W}$-algebras are typically nonlinear, but a few examples in which the algebra is isomorphic to a linear one are known. In what follows, I will focus on a particular example, namely the Pope-Romans-Shen $\mathcal{W}_\infty$ algebra [155, 156]. The explicit contraction and the resulting $\mathcal{FW}_\infty$ algebra can be found in Appendix D.5. In the following I will show how this $\mathcal{FW}_\infty$ algebra can be used to evade the NO-GO theorem of Section 11.3.2.

### Unitarity of $\mathcal{FW}_\infty$

The $\mathcal{FW}_\infty$ algebra, as shown in Appendix D.5, is generated by operators $V^s_n$ and $U^s_n$ and the following commutation relations

$$[V^s_n, V^t_m] = \sum_{\substack{u=2\\ \text{even}}}^{s+t-1} g^{st}_u(m,n;\lambda)V^{s+t-u}_{n+m} + c^s_L(m)\delta^{st}\delta_{n+m,0}, \tag{11.36a}$$

$$[V^s_n, U^t_m] = \sum_{\substack{u=2\\ \text{even}}}^{s+t-1} g^{st}_u(m,n;\lambda)U^{s+t-u}_{n+m} + c^s_M(m)\delta^{st}\delta_{n+m,0}, \tag{11.36b}$$

$$[U^s_n, U^t_m] = 0, \tag{11.36c}$$

where $c_L^s(m)$ and $c_M^s(m)$ are defined in the same way as (D.21), i.e. some combinatorial factors of $s$ and $m$ times $c_L$ and $c_M$, respectively.
In order to discuss unitarity, one can again use the highest-weight representation, defined by the vacuum annihilation conditions

$$\mathcal{V}_m^i|0\rangle = 0 \qquad \text{and} \qquad \mathcal{W}_m^i|0\rangle = 0\,, \qquad m \geq -i-1\,, \tag{11.37}$$

and the following definition of the hermitian conjugate

$$\left(\mathcal{V}_m^i\right)^\dagger = \mathcal{V}_{-m}^i\,, \qquad \left(\mathcal{W}_m^i\right)^\dagger = \mathcal{W}_{-m}^i. \tag{11.38}$$

Unitarity can again only be achieved when $c_M \to 0$. In contrast to the other, nonlinear cases treated before, this limit can be taken without any rescaling but still renders the states created by $\mathcal{W}_m^i$ generators having zero norm. Therefore one can mod them out of the module. For $c_L \neq 0$, the only non-trivial states left are the ones created by $\mathcal{V}_n^s$. One is therefore left with states that are descendants of a single copy of a $\mathcal{W}_\infty$ algebra that forms a unitary representation of the flat space higher-spin algebra $\mathcal{FW}_\infty$, thus circumventing the NO-GO theorem of Section 11.3.2.
After having found an explicit example of a flat space higher-spin algebra that allows for unitary representations without rendering the higher-spin states unphysical it is a natural question to look for a candidate theory realizing these symmetries as asymptotic symmetries. Since such a theory should have $c_L \neq 0$ and $c_M = 0$, a natural candidate would be a higher-spin extension of flat space chiral gravity. Such a theory would be a very interesting playground to explicitly explore higher-spin symmetries in flat space in combination with unitarity.

# Flat Space Higher-Spin Gravity 12

> *Once upon a midnight dreary, while I pondered, weak and weary,*
> *Over many a quaint and curious volume of forgotten lore—*
> *While I nodded, nearly napping, suddenly there came a tapping,*
> *As of some one gently rapping, rapping at my chamber door.*
> *"Tis some visitor," I muttered, "tapping at my chamber door"—*
> *Only this and nothing more."*

**– Edgar Allen Poe**
The Raven

In the previous chapter I looked for unitary highest-weight representations of $\mathcal{FW}$-algebras that are the flat space analogues of $\mathcal{W}$-algebras, which play an important role as the asymptotic symmetry algebras of (non-)AdS higher-spin gravity theories. In this chapter I review how these $\mathcal{FW}$-algebras arise as asymptotic symmetries of asymptotically flat spacetimes using spin-3 gravity in flat space as an explicit example. This will serve as a basis for Chapter 13 where I will extend the higher-spin description in flat space by including chemical potentials. On a technical level, especially in terms of a Chern-Simons description, higher-spin theories in three dimensional flat space are actually not very different from the description in $\text{AdS}_3$. I have already shown how to describe spin-2 gravity in flat space using an $\mathfrak{isl}(2,\mathbb{R})$ valued Chern-Simons gauge field previously in Section 10.1. The higher-spin extension of this formalism works in complete analogy to the AdS case by simply replacing the gauge algebra $\mathfrak{isl}(2,\mathbb{R}) \to \mathfrak{isl}(N,\mathbb{R})$. Depending on the choice of embedding of $\mathfrak{isl}(2,\mathbb{R}) \hookrightarrow \mathfrak{isl}(N,\mathbb{R})$ one again obtains different theories with different spectra of spins. The biggest subtlety one has to take care of is the invariant bilinear form which is different for $\mathfrak{isl}(N,\mathbb{R})$ in comparison to $\mathfrak{sl}(N,\mathbb{R})$. Aside from that subtlety one can apply most of the techniques and intuition already known from $\text{AdS}_3$ holography.

## 12.1 Spin-3 Gravity in Flat Space

Similar to the AdS case, spin-3 gravity in the principal embedding[1] is the simplest higher-spin extension one can consider in flat space and has been first described in [150].

[1]Here principal embedding refers to the principal embedding of $\mathfrak{isl}(2,\mathbb{R}) \hookrightarrow \mathfrak{isl}(3,\mathbb{R})$

The starting point is again the Chern-Simons action (2.3) where $\mathcal{A}$ takes values in the principal embedding of $\mathfrak{isl}(3,\mathbb{R})$ which is generated by the generators $L_n,\ M_n,\ U_n,\ V_n$ as

$$[L_n,\, L_m] = (n-m)L_{n+m}, \tag{12.1a}$$
$$[L_n,\, M_m] = (n-m)M_{n+m}, \tag{12.1b}$$
$$[L_n,\, U_m] = (2n-m)U_{n+m}, \tag{12.1c}$$
$$[L_n,\, V_m] = (2n-m)V_{n+m}, \tag{12.1d}$$
$$[M_n,\, U_m] = (2n-m)V_{n+m}, \tag{12.1e}$$
$$[U_n,\, U_m] = \sigma(n-m)(2n^2+2m^2-nm-8)L_{n+m}, \tag{12.1f}$$
$$[U_n,\, V_m] = \sigma(n-m)(2n^2+2m^2-nm-8)M_{n+m}, \tag{12.1g}$$

where the factor $\sigma$ fixes the overall normalization of the spin-3 generators $U_n$ and $V_n$ and can be chosen at will. I will fix this normalization constant to $\sigma = -\frac{1}{3}$.
The algebra (12.1) naturally comes with a $\mathbb{Z}_2$ grading so that the generators $L_n$, $U_n$ are even and $M_n$, $V_n$ are odd, respectively. Then even with even gives even, even with odd gives odd and odd with odd vanishes. In terms of İnönü–Wigner contractions, the $\epsilon$-independent generators are the even generators and the generators linear in $\epsilon$ are the odd generators [70].
There are various ways of constructing matrix representations of $\mathfrak{isl}(N,\mathbb{R})$ algebras. One particular neat way to do this is outlined in Appendix A.2. The basic idea is to combine fundamental matrix representations of $\mathfrak{sl}(2,\mathbb{R})$ and the Grassmann parameter $\epsilon$ in a block diagonal matrix in a certain way and then use $\epsilon^2 = 0$ for all computations involving the matrix representation constructed this way.
Moreover, this construction can be used to define various traces that correspond to different bilinear forms on (subalgebras of) $\mathfrak{isl}(N,\mathbb{R})$. For example the invariant bilinear form $\langle\cdot,\cdot\rangle$, which is relevant for a flat space description in terms of the Chern-Simons action (2.3) can be obtained from the representation I described above by defining a (hatted) trace

$$\langle\mathcal{G}_a\mathcal{G}_b\rangle = \widehat{\operatorname{Tr}}(\mathcal{G}_a\mathcal{G}_b) := \frac{\mathrm{d}}{\mathrm{d}\epsilon}\,\frac{1}{4}\operatorname{Tr}(\mathcal{G}_a\mathcal{G}_b\gamma^*_{(D)})\big|_{\epsilon=0}, \tag{12.2}$$

with

$$\gamma^\star_{(D)} = \begin{pmatrix} \mathbb{1}_{D\times D} & 0 \\ 0 & -\mathbb{1}_{D\times D} \end{pmatrix}, \tag{12.3}$$

and for two generators $\mathcal{G}_a,\ \mathcal{G}_b \in \mathfrak{isl}(N,\mathbb{R})$. The subscript in $\gamma^\star_{(D)}$ is the dimension of the matrix representation of $\mathfrak{sl}(N,\mathbb{R})$ used to construct the $\mathfrak{isl}(N,\mathbb{R})$ matrix representation. Since the matrix representations used in this thesis are based on the fundamental representation of $\mathfrak{sl}(N,\mathbb{R})$ one has for all practical purposes $D = N$.

Another invariant bilinear form[2] can be obtained from the representation I described above by defining a (twisted) trace over a product of $n$ $\mathfrak{isl}(N,\mathbb{R})$ generators $\mathcal{G}_i$ as

$$\widetilde{\mathrm{Tr}}\left(\prod_{i=1}^{n}\mathcal{G}_i\right)=\frac{1}{2}\mathrm{Tr}\left(\prod_{i=1}^{n}\frac{\mathrm{d}}{\mathrm{d}\epsilon}\mathcal{G}_i\gamma^{\star}_{(D)}\right). \tag{12.4}$$

Thus, there are basically three notions of traces available, the normal, twisted and hatted trace. This allows one to use these traces as tools to pick out information which is related to even (odd) [mixed] contributions by using the trace (twisted trace) [hatted trace].
The twisted trace (12.4) for example can be used to determine the spin-2 and spin-3 fields since those correspond to symmetric products of powers of the zuvielbein which is a purely odd quantity. Hence the metric can equivalently be determined by

$$g_{\mu\nu}=\frac{1}{2}\widetilde{\mathrm{Tr}}(\mathcal{A}_\mu\mathcal{A}_\nu)=\langle e_\mu e_\nu\rangle\ . \tag{12.5}$$

The spin-3 field is similarly defined from the cubic $\mathfrak{sl}(3,\mathbb{R})$-Casimir or, equivalently, by using again the twisted trace

$$\Phi_{\mu\nu\lambda}=\frac{1}{6}\widetilde{\mathrm{Tr}}\left(\mathcal{A}_\mu\mathcal{A}_\nu\mathcal{A}_\lambda\right)=\langle e_\mu e_\nu e_\lambda\rangle\,. \tag{12.6}$$

Explicit expressions for $\mathfrak{isl}(3,\mathbb{R})$ valued connections that obey asymptotically flat boundary conditions were established independently in [150] and [154] (see also [157]). As I have shown in Section 10.1 (see also [74]) one can, similar to the AdS case, gauge away the radial dependence of the action using

$$\mathcal{A}=b^{-1}\,\mathrm{d}b+b^{-1}a(u,\varphi)b,\qquad b=e^{\frac{r}{2}M_{-1}}. \tag{12.7}$$

In the spin-3 gravity setup one can then choose the following boundary conditions for $a(u,\varphi)$ [150, 154]

$$a(u,\varphi)=a_\varphi(u,\varphi)\,\mathrm{d}\varphi+a_u(u,\varphi)\,\mathrm{d}u, \tag{12.8}$$

where

$$a_\varphi(u,\varphi)=L_1-\frac{\mathcal{M}}{4}L_{-1}-\frac{\mathcal{N}}{2}M_{-1}+\frac{\mathcal{V}}{2}U_{-2}+\mathcal{Z}V_{-2}, \tag{12.9a}$$

$$a_u(u,\varphi)=M_1-\frac{\mathcal{M}}{4}M_{-1}+\frac{\mathcal{V}}{2}V_{-2}, \tag{12.9b}$$

with

$$\mathcal{N}=\mathcal{L}(\varphi)+\frac{u}{2}\mathcal{M}'(\varphi),\qquad \mathcal{Z}=\mathcal{U}(\varphi)+\frac{u}{2}\mathcal{V}'(\varphi), \tag{12.10}$$

[2] This is actually only an invariant bilinear form on the subalgebra of odd generators.

where all other components have to vanish at the asymptotic boundary $r \to \infty$. This form of the state dependent functions is a result of the equations of motion which require $F = \mathrm{d}\mathcal{A} + \mathcal{A} \wedge \mathcal{A} = 0$ which is tantamount to

$$\partial_u \mathcal{M} = \partial_u \mathcal{V} = 0, \qquad \partial_u \mathcal{N} = \tfrac{1}{2}\,\partial_\varphi \mathcal{M}, \qquad \partial_u \mathcal{Z} = \tfrac{1}{2}\,\partial_\varphi \mathcal{V}. \tag{12.11}$$

The canonical analysis for the connection (12.7) in combination with the boundary conditions (12.8) follows exactly the same lines as already outlined in Chapter 4. Thus, the first step is to determine the gauge transformations which preserve the boundary conditions (12.8). In analogy to the AdS case one can write the gauge transformations as $\varepsilon = b^{-1}\left(\sum_{n=-1}^{1} \epsilon^n L_n + \sum_{n=-1}^{1} \tau^n M_n + \sum_{n=-2}^{2} \chi^n U_n + \sum_{n=-2}^{2} \kappa^n V_n\right) b$, with $\epsilon^1 \equiv \epsilon$, $\tau^1 \equiv \tau = (\sigma + \frac{u}{2}\epsilon')$, $\chi^2 \equiv \chi$, $\kappa^2 \equiv \kappa = \rho + u\chi'$ and

$$\epsilon^0 = -\,\epsilon', \qquad \tau^0 = -\tau', \qquad \chi^1 = -\chi', \qquad \kappa^1 = -\kappa', \tag{12.12a}$$

$$\epsilon^{-1} = \frac{\epsilon''}{2} - \frac{\mathcal{M}}{4}\epsilon - 4\mathcal{V}\chi, \quad \tau^{-1} = \frac{\tau''}{2} - \frac{\mathcal{M}}{4}\tau - \frac{\mathcal{N}}{2}\epsilon - 4\mathcal{V}\kappa - 8\chi\mathcal{Z} \tag{12.12b}$$

$$\chi^0 = \frac{\chi''}{2} - \frac{\mathcal{M}}{2}\chi, \quad \chi^{-1} = -\frac{\chi'''}{6} + \frac{\mathcal{M}'}{6}\chi + \frac{5}{12}\mathcal{M}\chi', \tag{12.12c}$$

$$\chi^{-2} = \frac{\chi''''}{24} - \frac{\mathcal{M}''}{24}\chi - \frac{7}{48}\mathcal{M}'\chi' - \frac{\mathcal{M}}{6}\chi'' + \frac{\mathcal{M}^2}{16}\chi + \frac{\mathcal{V}}{2}\epsilon, \tag{12.12d}$$

$$\kappa^0 = \frac{\kappa''}{2} - \frac{\mathcal{M}}{2}\kappa - \mathcal{N}\chi, \;\; \kappa^{-1} = -\frac{\kappa'''}{6} + \frac{\mathcal{M}'}{6}\kappa + \frac{5}{12}\mathcal{M}\kappa' + \frac{\mathcal{N}'}{3}\chi + \frac{5}{6}\mathcal{N}\chi', \tag{12.12e}$$

$$\begin{aligned}\kappa^{-2} = &\frac{\kappa''''}{24} - \frac{\mathcal{M}''}{24}\kappa - \frac{7}{48}\mathcal{M}'\kappa' - \frac{\mathcal{M}}{6}\kappa'' - \frac{\mathcal{N}''}{12}\chi - \frac{7}{24}\mathcal{N}'\chi' - \frac{\mathcal{N}}{3}\chi'' \\ &+ \frac{\mathcal{M}^2}{16}\kappa + \frac{\mathcal{M}\mathcal{N}}{4}\chi + \mathcal{Z}\epsilon + \frac{\mathcal{V}}{2}\tau,\end{aligned} \tag{12.12f}$$

which only depend on the angular coordinate $\varphi$. This leads to the following transformation behavior of the state dependent functions under infinitesimal gauge transformations

$$\delta_\epsilon \mathcal{L} = \epsilon\mathcal{L}' + 2\epsilon'\mathcal{L}, \qquad \delta_\epsilon \mathcal{M} = \epsilon\mathcal{M}' + 2\epsilon'\mathcal{M} - 2\epsilon''', \tag{12.13a}$$

$$\delta_\epsilon \mathcal{U} = \epsilon\mathcal{U}' + 2\epsilon'\mathcal{U}, \qquad \delta_\epsilon \mathcal{V} = \epsilon\mathcal{V}' + 2\epsilon'\mathcal{V}, \tag{12.13b}$$

$$\delta_\sigma \mathcal{L} = \sigma\mathcal{M}' + 2\sigma'\mathcal{M} - 2\sigma''', \quad \delta_\sigma \mathcal{U} = \sigma\mathcal{V}' + 3\sigma'\mathcal{V}, \tag{12.13c}$$

$$\delta_\chi \mathcal{L} = -\,48\chi\mathcal{U}' - 72\chi'\mathcal{U}, \qquad \delta_\chi \mathcal{M} = -48\chi\mathcal{V}' - 72\chi'\mathcal{V}, \tag{12.13d}$$

$$\delta_\chi \mathcal{U} = -\frac{\mathcal{L}'''}{12}\chi - \frac{3}{8}\mathcal{L}''\chi' - \frac{5}{8}\mathcal{L}'\chi'' - \frac{5}{12}\mathcal{L}\chi''' + \frac{(\mathcal{M}\mathcal{L})'}{3}\chi + \frac{2}{3}\mathcal{M}\mathcal{L}\chi', \tag{12.13e}$$

$$\delta_\chi \mathcal{V} = \frac{\chi'''''}{12} - \frac{\mathcal{M}'''}{12}\chi - \frac{3}{8}\mathcal{M}''\chi' - \frac{5}{8}\mathcal{M}'\chi'' - \frac{5}{12}\mathcal{M}\chi''' + \frac{\mathcal{M}\mathcal{M}'}{6}\chi + \frac{1}{3}\mathcal{M}^2\chi', \tag{12.13f}$$

$$\delta_\rho \mathcal{L} = -\,24\rho\mathcal{V}' - 36\rho'\mathcal{V}, \tag{12.13g}$$

$$\delta_\rho \mathcal{U} = \frac{\rho'''''}{24} - \frac{\mathcal{M}'''}{24}\rho - \frac{3}{16}\mathcal{M}''\rho' - \frac{5}{16}\mathcal{M}'\rho'' - \frac{5}{24}\mathcal{M}\rho''' + \frac{\mathcal{M}\mathcal{M}'}{6}\rho + \frac{\mathcal{M}^2}{6}\rho'. \tag{12.13h}$$

In this way, one obtains as the variation of the canonical boundary charge

$$\delta\mathcal{Q}[\varepsilon] = \frac{k}{2\pi}\int \mathrm{d}\varphi \left(\epsilon\,\delta\mathcal{L} + \frac{\sigma}{2}\,\delta\mathcal{M} + 8\chi\,\delta\mathcal{U} + 4\rho\,\delta\mathcal{V}\right), \tag{12.14}$$

which is finite and integrable in field space. After integration one obtains the canonical boundary charge

$$\mathcal{Q}[\varepsilon] = \frac{k}{2\pi}\int \mathrm{d}\varphi \left(\epsilon\,\mathcal{L} + \frac{\sigma}{2}\,\mathcal{M} + 8\chi\,\mathcal{U} + 4\rho\,\mathcal{V}\right), \tag{12.15}$$

which is also conserved in retarded time, i.e. $\partial_u\mathcal{Q} = 0$.

The conserved charge (12.15) in combination with (12.13) can again be used to first determine the Dirac bracket algebra of the asymptotic symmetries and in turn also the semiclassical asymptotic symmetry algebra as

$$[L_n, L_m] = (n-m)L_{n-m} + \frac{c_L}{12}n(n^2-1)\delta_{n+m,0}, \tag{12.16a}$$

$$[L_n, M_m] = (n-m)M_{n-m} + \frac{c_M}{12}n(n^2-1)\delta_{n+m,0}, \tag{12.16b}$$

$$[L_n, U_m] = (2n-m)U_{n+m}, \tag{12.16c}$$

$$[L_n, V_m] = (2n-m)V_{n+m}, \tag{12.16d}$$

$$[M_n, U_m] = (2n-m)V_{n+m}, \tag{12.16e}$$

$$\begin{aligned}[U_n, U_m] =& (n-m)(2n^2+2m^2-nm-8)L_{n+m} + \frac{192}{c_M}(n-m)\Lambda_{n+m} \\ & - \frac{96c_L}{c_M^2}(n-m)\Theta_{n+m} + \frac{c_L}{12}n(n^2-1)(n^2-4)\delta_{n+m,0},\end{aligned} \tag{12.16f}$$

$$\begin{aligned}[U_n, V_m] =& (n-m)(2n^2+2m^2-nm-8)M_{n+m} + \frac{96}{c_M}(n-m)\Theta_{n+m} \\ & + \frac{c_M}{12}n(n^2-4)(n^2-1)\delta_{n+m,0},\end{aligned} \tag{12.16g}$$

with

$$\Lambda_n = \sum_{p\in\mathbb{Z}} L_p M_{n-p}, \qquad \Theta_n = \sum_{p\in\mathbb{Z}} M_p M_{n-p}, \tag{12.17}$$

and

$$c_L = 0, \qquad c_M = 12k. \tag{12.18}$$

This algebra can also be obtained as an ultrarelativistic limit of the asymptotic symmetries of spin-3 gravity in $\mathrm{AdS}_3$ and is the spin-3 extension of the $\mathfrak{bms}_3$ algebra encountered in spin-2 gravity. As in the AdS case, however, this algebra is only valid for large central charges $c_L$ and $c_M$ and therefore needs further modifications in order to also be valid for small values of $c_L$ and $c_M$ where products of operators have to be normal ordered.

## 12.2 Quantum Asymptotic Symmetries

Since $\Lambda_n$ and $\Theta_n$ are nonlinear operators one should introduce a notion of normal ordering for these operators. I will choose the same vacuum conditions as in (11.1) and (11.5) and define normal ordering accordingly as[3]

$$:\Lambda:_n := \sum_{p\geq -1} L_{n-p}M_p + \sum_{p\geq -1} M_p L_{n-p}. \tag{12.19}$$

In order to be compatible with the Jacobi identities one again has to modify the semiclassical algebra (12.16) in a suitable way. After performing some algebraic gymnastics one obtains

$$[L_n, L_m] = (n-m)L_{n+m} + \frac{c_L}{12}(n^3-n)\,\delta_{n+m,0}, \tag{12.20a}$$

$$[L_n, M_m] = (n-m)M_{n+m} + \frac{c_M}{12}(n^3-n)\,\delta_{n+m,0}, \tag{12.20b}$$

$$[L_n, U_m] = (2n-m)U_{n+m}, \tag{12.20c}$$

$$[L_n, V_m] = (2n-m)V_{n+m}, \tag{12.20d}$$

$$[M_n, U_m] = (2n-m)V_{n+m}, \tag{12.20e}$$

$$[U_n, U_m] = (n-m)(2n^2+2m^2-nm-8)L_{n+m} + \frac{192}{c_M}(n-m)\,:\tilde{\Lambda}:_{n+m} \tag{12.20f}$$

$$-\frac{96(c_L+\frac{44}{5})}{c_M^2}(n-m)\Theta_{n+m} + \frac{c_L}{12}n(n^2-1)(n^2-4)\,\delta_{n+m,0}, \tag{12.20g}$$

$$[U_n, V_m] = (n-m)(2n^2+2m^2-nm-8)M_{n+m} + \frac{96}{c_M}(n-m)\Theta_{n+m}$$

$$+\frac{c_M}{12}n(n^2-1)(n^2-4)\,\delta_{n+m,0}, \tag{12.20h}$$

where

$$:\tilde{\Lambda}:_n = :\Lambda:_n - \frac{3}{10}(n+2)(n+3)M_n, \tag{12.21}$$

which is nothing else than the contracted $\mathcal{W}_3$ algebra (10.30) already presented in Section 10.3. This example shows explicitly how the quantum $\mathcal{FW}_3$ algebra arises as the asymptotic symmetry algebra of spin-3 gravity in flat space.

Other flat space higher-spin gravity theories can be analyzed in the exact same way by starting out with a different embedding of $\mathfrak{isl}(2,\mathbb{R}) \hookrightarrow \mathfrak{isl}(N,\mathbb{R})$ and choosing appropriate boundary conditions for the gauge field $\mathcal{A}$.

[3] Since $\Theta_n$ is a sum of products of commuting operators it does not make a difference whether or not this operator is normal ordered, and thus one can omit the normal ordering altogether.

# 13 Flat Space Higher-Spin Gravity with Chemical Potentials

> 二兎を追う者は一兎をも得ず。
> *(If you run after two hares, you will catch neither.)*
>
> **– Japanese Proverb**

Chemical potentials were first introduced by Josiah Willard Gibbs at the end of the 19th century and have played an important role in both chemistry and physics since then. Mostly known from statistical physics one can also introduce chemical potentials $\mu$ in gauge theories by giving the 0-component of the gauge connection a vacuum expectation value (see e.g. [158])

$$A_0 \to A_0 + \mu. \tag{13.1}$$

Using holography to describe quantum field theories with non-zero (higher-spin) chemical potentials one principally has to think about how to implement them in a gravitational context. In three dimensions, where gravity can be reformulated as a Chern-Simons gauge theory this can be done in a very clear and straightforward way. Chemical potentials were introduced in spin-3 AdS gravity in the past few years, first in the form of new black hole solutions with spin-3 fields by Gutperle and Kraus [159] (see also [135]), next perturbatively in the spin-3 chemical potential [160], then to all orders by Compère, Jottar and Song [161] and independently by Henneaux, Perez, Tempo and Troncoso [162]. A comprehensive recent discussion of higher spin black holes with chemical potentials is provided in [163]. However, the discussion so far was focused mostly on AdS and holographic aspects thereof [32], see [IV, 33–35] for reviews.

In this chapter[1] I will describe how to extend the AdS considerations to flat space. In a similar manner as it is rewarding to study Bañados–Teitelboim–Zanelli (BTZ) black holes [12, 60] in $AdS_3/CFT_2$ it is also rewarding to study flat space cosmologies with higher-spin charges and chemical potentials in order to get a better understanding of flat space holography.

In Chapter 12 I have reviewed how to describe flat space higher-spin gravity in terms of a Chern-Simons gauge theory, with a special focus on spin-3 gravity. In this chapter

[1] Since this chapter is based on the publication [VII] for which I collaborated with my co-authors Mirah Gary, Daniel Grumiller and Jan Rosseel several parts of this chapter coincide with the content found in [VII].

I will generalize this discussion to flat space spin-3 gravity with additional chemical potentials $\mu_{\mathrm{M}}$, $\mu_{\mathrm{L}}$, $\mu_{\mathrm{V}}$, $\mu_{\mathrm{U}}$, one for each of the spin-2 and spin-3 fields, respectively.

## 13.1 Adding Chemical Potentials

The starting point is the connection (12.8), where following the procedure of [163] I also assume that the form of $a_\varphi$ remains unchanged by chemical potentials, in order to maintain the structure of the asymptotic canonical boundary charges.
Thus, I will make a general ansatz for $a_u$ with some arbitrary coefficients, which are then fixed by solving the equations of motion $F = \mathrm{d}\mathcal{A} + \mathcal{A} \wedge \mathcal{A} = 0$. Associating the coefficients of the highest weight components with the corresponding chemical potentials i.e. $\alpha M_1 \to \mu_{\mathrm{M}} M_1$ one obtains

$$a_u = a_u^{(0)} + a_u^{(\mu_{\mathrm{M}})} + a_u^{(\mu_{\mathrm{L}})} + a_u^{(\mu_{\mathrm{V}})} + a_u^{(\mu_{\mathrm{U}})}, \qquad a_\varphi = a_\varphi^{(0)}, \tag{13.2}$$

with $a_u^{(0)}$, $a_\varphi^{(0)}$ being the $u$ part of the connection (12.8) and

$$a_u^{(\mu_{\mathrm{M}})} = \mu_{\mathrm{M}}\, M_1 - \mu_{\mathrm{M}}'\, M_0 + \tfrac{1}{2}\,(\mu_{\mathrm{M}}'' - \tfrac{1}{2}\mathcal{M}\mu_{\mathrm{M}})\, M_{-1} + \tfrac{1}{2}\,\mathcal{V}\,\mu_{\mathrm{M}}\, V_{-2}, \tag{13.3a}$$

$$a_u^{(\mu_{\mathrm{L}})} = a_u^{(\mu_{\mathrm{M}})}\big|_{M\to L} - \tfrac{1}{2}\,\mathcal{N}\,\mu_{\mathrm{L}}\, M_{-1} + \mathcal{Z}\,\mu_{\mathrm{L}}\, V_{-2}, \tag{13.3b}$$

$$\begin{aligned} a_u^{(\mu_{\mathrm{V}})} = {} & \mu_{\mathrm{V}}\, V_2 - \mu_{\mathrm{V}}'\, V_1 + \tfrac{1}{2}\,(\mu_{\mathrm{V}}'' - \mathcal{M}\mu_{\mathrm{V}})\, V_0 + \tfrac{1}{6}\,(-\mu_{\mathrm{V}}''' + \mathcal{M}'\mu_{\mathrm{V}} + \tfrac{5}{2}\mathcal{M}\mu_{\mathrm{V}}')\, V_{-1} \\ & + \tfrac{1}{24}\,(\mu_{\mathrm{V}}'''' - 4\mathcal{M}\mu_{\mathrm{V}}'' - \tfrac{7}{2}\mathcal{M}'\mu_{\mathrm{V}}' + \tfrac{3}{2}\mathcal{M}^2\mu_{\mathrm{V}} - \mathcal{M}''\mu_{\mathrm{V}})\, V_{-2} - 4\mathcal{V}\,\mu_{\mathrm{V}}\, M_{-1}, \end{aligned} \tag{13.3c}$$

$$\begin{aligned} a_u^{(\mu_{\mathrm{U}})} = {} & a_u^{(\mu_{\mathrm{V}})}\big|_{M\to L} - 8\mathcal{Z}\,\mu_{\mathrm{U}}\, M_{-1} - \mathcal{N}\,\mu_{\mathrm{U}}\, V_0 + (\tfrac{5}{6}\mathcal{N}\mu_{\mathrm{U}}' + \tfrac{1}{3}\mathcal{N}'\mu_{\mathrm{U}})\, V_{-1} \\ & + (-\tfrac{1}{3}\mathcal{N}\mu_{\mathrm{U}}'' - \tfrac{7}{24}\mathcal{N}'\mu_{\mathrm{U}}' - \tfrac{1}{12}\mathcal{N}''\mu_{\mathrm{U}} + \tfrac{1}{4}\mathcal{M}\mathcal{N}\mu_{\mathrm{U}})\, V_{-2}, \end{aligned} \tag{13.3d}$$

where the subscript $M \to L$ denotes that in the corresponding quantity all odd generators and chemical potentials are replaced by corresponding even ones, $M_n \to L_n$, $V_n \to U_n$, $\mu_{\mathrm{M}} \to \mu_{\mathrm{L}}$ and $\mu_{\mathrm{V}} \to \mu_{\mathrm{U}}$, i.e.

$$a_u^{(\mu_{\mathrm{M}})}\big|_{M\to L} = \mu_{\mathrm{L}}\, L_1 - \mu_{\mathrm{L}}'\, L_0 + \tfrac{1}{2}\,(\mu_{\mathrm{L}}'' - \tfrac{1}{2}\mathcal{M}\mu_{\mathrm{L}})\, L_{-1} + \tfrac{1}{2}\,\mathcal{V}\,\mu_{\mathrm{L}}\, U_{-2}, \tag{13.3e}$$

$$\begin{aligned} a_u^{(\mu_{\mathrm{V}})}\big|_{M\to L} = {} & \mu_{\mathrm{U}}\, U_2 - \mu_{\mathrm{U}}'\, U_1 + \tfrac{1}{2}\,(\mu_{\mathrm{U}}'' - \mathcal{M}\mu_{\mathrm{U}})\, U_0 + \tfrac{1}{6}\,(-\mu_{\mathrm{U}}''' + \mathcal{M}'\mu_{\mathrm{U}} + \tfrac{5}{2}\mathcal{M}\mu_{\mathrm{U}}')\, U_{-1} \\ & + \tfrac{1}{24}\,(\mu_{\mathrm{U}}'''' - 4\mathcal{M}\mu_{\mathrm{U}}'' - \tfrac{7}{2}\mathcal{M}'\mu_{\mathrm{U}}' + \tfrac{3}{2}\mathcal{M}^2\mu_{\mathrm{U}} - \mathcal{M}''\mu_{\mathrm{U}})\, U_{-2} - 4\mathcal{V}\,\mu_{\mathrm{U}}\, L_{-1}. \end{aligned} \tag{13.3f}$$

Here, dots (primes) denote derivatives with respect to retarded time $u$ (angular coordinate $\varphi$).

The equations of motion (12.11) impose the additional conditions

$$\dot{\mathcal{M}} = -2\mu_{\mathrm{L}}''' + 2\mathcal{M}\mu_{\mathrm{L}}' + \mathcal{M}'\mu_{\mathrm{L}} + 24\mathcal{V}\mu_{\mathrm{U}}' + 16\mathcal{V}'\mu_{\mathrm{U}}, \tag{13.4a}$$

$$\dot{\mathcal{N}} = \tfrac{1}{2}\,\dot{\mathcal{M}}\big|_{L\to M} + 2\mathcal{N}\mu_{\mathrm{L}}' + \mathcal{N}'\mu_{\mathrm{L}} + 24\mathcal{Z}\mu_{\mathrm{U}}' + 16\mathcal{Z}'\mu_{\mathrm{U}}, \tag{13.4b}$$

$$\begin{aligned}\dot{\mathcal{V}} &= \tfrac{1}{12}\,\mu_{\mathrm{U}}''''' - \tfrac{5}{12}\,\mathcal{M}\mu_{\mathrm{U}}''' - \tfrac{5}{8}\,\mathcal{M}'\mu_{\mathrm{U}}'' - \tfrac{3}{8}\,\mathcal{M}''\mu_{\mathrm{U}}' + \tfrac{1}{3}\,\mathcal{M}^2\mu_{\mathrm{U}}' \\ &\quad - \tfrac{1}{12}\mathcal{M}'''\mu_{\mathrm{U}} + \tfrac{1}{3}\,\mathcal{M}\mathcal{M}'\mu_{\mathrm{U}} + 3\mathcal{V}\mu_{\mathrm{L}}' + \mathcal{V}'\mu_{\mathrm{L}},\end{aligned} \tag{13.4c}$$

$$\begin{aligned}\dot{\mathcal{Z}} &= \tfrac{1}{2}\,\dot{\mathcal{V}}\big|_{L\to M} - \tfrac{5}{12}\,\mathcal{N}\mu_{\mathrm{U}}''' - \tfrac{5}{8}\,\mathcal{N}'\mu_{\mathrm{U}}'' - \tfrac{3}{8}\,\mathcal{N}''\mu_{\mathrm{U}}' + \tfrac{2}{3}\,\mathcal{M}\mathcal{N}\mu_{\mathrm{U}}' \\ &\quad - \tfrac{1}{12}\,\mathcal{N}'''\mu_{\mathrm{U}} + \tfrac{1}{3}\,(\mathcal{M}\mathcal{N})'\mu_{\mathrm{U}} + 3\mathcal{Z}\mu_{\mathrm{L}}' + \mathcal{Z}'\mu_{\mathrm{L}},\end{aligned} \tag{13.4d}$$

with

$$\tfrac{1}{2}\,\dot{\mathcal{M}}\big|_{L\to M} = -\mu_{\mathrm{M}}''' + \mathcal{M}\mu_{\mathrm{M}}' + \tfrac{1}{2}\,\mathcal{M}'\mu_{\mathrm{M}} + 12\mathcal{V}\mu_{\mathrm{V}}' + 8\mathcal{V}'\mu_{\mathrm{V}}, \tag{13.4e}$$

$$\begin{aligned}\tfrac{1}{2}\,\dot{\mathcal{V}}\big|_{L\to M} &= \tfrac{1}{24}\,\mu_{\mathrm{V}}''''' - \tfrac{5}{24}\,\mathcal{M}\mu_{\mathrm{V}}''' - \tfrac{5}{16}\,\mathcal{M}'\mu_{\mathrm{V}}'' - \tfrac{3}{16}\,\mathcal{M}''\mu_{\mathrm{V}}' + \tfrac{1}{6}\,\mathcal{M}^2\mu_{\mathrm{V}}' \\ &\quad - \tfrac{1}{24}\,\mathcal{M}'''\mu_{\mathrm{V}} + \tfrac{1}{6}\,\mathcal{M}\mathcal{M}'\mu_{\mathrm{V}} + \tfrac{3}{2}\,\mathcal{V}\mu_{\mathrm{M}}' + \tfrac{1}{2}\,\mathcal{V}'\mu_{\mathrm{M}}.\end{aligned} \tag{13.4f}$$

The chemical potentials $\mu_{\mathrm{M}}$, $\mu_{\mathrm{L}}$, $\mu_{\mathrm{V}}$ and $\mu_{\mathrm{U}}$ can be in principal arbitrary functions of the angular coordinate $\varphi$ and the retarded time $u$. In many applications, however, they are constant, which simplifies most of the formulas considerably.

## 13.2 Consistency Checks

After having added the chemical potentials I will perform some consistency checks before proceeding.

- In the absence of chemical potentials, $\mu_{\mathrm{M}} = \mu_{\mathrm{L}} = \mu_{\mathrm{V}} = \mu_{\mathrm{U}} = 0$ one should recover the results from Chapter 12. This is indeed true. In particular, the on-shell conditions (13.4) simplify to (12.11).

- In the presence of chemical potentials the on-shell conditions (13.4) should contain information about the asymptotic symmetry algebra (10.30). For example, the $\mu_{\mathrm{L}}$-terms in (13.4a) are an infinitesimal Schwarzian derivative, while the $\mu_{\mathrm{U}}$-terms exhibit transformation behavior of a spin-3 field.

- Since any solution to the field equations $F = 0$ must be locally pure gauge, and any solution that obeys the boundary conditions (12.8) can be generated by the boundary condition preserving gauge transformations (12.12), it should be possible to obtain (13.3) directly from a gauge transformation. Indeed, comparing the expressions for (12.12) with (13.3) one can see that they coincide upon identifying $\epsilon \to \mu_{\mathrm{L}}$, $\tau \to \mu_{\mathrm{M}}$, $\kappa \to \mu_{\mathrm{V}}$ and $\chi \to \mu_{\mathrm{U}}$.

It is possible to derive the results of section 13.1 also in a different way. One could start from equation (3.7)-(3.12) in [163] and use the Grassmann-approach of [70] to derive the flat space connection with chemical potentials along similar lines as in Section 10.1, dropping in the end all terms quadratic in the Grassmann-parameter $\epsilon$. The map that leads from (3.7)-(3.12) in [163] (left hand side) to the results presented in section 13.1 (right hand side) is given by

| | | |
|---|---|---|
| Coordinates: | $x^{\pm} = \epsilon\, u \pm \varphi,$ | (13.5a) |
| Connection 1-form: | $2a_{\pm}(x^+, x^-) = \frac{1}{\epsilon} a_u(u,\varphi) \pm a_{\varphi}(u,\varphi),$ | (13.5b) |
| Spin-2 Generators: | $2L_n^{\pm} = L_n \pm \frac{1}{\epsilon} M_n,$ | (13.5c) |
| Spin-3 Generators: | $2W_n^{\pm} = U_n \pm \frac{1}{\epsilon} V_n,$ | (13.5d) |
| Spin-2 Fields: | $\frac{24}{c_{\pm}} \mathcal{L}^{\pm}(x^{\pm}) = \mathcal{M}(u,\varphi) \pm 2\epsilon \mathcal{N}(u,\varphi),$ | (13.5e) |
| Spin-3 Fields: | $-\frac{3}{c_{\pm}} \mathcal{W}^{\pm}(x^{\pm}) = \mathcal{V}(u,\varphi) \pm 2\epsilon \mathcal{Z}(u,\varphi),$ | (13.5f) |
| Spin-2 Chemical Potentials: | $\frac{1}{4}\xi^{\pm}(x^+, x^-) = 1 + \mu_{\mathrm{M}}(u,\varphi) \pm \frac{1}{\epsilon}\mu_{\mathrm{L}}(u,\varphi),$ | (13.5g) |
| Spin-3 Chemical Potentials: | $\frac{1}{4}\eta^{\pm}(x^+, x^-) = \mu_{\mathrm{V}}(u,\varphi) \pm \frac{1}{\epsilon}\mu_{\mathrm{U}}(u,\varphi).$ | (13.5h) |

As expected, this procedure leads to the same results as displayed above in Section 13.1.

## 13.3 Canonical Charges and Chemical Potentials

Since I have not changed $a_{\varphi}$ the results for the canonical charges remain unchanged and all expressions displayed in Chapter 12 are also valid for non-vanishing $\mu_{\mathrm{M}}$, $\mu_{\mathrm{L}}$, $\mu_{\mathrm{V}}$ and $\mu_{\mathrm{U}}$. In particular, from (12.15) one can read off the following four zero mode charges

$$\mathcal{Q}_{\mathcal{M}} = \frac{k}{2}\mathcal{M}, \qquad \mathcal{Q}_{\mathcal{L}} = k\,\mathcal{L}, \qquad \mathcal{Q}_{\mathcal{V}} = 4k\,\mathcal{V}, \qquad \mathcal{Q}_{\mathcal{U}} = 8k\,\mathcal{U}, \tag{13.6}$$

which can be interpreted as mass, angular momentum, odd and even spin-3 charges, respectively. These zero-mode charges play a prominent role in the variational principle and the calculation of entropy of flat space cosmologies with higher-spin hair. But before proceeding in calculating physical observables one has to check whether or not introducing chemical potentials spoils the variational principle of the Chern-Simons action.
As shown in Chapter 2, varying the Chern-Simons action (2.3) in general yields a boundary term of the form $\frac{k}{4\pi}\int\langle \mathcal{A} \wedge \delta\mathcal{A}\rangle$. Evaluating this term explicitly for the connection (12.8) including chemical potentials as in (13.2) one finds

$$\langle A_{\varphi}\delta A_u - A_u \delta A_{\varphi}\rangle \simeq \mathcal{M}\delta\mu_{\mathrm{M}} + 2\mathcal{N}\delta\mu_{\mathrm{L}} + 12\mathcal{V}\delta\mu_{\mathrm{V}} + 24\mathcal{Z}\delta\mu_{\mathrm{U}} + 4\mu_{\mathrm{V}}\delta\mathcal{V} + 8\mu_{\mathrm{U}}\delta\mathcal{Z}\,. \tag{13.7}$$

For vanishing spin-3 potentials this means that the bulk Chern-Simons action has a well defined variational principle. For non-vanishing spin-3 chemical potentials, however, the last two terms are incompatible with a well defined variational principle. This can be fixed by adding an appropriate boundary term by hand. In this case this term has the form

$$\Gamma[A] = S_{\mathrm{CS}}[\mathcal{A}] - S_{\mathrm{B}}[\mathcal{A}], \qquad \text{with} \qquad S_{\mathrm{B}}[\mathcal{A}] = \frac{k}{4\pi}\int \mathrm{d}u\,\mathrm{d}\varphi\,\langle \bar{A}_u A_\varphi\rangle, \tag{13.8}$$

where

$$\bar{a}_u = a_u - 2(1+\mu_{\mathrm{M}})\,M_1 - 2\mu_{\mathrm{L}}\,L_1 - 2\mu_{\mathrm{V}}\,V_2 - 2\mu_{\mathrm{U}}\,U_2\,. \tag{13.9}$$

Thus, after adding this term one obtains

$$\begin{aligned}\delta\Gamma\big|_{\mathrm{EOM}} =& \frac{k}{4\pi}\int \mathrm{d}u\,\mathrm{d}\varphi\,\left(\langle A_\varphi\delta A_u - A_u\delta A_\varphi\rangle - \delta\langle \bar{A}_u A_\varphi\rangle\right)\\ =& \int \mathrm{d}u\,\left(\mathcal{Q}_{\mathcal{M}}\,\delta\mu_{\mathrm{M}} + \mathcal{Q}_{\mathcal{N}}\,\delta\mu_{\mathrm{L}} + \mathcal{Q}_{\mathcal{V}}\,\delta\mu_{\mathrm{V}} + \mathcal{Q}_{\mathcal{Z}}\,\delta\mu_{\mathrm{U}}\right).\end{aligned} \tag{13.10}$$

This small modification is already sufficient for the action (13.8) to have a well-defined variational principle, in the sense that the first variation of the full action vanishes on-shell for arbitrary (but fixed) chemical potentials. As expected, the response functions (13.10) are determined by the canonical charges, and the chemical potentials act as sources.

## Metric and Spin-3 Field with Chemical Potentials

Plugging the results for the connection with chemical potentials, (13.2)-(13.4) together with (12.8) into the definitions of the metric (12.5) one obtains

$$\mathrm{d}s^2 = g_{uu}\,\mathrm{d}u^2 + g_{u\varphi}\,2\,\mathrm{d}u\,\mathrm{d}\varphi - (1+\mu_{\mathrm{M}})\,2\,\mathrm{d}r\,\mathrm{d}u + r^2\,\mathrm{d}\varphi^2, \tag{13.11}$$

with

$$g_{uu} = r^2\left(\mu_{\mathrm{L}}^2 - 4\mu_{\mathrm{U}}''\mu_{\mathrm{U}} + 3\mu_{\mathrm{U}}'^2 + 4\mathcal{M}\mu_{\mathrm{U}}^2\right) + r\,g_{uu}^{(r)} + g_{uu}^{(0)} + g_{uu}^{(0')}, \tag{13.12a}$$

$$g_{u\varphi} = r^2\mu_{\mathrm{L}} - r\mu_{\mathrm{M}}' + \mathcal{N}(1+\mu_{\mathrm{M}}) + 8\mathcal{Z}\mu_{\mathrm{V}}, \tag{13.12b}$$

where

$$\begin{aligned}g_{uu}^{(0)} =& \mathcal{M}(1+\mu_{\mathrm{M}})^2 + 2(1+\mu_{\mathrm{M}})(\mathcal{N}\mu_{\mathrm{L}} + 12\mathcal{V}\mu_{\mathrm{V}} + 16\mathcal{Z}\mu_{\mathrm{U}})\\ &+ 16\mathcal{Z}\mu_{\mathrm{L}}\mu_{\mathrm{V}} + \tfrac{4}{3}(\mathcal{M}^2\mu_{\mathrm{V}}^2 + 4\mathcal{M}\mathcal{N}\mu_{\mathrm{U}}\mu_{\mathrm{V}} + \mathcal{N}^2\mu_{\mathrm{U}}^2),\end{aligned} \tag{13.12c}$$

and the contributions $g_{uu}^{(r)}$ and $g_{uu}^{(0')}$ are presented in (E.1) in Appendix E.

Similarly, one obtains from the definition of the spin-3 field (12.6)

$$\Phi_{\mu\nu\lambda}\,\mathrm{d}x^\mu\,\mathrm{d}x^\nu\,\mathrm{d}x^\lambda = \Phi_{uuu}\,\mathrm{d}u^3 + \Phi_{ruu}\,\mathrm{d}r\,\mathrm{d}u^2 + \Phi_{uu\varphi}\,\mathrm{d}u^2\,\mathrm{d}\varphi - (2\mu_{\mathrm{U}}r^2 - r\mu_{\mathrm{V}}' + 2\mathcal{N}\mu_{\mathrm{V}})\mathrm{d}r\,\mathrm{d}u\,\mathrm{d}\varphi + \mu_{\mathrm{V}}\,\mathrm{d}r^2\,\mathrm{d}u - \left(\mu_{\mathrm{U}}'r^3 - \tfrac{1}{3}r^2(\mu_{\mathrm{V}}'' - \mathcal{M}\mu_{\mathrm{V}} + 4\mathcal{N}\mu_{\mathrm{U}}) + r\mathcal{N}\mu_{\mathrm{V}}' - \mathcal{N}^2\mu_{\mathrm{V}}\right)\mathrm{d}u\,\mathrm{d}\varphi^2, \tag{13.13}$$

with

$$\begin{aligned}\Phi_{uuu} = {} & r^2\left[2(1+\mu_{\mathrm{M}})\mu_{\mathrm{U}}(\mathcal{M}\mu_{\mathrm{L}} - 4\mathcal{V}\mu_{\mathrm{U}}) - \tfrac{1}{3}\mu_{\mathrm{L}}^2(\mathcal{M}\mu_{\mathrm{V}} - 4\mathcal{N}\mu_{\mathrm{U}}) + 16\mu_{\mathrm{L}}\mu_{\mathrm{U}}(\mathcal{V}\mu_{\mathrm{V}} + \mathcal{Z}\mu_{\mathrm{U}}) - \tfrac{4}{3}\mathcal{M}\mu_{\mathrm{U}}^2(\mathcal{M}\mu_{\mathrm{V}} + 2\mathcal{N}\mu_{\mathrm{U}})\right] \\ & + 2\mathcal{V}(1+\mu_{\mathrm{M}})^3 + \tfrac{2}{3}(1+\mu_{\mathrm{M}})^2(6\mathcal{Z}\mu_{\mathrm{L}} + \mathcal{M}^2\mu_{\mathrm{V}} + 2\mathcal{M}\mathcal{N}\mu_{\mathrm{U}}) \\ & + \tfrac{2}{3}(1+\mu_{\mathrm{M}})\left((\mathcal{N}\mu_{\mathrm{L}} + 16\mathcal{Z}\mu_{\mathrm{U}})(2\mathcal{M}\mu_{\mathrm{V}} + \mathcal{N}\mu_{\mathrm{U}}) + 12\mathcal{M}\mathcal{V}\mu_{\mathrm{V}}^2\right) + \mathcal{N}^2\mu_{\mathrm{L}}^2\mu_{\mathrm{V}} \\ & + 16\mu_{\mathrm{L}}\mu_{\mathrm{V}}^2(\mathcal{N}\mathcal{V} - \tfrac{1}{3}\mathcal{M}\mathcal{Z}) + \tfrac{64}{3}\mathcal{Z}\mu_{\mathrm{U}}\mu_{\mathrm{V}}(\mathcal{N}\mu_{\mathrm{L}} + 12\mathcal{V}\mu_{\mathrm{V}} + 12\mathcal{Z}\mu_{\mathrm{U}}) + 64\mathcal{V}^2\mu_{\mathrm{V}}^3 \\ & - \tfrac{8}{27}(\mathcal{M}^3\mu_{\mathrm{V}}^3 - \mathcal{N}^3\mu_{\mathrm{U}}^3) - \tfrac{4}{9}\mathcal{M}\mathcal{N}\mu_{\mathrm{U}}\mu_{\mathrm{V}}(4\mathcal{M}\mu_{\mathrm{V}} + 5\mathcal{N}\mu_{\mathrm{U}}) \\ & + r^3\,\Phi_{uuu}^{(r^3)} + r^2\,\Phi_{uuu}^{(r^2)} + r\,\Phi_{uuu}^{(r)} + \Phi_{uuu}^{(0)}, \end{aligned} \tag{13.14a}$$

$$\Phi_{ruu} = -2r^2\mu_{\mathrm{L}}\mu_{\mathrm{U}} - \tfrac{2}{3}(1+\mu_{\mathrm{M}})(2\mathcal{M}\mu_{\mathrm{V}} + \mathcal{N}\mu_{\mathrm{U}}) - 2\mathcal{N}\mu_{\mathrm{L}}\mu_{\mathrm{V}} - 16\mu_{\mathrm{V}}(\mathcal{V}\mu_{\mathrm{V}} + 2\mathcal{Z}\mu_{\mathrm{U}}) + r\,\Phi_{ruu}^{(r)} + \Phi_{ruu}^{(0)}, \tag{13.14b}$$

$$\begin{aligned}\Phi_{uu\varphi} = {} & r^2\left[2\mathcal{M}(1+\mu_{\mathrm{M}})\mu_{\mathrm{U}} - \tfrac{2}{3}\mu_{\mathrm{L}}(\mathcal{M}\mu_{\mathrm{V}} - 4\mathcal{N}\mu_{\mathrm{U}}) + 16\mu_{\mathrm{U}}(\mathcal{V}\mu_{\mathrm{V}} + \mathcal{Z}\mu_{\mathrm{U}})\right] \\ & + 4\mathcal{Z}(1+\mu_{\mathrm{M}})^2 + \tfrac{2}{3}\mathcal{N}(1+\mu_{\mathrm{M}})(2\mathcal{M}\mu_{\mathrm{V}} + \mathcal{N}\mu_{\mathrm{U}}) + 2\mathcal{N}\mu_{\mathrm{V}}(\mathcal{N}\mu_{\mathrm{L}} + \tfrac{32}{3}\mathcal{Z}\mu_{\mathrm{U}}) \\ & - \tfrac{16}{3}(\mathcal{M}\mathcal{Z} - 3\mathcal{V}\mathcal{N})\mu_{\mathrm{V}}^2 + r^3\,\Phi_{uu\varphi}^{(r^3)} + r^2\,\Phi_{uu\varphi}^{(r^2)} + r\,\Phi_{uu\varphi}^{(r)} + \Phi_{uu\varphi}^{(0)}, \end{aligned} \tag{13.14c}$$

where the contributions $\Phi_{uuu}^{(r^3)}$, $\Phi_{uuu}^{(r^2)}$, $\Phi_{uuu}^{(r)}$, $\Phi_{uuu}^{(0)}$, $\Phi_{ruu}^{(r)}$, $\Phi_{ruu}^{(0)}$, $\Phi_{uu\varphi}^{(r^3)}$, $\Phi_{uu\varphi}^{(r^2)}$, $\Phi_{uu\varphi}^{(r)}$ and $\Phi_{uu\varphi}^{(0)}$ are collected in Appendix E.

Note that for zero mode solutions of the charges with constant chemical potentials i.e. $\mathcal{M}' = \mathcal{N}' = \mu_{\mathrm{M}}' = \mu_{\mathrm{L}}' = \mu_{\mathrm{V}}' = \mu_{\mathrm{U}}' = 0$, all the expressions in Appendix E vanish and thus the spin-2 and spin-3 fields simplify considerably in this case.

## 13.4 Flat Space Einstein Gravity with Chemical Potentials

Flat space Einstein gravity with chemical potentials can be obtained by setting the spin-3 charges and spin-3 chemical potentials, $\mathcal{V} = \mathcal{Z} = \mu_{\mathrm{V}} = \mu_{\mathrm{U}} = 0$ in the results previously obtained. Even though this is just a special case, it is a good starting point to see how adding chemical potentials influences Einstein gravity before moving on to the spin-3 case

### 13.4.1 General Solution

The connection for spin-2 Einstein gravity is given by (12.8) with

$$\begin{aligned} a_u &= (1+\mu_{\text{M}})\, M_1 - \mu'_{\text{M}}\, M_0 + \tfrac{1}{2}\left(\mu''_{\text{M}} - \tfrac{1}{2}\mathcal{M}(1+\mu_{\text{M}}) - \mathcal{N}\,\mu_{\text{L}}\right) M_{-1} \\ &\quad + \mu_{\text{L}}\, L_1 - \mu'_{\text{L}}\, L_0 + \tfrac{1}{2}\left(\mu''_{\text{L}} - \tfrac{1}{2}\mathcal{M}\mu_{\text{L}}\right) L_{-1}, \qquad (13.15\text{a}) \\ a_\varphi &= L_1 - \frac{\mathcal{M}}{4}\, L_{-1} - \frac{\mathcal{N}}{2}\, M_{-1}\,. \qquad (13.15\text{b}) \end{aligned}$$

The corresponding line-element reads

$$\begin{aligned} \mathrm{d}s^2 =& \big[r^2\mu_{\text{L}}^2 + 2r\left(\mu'_{\text{L}}(1+\mu_{\text{M}}) - \mu_{\text{L}}\mu'_{\text{M}}\right) + \mathcal{M}(1+\mu_{\text{M}})^2 + 2(1+\mu_{\text{M}})(\mathcal{N}\mu_{\text{L}} - \mu''_{\text{M}}) + \mu'^2_{\text{M}}\big]\,\mathrm{d}u^2 \\ &+ \left(r^2\mu_{\text{L}} - r\mu'_{\text{M}} + \mathcal{N}(1+\mu_{\text{M}})\right) 2\,\mathrm{d}u\,\mathrm{d}\varphi - (1+\mu_{\text{M}})\, 2\,\mathrm{d}r\,\mathrm{d}u + r^2\,\mathrm{d}\varphi^2, \qquad (13.16) \end{aligned}$$

with the on-shell conditions

$$\begin{aligned} \dot{\mathcal{M}} &= -2\mu'''_{\text{L}} + 2\mathcal{M}\mu'_{\text{L}} + \mathcal{M}'\mu_{\text{L}}, \qquad (13.17\text{a}) \\ \dot{\mathcal{N}} &= -\mu'''_{\text{M}} + \mathcal{M}\mu'_{\text{M}} + \tfrac{1}{2}\mathcal{M}'\mu_{\text{M}} + 2\mathcal{N}\mu'_{\text{L}} + \mathcal{N}'\mu_{\text{L}}\,. \qquad (13.17\text{b}) \end{aligned}$$

### 13.4.2 Zero Mode Solutions with Constant Chemical Potentials

In order to simplify the results from before even further and to gain an intuition of the geometric meaning of the spin-2 chemical potentials $\mu_{\text{M}}$, $\mu_{\text{L}}$ I will consider now the case of constant $\mathcal{M}$, $\mathcal{N}$, $\mu_{\text{M}}$ and $\mu_{\text{L}}$. For this choice of state dependent functions and chemical potentials the line-element reads

$$\begin{aligned} \mathrm{d}s^2 =& \left[r^2\mu_{\text{L}}^2 + \mathcal{M}(1+\mu_{\text{M}})^2 + 2\mathcal{N}(1+\mu_{\text{M}})\mu_{\text{L}}\right]\mathrm{d}u^2 \\ &+ \left(r^2\mu_{\text{L}} + \mathcal{N}(1+\mu_{\text{M}})\right) 2\,\mathrm{d}u\,\mathrm{d}\varphi - (1+\mu_{\text{M}})\, 2\,\mathrm{d}r\,\mathrm{d}u + r^2\,\mathrm{d}\varphi^2. \qquad (13.18) \end{aligned}$$

Setting the even chemical potential, $\mu_{\text{L}} = 0$, the line-element (13.18) simplifies to the FSC solution (10.1) but with $\tilde{u} = (1+\mu_{\text{M}})u$. Therefore, a constant odd chemical potential $\mu_{\text{M}}$ effectively rescales the retarded time coordinate. Speaking in terms of canonical general relativity language, the odd chemical potential $\mu_{\text{M}}$ rescales the lapse function.
If instead one sets to zero the odd chemical potential $\mu_{\text{M}} = 0$, then the line-element (13.18) simplifies to

$$\mathrm{d}s^2 = \left(\mathcal{M} - \frac{\mathcal{N}^2}{r^2}\right)\mathrm{d}u^2 - 2\,\mathrm{d}r\,\mathrm{d}u + r^2\left(\mathrm{d}\varphi + \frac{\mathcal{N}}{r^2}\,\mathrm{d}u + \mu_{\text{L}}\,\mathrm{d}u\right)^2. \qquad (13.19)$$

Comparing this result with the FSC solution (10.1) in ADM-like form,

$$\mathrm{d}s^2 = \left(\mathcal{M} - \frac{\mathcal{N}^\in}{r^2}\right)\mathrm{d}u^2 - 2\,\mathrm{d}r\,\mathrm{d}u + r^2\left(\mathrm{d}\varphi + \frac{\mathcal{N}}{r^2}\,\mathrm{d}u\right)^2 \qquad (13.20)$$

one finds that the even chemical potential $\mu_{\text{L}}$ only changes the last term. Thus, again speaking in canonical general relativity language, the even chemical potential $\mu_{\text{L}}$ changes the shift vector.

## 13.4.3 Perturbative Solutions Linearized in Chemical Potentials

A different kind of simplification arises when linearizing in the chemical potentials. Expanding the metric (13.11) in terms of the chemical potentials,

$$g_{\mu\nu} = \bar{g}_{\mu\nu} + h_{\mu\nu} + \mathcal{O}(\mu_{\text{M}}^2, \mu_{\text{L}}^2, \mu_{\text{M}}\mu_{\text{L}}) \tag{13.21}$$

with the background line-element $\bar{g}_{\mu\nu}\,\mathrm{d}x^\mu\,\mathrm{d}x^\nu$ given by the right hand side of (10.1), yields for the linear terms

$$\begin{aligned} h_{\mu\nu}\,\mathrm{d}x^\mu\,\mathrm{d}x^\nu =& 2(\mathcal{M}\,\mu_{\text{M}} + \mathcal{N}\,\mu_{\text{L}})\,\mathrm{d}u^2 + (r^2\,\mu_{\text{L}} + \mathcal{N}\,\mu_{\text{M}})2\,\mathrm{d}u\,\mathrm{d}\varphi - 2\mu_{\text{M}}\,\mathrm{d}r\,\mathrm{d}u \\ &+ 2(r\,\mu_{\text{L}}' - \mu_{\text{M}}'')\,\mathrm{d}u^2 - 2r\,\mu_{\text{M}}'\,\mathrm{d}u\,\mathrm{d}\varphi\,. \end{aligned} \tag{13.22}$$

The terms in the second line vanish for constant chemical potentials.

#### Comparison with Holographic Dictionary

From a holographic perspective, the first two terms in the linearized solution (13.22) show the typical coupling between sources (chemical potentials) and vacuum expectation values (canonical charges). The $r^2\mu_{\text{L}}\,\mathrm{d}u\,\mathrm{d}\varphi$ term and the $\mu_{\text{M}}\,\mathrm{d}r\,\mathrm{d}u$ term correspond to the essential terms in the two towers of non-normalizable[2] solutions to the linearized equations of motion.

In the holographic dictionary, these non-normalizable contributions should be dual to sources of the corresponding operators in the dual field theory. Indeed, this is what happens as shown in [72]. Note, however, that [72] worked in Euclidean signature, restricted to zero mode solutions and imposed axial gauge for the non-normalizable solutions to the linearized Einstein equations on a flat space background, so a direct comparison is not straightforward. Exploiting the interpretation of constant chemical potentials as modifications of lapse and shift (see section 13.4.2) one can interpret the results of [72] as follows (see their section 3.4): their quantity $\delta\xi_J$ corresponds precisely to the (linearized) even chemical potential $\delta\xi_J \sim \mu_{\text{L}}$, and their quantity $\delta\xi_M$ corresponds to twice the (linearized) odd chemical potential, $\delta\xi_M \sim 2\mu_{\text{M}}$. This identification is perfectly consistent with the holographic interpretation summarized above.

[2]Here and in the following the attribute “non-normalizable” always means “breaking the Barnich–Compère boundary conditions” [51] or the corresponding spin-3 version [150, 154].

## 13.5 Applications

After having introduced chemical potentials in spin-2 and spin-3 gravity in three dimensional flat space I now want to present some applications thereof, such as the entropy of flat space cosmologies with spin-3 charges and the corresponding free energies in the following section[3].

### 13.5.1 Entropy

In this section I will determine the entropy of flat space cosmologies including spin-3 charges, by solving holonomy conditions of the connection $\mathcal{A}$ around the non-contractible $\varphi$-cycle. In addition, I will restrict myself to solutions with constant chemical potentials in order to carry out this calculation.

Using the hatted trace introduced in (12.2) one can write the entropy of a spin-3 charged flat space cosmology, similar to a spin-3 charged BTZ black hole [164] as

$$S = 2k\beta_L \,\widehat{\mathrm{Tr}}(a_u a_\varphi)\Big|_{\mathrm{EOM}} = \beta_L \left(2(1+\mu_{\mathrm{M}})\mathcal{Q}_{\mathcal{M}} + 2\mu_{\mathrm{L}}\mathcal{Q}_{\mathcal{L}} + 3\mu_{\mathrm{V}}\mathcal{Q}_{\mathcal{V}} + 3\mu_{\mathrm{U}}\mathcal{Q}_{\mathcal{U}}\right). \quad (13.23)$$

The quantity $\beta_L$ is not necessarily the inverse temperature, but rather the length of the relevant cycle appearing in the holonomy condition below. The zero mode charges $\mathcal{Q}_i$ are displayed in (13.6).

The holonomy condition I want to solve is given by

$$\exp\left(i\beta_L a_u\right) = \mathbb{1}. \quad (13.24)$$

This condition is completely analogous to the holonomy conditions for higher spin black holes in AdS [159]. To solve the holonomy condition (13.24) one can use the representation summarized in Appendix A.2.3 in terms of $9\times 9$ matrices. By a similarity transformation one can diagonalize the ad-part of a generic matrix of the form (A.22).

$$\begin{pmatrix} \mathcal{A}^{-1}_{8\times 8} & \mathbb{0}_{8\times 1} \\ \mathbb{0}_{1\times 8} & 1 \end{pmatrix} \begin{pmatrix} \mathrm{ad}_{8\times 8} & \mathrm{odd}_{8\times 1} \\ \mathbb{0}_{1\times 8} & 0 \end{pmatrix} \begin{pmatrix} \mathcal{A}_{8\times 8} & \mathbb{0}_{8\times 1} \\ \mathbb{0}_{1\times 8} & 1 \end{pmatrix} = \begin{pmatrix} (\mathcal{A}^{-1}\mathrm{ad}\mathcal{A})_{8\times 8} & (\mathcal{A}^{-1}\mathrm{odd})_{8\times 1} \\ \mathbb{0}_{1\times 8} & 0 \end{pmatrix} \quad (13.25)$$

A matrix of this form is easily exponentiated. Assuming that ad has zero as eigenvalue with geometric and algebraic multiplicity $n$ and denoting $v = \mathcal{A}^{-1}\mathrm{odd}$ yields

$$\exp\begin{pmatrix} (\mathcal{A}^{-1}\mathrm{ad}\mathcal{A})_{8\times 8} & (\mathcal{A}^{-1}\mathrm{odd})_{8\times 1} \\ \mathbb{0}_{1\times 8} & 0 \end{pmatrix} = \begin{pmatrix} 1 & & & & & & v_1 \\ & \ddots & & & & & \vdots \\ & & 1 & & & & v_n \\ & & & e^{\lambda_1} & & & v_{n+1}\frac{e^{\lambda_1}-1}{\lambda_1} \\ & & & & \ddots & & \vdots \\ & & & & & e^{\lambda_{8-n}} & v_8\frac{e^{\lambda_{8-n}}-1}{\lambda_{8-n}} \\ & & & & & & 1 \end{pmatrix}. \quad (13.26)$$

[3] The following section of applications is based on a collaboration with Grumiller, Gary and Rossel first presented in [III].

In the case at hand one has $n = 2$ which is the rank of $\mathfrak{sl}(3,\mathbb{R})$[4]. The holonomy condition (13.24) is then solved by the relations

$$\lambda_a = 0 \bmod \frac{2\pi}{\beta_L}\,, \quad a = 1\ldots 6\,; \qquad v_m = 0\,, \quad m = 1\ldots 2\,. \tag{13.27}$$

The first set of relations (13.27) is precisely the same as for one chiral half of AdS spin-3 gravity. Therefore, one must be able to represent these conditions in the same way as it was done in AdS. In fact, a plausible guess for the two holonomy conditions that follow from the first set of relations (13.27) is given by

$$\tfrac{1}{4}\mathrm{Tr}(a_u a_u)\Big|_{\epsilon=0} = \mathcal{M}\mu_{\mathrm{L}}^2 + 24\mathcal{V}\mu_{\mathrm{L}}\mu_{\mathrm{U}} + \tfrac{4}{3}\mathcal{M}^2\mu_{\mathrm{U}}^2 = \frac{4\pi^2}{\beta_L^2}\,, \tag{13.28}$$

$$\tfrac{1}{4}\sqrt{\det a_u}\Big|_{\epsilon=0} = |\mathcal{V}\mu_{\mathrm{L}}^3 + \tfrac{1}{3}\mathcal{M}^2\mu_{\mathrm{L}}^2\mu_{\mathrm{U}} + 4\mathcal{M}\mathcal{V}\mu_{\mathrm{L}}\mu_{\mathrm{U}}^2 - \tfrac{4}{27}\mathcal{M}^3\mu_{\mathrm{U}}^3 + 32\mathcal{V}^2\mu_{\mathrm{U}}^3| = 0. \tag{13.29}$$

Since the matrix $\mathcal{A}^{-1}\mathrm{ad}\mathcal{A}$ is diagonal, it must lie in the Cartan subalgebra of $\mathfrak{sl}(3,\mathbb{R})$. Diagonalizing simultaneously $L_0$ and $U_0$ one finds

$$\mathcal{A}^{-1}\mathrm{ad}\mathcal{A} = \mathrm{diag}\left(0, 0, f_L{+}2f_U, f_L{-}2f_U, -f_L{+}2f_U, -f_L{-}2f_U, 2f_L, -2f_L\right), \tag{13.30}$$

with some functions $f_L$, $f_U$ of the charges and chemical potentials that can be determined by explicitly calculating the characteristic polynomial of the matrix $i\beta_L a_u$ for the eigenvalues $\lambda$ as derived from the solution (13.3) (with constant charges and chemical potentials) and comparing it with the characteristic polynomial that follows from (13.30). The first set of relations (13.27) yields the conditions

$$f_L = \frac{m\pi}{\beta_L}, \qquad f_U = \frac{(n - \frac{m}{2})\pi}{\beta_L}, \qquad n, m \in \mathbb{Z}\,. \tag{13.31}$$

Thus, the first half of the holonomy conditions leads to a discrete family of solutions parametrized by two integers $n$ and $m$. For the choice $m = 2$ and $n = 1$ these conditions reproduce precisely the guess (13.28) and (13.29). This choice is unique by requiring that in the absence of spin-3 chemical potentials and spin-3 charges the holonomy conditions reduce to the ones for flat space cosmologies. I will therefore always make this choice in the following.

So far only half of the holonomy conditions have been solved. The other half emerges from imposing the second set of relations (13.27). After a straightforward calculation[5] one finds that one of these conditions is linear in the charges and

[4] Since the even part in this representation corresponds to $\mathfrak{sl}(3,\mathbb{R})$.

[5] There are numerous different ways to obtain these results, but it is not always easy to extract the simple conditions (13.32) and (13.33). For instance, one can contract the AdS holonomy conditions using the map (13.5), but this leads naturally to nonlinear relations between charges and chemical potentials. Two combinations of these relations immediately provide the holonomy conditions (13.28) and (13.29), but it takes a bit of work to extract the other two conditions in their simplest form. Alternatively, one can explicitly construct the matrix $\mathcal{A}$ in (13.25) that diagonalizes the $\mathfrak{sl}(3,\mathbb{R})$ part of the generators and then determine the two eigenvectors associated with the two

chemical potentials, while the other is quadratic in the charges and linear in the chemical potentials

$$\mathcal{M}(1+\mu_{\text{M}})+\mathcal{L}\mu_{\text{L}}+12\mathcal{V}\mu_{\text{V}}+16\mathcal{U}\mu_{\text{U}}=0, \tag{13.32}$$

$$9\mathcal{V}(1+\mu_{\text{M}})+6\mathcal{U}\mu_{\text{L}}+\mathcal{M}^2\mu_{\text{V}}+2\mathcal{L}\mathcal{M}\mu_{\text{U}}=0\,. \tag{13.33}$$

These results are considerably simpler than the corresponding holonomy conditions in AdS, which are at least quadratic in chemical potentials and charges.
The linear holonomy condition (13.32) simplifies the entropy (13.23) to

$$S=\beta_L\left(\mu_{\text{L}}\mathcal{Q}_{\mathcal{L}}+\mu_{\text{U}}\mathcal{Q}_{\mathcal{U}}\right). \tag{13.34}$$

For the special case $\mu_{\text{U}}=0$, the entropy (13.34) depends only on spin-2 charges and chemical potentials. Moreover, the solution to the four holonomy conditions (13.28), (13.29), (13.32), (13.33) is given by

$$\mathcal{M}=\frac{4\pi^2}{\beta_L^2\mu_{\text{L}}^2},\qquad \mathcal{L}=-\mathcal{M}\frac{1+\mu_{\text{M}}}{\mu_{\text{L}}},\qquad \mathcal{V}=0,\qquad \mathcal{U}=-\mathcal{M}^2\,\frac{\mu_{\text{V}}}{6\mu_{\text{L}}}. \tag{13.35}$$

For that case entropy is given by the Bekenstein–Hawking area law with $k=\frac{1}{4G_N}$

$$S\big|_{\mu_{\text{U}}=0}=k\beta_L\left|\mu_{\text{L}}\mathcal{L}\right|=k\,\frac{2\pi|\mathcal{L}|}{\sqrt{\mathcal{M}}}. \tag{13.36}$$

The absolute values used above ensure that entropy is positive regardless of the sign of the charge $\mathcal{L}$. The inverse temperature

$$\beta=-\frac{\partial S}{\partial\mathcal{Q}_{\mathcal{M}}}\Big|_{\mathcal{Q}_{\mathcal{L}}}=-\frac{2\partial S}{k\partial\mathcal{M}}\Big|_{\mathcal{L}}=2\pi\,\frac{|\mathcal{L}|}{\mathcal{M}^{3/2}}, \tag{13.37}$$

then coincides with the spin-2 result (see e.g. [63]; note that in their conventions $\mathcal{M}=r_+^2$ and $|\mathcal{L}|=|r_0r_+|$)

$$T=\frac{1}{2\pi}\frac{\mathcal{M}^{3/2}}{|\mathcal{L}|}\,. \tag{13.38}$$

The minus sign in the definition (13.37) can be seen as a remnant of the inner horizon first law of black hole mechanics [148, 149, 165–167] as explained in [63] when thinking about a flat space cosmology as a flat space limit of a non-extremal BTZ black hole as elaborated on at the beginning of Chapter 10. From the corresponding first law

$$-\,\mathrm{d}\mathcal{Q}_{\mathcal{M}}=T\;\mathrm{d}S+\Omega\;\mathrm{d}\mathcal{Q}_{\mathcal{L}}, \tag{13.39}$$

one can deduce the angular potential

$$\Omega=-T\,\frac{\partial S}{\partial\mathcal{Q}_{\mathcal{L}}}\Big|_{\mathcal{Q}_{\mathcal{M}}}=-T\,\frac{\partial S}{k\partial\mathcal{L}}\Big|_{\mathcal{M}}=\frac{\mathcal{M}}{\mathcal{L}}, \tag{13.40}$$

zero eigenvalues. This approach makes it clear from the start that the remaining two holonomy conditions must be linear in the chemical potentials.

which again coincides with the spin-2 result [63].

In the general case $\mu_{\rm U} \neq 0$ not all holonomy conditions are linear. Instead, one has to solve one quadratic and one cubic equation, similar to the AdS case. Defining $\mu = \mu_{\rm L}\mu_{\rm U}$ and $\eta = \mu_{\rm L}/\mu_{\rm U} + \frac{1}{9}\mathcal{M}^2/\mathcal{V}$ the holonomy conditions (13.28), (13.29) simplify to

$$\eta^3 + \eta\Big(4\mathcal{M} - \frac{\mathcal{M}^4}{27\mathcal{V}^2}\Big) + 32\mathcal{V} - \frac{16\mathcal{M}^3}{27\mathcal{V}} + \frac{2\mathcal{M}^6}{729\mathcal{V}^3} = 0, \tag{13.41}$$

$$\mu = \frac{4\pi^2}{\beta_L^2}\Big(\frac{\mu_{\rm L}}{\mu_{\rm U}}\mathcal{M} + 24\mathcal{V} + \frac{4\mu_{\rm U}}{3\mu_{\rm L}}\mathcal{M}^2\Big)^{-1}. \tag{13.42}$$

Solving the cubic equation (13.41) yields a result for the ratio $\mu_{\rm L}/\mu_{\rm U}$, which can then be plugged into the linear equation (13.42) to determine the product of the chemical potentials. The sign of the discriminant $D$ of the cubic equation (13.41) is given by

$$\text{sign}\, D = \text{sign}(\mathcal{M}^3 - 108\mathcal{V}^2)\,. \tag{13.43}$$

If $D$ is negative there is exactly one real solution. This happens only if the spin-3 charge $\mathcal{V}$ is sufficiently large or if the mass $\mathcal{M}$ is negative. For a critical tuning of the charges,

$$\text{criticality:} \qquad 108\mathcal{V}^2 = \mathcal{M}^3, \tag{13.44}$$

the discriminant vanishes, $D = 0$, and there is a unique real solution $\eta = 0$. However, the linear equation (13.43) has no finite solution for $\mu$ in this case. Therefore, starting from finite and positive $\mathcal{M}$ it is not possible to smoothly increase the spin-3 charge $\mathcal{V}$ beyond the critical value (13.44).

Henceforth, the following inequality will be assumed to hold

$$\mathcal{M} > \big(108\mathcal{V}^2\big)^{1/3} \geq 0\,. \tag{13.45}$$

In other words, from now on exclusively the case of positive discriminant $D > 0$ will be considered. In this case there are three real solutions for $\eta$. The resulting entropy is real for all three branches. However, only one branch recovers the same entropy (13.34) as for the spin-2 case in the limit $\mathcal{V} \to 0$. It is exactly this branch that will be of interest for the following discussion.

On this particular branch, there is a neat way to express all results in terms of the charges $\mathcal{M}, \mathcal{L}, \mathcal{U}$ and a new parameter $\mathcal{R}$ that depends on the ratio of spin-3 and spin-2 charges $\frac{\mathcal{V}^2}{\mathcal{M}^3}$, just like in the AdS case [159]

$$\frac{\mathcal{R}-1}{4\mathcal{R}^{3/2}} = \frac{|\mathcal{V}|}{\mathcal{M}^{3/2}}, \qquad \mathcal{R} > 3. \tag{13.46}$$

The restriction to $\mathcal{R} > 3$ guarantees that indeed the correct branch is chosen.

The chemical potentials then read

$$1+\mu_{\mathrm{M}} = -\frac{2\pi|\mathcal{L}|}{\mathcal{M}\sqrt{\mathcal{M}}\beta_L}\cdot\frac{4\mathcal{R}(2\mathcal{R}^2+6\mathcal{R}-9)-24\mathcal{P}\sqrt{\mathcal{R}}(10\mathcal{R}^2-15\mathcal{R}+9)}{(\mathcal{R}-3)^3(4-3/\mathcal{R})^{3/2}}, \tag{13.47}$$

$$\mu_{\mathrm{L}} = \frac{2\pi\,\mathrm{sign}\mathcal{L}}{\sqrt{\mathcal{M}}\beta_L}\cdot\frac{2\mathcal{R}-3}{(\mathcal{R}-3)\sqrt{4-3/\mathcal{R}}}, \tag{13.48}$$

$$\mu_{\mathrm{U}} = -\frac{3\pi\,\mathrm{sign}\mathcal{L}}{\mathcal{M}\beta_L}\cdot\frac{\sqrt{\mathcal{R}}}{(\mathcal{R}-3)\sqrt{4-3/\mathcal{R}}}, \tag{13.49}$$

$$\mu_{\mathrm{V}} = \frac{3\pi|\mathcal{L}|}{\mathcal{M}^2\beta_L}\cdot\frac{2\sqrt{\mathcal{R}}(10\mathcal{R}^2-15\mathcal{R}+9)-16\mathcal{P}\mathcal{R}(2\mathcal{R}^2+6\mathcal{R}-9)}{(\mathcal{R}-3)^3(4-3/\mathcal{R})^{3/2}}, \tag{13.50}$$

while the entropy is given by

$$S(\mathcal{M},\mathcal{L},\mathcal{R},\mathcal{P}) = 2\pi k\,\frac{|\mathcal{L}|}{\sqrt{\mathcal{M}}}\cdot\frac{2\mathcal{R}-3-12\mathcal{P}\sqrt{\mathcal{R}}}{(\mathcal{R}-3)\sqrt{4-3/\mathcal{R}}}, \tag{13.51}$$

with the dimensionless ratio

$$\mathcal{P} = \frac{\mathcal{U}}{\sqrt{\mathcal{M}\mathcal{L}}}. \tag{13.52}$$

The expression for entropy (13.51) is the main result of this section. The pre-factor containing the spin-2 charges $\mathcal{M},\mathcal{L}$ coincides with the spin-2 result (13.36). The spin-3 correction depends nonlinearly on one of the combinations of spin-3 charges, $\mathcal{R}$, and linearly on the other, $\mathcal{P}$.

For some purposes it can be useful to have a simpler perturbative result for entropy in the limit of small spin-3 charge $\mathcal{V}$ (large $\mathcal{R}$) given by

$$S(\mathcal{M},\mathcal{L},\mathcal{V},\mathcal{U}) = 2\pi k\,\frac{|\mathcal{L}|}{\sqrt{\mathcal{M}}}\Big(1+\frac{15\mathcal{V}^2}{8\mathcal{M}^3}-\frac{6\mathcal{U}|\mathcal{V}|}{\mathcal{M}^2\mathcal{L}}\Big)+\mathcal{O}(\mathcal{V}^3). \tag{13.53}$$

I will close this discussion on entropy of higher-spin charged flat space cosmologies by addressing sign issues. The mass should be positive, $\mathcal{M}>0$, motivated by the necessity of this condition in the spin-2 case. The sign of $\mathcal{L}$ does not matter, which is why absolute values in the final result for entropy (13.51) have been used. Suppose that $\mathcal{L}>0$ ($\mathcal{L}<0$). Then one can exploit the sign ambiguity in the definitions of $\mu_{\mathrm{L}}$, $\mu_{\mathrm{U}}$ by choosing $\mu_{\mathrm{L}}>0$ ($\mu_{\mathrm{L}}<0$) so that the first term in (13.34) is always positive and thus entropy is positive in the limit of vanishing spin-3 fields. The sign of $\mathcal{V}$ is taken care of by the definition (13.46), which ensures positive $\mathcal{R}$ regardless of the sign of $\mathcal{V}$. Thus, the only remaining signs of potential relevance are the signs of the spin-3 charge $\mathcal{U}$ and the corresponding chemical potential $\mu_{\mathrm{U}}$. The latter is fixed through the sign choice of $\mu_{\mathrm{L}}$ explained above, but the former is free to change, and this change is physically relevant. This implies that the quantity $\mathcal{P}$ defined in (13.52) can have either sign, so that the last term in the entropy (13.51) can have either sign. Demanding positivity of entropy then establishes an upper bound on $\mathcal{U}$.

### 13.5.2 Grand Canonical Free Energy and Phase Transitions

There are three branches of solutions of all the holonomy conditions as shown in the previous section. The proposal for the correct branch to choose was the branch which connects continuously to the spin-2 results in the limit of vanishing spin-3 charges. However, it is not guaranteed that this branch is actually the correct one from a thermodynamical perspective in the whole parameter space. A way to check whether or not the chosen branch is thermodynamically sensible is to compare the free energies of all branches for given values of the chemical potentials and check which of the branches leads to the lowest free energy.

The first step will be to write the general result for the (grand canonical) free energy, regardless of the specific branch[6]. Since the entropy has been determined previously, which is a thermodynamic potential in terms of extensive quantities (charges), the only thing to do is to perform a Legendre transformation with respect to all pairs of charges and chemical potentials.[7]

$$F(T,\,\Omega,\,\Omega_{\rm V},\,\Omega_{\rm U}) = -\mathcal{Q}_{\mathcal{M}} - TS - \Omega\,\mathcal{Q}_{\mathcal{L}} - \Omega_{\rm V}\,\mathcal{Q}_{\mathcal{V}} - \Omega_{\rm U}\,\mathcal{Q}_{\mathcal{U}}. \tag{13.54}$$

The zero mode charges are given by (13.6) and the intensive quantities by the chemical potentials.

$$T^{-1} = \beta = -\frac{\partial S}{\partial \mathcal{Q}_{\mathcal{M}}}\Big|_{\mathcal{L},\mathcal{V},\mathcal{U}} = -\beta_L\,(1+\mu_{\rm M}), \tag{13.55}$$

$$\beta\,\Omega = -\frac{\partial S}{\partial \mathcal{Q}_{\mathcal{L}}}\Big|_{\mathcal{M},\mathcal{V},\mathcal{U}} = -\beta_L\,\mu_{\rm L}, \tag{13.56}$$

$$\beta\,\Omega_{\rm V} = -\frac{\partial S}{\partial \mathcal{Q}_{\mathcal{V}}}\Big|_{\mathcal{M},\mathcal{L},\mathcal{U}} = -\beta_L\,\mu_{\rm V}, \tag{13.57}$$

$$\beta\,\Omega_{\rm U} = -\frac{\partial S}{\partial \mathcal{Q}_{\mathcal{U}}}\Big|_{\mathcal{M},\mathcal{L},\mathcal{V}} = -\beta_L\,\mu_{\rm U}. \tag{13.58}$$

In order to express free energy in terms of intensive variables one has to invert the holonomy conditions (13.28),(13.29),(13.32) and (13.33) and solve for the charges in terms of chemical potentials.

Before doing so, however, it is instructive to consider the free energy expressed in terms of charges in certain limits. In the large $\mathcal{R}$ limit (weak contribution from spin-3 charges) one recovers the spin-2 result

$$F_{\rm weak} = -\frac{\mathcal{M}}{2} + \mathcal{O}(\mathcal{P}/\sqrt{\mathcal{R}}) + \mathcal{O}(1/\mathcal{R})\,. \tag{13.59}$$

In the $\mathcal{R}\to 3$ limit (strong contribution from spin-3 charges) one obtains

$$F_{\rm strong} = -\frac{\mathcal{M}}{6} + \mathcal{O}(\mathcal{R}-3)^2\,. \tag{13.60}$$

[6] I set $k=1$ in this subsection.
[7] Alternatively, one could use the on-shell action method by Bañados, Canto and Theisen [168].

Thus, one finds a universal ratio

$$\frac{F_{\text{weak}}}{F_{\text{strong}}} = 3\,. \tag{13.61}$$

The results (13.59)-(13.61) are valid on all branches and show that the free energy approaches the correct spin-2 value.
Performing the Legendre transformation (13.54) with the entropy (13.34) yields

$$F = -\mathcal{Q}_{\mathcal{M}} + T\beta_L\,\mu_{\text{V}}\mathcal{Q}_{\mathcal{V}} = -\frac{\mathcal{M}}{2} - 4\Omega_{\text{V}}\mathcal{V}\,. \tag{13.62}$$

In order to obtain the free energy as function of intensive variables one has to solve the nonlinear holonomy conditions (13.28), (13.29) for the charges in terms of the chemical potentials. Solving (13.28) for $\mathcal{V}$ allows one to express free energy in terms of the mass $\mathcal{M}$ and chemical potentials.

$$F = -\frac{\mathcal{M}}{2} + \frac{\mathcal{M}\Omega\Omega_{\text{V}}}{6\Omega_{\text{U}}} + \frac{2\mathcal{M}^2\Omega_{\text{U}}\Omega_{\text{V}}}{9\Omega} - \frac{2\pi^2T^2\Omega_{\text{V}}}{3\Omega\Omega_{\text{U}}}. \tag{13.63}$$

Plugging the solution for the spin-3 charge $\mathcal{V}$ in terms of the mass $\mathcal{M}$ into the other holonomy condition (13.29) establishes a quartic equation for the mass $\mathcal{M}$, which leads to four branches of solutions for the free energy. The discriminant of that equation is positive, provided the spin-3 chemical potential obeys the bound

$$\Omega_{\text{U}}^2 < \frac{9(2\sqrt{3}-3)}{64}\,\frac{\Omega^4}{4\pi^2T^2} \approx 0.065\,\frac{\Omega^4}{4\pi^2T^2}\,. \tag{13.64}$$

Another way to read the inequality (13.64) is that it provides an upper bound on the temperature for given spin-3 chemical potential $\Omega_{\text{U}}$. The maximal temperature is given by

$$T_{\text{max}} = \frac{3\sqrt{2\sqrt{3}-3}}{8}\,\frac{\Omega^2}{2\pi|\Omega_{\text{U}}|}\,. \tag{13.65}$$

In the limit of small $\Omega_{\text{U}}$ it turns out that only one of the branches has finite free energy. This is the branch that continuously connects with spin-2 results, on which free energy yields

$$F = -\frac{2\pi^2T^2}{\Omega^2}\Big(1 - \frac{32\pi^2T^2\Omega_{\text{V}}\Omega_{\text{U}}}{3\Omega^3} + \frac{80\pi^2T^2\Omega_{\text{U}}^2}{3\Omega^4} + \mathcal{O}(\Omega_{\text{U}}^3)\Big)\,. \tag{13.66}$$

The term before the parentheses reproduces the spin-2 result for free energy. The term in the parentheses depends only on two linear combinations of the chemical potentials[8]. As in the spin-2 case [63] there will be a phase transition between flat space cosmologies and hot flat space at some critical temperature.
A novel feature of the spin-3 case is that there are additional phase transitions between the various flat space cosmology branches. To see this, I consider the difference between the free energies of two branches

$$\Delta F_{12} = \frac{2\Omega_{\text{U}}\Omega_{\text{V}}}{9\Omega}\left(\mathcal{M}_1 - \mathcal{M}_2\right)\Big(\mathcal{M}_1 + \mathcal{M}_2 + \frac{3\Omega(\Omega\Omega_{\text{V}} - 3\Omega_{\text{U}})}{4\Omega_{\text{U}}^2\Omega_{\text{V}}}\Big). \tag{13.67}$$

[8] To be more precise, on $\mathfrak{t}$ and $\mathfrak{v}$ introduced in (13.68).

There are two zeros in the difference (13.67), an obvious one when the masses of the two branches coincide, $\mathcal{M}_1 = \mathcal{M}_2$, and a non-obvious one when the expression in the last parentheses in (13.67) vanishes. In the following the focus will be on the difference between the branch that continuously connects to spin-2 results (branch 1) and the other branch that ceases to exist if the bound (13.64) is violated (branch 2). The other two branches are then branch 3 and 4, which will only play minor roles.

To reduce clutter it will be assumed from now on that temperature and the chemical potentials are non-negative. Moreover, it is convenient to introduce dimensionless combinations of chemical potentials as

$$\mathfrak{t} = 2\pi T\,\frac{\Omega_{\text{U}}}{\Omega^2}, \qquad \mathfrak{v} = \Omega_{\text{V}}\,\frac{\Omega}{\Omega_{\text{U}}}\,. \tag{13.68}$$

The quantity $\mathfrak{t}$ is a dimensionless temperature, while $\mathfrak{v}$ is essentially a ratio of odd over even spin-3 chemical potential. Expressing the difference of free energies (13.67) between branches 1 and 2 as function of these two combinations, up to a non-negative overall constant, yields

$$\Delta F_{12} \propto 15\mathfrak{v} - 18 - \mathfrak{v}\sqrt{64\mathfrak{t}^2 + 9 + \frac{8\mathfrak{t}(64\mathfrak{t}^2+27)}{N(\mathfrak{t})} + 8\mathfrak{t}N(\mathfrak{t})}\,. \tag{13.69}$$

with

$$N(\mathfrak{t}) = \left(512\mathfrak{t}^3 + 648\mathfrak{t} + 9\sqrt{4096\mathfrak{t}^4 + 3456\mathfrak{t}^2 - 243}\right)^{1/3}. \tag{13.70}$$

The positive real zero of the term under the square-root in (13.70) corresponds precisely to the critical temperature (13.65). For each value of dimensionless temperature $\mathfrak{t}$ there is a simple zero in $\Delta F_{12}$ since it depends linearly on $\mathfrak{v}$. The corresponding value of $\mathfrak{v}$ will be called "critical" and denoted by a subscript "c". For vanishing temperature one finds from setting (13.69) to zero

$$\mathfrak{v}_c|_{\mathfrak{t}=0} = \frac{3}{2}, \tag{13.71}$$

while at the critical temperature (13.65) one finds similarly

$$\mathfrak{v}_c\Big|_{\mathfrak{t}=\mathfrak{t}_c=\frac{3}{8}\sqrt{2\sqrt{3}-3}} = 2\,. \tag{13.72}$$

The corresponding free energy differences near these temperatures read, respectively

$$\Delta F_{12} \propto 12\mathfrak{v} - 18 - 12\mathfrak{t}\mathfrak{v} + \tfrac{8}{3}\mathfrak{t}^2\mathfrak{v} + \mathcal{O}(\mathfrak{t}^3), \tag{13.73}$$

$$\Delta F_{12} \propto 9\mathfrak{v} - 18 - 8\sqrt{1 + \tfrac{2}{\sqrt{3}}}\,(\mathfrak{t} - \mathfrak{t}_c)\mathfrak{v} - \tfrac{16}{27}\,(\mathfrak{t} - \mathfrak{t}_c)^2\,\mathfrak{v} + \mathcal{O}(\mathfrak{t} - \mathfrak{t}_c)^3\,. \tag{13.74}$$

Hence one arrives at the following picture, depending on the value of the parameter $\mathfrak{v}$:[9]

- $0 < \mathfrak{v} < \frac{3}{2}$: Branch 1 is thermodynamically unstable for all temperatures.
- $\mathfrak{v} = \frac{3}{2}$: Branch 1 degenerates with branch 2 at vanishing temperature and is thermodynamically unstable for all positive temperatures.
- $\frac{3}{2} < \mathfrak{v} < 2$: Branch 1 degenerates with branch 2 at some positive temperature. Below that temperature branch 1 is thermodynamically unstable. At that temperature there is a phase transition from branch 2 to branch 1. Above that temperature branch 1 is stable (modulo the phase transition to hot flat space [63]).
- $\mathfrak{v} = 2$: Branch 1 degenerates with branch 2 at the maximal temperature (13.65) and is thermodynamically stable for all temperatures (again modulo the phase transition to hot flat space).
- $\mathfrak{v} > 2$: Branch 1 is thermodynamically stable for all temperatures (with the same caveat as above).

To illustrate the results above I show an example in Figure 13.1. In all six graphs the thick line depicts free energy for branch 1 and the dashed line for branch 2[10]. The three upper plots show explicitly the phase transition between branches 1 and 2, depending on the choice of $\mathfrak{v}$. The three lower plots show that there are further phase transitions involving the branches 3 and 4 if branch 1 is unstable for all values of temperature. In addition to all these new phase transitions there is the 'usual' phase transition to hot flat space [63], which in the present case can be of zeroth, first or second order. Since there are several phase transitions possible, there exist also multi-critical points where three or four phases co-exist.
The most striking difference between the AdS results by David, Ferlaino and Kumar [169] and this flat space results is that one observes the possibility of first order phase transitions between various branches (see the right upper and middle lower plot in figure 13.1). In contrast, for AdS the only phase transitions (other than Hawking–Page like) arise because two of the branches end, at which point the free energy jumps (These zeroth order phase transitions are also recovered in flat space, see e.g. the left lower plot in figure 13.1).

[9] Positivity of entropy imposes additional constraints on the existence of branches. The existence of the first order phase transition between branches 1 and 2 described below is not influenced by such constraints.
[10] The other two branches are not essential for this discussion. If visible they are plotted as dotted lines.

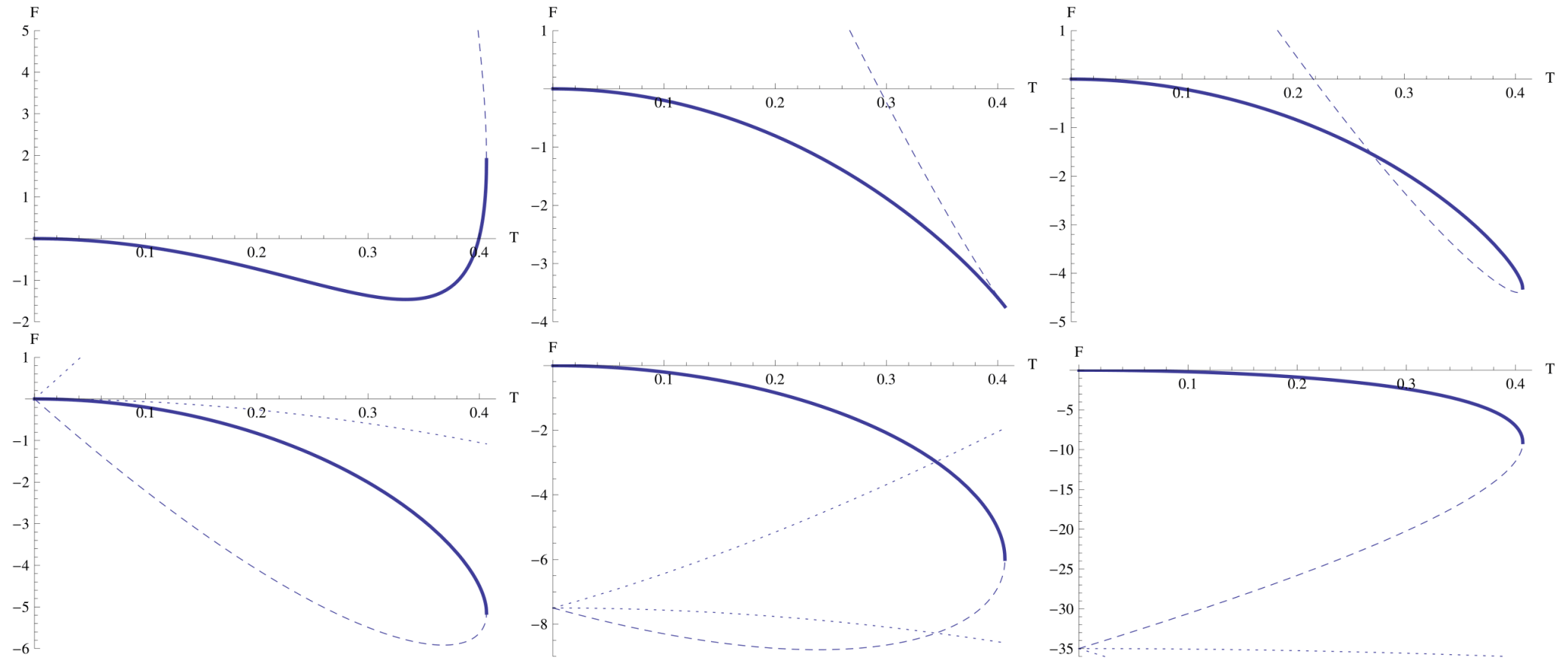

**Fig. 13.1.:** Plots of free energy as function of temperature. In all plots $\Omega = 1$, $\Omega_{\text{U}} = 0.1$. Upper Left: $\Omega_{\text{V}} = 0.4$. Upper Middle: $\Omega_{\text{V}} = 0.2$. Upper Right: $\Omega_{\text{V}} = 0.18$. Lower Left: $\Omega_{\text{V}} = 0.15$. Lower Middle: $\Omega_{\text{V}} = 0.12$. Lower Right: $\Omega_{\text{V}} = 0.01$. The branch with smooth spin-2 limit is displayed as thick line, the second branch as dashed line, the other two branches as dotted lines (in the upper plots these lines are at positive $F$).

## 13.6 Conclusions

This Part III was mainly focused on the gravity aspects of a flat space (higher-spin) holographic correspondence. In Chapter 10 I have shown how suitable limits from known results related to AdS$_3$ holography can be used to determine an efficient Chern-Simons description of flat space, the asymptotic symmetries of flat space (higher-spin) gravity theories and a flat space (higher-spin) Cardy formula. This shows that for certain instances the limit of vanishing cosmological constant of the AdS$_3$ results can yield physically sensible results which can be used for a holographic correspondence involving asymptotically flat spacetimes.

Chapter 11 discussed unitarity of linear and nonlinear $\mathcal{FW}$-algebras alike. Similar to the results obtained for $\mathcal{W}$-algebras, the requirement of unitary representations yields restrictions for both $c_L$ and $c_M$ under certain assumptions. In order to have unitary representations one has to set $c_M = 0$. This has severe consequences for the presence of higher-spin excitations of nonlinear $\mathcal{FW}$-algebras. I showed that $c_M = 0$ also renders all higher-spin excitations for general nonlinear $\mathcal{FW}$-algebras unphysical and thus concluded that this is a NO-GO result for having flat space, unitarity and higher-spins at the same time. Since I assumed that the $\mathcal{FW}$-algebras are realized by highest-weight representations, I also argued that there could be possible loopholes circumventing my NO-GO result. I then showed how to exploit a specific loophole, namely the nonlinearity of the $\mathcal{FW}$-algebras, by considering the linear $\mathcal{FW}_\infty$ algebra. Using this algebra I showed that for a linear $\mathcal{FW}$-algebra it is indeed possible to have flat space, unitarity and higher-spins at the same time.
Following up on this discussion I reviewed in Chapter 12 how to describe higher-spin gravity in flat space using a Chern-Simons formulation. Building up on this I showed how to add chemical potentials to flat space (higher-spin) gravity in Chapter 13. I also performed consistency checks and closely examined the special case of flat space Einstein gravity with chemical potentials. Furthermore, I determined the entropy of flat space cosmological solutions with chemical (higher-spin) potentials turned on. Following up on this I also showed how to determine the grand canonical free energy which then led to the discovery of new first order phase transitions between various flat space cosmological solutions.
Building up on these results from the gravity side I will provide explicit checks of a holographic correspondence in asymptotically flat spacetimes in the following Part IV of this thesis.

# Part IV

## Flat Space Holography–Field Theory Side

In this part of the thesis I will first give a brief introduction to Galilean conformal field theories and show why it is useful to study them in the context of flat space holography by calculating entanglement entropy for different Galilean conformal field theories. I will also mention early checks for a holographic correspondence involving asymptotically flat spacetimes. This part will conclude with the first holographic derivation of entanglement entropy and thermal entropy for $\mathfrak{bms}_3$ invariant field theories using a specialized Wilson line, which will also be generalized to higher-spin symmetries.

# 14 Galilean Conformal Field Theory in $1+1$ Dimensions

> *Eppur si muove.*
> *(And yet it moves.)*
>
> **– Galileo Galilei**
> Italian mathematician, physicist and philosopher

In this chapter[1] I will review the basics of Galilean conformal field theories (GCFTs) in $1+1$ dimensions and then use this knowledge to compute entanglement entropy first for a GCFT at zero temperature and then for a GCFT at finite temperature. More details on the mathematical tools used in the following sections can be found in [77].

This chapter is organized as follows. First I will give a brief introduction on Galilean conformal symmetries. I will then proceed by reviewing a highest-weight representation for the Galilean conformal algebra, energy-momentum tensors and the corresponding transformation properties under Galilean conformal transformations. After having introduced the basics of Galilean conformal field theory, I will then continue by reviewing how to employ the replica trick for Galilean conformal field theories and determine the entanglement entropy for a Galilean conformal field theory. In a similar manner I will also review how to determine the thermal entropy of a Galilean conformal field theory at non-zero temperature. I will finish this chapter with a short overview of early checks of a holographic correspondence in asymptotically flat spacetimes.

## 14.1 Galilean Conformal Symmetries

First let us consider a system in $1+1$ dimensions with a spatial coordinate $x$ and a timelike coordinate $t$ invariant under Galilei transformations. By also demanding scaling invariance of the system generated by a dilatation operator, one obtains the finite dimensional Galilean conformal group in two dimensions.

The (finite) Galilean conformal algebra generating this group has already been

[1] This review section and the following Chapter 15 is based on [VIII], which I published together with my collaborator Rudranil Basu. Thus, the general structure and several parts of this review coincide with the contents found in [VIII].

encountered in Chapter 5. It is generated by six generators $L_n$ and $M_n$ with $n = \pm 1, 0$ whose vector field representation is given by (5.6) and obeys the algebra (5.10). The generators correspond to the following symmetries

| Symmetry | Generators | Change of $x$ | Change of $t$ |
|---|---|---|---|
| Time Translations | $L_{-1}$ | $x \to x$ | $t \to t + a$ |
| Space Translations | $M_{-1}$ | $x \to x + a$ | $t \to t$ |
| Galilean Boosts | $M_0$ | $x \to x + vt$ | $t \to t$ |
| Dilatations | $L_0$ | $x \to \lambda x$ | $t \to \lambda t$ |
| Spec. Conf. Trans. | $L_1$ | $x \to x + 2\kappa x t$ | $t \to t + \kappa t^2$ |
| Constant Acceleration | $M_1$ | $x \to x + bt^2$ | $t \to t$ |

and can also be interpreted as a nonrelativistic contraction of the global part of relativistic conformal symmetries as shown in Chapter 5. For the purpose of studying flat space holography, however, the relevant global symmetries correspond to the ultrarelativistic limit of the conformal symmetries whose global part can be summarized as

| Symmetry | Generators | Change of $x$ | Change of $t$ |
|---|---|---|---|
| Time Translations | $M_{-1}$ | $x \to x$ | $t \to t + a$ |
| Space Translations | $L_{-1}$ | $x \to x + a$ | $t \to t$ |
| Galilean Boosts | $M_0$ | $x \to x$ | $t \to t + vx$ |
| Dilatations | $L_0$ | $x \to \lambda x$ | $t \to \lambda t$ |
| Spec. Conf. Trans. | $L_1$ | $x \to x + \kappa x^2$ | $t \to t + 2\kappa x t$ |
| Constant Acceleration | $M_1$ | $x \to x$ | $t \to t + bx^2$ |

Now why is it helpful to study GCFTs in the context of flat space holography? Looking at the expressions of the corresponding symmetry generators in terms of vector fields (5.9) and (5.17) or alternatively at the two lists above, one can see that at least at the level of the symmetries in $1+1$ dimensions the only difference between the nonrelativistic and ultrarelativistic symmetries is an exchange of time and space. Thus, at least in $1+1$ dimensions, one can for some instances use the knowhow one has gained from Galilean conformally invariant quantum field theories in order to learn something about flat space holography.

The six generators mentioned above generate the finite Galilean conformal algebra. It is worth noting that this algebra still closes under the Lie bracket even if one does not restrict the mode indices to $\{-1, 0, 1\}$.

## 14.2 Quantization and Highest Weight Representation

Similar to the Virasoro algebra (5.4), which is a centrally extended version of the Witt algebra (5.5), the GCA also admits central extensions of the following form

$$\begin{aligned} [L_n, L_m] &= (n-m)L_{n+m} + \frac{c_L}{12}(n^3-n)\delta_{m+n,0}, \\ [L_n, M_m] &= (n-m)M_{n+m} + \frac{c_M}{12}(n^3-n)\delta_{m+n,0}, \\ [M_n, M_m] &= 0. \end{aligned} \tag{14.1}$$

I want to stress at this point that the algebra (14.1) and its highest weight representations are independent of any limit. Whether one is looking at a GCFT or a $\mathfrak{bms}_3$ invariant quantum field theory cannot be seen at the level of the algebra (14.1) alone.

In order to compute the entanglement entropy of a GCFT using similar methods as introduced in Chapter 6 one first needs to introduce the notion of a highest weight representation. This representation is fixed by the highest weight state $|h_L, h_M\rangle$ defined as

$$\begin{aligned} L_0|h_L, h_M\rangle &= h_L|h_L, h_M\rangle, \\ M_0|h_L, h_M\rangle &= h_M|h_L, h_M\rangle, \\ L_n|h_L, h_M\rangle &= M_n|h_L, h_M\rangle = 0 \ \text{ for } n > 0. \end{aligned} \tag{14.2}$$

Repeated application of $L_{-n}$ and $M_{-n}$ for $n > 0$ creates new states in this representation. In analogy to a CFT one can introduce operators corresponding to each of these states. The GCA primaries are local operators $\Phi_{h_L,h_M}(x,t)$ which map the vacuum state to the highest weight state

$$\Phi_{h_L,h_M}(0,0)|0\rangle = |h_L, h_M\rangle. \tag{14.3}$$

The transformation properties of the primaries under the Galilean conformal transformations can be easily derived from first principles by using

$$[L_0, \Phi_{h_L,h_M}(0,0)] = h_L \Phi_{h_L,h_M}(0,0), \tag{14.4}$$

and similarly for $M_0$. Abbreviating $\Phi_{h_L,h_M}(x,t) \equiv \Phi$ one thus obtains

$$\delta_{L_n}\Phi = \Big[t^{n+1}\partial_t + (n+1)t^n x\partial_x + (n+1)t^{n-2}(h_L t - n\, h_M x)\Big]\Phi, \tag{14.5a}$$

$$\delta_{M_n}\Phi = \Big[-t^{n+1}\partial_x + (n+1)t^n h_M\Big]\,\Phi. \tag{14.5b}$$

These relations can be more conveniently encoded in a pair of fields that one can interpret as Galilean energy-momentum tensors [170]

$$T_{(1)}(x,t) = \sum_n t^{-n-2}\left[L_n + (n+2)\frac{x}{t}M_n\right], \tag{14.6a}$$

$$T_{(2)}(x,t) = \sum_n t^{-n-2}M_n, \tag{14.6b}$$

which is analogous to the mode expansion of the energy-momentum tensor in terms of Virasoro generators in a CFT.

With these definitions at hand, one can now determine Galilean conformal Ward identities, which will play an important role in determining entanglement entropy. For the purpose of calculating entanglement entropy, the Ward identities involving two primary fields $\Phi_{h_L^{(i)},h_M^{(i)}}(x_i,t_i) \equiv \Phi^{(i)}$, with $i=1,2$ will be of main interest. One can for example determine the Ward identities for $T_{(2)}(x,t)$ via

$$\begin{aligned}\langle T_{(2)}\Phi^{(1)}\Phi^{(2)}\rangle &= \sum_{n=-1}^{\infty} t^{-n-2}\langle 0|\left[M_n, \Phi^{(1)}\Phi^{(2)}\right]|0\rangle \\ &= \sum_{i=1,2}\sum_{n=-1}^{\infty} t^{-n-2}\left[-t_i^{n+1}\partial_{x_i} + (n+1)h_M^{(i)}t_i^n\right]\langle\Phi^{(1)}\Phi^{(2)}\rangle \\ &= \sum_{i=1,2}\left(\frac{h_M^{(i)}}{(t-t_i)^2} - \frac{1}{t-t_i}\partial_{x_i}\right)\langle\Phi^{(1)}\Phi^{(2)}\rangle. \end{aligned}\tag{14.7}$$

The Ward identity involving $T_{(1)}(x,t)$ can be derived similarly and is given by

$$\begin{aligned}\langle T_{(1)}\Phi^{(1)}\Phi^{(2)}\rangle = \sum_{i=1,2}\Bigg[&\frac{1}{(t-t_i)}\partial_{t_i} + 2h_M^{(i)}\frac{x-x_i}{(t-t_i)^3} \\ &+ \frac{1}{(t-t_i)^2}\left(h_L^{(i)} - (x-x_i)\partial_{x_i}\right)\Bigg]\langle\Phi^{(1)}\Phi^{(2)}\rangle. \end{aligned}\tag{14.8}$$

In order to proceed one needs to know the exact form of the two-point correlation function for GCFT primaries which is given up to some normalization constant by

$$\langle\Phi^{(1)}\Phi^{(2)}\rangle \sim \delta_{h_L^{(1)}h_L^{(2)}}\delta_{h_M^{(1)}h_M^{(2)}}\, t_{12}^{-2h_L^{(1)}}\exp\left(-2h_M^{(1)}\frac{x_{12}}{t_{12}}\right). \tag{14.9}$$

Using this expression for the two-point function and inserting this into (14.7) and (14.8) one arrives at

$$\begin{aligned}\langle T_{(1)}\Phi^{(1)}\Phi^{(2)}\rangle = &\left(\frac{t_{12}}{t_{01}t_{02}}\right)^2 t_{12}^{-2h_L}\exp\left(-2h_M\frac{x_{12}}{t_{12}}\right)\times \\ &\times\left[h_L - 2h_M\left(\frac{x_{12}}{t_{12}} - \frac{x_{01}}{t_{01}} - \frac{x_{02}}{t_{02}}\right)\right], \end{aligned}\tag{14.10a}$$

$$\langle T_{(2)}\Phi^{(1)}\Phi^{(2)}\rangle = h_M\left(\frac{t_{12}}{t_{01}t_{02}}\right)^2 t_{12}^{-2h_L}e^{-2h_M\frac{x_{12}}{t_{12}}}, \tag{14.10b}$$

where $t_{ab} = t_a - t_b$, and the energy-momentum tensor has been inserted at $(t_0, x_0)$.

## 14.3 Transformation Properties of the Energy Momentum Tensor

The Galilean conformal Ward identities (14.7) and (14.8) derived in the previous section are intimately related with the transformation properties of primary fields in a GCFT. In this section I will review how to determine the transformation rules of the components of the energy-momentum tensor, $T_{(1)}$ and $T_{(2)}$ under Galilean conformal transformations.

First, note that arbitrary diffeomorphisms $(t,x) \to (t',x')$ are not compatible with Galilean conformal transformations. The form of the most general transformation of the coordinates $(t,x)$, which are compatible with Galilean conformal transformations can be determined via (5.9). These coordinate transformations are given by [10]

$$t = f(t'), \quad \text{and} \quad x = \frac{df(t')}{dt'}x' + g(t'), \tag{14.11}$$

where $f, g$ are arbitrary functions of $t'$. These transformations can be seen as the Galilean conformal analogues of the holomorphic and anti-holomorphic transformations generated by the Virasoro vector fields in a relativistic CFT. From (14.11) one can also straightforwardly determine the relations

$$\frac{\partial t}{\partial t'} = \frac{\partial x}{\partial x'}, \qquad \frac{\partial t}{\partial x'} = 0 \tag{14.12}$$

whose structure resembles the Cauchy–Riemann equations encountered in complex analysis.

Similar to a relativistic CFT one can determine the transformation properties of $T_{(1)}$ and $T_{(2)}$ under Galilean conformal transformations by integrating the infinitesimal transformation relations which are determined by the two-point correlators [171] of the energy-momentum tensor with itself. A very useful cross-check of the results obtained this way is given by taking the nonrelativistic limit of the corresponding CFT results. On the level of the energy-momentum tensor, this limit is performed as

$$T_{(1)}(t,x) = \lim_{\epsilon\to 0}\left(T(z) + \bar{T}(\bar{z})\right), \tag{14.13}$$

$$T_{(2)}(t,x) = \lim_{\epsilon\to 0}\epsilon\left(T(z) - \bar{T}(\bar{z})\right), \tag{14.14}$$

where $z = t + \epsilon x$ and $\bar{z} = t - \epsilon x$. In the limit $\epsilon \to 0$ this yields

$$\begin{aligned} T_{(1)}(t',x') \to \left(\frac{\mathrm{d}t}{\mathrm{d}t'}\right)^2 T_{(1)}(t,x) + 2\left(\frac{\mathrm{d}t}{\mathrm{d}t'}\right)\left(\frac{\mathrm{d}x}{\mathrm{d}t'}\right) T_{(2)}(t,x) \\ + \frac{c_L}{12}\{t,t'\} + \frac{c_M}{12}\left(\frac{\mathrm{d}t}{\mathrm{d}t'}\right)^{-1} [\![(t,x),t']\!], \end{aligned} \tag{14.15a}$$

$$T_{(2)}(t',x') \to \left(\frac{\mathrm{d}t}{\mathrm{d}t'}\right)^2 T_{(2)}(t,x) + \frac{c_M}{12}\{t,t'\}. \tag{14.15b}$$

where $\{,\}$ denotes the Schwarzian derivative given by

$$\{t, t'\} = \left[\left(\frac{\mathrm{d}^3 t}{\mathrm{d}t'^3}\right) - \frac{3}{2}\left(\frac{\mathrm{d}^2 t}{\mathrm{d}t'^2}\right)^2 \left(\frac{\mathrm{d}t}{\mathrm{d}t'}\right)^{-1}\right]\left(\frac{\mathrm{d}t}{\mathrm{d}t'}\right)^{-1}, \tag{14.16}$$

and $[\![,]\!]$ denotes the corresponding Galilean conformal equivalent (GCA Schwarzian) thereof which can be defined via

$$[\![(t,x), t']\!] := \{(t,x), t'\} - \left(\frac{\mathrm{d}x}{\mathrm{d}t'}\right)\{t, t'\}, \tag{14.17}$$

where

$$\{(t,x), t'\} = \left(\frac{\mathrm{d}^3 x}{\mathrm{d}t'^3}\right) + 3\left[\frac{1}{2}\left(\frac{\mathrm{d}^2 t}{\mathrm{d}t'^2}\right)^2 \left(\frac{\mathrm{d}x}{\mathrm{d}t'}\right)\left(\frac{\mathrm{d}t}{\mathrm{d}t'}\right)^{-1} - \left(\frac{\mathrm{d}^2 t}{\mathrm{d}t'^2}\right)\left(\frac{\mathrm{d}^2 x}{\mathrm{d}t'^2}\right)\right]\left(\frac{\mathrm{d}t}{\mathrm{d}t'}\right)^{-1}. \tag{14.18}$$

The exact form of $\{,\}$ and $[\![,]\!]$ and their appearance together with the central charges $c_L$ and $c_M$ can also be understood as follows. As in a relativistic CFT, the quantum corrections that the classical transformation law of the energy-momentum tensor obtains should vanish for $c_L = 0$, $c_M = 0$ and global ISL$(2,\mathbb{R})$ transformations. Thus, it is clear that the quantum corrections have to depend on $c_L$ and $c_M$. In addition, whatever $\{,\}$ and $[\![,]\!]$ are, they have to be compatible with the group composition law that two successive transformations $(t,x) \mapsto (t', x') \mapsto (t'', x'')$ yield the same result as mapping $(t,x) \mapsto (t'', x'')$ directly. This together with the invariance under global ISL$(2,\mathbb{R})$ transformations, i.e.

$$\begin{aligned} \{f[t], t\} &= \left\{\frac{af[t]+b}{cf[t]+d}, t\right\}, \\ [\![(f[t], g[x]), t]\!] &= \left[\!\!\left[\left(\frac{af[t]+b}{cf[t]+d}, \frac{g[x]}{(d+cf[t])^2}\right), t\right]\!\!\right], \end{aligned} \tag{14.19}$$

where $a, b, c, d$ are some constants with $ad - bc = 1$, determines the form of $\{,\}$ and $[\![,]\!]$ uniquely.

As an addendum it is noteworthy that in the same sense that the Schwarzian derivative measures the degree to which a function fails to be a fractional linear transformation, i.e.

$$\{t, t'\} = 0 \quad \Leftrightarrow \quad t(t') = \frac{at'+b}{ct'+d}, \tag{14.20}$$

the GCA Schwarzian also measures the degree to which two functions fail to be Galilean conformal, i.e.

$$[\![(t,x), t']\!] = 0 \quad \Leftrightarrow \quad t(t') = \frac{at'+b}{ct'+d}, \quad x(t', x') = \frac{x'}{(d+ct')^2}. \tag{14.21}$$

## 14.4 Entanglement Entropy in GCFTs

In this section I will review how to calculate the entanglement entropy of a one-dimensional subsystem in a $1+1$ dimensional GCFT in close analogy to [107]. Consider a subsystem $A$, given by a line connecting the points $(t_1, x_1)$ and $(t_2, x_2)$ and its complement, which will be called $B$ as shown in Figure 14.1.
The motivation for considering such an interval lies in the nonrelativistic nature of a GCFT. In a Lorentz invariant theory, observables are not sensitive to a certain choice of frame. Hence, assuming one quantized a Lorentz invariant theory with respect to some time coordinate $t$, entanglement entropy can be simply computed on a $t=0$ slice. A GCFT, however, is not a Lorentz invariant theory and therefore observables are sensitive to a choice of frame. Thus, in order to determine how the entanglement entropy in a GCFT depends on the choice of frame one should use a (Galilean) boosted interval (A) bounded by the points $(t_1, x_1)$ and $(t_2, x_2)$ instead of an equal time interval (A') which would be bounded by $(x_1, t_2)$ and $(t_2, x_2)$ as depicted in Figure 14.1.

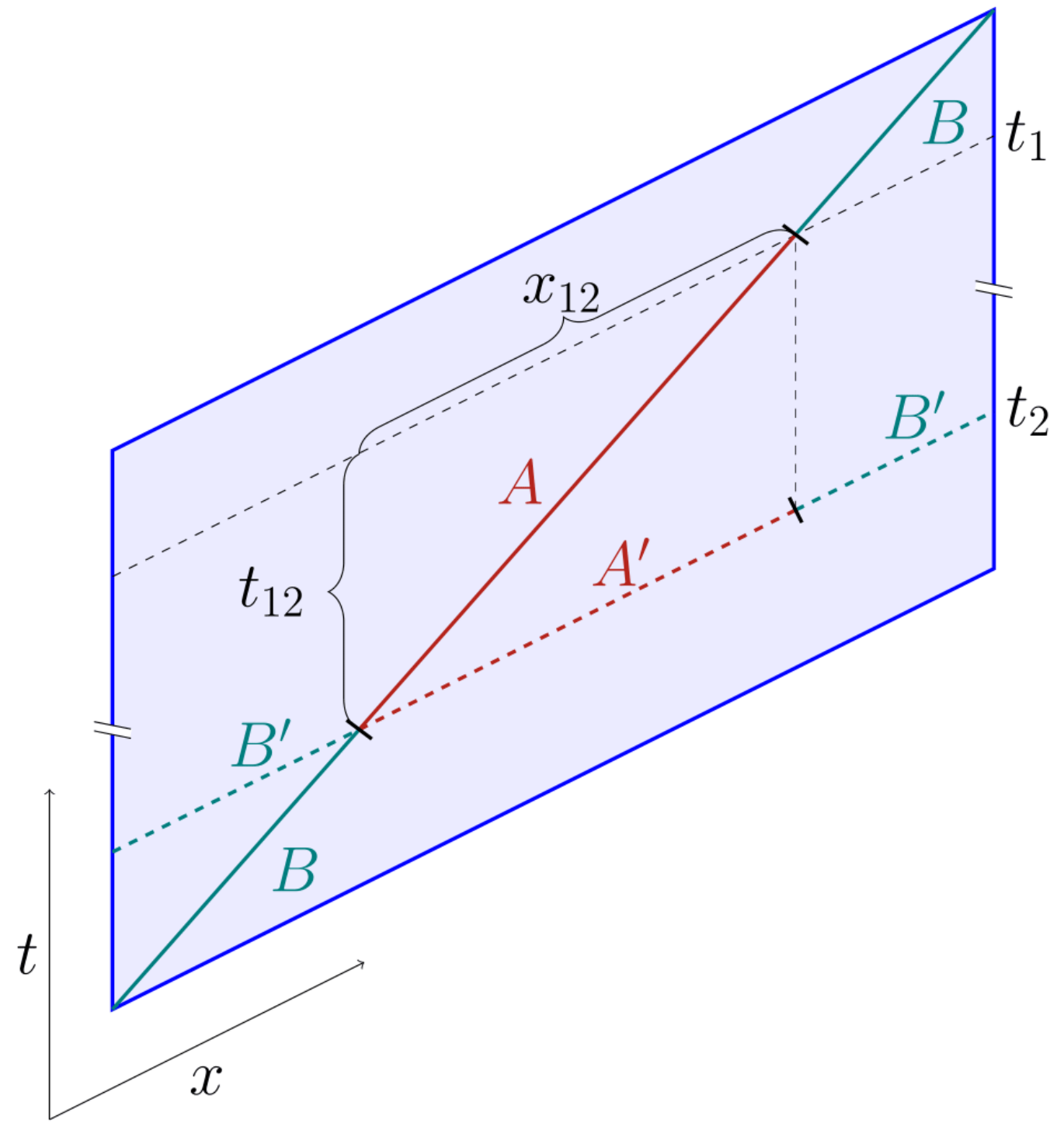

**Fig. 14.1.:** Boosted ($A$, $B$) and equal time ($A'$, $B'$) entangled intervals used to determine entanglement entropy in GCFTs.

In close analogy to the computations in [107] and the review in Chapter 6, the first step is to calculate the $n^{\text{th}}$ Tsallis entropies $S_n^{\text{Tsallis}} = \frac{\text{Tr}_A \rho_A^n - 1}{1-n}$ where $\rho_A$ is the reduced density matrix of the state of the system. The interpretation of rewriting the trace of the $n^{\text{th}}$ power of the reduced density matrix $\rho_A$ in terms of $n$-sheeted Riemann surfaces $\Sigma_n$ as in the CFT case will also apply in the GCFT case.

A map between $\Sigma_n$ $(t, x)$ and the GCFT plane $(t_a, x_b)$ can be established by the coordinate transformations,

$$t = \left(\frac{t_0 - t_1}{t_0 - t_2}\right)^{1/n}, \tag{14.22a}$$

$$x = \frac{1}{n}\left(\frac{t_0 - t_1}{t_0 - t_2}\right)^{1/n}\left(\frac{x_0 - x_1}{t_0 - t_1} - \frac{x_0 - x_2}{t_0 - t_2}\right). \tag{14.22b}$$

The form of the transformed energy-momentum tensor components on $\Sigma_n$ can then be determined using (14.15). The vacuum expectation values of the energy-momentum tensors thus take the following form

$$\langle T_{(1)}(t_0, x_0)\rangle_{\Sigma_n} = \left(1 - \frac{1}{n^2}\right)\left(\frac{t_{12}}{t_{01}t_{02}}\right)^2\left[\frac{c_L}{24} - \frac{c_M}{12}\left(\frac{x_{12}}{t_{12}} - \frac{x_{01}}{t_{01}} - \frac{x_{02}}{t_{02}}\right)\right], \tag{14.23a}$$

$$\langle T_{(2)}(t_0, x_0)\rangle_{\Sigma_n} = \left(1 - \frac{1}{n^2}\right)\left(\frac{t_{12}}{t_{01}t_{02}}\right)^2\frac{c_M}{24}. \tag{14.23b}$$

Similar to the CFT case the planar energy-momentum tensor has vanishing vacuum expectation value due to its symmetries, which reduces (14.23) basically to the (GCA) Schwarzian. One can now compare the explicitly evaluated Ward identities on the GCFT plane (14.10) and the above vacuum expectation values (14.23) on $\Sigma_n$. This comparison yields

$$\langle T_{(i)}(t_0, x_0)\rangle_{\Sigma_n} = \frac{\langle T_{(i)}(t_0, x_0)\Phi_n^{(1)}\Phi_{-n}^{(2)}\rangle_{\mathbb{C}}}{\langle\Phi_n^{(1)}\Phi_{-n}^{(2)}\rangle_{\mathbb{C}}}, \tag{14.24}$$

for $i = 1, 2$, provided one identifies the weights of the twist primaries $\Phi_n$ as $h_L = \frac{c_L}{24}\left(1 - \frac{1}{n^2}\right)$ and $h_M = \frac{c_M}{24}\left(1 - \frac{1}{n^2}\right)$.
As mentioned in the beginning of this section, evaluating the quantity of interest, i.e. $\mathrm{Tr}\rho_A^n$, is the same as doing a path integral over $\Sigma_n$. The left hand side of the identity (14.24) then corresponds to a $T_{(i)}$ insertion in that functional integral. This insertion is equivalent to a Galilean conformal transformation defined by the Ward identities (14.7), (14.8). The functional integral on $\Sigma_n$ is therefore proportional to $n$ products of the GCFT plane 2-point correlators of the twist fields evaluated at the end-points of $A$: $(\langle\Phi_n(t_1, x_1)\Phi_{-n}(t_2, x_2)\rangle)^n$. This allows one to infer

$$\begin{aligned}\mathrm{Tr}\rho_A^n =& \alpha_n\left(\langle\Phi_n(t_1, x_1)\Phi_{-n}(t_2, x_2)\rangle\right)_{\mathbb{C}}^n \\ =& \alpha_n t_{12}^{-\frac{c_L}{12}\left(n - \frac{1}{n}\right)}\exp\left[\frac{c_M}{12}\left(n - \frac{1}{n}\right)\frac{x_{12}}{t_{12}}\right],\end{aligned} \tag{14.25}$$

where the $\alpha_n$ are some normalization constants which can be chosen in a convenient way[2].

[2]This means in particular that $\alpha_1 = 1$.

The final step is to determine the entanglement entropy $\mathrm{Tr}(\rho_A \ln \rho_A)$ for the segment $A$ as a limit of Tsallis entropies. Taking the limit $n \to 1$ of $\frac{\partial}{\partial n}\mathrm{Tr}\rho_A^n$ one obtains

$$S_E = -\lim_{n\to 1}\frac{\partial}{\partial n}\mathrm{Tr}\rho_A^n = \frac{c_L}{6}\ln\left(\frac{t_{12}}{a}\right) + \frac{c_M}{6}\left(\frac{x_{12}}{t_{12}}\right), \tag{14.26}$$

where $a$ is interpreted as a small scale cut-off or lattice spacing of the underlying GCFT. An interpretation of this result and its physical meaning in light of holography will be discussed in the next section.

## The BMS/GCA Correspondence and Entanglement Entropy

In section 14 I reviewed the basics for calculations in a $1+1$ dimensional quantum field theory, which is invariant under Galilean transformations in addition to having scale invariance. I will now elaborate on how the results obtained previously are related to flat space holography in $2+1$ dimensions.
Taking this BMS/GCA correspondence [65] into account all the field theory results presented earlier in this chapter in the context of GCFT can be used for calculations for ultrarelativistic field theories invariant under $\mathfrak{bms}_3$ symmetries at null infinity of asymptotically flat spacetimes by only exchanging the role of time and space.
Using this argument one can immediately determine entanglement entropy in a planar field theory having ultrarelativistic conformal/$\mathfrak{bms}_3$ symmetry. Once again assuming a rectilinear segment $A$ with end points $(t_1, x_1)$ and $(t_2, x_2)$ as the entangling region, one can readily determine the entanglement entropy of that region by using (14.26) and exchanging time and space

$$S_E = \frac{c_L}{6}\ln\left(\frac{x_{12}}{a}\right) + \frac{c_M}{6}\left(\frac{t_{12}}{x_{12}}\right). \tag{14.27}$$

This interval can again be interpreted as a boosted version of a purely spatial (or equal time) interval. For $t_{12} = 0$ (14.27) reduces to

$$S_E = \frac{c_L}{6}\ln\left(\frac{x_{12}}{a}\right). \tag{14.28}$$

The entanglement entropy calculated above corresponds to a $1+1$ dimensional system of infinite spatial extent at zero temperature. It is also of interest to see what happens with the entanglement entropy when dealing with a system at finite temperature $T = \beta^{-1}$ and/or finite spatial extent.
This generalization can be achieved by using geometric properties of the $1+1$ dimensional field theory in question very much alike to the CFT case, where one can use conformal maps to map between the entanglement entropy of different systems.

To elaborate on this further, note that one can map the $1+1$ dimensional GCFT on the plane to a cylinder by

$$x = e^{2\pi\xi/\beta}, \quad t = \frac{2\pi\tau}{\beta} e^{2\pi\xi/\beta}, \tag{14.29}$$

where $\xi$ and $\tau$ denote the coordinates on the cylinder. This means effectively that one dimension gets compactified in the construction of $\Sigma_n$. As shown in [71] this induces a transformation of the GCFT primaries as

$$\tilde{\Phi}(\xi,\tau) = e^{\frac{2\pi}{\beta}(\xi h_L + \tau h_M)} \Phi(x(\xi,\tau), t(\xi,\tau)). \tag{14.30}$$

The two point function evaluated in this geometry is then given by

$$\langle \tilde{\Phi}(\xi_1,\tau_1) \tilde{\Phi}(\xi_2,\tau_2) \rangle = \left[ 2 \sinh\left( \frac{\pi \xi_{12}}{\beta} \right) \right]^{-2h_L} e^{-h_M \frac{\pi\tau_{12}}{\beta} \coth\left(\frac{\pi\xi_{12}}{\beta}\right)}. \tag{14.31}$$

The following steps which are necessary for calculating entanglement entropy in a thermal state for a subsystem with endpoints $\xi_1, \tau_1$ and $\xi_2, \tau_2$ are the same as in the zero-temperature case. Thus one obtains for the entanglement entropy for a system at finite temperature the following expression

$$S_E = \frac{c_L}{6} \ln\left[ \frac{\beta}{\pi a} \sinh\left( \frac{\pi\xi_{12}}{\beta} \right) \right] + \frac{\pi}{6\beta} c_M \tau_{12} \coth\left( \frac{\pi \xi_{12}}{\beta} \right). \tag{14.32}$$

At leading order the expansion of the right hand side of (14.32) in $\beta^{-1}$ yields again the zero-temperature answer (14.27) with the identification of $\tau_{12} \sim t_{12}/a$ and $\xi_{12} \sim x_{12}/a$. In the high-temperature limit on the other hand, i.e. for $\xi_{12} \gg \beta$, one obtains

$$S_E = \frac{\pi}{6\beta} (c_L \xi_{12} + c_M \tau_{12}) + \frac{c_L}{6} \ln \beta + \mathcal{O}(\beta). \tag{14.33}$$

A very similar analysis works when considering the spatial extent of the system to be of finite length $L$ in the ground state. The only difference in comparison to the analysis before lies in the direction of the compactification to the cylinder along the spatial cycle of length $L \sim \beta$ which is perpendicular to the previous case. The entanglement entropy for that system then turns out to be

$$S_E = \frac{c_L}{6} \ln\left[ \frac{L}{\pi a} \sin\left( \frac{\pi\xi_{12}}{L} \right) \right] + \frac{\pi}{6L} c_M \tau_{12} \cot\left( \frac{\pi \xi_{12}}{L} \right). \tag{14.34}$$

## 14.5 Thermal Entropy in GCFTs

In this section I will briefly review how to derive the high-temperature density of states and the corresponding entropy for ordinary $1+1$ dimensional GCFTs (for more details see e.g. [61, 62, 71]) in order to make contact with the holographic results for the thermal entropy of FSCs found in section 15.4.
The partition function for a $1+1$ dimensional GCFT on a torus is given by

$$\begin{aligned} Z^0_{GCFT}(\eta,\rho) =& \mathrm{Tr}\left(e^{2\pi i\eta\left(L_0-\frac{c_L}{24}\right)}e^{2\pi i\rho\left(M_0-\frac{c_M}{24}\right)}\right) \\ =& e^{\frac{\pi}{12}i(\eta c_L+\rho c_M)}Z_{GCFT}(\eta,\rho), \end{aligned} \tag{14.35}$$

where $\eta$ and $\rho$ are the Galilean conformal equivalents of the modular parameters of a CFT. In the same spirit as in a relativistic CFT one demands that (14.35) is invariant under the Galilean conformal equivalent of $S$ modular transformations given by

$$(\eta,\rho)\to\left(-\frac{1}{\eta},\frac{\rho}{\eta^2}\right), \tag{14.36}$$

i.e.

$$Z^0_{GCFT}(\eta,\rho) = Z^0_{GCFT}(-\frac{1}{\eta},\frac{\rho}{\eta^2}). \tag{14.37}$$

This is tantamount to requiring

$$Z_{GCFT}(\eta,\rho) = e^{2\pi i\left(\tilde{f}(\eta,\rho)+h_L\eta+h_M\rho\right)}Z_{GCFT}(-\frac{1}{\eta},\frac{\rho}{\eta^2}), \tag{14.38}$$

where

$$\tilde{f}\left(\eta,\rho\right) = \frac{c_L\eta}{24}+\frac{c_M\rho}{24}+\frac{c_L}{24\eta}-\frac{c_M\rho}{24\eta^2}-h_L\eta-h_M\rho. \tag{14.39}$$

In order to proceed, one has first to rewrite the density of states $d\left(h_L,h_M\right)$ in terms of the GCA partition function. This can be done by using

$$Z_{GCFT}(\eta,\rho) = \mathrm{Tr}\left(e^{2\pi i\eta L_0}e^{2\pi i\rho M_0}\right) = \sum d\left(h_L,h_M\right)e^{2\pi i\eta h_L}e^{2\pi i\rho h_M}, \tag{14.40}$$

and performing an inverse Laplace transformation

$$d\left(h_L,h_M\right) = \int \mathrm{d}\eta\,\mathrm{d}\rho e^{2\pi i\tilde{f}(\eta,\rho)}Z_{GCFT}(-\frac{1}{\eta},\frac{\rho}{\eta^2}). \tag{14.41}$$

In the limit of large central charges the density of states (14.41) can by approximated by the value of the integrand, when the exponential factor is extremal. Using this approximation the density of states is given by

$$d\left(h_L,h_M\right)\sim e^{\pi\sqrt{\frac{c_Mh_M}{6}}\left(\frac{h_L}{h_M}+\frac{c_L}{c_M}\right)}. \tag{14.42}$$

The corresponding entropy is then given by the logarithm of the density of states

$$S=\ln\left(d\left(h_L,h_M\right)\right) = \pi\sqrt{\frac{c_Mh_M}{6}}\left(\frac{h_L}{h_M}+\frac{c_L}{c_M}\right). \tag{14.43}$$

For Einstein gravity, where $c_L = 0$ the entropy (14.43) agrees precisely with the thermal entropy of a flat space cosmology [61, 62]. Thus, the 2d GCFT state counting reproduces exactly the entropy of the cosmological horizon of a FSC. This is one particular example of a specific check that one can indeed establish a holographic principle for asymptotically flat spacetimes.

## 14.6 Early Checks of Flat Space Holography

Further selected checks, some of which I also mentioned briefly in the introduction include:

- Barnich and Compère proposed in 2006 [51] asymptotic boundary conditions for flat spacetimes in three dimensional Einstein gravity and showed that the asymptotic symmetries corresponding to these boundary conditions are given by the $\mathfrak{bms}_3$ algebra (10.3) with central charges $c_L = 0$ and $c_M = \frac{3}{G_N}$.
- Bagchi then showed in 2010 [56] (and also together with Fareghbal in [65]) that the $\mathfrak{bms}_3$ algebra is isomorphic to the Galilean conformal algebra $\mathfrak{gca}_2$, which he coined the BMS/GCA correspondence, and thus was able to propose a framework for the dual field theory of asymptotically flat spacetimes.
- Even though the framework for a dual theory was found in [56, 65], a concrete proposal for a specific theory with $c_L = 0$ and $c_M \neq 0$ was missing. For $c_L \neq 0$ and $c_M = 0$, however, the situation was the complete opposite. The field theory dual was clear, as it would be just a chiral half of a CFT, whereas the corresponding gravity dual was unknown. In [57] Bagchi, Detournay and Grumiller found the gravity dual whose asymptotic symmetry algebra is $\mathfrak{bms}_3$ with $c_L \neq 0$ and $c_M = 0$. This theory is known as flat space chiral gravity and can be obtained as a flat space limit from Topologically Massive Gravity [152] whose action is given by (10.49).
- In 2013 Bagchi et al. showed that there exist Hawking-Page like phase transitions between flat space cosmologies and hot flat space[3] at a critical temperature $T_c = \frac{1}{2\pi r_0}$, where $r_0$ is the radius of the cosmological horizon.

All these explicit checks, and many more that I did not explicitly mention, strongly suggest that doing flat space holography is indeed possible.
In the next chapter I will provide an additional explicit check of flat space holography by computing the entanglement entropy of a $\mathfrak{bms}$ invariant quantum field theory holographically.

[3] This is basically the Euclidean version of the null orbifold.

# 15 Flat Space Holographic Entanglement Entropy

七転び八起き。
*(Fall down seven times and get up eight times.)*

**– Japanese proverb**

Explicit checks are of vital importance for establishing a holographic dictionary and also a given correspondence itself. This chapter will provide a novel and explicit check of a holographic correspondence in flat space by calculating the entanglement entropy as well as the thermal entropy of a $\mathfrak{bms}$ ($\mathcal{FW}$) invariant quantum field theory using a Wilson line.

This chapter is organized as follows. First I will review the Wilson line construction used in $\text{AdS}_3$ to determine entanglement entropy holographically. I will then proceed with constructing a suitable topological probe for flat space and apply this construction explicitly to spin-2 gravity in flat space. Furthermore, I will show how to generalize this construction to include higher-spin symmetries. As a further check of a holographic correspondence in flat space I will use this Wilson line approach to holographically determine the thermal entropy of flat space cosmologies with spin-2 and spin-3 charges and show that all the results obtained in this chapter precisely agree with the results obtained previously in this thesis.

## 15.1 Wilson Lines in $\text{AdS}_3$ Representing a *Massive* and *Spinning* Particle

In order to find a suitable proposal for holographic entanglement entropy in flat space I will modify the proposal made by Alejandra Castro, Nabil Iqbal and Martin Ammon in [81, 172][1] for $\text{AdS}_3$. In the following section I will review the main concepts underlying this proposal.

The basic idea of these proposals relies on finding a gauge invariant object which generalizes a geodesic for the spin-2 gravity case but is also invariant under higher-spin symmetries. An object which satisfies these properties in $\mathrm{AdS}_3$ is given by a Wilson line which takes as an argument a given gauge field $\mathcal{A}$. In [81, 172] the authors argued that indeed a Wilson line $W_{\mathcal{R}}(C)$, for an appropriate choice of representation $\mathcal{R}$, attached to the boundary of $\mathrm{AdS}_3$ can be used to determine the entanglement of the region bounded by the endpoints of the Wilson line holographically as

$$S_{\mathrm{EE}} = -\log\left[W_{\mathcal{R}}(C)\right]. \tag{15.1}$$

[1] Another proposal for holographic entanglement entropy using Wilson lines was also made in [82] by Jan de Boer and Juan Jottar. In [173] it was shown that this proposal is equivalent to the proposal in [81, 172].

In case the central charges $c$ and $\bar{c}$ of the two copies of the Virasoro algebra at the boundary of $\mathrm{AdS}_3$ are not equal, i.e. $c \neq \bar{c}$, this Wilson line describes a massive and spinning particle[2] probing the bulk geometry. Thus, for a massive and spinning particle in $\mathrm{AdS}_3$, where one can split $\mathcal{A}$ into a left moving $A_L$ and a right moving part $A_R$ one can also split the Wilson line accordingly as

$$W_{\mathcal{R}}(C) = W^L_{\mathcal{R}}(C) \times W^R_{\mathcal{R}}(C), \tag{15.2}$$

where

$$W^L_{\mathcal{R}}(C) = \mathrm{Tr}_{\mathcal{R}}\left[\mathcal{P}\exp\left(\int_C A_L\right)\right] = \int \mathcal{D}U_L \exp\left[-S_L(U_L;A_L)_C\right], \tag{15.3}$$

and $A_L$ is the pullback of the connections along the curve $C$, i.e. $A = A_\mu \dot{x}^\mu$. The relevant expressions for $W^L_{\mathcal{R}}$ can be obtained by a simple exchange of the labels as $L \leftrightarrow R$. The corresponding actions, which describe the topological probe, for left and right movers are given by

$$S_L(U_L;A_L)_C = \int_C \mathrm{d}s \left\langle P_L D_L U_L U_L^{-1}\right\rangle + \lambda_L\left(\left\langle P_L^2\right\rangle - c_2\right), \tag{15.4a}$$

$$S_R(U_R;A_R)_C = \int_C \mathrm{d}s \left\langle P_R U_R^{-1} D_R U_R\right\rangle + \lambda_R\left(\left\langle P_R^2\right\rangle - \bar{c}_2\right), \tag{15.4b}$$

where $c_2$ and $\bar{c_2}$ are the quadratic casimirs of the two $\mathfrak{sl}(2,\mathbb{R})$ copies, $\langle\ldots\rangle$ corresponds to the invariant bilinear form on each of the $\mathfrak{sl}(2,\mathbb{R})$ algebras, $U_L$ ($U_R$) describes the probe and takes values in the group manifold $\mathrm{SL}(2,\mathbb{R})$ and $P_L$ ($P_R$) are the canonical momenta associated with $U_L$ ($U_R$) and take values in the Lie algebra $\mathfrak{sl}(2,\mathbb{R})$. The general strategy to determine the holographic entanglement entropy using these ingredients can be roughly summarized as follows:

- Determine the equations of motion (EOM) of (15.4).
- Solve EOM with $A_L$ ($A_R$) set to zero ("nothingness trick").
- Use a suitable (large) gauge transformation in order to generate a non-trivial solution of interest.

[2] If $c = \bar{c}$ then the Wilson line describes a massive particle without spin.

- Determine the path integral in (15.3) using a saddle point approximation.
- Use (15.1) to determine the holographic entanglement entropy.

## 15.2 Constructing a Topological Probe for Flat Space

Having recapitulated the basic ingredients of the holographic entanglement entropy proposal using Wilson lines in $\mathrm{AdS}_3$ in the previous subsection, I will now proceed in constructing a topological probe for flat space.

The biggest difference between AdS and flat space, formulated as a Chern-Simons theory, lies in the different structure of the underlying symmetry algebras. Whereas $\mathfrak{so}(2,2)$ can be written as a direct sum of two $\mathfrak{sl}(2,\mathbb{R})$ algebras, flat space is formulated in terms of a *semi*-direct sum of $\mathfrak{sl}(2,\mathbb{R})$ and translations in three dimensions. Thus, it is at first sight not completely clear how to implement the prescription (15.3) for flat space.

The calculations performed in section 14.4 show that the entanglement entropy for GCFTs splits into two different parts which are proportional to the central charges $c_L$ and $c_M$. Thus, it seems natural that, similar to the $\mathrm{AdS}_3$ case, one mimics that behavior by splitting the action $S(U;\mathcal{A})_C$ appearing in the path integral

$$W_{\mathcal{R}}(C) = \mathrm{Tr}_{\mathcal{R}}\left[\mathcal{P}\exp\left(\int_C \mathcal{A}\right)\right] = \int \mathcal{D}U \exp\left[-S(U;\mathcal{A})_C\right], \tag{15.5}$$

which determines the Wilson line and is used to construct an auxiliary quantum system[3], into even and odd parts labeled by $L$ and $M$ respectively and also fix the norm of the canonical momenta in a similar manner. This in turn also means that the holographic entanglement entropy written in terms of Wilson lines should be given by

$$S_{\mathrm{E}} = -\log\left[W^L_{\mathcal{R}}(C)\right] - \log\left[W^M_{\mathcal{R}}(C)\right]. \tag{15.6}$$

In order to proceed with this split I also assume that the topological probe $U$ and $S(U;\mathcal{A})_C$ can be written as

$$\begin{gathered} U \in \mathrm{ISL}(2,\mathbb{R}), \quad \text{and} \quad U_{L+M} = U_L U_M, \\ S(U;\mathcal{A})_C = S_L(U_L;A_L)_C + S_M(U_M;A_M)_C, \end{gathered} \tag{15.7}$$

[3] In Chapter 14 I mentioned that aside from a curve $C$ one has also to choose an appropriate representation $\mathcal{R}$. For $\mathrm{AdS}_3$ the representation has to be chosen in such a way that the Wilson line corresponds to a massive and spinning particle moving in the $\mathrm{AdS}_3$ bulk. As argued in [81] one possible choice for this representation is an infinite dimensional highest-weight representation of $\mathrm{SL}(2,\mathbb{R})\times\mathrm{SL}(2,\mathbb{R})$ characterized by the conformal weights $(h,\bar{h})$. In close analogy to this I claim that for flat space the correct choice of representations is an infinite dimensional representation of $\mathrm{ISL}(2,\mathbb{R})$ characterized by the Galilean conformal weights $(h_L,h_M)$. I will implement these representations in a similar way as described in [81], i.e. by constructing an auxiliary quantum mechanical system defined on the Wilson line whose Hilbert space will be exactly the representation $\mathcal{R}$ needed to compute entanglement entropy.

with

$$S_L = \int_C \mathrm{d}s \left\langle P_L D_L U_L U_L^{-1} \right\rangle_L + \lambda_L \left( \left\langle P_L^2 \right\rangle_L - c_2 \right), \tag{15.8a}$$

$$S_M = \int_C \mathrm{d}s \left\langle P_M D_M U_M U_M^{-1} \right\rangle_M + \lambda_M \left( \left\langle P_M^2 \right\rangle_M - \bar{c}_2 \right), \tag{15.8b}$$

where $s \in [0,1]$ parametrizes the curve $C$ and

$$D_L = \partial_s + A_L, \quad D_M = \partial_s + A_M. \tag{15.9}$$

$P_L$ $(P_M)$ is the canonical momentum conjugate to $U_L$ $(U_M)$ and $\lambda_L$ $(\lambda_M)$ is a lagrange multiplier that constrains the norm of $P_L$ $(P_M)$ to $c_2$ or $\bar{c}_2$ respectively[4]. The invariant bilinear forms $\langle \ldots \rangle_L$ ($\langle \ldots \rangle_M$) can also be written in terms of Lie algebra metrics $\omega_{ab}$ and $\bar{\omega}_{ab}$, where $\omega_{ab}$ is restricted to the even and $\bar{\omega}_{ab}$ to odd generators as follows

$$\left\langle P^2 \right\rangle_L = P_a P_b \omega^{ab} = 2P_0^2 - \left(P_{-1}P_1 + P_1 P_{-1}\right), \tag{15.10a}$$

$$\left\langle \bar{P}^2 \right\rangle_M = \bar{P}_a \bar{P}_b \bar{\omega}^{ab} = 2\bar{P}_0^2 - \left(\bar{P}_{-1}\bar{P}_1 + \bar{P}_1 \bar{P}_{-1}\right), \tag{15.10b}$$

for $P = P_a L^a$ and $\bar{P} = \bar{P}_a M^a$. The even metric $\omega_{ab}$ can be determined using the ordinary trace and the matrix representation found in appendix A.2.1 (with a factor of 1/2), i.e. $\omega_{ab} = \frac{1}{2}\mathrm{Tr}\left(L_a L_b\right)$. In order to determine $\bar{\omega}_{ab}$ one can use the twisted trace as defined in (12.4) i.e. $\bar{\omega}_{ab} = \bar{\mathrm{Tr}}\left(M_a M_b\right)$.
The EOM for the even part of (15.8) are given by

$$D_L U_L U_L^{-1} + 2\lambda_L P_L = 0, \qquad \frac{\mathrm{d}}{\mathrm{d}s} P_L = 0, \tag{15.11}$$

in addition to the constraints $\langle P_L^2 \rangle_L = c_2$. The EOM for the odd part are the same as in (15.11) upon replacing $L \leftrightarrow M$ in addition to the constraint $\langle P_M^2 \rangle_M = \bar{c}_2$. In order to solve these EOM one can use the same "nothingness trick" as in the $\mathrm{AdS}_3$ case by finding a solution of these equations first for $A_L = 0$ $(A_M = 0)$ and then generating a non-trivial solution by using a (large) gauge transformation. For $A_L = 0$ $(A_M = 0)$ solutions of (15.11) are given by

$$U_L^{(0)} = u_L^{(0)} \exp\left(-2\alpha_L(s) P_L^{(0)}\right), \qquad \frac{d\alpha_L(s)}{ds} = \lambda_L(s), \quad (L \leftrightarrow M), \tag{15.12}$$

where $u_L^{(0)}$ $(u_M^{(0)})$ are constant group elements chosen in such a way that they are compatible with (15.7). Looking at (15.12) and the assumption (15.7) one finds that also $\left[P_L^{(0)}, P_M^{(0)}\right] = 0$ has to be satisfied.

[4]For flat space $c_2$ is the value of the quadratic casimir operator of the $\mathfrak{sl}(2,\mathbb{R})$ part of $\mathfrak{isl}(2,\mathbb{R})$, i.e. $c_2 = 2L_0^2 - (L_{-1}L_1 + L_1 L_{-1})$, while $\bar{c}_2$ is one of the quadratic casimirs of the *full* $\mathfrak{isl}(2,\mathbb{R})$ algebra (the other one would be the helicity), i.e. $\bar{c}_2 = 2M_0^2 - (M_{-1}M_1 + M_1 M_{-1})$, which label the representation $\mathcal{R}$ via the highest weights $h_L$ and $h_M$.

Using this one obtains the following on-shell actions

$$S_L^{\text{on-shell}} = -2\Delta\alpha_L c_2, \qquad S_M^{\text{on-shell}} = -2\Delta\alpha_M \bar{c}_2, \tag{15.13}$$

where $\Delta\alpha_L = \alpha_L(1) - \alpha_L(0)$ and equivalently for $\Delta\alpha_M$. By using a saddle point approximation for the path integral

$$\int \mathcal{D}U e^{-S(U;\mathcal{A})_C} \sim e^{-S_{\text{on-shell}}(U;\mathcal{A})_C}, \tag{15.14}$$

one can write (15.6) as

$$S_{\text{EE}} = -2\Delta\alpha_L c_2 - 2\Delta\alpha_M \bar{c}_2. \tag{15.15}$$

Thus the calculation of holographic entanglement entropy using Wilson lines reduces to calculating $\Delta\alpha_L$ and $\Delta\alpha_M$ for the relevant theories in question.

## 15.3 Calculating Holographic Entanglement Entropy for Flat Space

Having constructed a suitable topological probe for flat space in the previous subsection I will now calculate the entanglement entropy for various different flat spacetimes holographically.

### 15.3.1 Spin-2

As in the $AdS_3$ case it is convenient to formulate flat space gravity in terms of a Chern-Simons action and the corresponding gauge connection $\mathcal{A}$, which is given by (10.21a). As in the previous section I will split this connection into an even and odd part respectively as $\mathcal{A} = A_L + A_M$ with

$$A_L = \left(L_1 - \frac{\mathcal{M}}{4}L_{-1}\right)\mathrm{d}\varphi, \tag{15.16a}$$

$$A_M = \frac{1}{2}M_{-1}\,\mathrm{d}r + \left(M_1 - \frac{\mathcal{M}}{4}M_{-1}\right)\mathrm{d}u + \left(rM_0 - \frac{\mathcal{N}}{2}M_{-1}\right)\mathrm{d}\varphi. \tag{15.16b}$$

One can now use the connections $A_L$ and $A_M$ and perform a large gauge transformation on the trivial solution (15.12) in order to obtain a solution for (15.11) with $A_L$ and $A_M$ given by (15.16). This gauge transformation can be compactly written as

$$A_L + A_M = A\,\mathrm{d}A^{-1} \quad \text{with} \quad A = b^{-1}e^{-\int a_i\,\mathrm{d}x^i}, \tag{15.17}$$

where $b$ is the same group element that is used to gauge away the radial dependence as in (10.21a). The topological probe $U(s)$ transforms under this gauge transformation as

$$U(s) = (U_L U_M)(s) = A(s)U_L^{(0)}U_M^{(0)}A^{-1}(s). \tag{15.18}$$

Up until this point of the calculation it was not necessary to specify the exact points at which the Wilson line is attached to. However, since one will have to fix boundary conditions for the probe at some point during the calculations, I will now specify where exactly the Wilson line is attached to and which entangling interval it is bounding, see Figure 15.1.

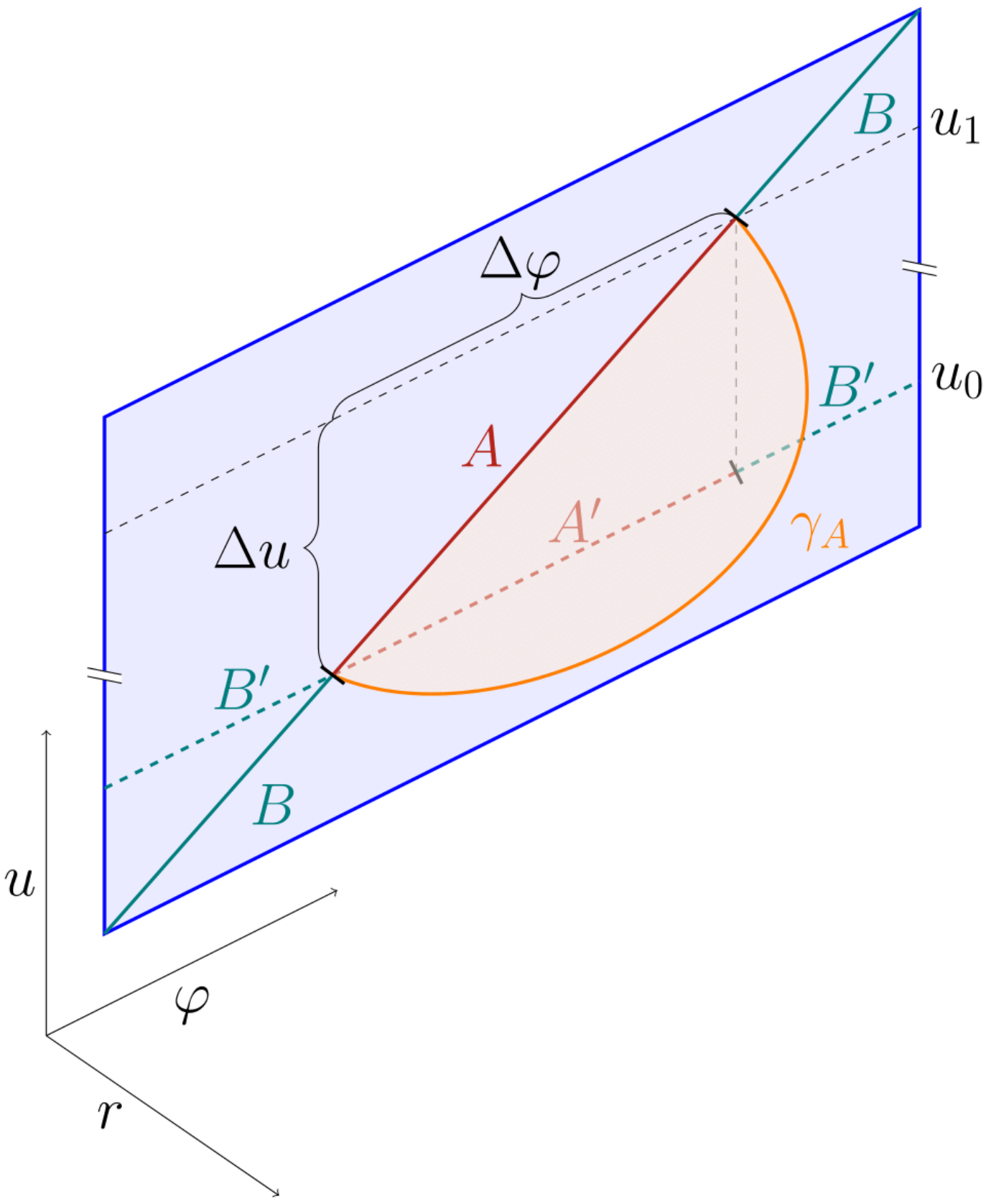

**Fig. 15.1.:** Boosted ($A$, $B$) and equal time ($A'$, $B'$) entangled intervals and the corresponding Wilson line ($\gamma_A$) used to determine holographic entanglement entropy in flat space.

First one introduces a radial cut-off $r_0$ which is placed very close to the boundary $r = \infty$ in order to regulate infinities when approaching the boundary. The Wilson line will then be attached at the hypersurface with $r = r_0$ at the points $x_i^\mu = (r_0, u_i, \varphi_i)$ and $x_f^\mu = (r_0, u_f, \varphi_f)$. Denoting

$$
\begin{aligned}
U(0) &= U_i, & U(1) &= U_f, \\
A(0) &= A|_{x=x_i} = A_i, & A(1) &= A|_{x=x_f} = A_f, \\
\alpha_L(0) &= \alpha_L^i, & \alpha_L(1) &= \alpha_L^f, \\
\alpha_M(0) &= \alpha_M^i, & \alpha_M(1) &= \alpha_M^f,
\end{aligned}
\tag{15.19}
$$

one can use (15.18) to write

$$U_i = A_i u_L^{(0)} u_M^{(0)} \exp\left(-2\alpha_L^i P_L^{(0)} - 2\alpha_M^i P_M^{(0)}\right) A_i^{-1}, \tag{15.20a}$$

$$U_f = A_f u_L^{(0)} u_M^{(0)} \exp\left(-2\alpha_L^f P_L^{(0)} - 2\alpha_M^f P_M^{(0)}\right) A_f^{-1}. \tag{15.20b}$$

Next solving for $u_L^{(0)} u_M^{(0)}$ in one of the two equations and replacing the expression in the remaining equation one obtains

$$e^{-2\Delta\alpha_L P_L^{(0)} - 2\Delta\alpha_M P_M^{(0)}} = A_i^{-1} U_i^{-1} A_i A_f^{-1} U_f A_f = \Omega. \tag{15.21}$$

With this equation one can now almost determine $\Delta\alpha_L$ and $\Delta\alpha_M$. The only thing left to do is to choose appropriate boundary conditions for the topological probe at the initial and final point of the Wilson line. As in the AdS$_3$ case it is, as of yet, not known how unique such a choice of boundary conditions actually is, i.e. if there is only one set of boundary conditions that yields the correct entanglement entropy or if there is a whole family thereof. In the pure AdS$_3$ spin-2 case one can employ boundary conditions for example in such a way that the curve the Wilson line is describing is actually a geodesic [81], in accordance with the Ryu-Takayanagi proposal. For other cases like the ones described in [172] where one has to deal with gravitational anomalies which render the theory non-Lorentz invariant, the guiding principle is not so clear. In the case at hand I will choose the boundary conditions in such a way that they are as simple as possible and analogous to the ones for theories with gravitational anomalies. The reason for this is that looking at flat space as a limit from AdS theories with gravitational anomalies can be seen as the "parent" theories for GCFTs with $c_M \neq 0$. Following this reasoning I propose the following boundary conditions for the topological probe $U$ at the initial and final point

$$U_i^{-1} = e^{\frac{r}{2}L_{-1}} b, \quad U_f = e^{-\frac{r}{2}L_{-1}} b. \tag{15.22}$$

After fixing the boundary conditions one can solve (15.21) for $\Delta\alpha_L$ and $\Delta\alpha_M$. In order to proceed, it makes sense to first take a closer look at (15.21) and use the fact that $\mathfrak{isl}(2,\mathbb{R})$ has a nilpotent subalgebra. Since $[M_n, M_n] = 0$ and we assumed $\left[P_L^{(0)}, P_M^{(0)}\right] = 0$, (15.21) simplifies to

$$e^{-2\Delta\alpha_L P_L^{(0)}} \left(\mathbb{1} - 2\Delta\alpha_M P_M^{(0)}\right) = \Omega. \tag{15.23}$$

At this point the way the $\mathfrak{isl}(2,\mathbb{R})$ matrix representation used is constructed and categorized in even and odd parts is again very convenient as one can schematically write the left hand side of this equation as

$$\begin{pmatrix} e^{\gamma_L} & 0 \\ 0 & e^{-\gamma_L} \end{pmatrix} \otimes \mathbb{1}_{2\times 2} + \epsilon\gamma_M \begin{pmatrix} e^{\gamma_L} & 0 \\ 0 & -e^{\gamma_L} \end{pmatrix} \otimes \gamma^\star_{(1)} \tag{15.24}$$

where $e^{\pm\gamma_L}$ and $\pm\epsilon\gamma_M$ are the eigenvalues of $e^{-2\Delta\alpha_L P_L^{(0)}}$ and $-2\Delta\alpha_M P_M^{(0)}$, respectively and $\gamma^\star_{(1)}$ is given by (12.3). Thus, one can conveniently distinguish between even and odd eigenvalues.

One could of course just determine the eigenvalues of the matrices on both sides of but there is a more efficient way of doing things, i.e. taking two different traces of (15.21) in such a way that one trace picks out the purely even part and the other one the mixed even-odd part. The ordinary matrix trace used for determining $\omega_{ab}$ does the trick for the even part, as can be seen from (15.24). For the mixed part one uses the hatted trace as defined in (12.2). Using this trick one obtains the following two equations

$$2\cosh\left(\sqrt{2c_2}\Delta\alpha_L\right) = \text{Tr}\,(\Omega)|_{r_0\to\infty}\,, \tag{15.25a}$$

$$2\sinh\left(\sqrt{2c_2}\Delta\alpha_L\right)\sqrt{2\bar{c}_2}\Delta\alpha_M = \widehat{\text{Tr}}\,(\Omega)\Big|_{r_0\to\infty}\,. \tag{15.25b}$$

Since the Wilson line is pushed to the boundary, Tr $(\Omega)$ and thus also the left hand side of (15.25a) will be very large and positive. As the cosh is an even function, there are two branches to solve for $\Delta\alpha_L$, depending on whether $\Delta\alpha_L$ is bigger or smaller than zero. This part of the calculation is identical to the AdS$_3$ case and thus one can use it as a pointer to choose the right branch which in this case is

$$e^{-\sqrt{2c_2}\Delta\alpha_L} = \text{Tr}\,(\Omega)|_{r_0\to\infty}\,. \tag{15.26}$$

Using this (15.25b) simplifies to

$$-\sqrt{2\bar{c}_2}\Delta\alpha_M\,\text{Tr}\,(\Omega)|_{r_0\to\infty} = \widehat{\text{Tr}}\,(\Omega)\Big|_{r_0\to\infty}\,. \tag{15.27}$$

$\Delta\alpha_L$ and $\Delta\alpha_M$ can now be determined as

$$\Delta\alpha_L = -\frac{\ln\left(\text{Tr}\,(\Omega)|_{r_0\to\infty}\right)}{\sqrt{2c_2}}\,, \tag{15.28a}$$

$$\Delta\alpha_M = -\frac{\widehat{\text{Tr}}\,(\Omega)\Big|_{r_0\to\infty}}{\sqrt{2\bar{c}_2}\,\text{Tr}\,(\Omega)|_{r_0\to\infty}}\,. \tag{15.28b}$$

The entanglement entropy can thus equivalently be written as

$$S_E = \sqrt{2c_2}\ln\left(\widehat{\text{Tr}}\,(\Omega)\Big|_{r_0\to\infty}\right) + \sqrt{2\bar{c}_2}\frac{\widehat{\text{Tr}}\,(\Omega)\Big|_{r_0\to\infty}}{\text{Tr}\,(\Omega)|_{r_0\to\infty}}.\,, \tag{15.29}$$

Writing $u_f - u_i = \Delta u$ and $\varphi_f - \varphi_i = \Delta\varphi$ then the holographic entanglement entropy for an interval with spatial extension $\Delta\varphi$ and timelike extension $\Delta u$ for the null-orbifold ($\mathcal{M} = \mathcal{N} = 0$), (global) flat space ($\mathcal{M} = -1, \mathcal{N} = 0$) and FSCs ($\mathcal{M} \geq 0, \mathcal{N} \neq 0$ is given by

$$S_E^{\text{NO}} = 2\sqrt{2c_2}\ln\left[\frac{r_0\Delta\phi}{2}\right] + 2\sqrt{2\bar{c}_2}\frac{\Delta u}{\Delta\phi}, \tag{15.30a}$$

$$S_E^{\text{GFS}} = 2\sqrt{2c_2}\ln\left[r_0\sin\left(\frac{\Delta\phi}{2}\right)\right] + \sqrt{2\bar{c}_2}\cot\left(\frac{\Delta\phi}{2}\right)\Delta u, \tag{15.30b}$$

$$\begin{aligned} S_E^{\text{FSC}} = & 2\sqrt{2c_2}\ln\left[\frac{r_0\sinh\left(\frac{\sqrt{\mathcal{M}}\Delta\phi}{2}\right)}{\sqrt{\mathcal{M}}}\right] + \sqrt{2\bar{c}_2}\left(-\frac{2\mathcal{N}}{\mathcal{M}}\right. \\ & \left. + \sqrt{\mathcal{M}}\coth\left(\frac{\sqrt{\mathcal{M}}\Delta\phi}{2}\right)\left(\Delta u + \frac{\mathcal{N}}{\mathcal{M}}\Delta\phi\right)\right). \end{aligned} \tag{15.30c}$$

Relating the quadratic casimirs and central charges in a similar way to the $AdS_3$ case, i.e. $\sqrt{2c_2} = \frac{c_L}{12}$ and $\sqrt{2\bar{c}_2} = \frac{c_M}{12}$ one obtains the following final results

$$S_E^{\text{NO}} = \frac{c_L}{6}\ln\left[\frac{r_0\Delta\phi}{2}\right] + \frac{c_M}{6}\frac{\Delta u}{\Delta\phi}, \tag{15.31a}$$

$$S_E^{\text{GFS}} = \frac{c_L}{6}\ln\left[r_0\sin\left(\frac{\Delta\phi}{2}\right)\right] + \frac{c_M}{12}\cot\left(\frac{\Delta\phi}{2}\right)\Delta u, \tag{15.31b}$$

$$\begin{aligned} S_E^{\text{FSC}} = & \frac{c_L}{6}\ln\left[\frac{r_0\sinh\left(\frac{\sqrt{\mathcal{M}}\Delta\phi}{2}\right)}{\sqrt{\mathcal{M}}}\right] + \frac{c_M}{12}\left(-\frac{2\mathcal{N}}{\mathcal{M}}\right. \\ & \left. + \sqrt{\mathcal{M}}\coth\left(\frac{\sqrt{\mathcal{M}}\Delta\phi}{2}\right)\left(\Delta u + \frac{\mathcal{N}}{\mathcal{M}}\Delta\phi\right)\right), \end{aligned} \tag{15.31c}$$

which precisely coincide with the calculations done for GCFTs in Section 14.4 (where the UV cut-off $a$ is related to $r_0$ as $a = \frac{1}{r_0}$) and the results in [174], which were obtained as a limiting procedure from the $AdS_3$ results.

### 15.3.2 Spin-3

Having developed the flat space equivalent of the Wilson line proposal for holographic entanglement entropy in $AdS_3$ one can now also straightforwardly extend the formalism to higher-spin theories in flat space in analogy to the $AdS_3$ formulations. I will illustrate how to extend this construction for the case of spin-3 flat space gravity.

For flat space spin-3 gravity I will make the following generalizations to the ansatz (15.8) used before. First, take as a gauge algebra now the principal embedding of $\mathfrak{isl}(2,\mathbb{R})$ into $\mathfrak{isl}(3,\mathbb{R})$ with generators[5] $L_n, M_n, U_n, V_n$ which obey (12.1). Next,

[5]For more details on an appropriate matrix representation see appendix A.2.2.

modify the actions (15.8a) and (15.8b) in such a way that $P_L \in \{L_n, U_n\}$ and $P_M \in \{M_n, V_n\}$ and add the following constraints to the actions

$$S_L = \int_C \mathrm{d}s \left\langle P_L D_L U_L U_L^{-1} \right\rangle_L + \lambda_L \left( \left\langle P_L^2 \right\rangle_L - c_2 \right) + \lambda_L^{(3)} \left( \left\langle P_L^3 \right\rangle_L - c_3 \right), \quad (15.32a)$$

$$S_M = \int_C \mathrm{d}s \left\langle P_M U_M^{-1} D_M U_M \right\rangle_M + \lambda_M \left( \left\langle P_M^2 \right\rangle_M - \bar{c}_2 \right) + \lambda_M^{(3)} \left( \left\langle P_M^3 \right\rangle_M - \bar{c}_3 \right), \quad (15.32b)$$

where $\lambda_L^{(3)}$ ($\lambda_M^{(3)}$) are again lagrange multipliers, $c_3$ and $\bar{c}_3$ are the cubic even and odd casimirs[6] and $\langle P_L^3 \rangle_L$ ($\langle P_M^3 \rangle_M$) is a short hand notation for $\langle P_L^3 \rangle_L = h_{abc} P_L^a P_L^b P_L^c$ ($\langle P_M^3 \rangle_M = \bar{h}_{abc} P_M^a P_M^b P_M^c$). The tensors $h_{abc}$ and $\bar{h}_{abc}$ coincide with the $\mathfrak{sl}(3, \mathbb{R})$ Killing form that defines the cubic casimir with the only difference being that $h_{abc}$ can be obtained via

$$\frac{1}{2} \mathrm{Tr}(G_a G_b G_c) = h_{abc}, \quad (15.33)$$

with $G_a \in \{L_n, U_n\}$ and $\bar{h}_{abc}$ via

$$\widetilde{\mathrm{Tr}}(\bar{G}_a \bar{G}_b \bar{G}_c) = \bar{h}_{abc}, \quad (15.34)$$

with $\bar{G}_a \in \{M_n, V_n\}$. The EOM of (15.32) are given by

$$D_L U_L U_L^{-1} + 2\lambda_L P_L + 3\lambda_L^{(3)} P_L \times P_L = 0, \qquad \frac{\mathrm{d}}{\mathrm{d}s} P_L = 0, \qquad (L \leftrightarrow M) \quad (15.35)$$

in addition to the constraints $\langle P_L^2 \rangle_L = c_2$, $\langle P_M^2 \rangle_M = \bar{c}_2$, $\langle P_L^3 \rangle_L = c_3$ and $\langle P_M^3 \rangle_M = \bar{c}_3$. $P_L \times P_L = 0$ and $P_M \times P_M = 0$ are shorthand notations for $P_L \times P_L = h_{abc} G^a P_L^b P_L^c$ and $P_M \times P_M = \bar{h}_{abc} \bar{G}^a P_M^b P_M^c$. Using again the "nothingness" trick one obtains the following solution for $A_L = A_M = 0$

$$U_L^{(0)} = u_L^{(0)} e^{\left(-2\alpha_L(s) P_L^{(0)} - 3\alpha_L^{(3)}(s) P_L^{(0)} \times P_L^{(0)}\right)},$$

$$\frac{d\alpha_L(s)}{ds} = \lambda_L(s), \qquad \frac{d\alpha_L^{(3)}(s)}{ds} = \lambda_L^{(3)}(s), \qquad (L \leftrightarrow M). \quad (15.36)$$

The on-shell action is given by

$$S_L^{\text{on-shell}} = -\, 2\Delta\alpha_L c_2 - 3\Delta\alpha_L^{(3)} c_3, \quad (15.37a)$$

$$S_M^{\text{on-shell}} = -\, 2\Delta\alpha_M \bar{c}_2 - 3\Delta\alpha_M^{(3)} \bar{c}_3. \quad (15.37b)$$

Now one can define

$$\mathbb{P}_L := -\, 2\Delta\alpha_L(s) P_L^{(0)} - 3\Delta\alpha_L^{(3)}(s) P_L^{(0)} \times P_L^{(0)}, \quad (15.38a)$$

$$\mathbb{P}_M := -\, 2\Delta\alpha_M(s) P_M^{(0)} - 3\Delta\alpha_M^{(3)}(s) P_M^{(0)} \times P_M^{(0)}, \quad (15.38b)$$

[6] In the same sense as in the spin-2 case, i.e. $c_3$ is the $\mathfrak{sl}(3, \mathbb{R})$ casimir and $\bar{c}_3$ the cubic casimir of the *full* $\mathfrak{isl}(3, \mathbb{R})$.

and perform the same steps as in the spin-2 case in order to obtain the spin-3 analogue of (15.21)

$$e^{\mathbb{P}_L+\mathbb{P}_M}=\Omega, \tag{15.39}$$

where $\Omega$ is the same expression as in (15.21) with the exception that $U$ now takes values in $\mathfrak{isl}(3,\mathbb{R})$ and $A_{i/f}$ are determined by the corresponding spin-3 Chern-Simons connection.

Using the EOM one can further simplify this equation to the following set of equations

$$S_L^{\text{on-shell}}=-2\Delta\alpha_L c_2-3\Delta\alpha_L^{(3)}c_3=\frac{1}{2}\mathrm{Tr}\left[\ln(\Omega)\,P_L^{(0)}\right], \tag{15.40a}$$

$$S_M^{\text{on-shell}}=-2\Delta\alpha_M\bar{c}_2-3\Delta\alpha_M^{(3)}\bar{c}_3=\bar{\mathrm{Tr}}\left[\ln(\Omega)\,P_M^{(0)}\right]. \tag{15.40b}$$

Since in the semiclassical limit the entanglement entropy is proportional to the on-shell action one can thus write the entanglement entropy as

$$S_E=S_L^{\text{on-shell}}+S_M^{\text{on-shell}}=\frac{1}{2}\mathrm{Tr}\left[\ln(\Omega)\,P_L^{(0)}\right]+\bar{\mathrm{Tr}}\left[\ln(\Omega)\,P_M^{(0)}\right], \tag{15.41}$$

or equivalently as

$$S_E=-2\Delta\alpha_L c_2-3\Delta\alpha_L^{(3)}c_3-2\Delta\alpha_M\bar{c}_2-3\Delta\alpha_M^{(3)}\bar{c}_3. \tag{15.42}$$

One can now use this expression and the spin-3 connection given by (12.8) and determine the holographic entanglement entropy of a spin-3 charged FSC along the same lines as in the spin-2 case in the previous subsection. Since the results are rather lengthy I will, however, not display them here explicitly.

## 15.4 Thermal Entropy of Flat Space Cosmologies

In this section I will show how to use Wilson lines to determine the thermal entropy of FSCs. In order to do this one has to consider a closed Wilson loop around the non-contractable cycle of the FSC, i.e. the $\varphi$ cycle, which tremendously simplifies things from a computational perspective.

### 15.4.1 Spin-2

I will start by determining the thermal entropy of a FSC with only mass $\mathcal{M}$ and angular momentum $\mathcal{N}$. Since one is now dealing with a Wilson *loop* instead of a Wilson *line* the topological probe should be continuous at the initial and final points and thus periodic, i.e.

$$U_i=U_f,\quad\text{and}\quad P_i=P_f. \tag{15.43}$$

As $P(s) = AP^{(0)}A^{-1}$, these boundary conditions imply that

$$\left[P^{(0)}, A_i^{-1}A_f\right] = \left[P_L^{(0)} + P_M^{(0)}, A_i^{-1}A_f\right] = 0. \tag{15.44}$$

$P_L^{(0)}$ and $P_M^{(0)}$ commute with each other and thus one can simultaneously diagonalize them. This in turn means that $\left[P_L^{(0)}, A_i^{-1}A_f\right]$ and $\left[P_M^{(0)}, A_i^{-1}A_f\right]$ vanish simultaneously. For the topological probe one finds the following relation

$$e^{-2\Lambda} = \left(u_L^{(0)}u_M^{(0)}\right)^{-1}\left(A_i^{-1}A_f\right)^{-1}u_L^{(0)}u_M^{(0)}A_i^{-1}A_f, \tag{15.45}$$

where $\Lambda = \Delta\alpha_L P_L^{(0)} + \Delta\alpha_M P_M^{(0)}$ and assuming that $P_L^{(0)}$, $P_M^{(0)}$ and $A_i^{-1}A_f$ have already been diagonalized. Since the non-contractible cycle is the $\varphi$ cycle $A_i^{-1}A_f$ reduces to the holonomy around this cycle, i.e.

$$A_i^{-1}A_f = e^{-2\pi\lambda_\varphi}, \tag{15.46}$$

where $\lambda_\varphi$ denotes the diagonalized form of $a_\varphi$ given in (10.21a). Since the r.h.s of (15.45) should be non-trivial one has to choose $u_L^{(0)}u_M^{(0)}$ in such a way that

$$\left(u_L^{(0)}u_M^{(0)}\right)^{-1}\left(A_i^{-1}A_f\right)^{-1}u_L^{(0)}u_M^{(0)} = A_i^{-1}A_f, \tag{15.47}$$

or equivalently

$$\left(u_L^{(0)}u_M^{(0)}\right)^{-1}e^{2\pi\lambda_\varphi}u_L^{(0)}u_M^{(0)} = e^{-2\pi\lambda_\varphi}. \tag{15.48}$$

Using (15.48) one finds that the following equation has to be satisfied

$$\Delta\alpha_L P_L^{(0)} + \Delta\alpha_M P_M^{(0)} = 2\pi\lambda_\varphi. \tag{15.49}$$

Since $P_L^{(0)}$ and $P_M^{(0)}$ are traceless and the constraints fix $\frac{1}{2}\mathrm{Tr}\left(P_L^2\right) = c_2$ and $\bar{\mathrm{Tr}}\left(P_M^2\right) = \bar{c}_2$ (and $\frac{1}{2}\mathrm{Tr}\left(P_L M_0\right) = 0$, $\bar{\mathrm{Tr}}\left(P_M L_0\right) = 0$) one immediately sees that the eigenvalues of $P_L^{(0)}$ and $P_M^{(0)}$ are $\pm\sqrt{\frac{c_2}{2}}$ and $\pm\epsilon\sqrt{\frac{\bar{c}_2}{2}}$ in each of the $\mathfrak{sl}(2,\mathbb{R})$ blocks which were employed in the construction of the $\mathfrak{isl}(2,\mathbb{R})$ matrix representation found in A.2.1. Phrased in terms of traces this means that

$$\frac{1}{2}\mathrm{Tr}\left(P_L L_0\right) = \sqrt{\frac{c_2}{2}}, \quad \text{and} \quad \bar{\mathrm{Tr}}\left(P_M M_0\right) = \sqrt{\frac{\bar{c}_2}{2}}. \tag{15.50}$$

Thus, by multiplying (15.49) with $L_0$ and $M_0$ and taking either the trace or twisted trace one can determine $\Delta\alpha_L$ and $\Delta\alpha_M$ via

$$\Delta\alpha_L = \pi\sqrt{\frac{2}{c_2}}\mathrm{Tr}\left(\lambda_\varphi L_0\right), \quad \Delta\alpha_M = 2\pi\sqrt{\frac{2}{\bar{c}_2}}\bar{\mathrm{Tr}}\left(\lambda_\varphi M_0\right). \tag{15.51}$$

For the FSC given by the connection (10.21a) one obtains the following values for $\Delta\alpha_L$ and $\Delta\alpha_M$

$$\Delta\alpha_L = -\pi\sqrt{\frac{2\mathcal{M}}{c_2}}, \quad \Delta\alpha_M = -\frac{\pi\mathcal{N}}{\sqrt{\mathcal{M}}}\sqrt{\frac{2}{\bar{c}_2}}. \tag{15.52}$$

Making also the same identifications of the quadratic casimirs as in the case of the thermal entropy this yields the following thermal entropy

$$S_{\text{Th}} = \frac{\pi}{6}\left(c_L\sqrt{\mathcal{M}} + c_M\frac{\mathcal{N}}{\sqrt{\mathcal{M}}}\right). \tag{15.53}$$

Taking into account

$$\mathcal{M} = \frac{24h_M}{c_M}, \quad \mathcal{N} = \frac{12\left(c_M h_L - c_L h_M\right)}{c_M^2}, \tag{15.54}$$

one can immediately check that this is exactly the same result as (14.43) obtained previously in section 14.5 or alternatively by performing an İnönü–Wigner contraction of the inner horizon thermal entropy of the BTZ black hole in $AdS_3$ [V].

## 15.4.2 Spin-3

As in the spin-2 case I will now determine the thermal entropy for a spin-3 charged FSC holographically. In order to proceed one has to perform the same steps as in the spin-2 case, i.e. using a closed Wilson loop around the $\varphi$ direction. With the notation used in section 15.3.2 this leads to the following equation which has to be solved

$$\mathbb{P}_L + \mathbb{P}_M = 2\pi\lambda_\varphi^{(3)}, \tag{15.55}$$

where $\lambda_\varphi^{(3)}$ denotes the diagonalized form of $a_\varphi^{(3)}$ which is given by [VII]

$$a_\varphi^{(3)} = L_1 - \frac{\mathcal{M}}{4}L_{-1} + \frac{\mathcal{V}}{2}U_{-2} - \frac{\mathcal{N}}{2}M_{-1} + \mathcal{Z}V_{-2}, \tag{15.56}$$

and the eigenvalues of $a_\varphi^{(3)}$ are ordered in such a way that they coincide with the spin-2 case for vanishing spin-3 charges $\mathcal{V}$ and $\mathcal{Z}$. Setting[7]

$$P_L^{(0)} = \sqrt{\frac{c_L}{2}}L_0, \quad P_M^{(0)} = \sqrt{\frac{c_M}{2}}M_0, \tag{15.57}$$

this equation simplifies to

$$2\pi\lambda_\varphi^{(3)} = -\sqrt{2c_2}\Delta\alpha_L L_0 - 2c_2\Delta\alpha_L^{(3)}U_0 - \sqrt{2\bar{c}_2}\Delta\alpha_M M_0 - 2\bar{c}_2\Delta\alpha_M^{(3)}V_0. \tag{15.58}$$

[7]This choice of $P_L^{(0)}$ and $P_M^{(0)}$ is tantamount to setting the cubic casimirs $c_3, \bar{c}_3$ to zero.

As in the spin-2 case one can now solve for $\Delta\alpha_L$ and $\Delta\alpha_M$ by multiplying either $L_0$, or $M_0$ on both sides of (15.58) and taking the (twisted) trace. This yields the following relations

$$\Delta\alpha_L = -\frac{\pi}{2\sqrt{2c_2}}\mathrm{Tr}\left(\lambda_\varphi^{(3)}L_0\right), \quad \Delta\alpha_M = -\frac{\pi}{\sqrt{2\bar{c}_2}}\bar{\mathrm{Tr}}\left(\lambda_\varphi^{(3)}M_0\right). \tag{15.59}$$

Thus, one can write the thermal entropy for the spin-3 charged FSC as

$$S_{\mathrm{Th}} = \pi\left(\sqrt{\frac{c_2}{2}}\mathrm{Tr}\left(\lambda_\varphi^{(3)}L_0\right) + \sqrt{2\bar{c}_2}\bar{\mathrm{Tr}}\left(\lambda_\varphi^{(3)}M_0\right)\right). \tag{15.60}$$

Replacing again the quadratic casimirs in the same fashion as in the section before and evaluting the traces one obtains the following expression

$$S_{\mathrm{Th}} = \frac{\pi}{6}\left(c_L\sqrt{\mathcal{M}}\sqrt{1-\frac{3}{4\mathcal{R}}} + c_M\frac{\mathcal{N}}{\sqrt{\mathcal{M}}}\frac{\left(2\mathcal{R}-3+12\mathcal{P}\sqrt{\mathcal{R}}\right)}{2\left(\mathcal{R}-3\right)\sqrt{1-\frac{3}{4\mathcal{R}}}}\right), \tag{15.61}$$

where I have rewritten $\mathcal{V}$ and $\mathcal{Z}$ in terms of the dimensionless parameters $\mathcal{R}$ and $\mathcal{P}$ as

$$\frac{|\mathcal{V}|}{\mathcal{M}^{\frac{3}{2}}} = \frac{\mathcal{R}-1}{4\mathcal{R}^{\frac{3}{2}}}, \quad \frac{\mathcal{Z}}{\mathcal{N}\sqrt{\mathcal{M}}} = \mathcal{P}, \tag{15.62}$$

which exactly coincides with the results obtained in [V, VII].

## 15.5 Conclusions

In this Part IV of my thesis I provided explicit checks of a holographic correspondence in asymptotically flat spacetimes by determining the entanglement entropy and thermal entropy of certain asymptotically flat spacetimes holographically. I started with a review of the basics of Galilean conformal field theories in Chapter 14 that also included the calculation of entanglement entropy and thermal entropy using Galilean conformal field theory methods. In Chapter 15 I used a Wilson line attached at the boundary of asymptotically flat spacetimes to holographically determine the entanglement entropy of $\mathfrak{bms}_3$ invariant quantum field theories as well as higher-spin versions thereof. Furthermore, I used a similar construction involving Wilson loops that wind around the horizon of cosmological solutions in flat space to determine the thermal entropy of the dual field theories at finite temperature. All the results in this chapter agreed perfectly with the results obtained previously in Chapters 10, 13 and 14 and thus provide strong evidence that, indeed, there exists a holographic correspondence for asymptotically flat spacetimes.

# Part V

## How General is Holography?

This part will conclude my thesis. I will summarize my results and give a conclusion on how general the holographic principle is, based on the results of my research. An outlook on possible follow-up projects and things which could not be addressed in full detail during my time as a PhD student will complete this concluding part.

# 16 The Frog's Perspective

> *"All right," said Deep Thought. "The Answer to the Great Question ..."*
> *"Yes ... !"*
> *"Of Life, the Universe and Everything ..." said Deep Thought.*
> *"Yes ... !"*
> *"Is ..." said Deep Thought, and paused. "Yes ... !"*
> *"Is ..."*
> *"Yes ... !!! ... ?"*
> *"Forty-two," said Deep Thought, with infinite majesty and calm.*
>
> **– Douglas Adams**
> The Hitchhiker's Guide to the Galaxy

In this thesis I tried to gain a better understanding of the holographic principle by developing various new aspects of non-AdS (higher-spin) holography and in particular flat space holography in $2+1$ dimensions. In the following I will give a summary of this thesis from a technical perspective (*frog's perspective*), but not from a broader perspective (*bird's perspective*). I will give such a summary from a broader perspective in Chapter 17.

For the purpose of studying non-AdS holography I mainly focused on higher-spin Lobachevsky holography where I explicitly determined the asymptotic symmetries for three different kinds of higher-spin theories, which can be characterized by different Lie algebra valued Chern-Simons gauge connections. These theories corresponded to all non-trivial embeddings of $\mathfrak{sl}(2,\mathbb{R}) \hookrightarrow \mathfrak{sl}(4,\mathbb{R})$. In addition, I analyzed the extension of the 3-1 embedding of $\mathfrak{sl}(2,\mathbb{R}) \hookrightarrow \mathfrak{sl}(4,\mathbb{R})$ to the $(N-1)-1$ embedding of $\mathfrak{sl}(2,\mathbb{R}) \hookrightarrow \mathfrak{sl}(N,\mathbb{R})$.

The results obtained in this way were then used to study unitary representations of the resulting quantum $\mathcal{W}$-algebras. This culminated in a family of $\mathcal{W}_N^{(2)}$ quantum field theories whose Virasoro central charge $c$ can take arbitrary (but not infinitely) large as well as very small values without violating unitarity as long as it satisfies the bounds

$$\text{For Odd } N: \qquad c \le \frac{N}{4} - \frac{1}{8} - \mathcal{O}(1/N), \tag{16.1}$$

$$\text{For Even } N: \qquad c = (1-\beta)\frac{1-2N+\beta N\frac{N-2}{N-1}}{1+\frac{\beta}{N-1}} \sim (\beta-1)(2-\beta)N, \tag{16.2}$$

where $\beta \in [1,2]$ for large $N$. Thus, the family of $\mathcal{W}_N^{(2)}$ quantum field theories can be studied both in the semi-classical regime as well as its ultra-quantum limit.

I then put my focus on flat space holography. First I showed how one can directly obtain a Chern-Simons formulation of flat space as a limiting procedure from the existing AdS prescription. In the same spirit I derived various $\mathcal{FW}$-algebras as İnönü–Wigner contractions from their $\mathcal{W}$-algebra counterparts. Furthermore, I used the flat space limit to derive a flat space analogue of a (higher-spin) Cardy formula for generic $\mathfrak{bms}_3$

$$S_{\mathfrak{bms}_3} = \frac{\pi}{6}\left|c_L\sqrt{\mathcal{M}} + c_M\frac{\mathcal{N}}{\sqrt{\mathcal{M}}}\right|, \tag{16.3}$$

with the $\mathfrak{bms}_3$ central charges $c_L$ and $c_M$ (10.24) and where the parameters $\mathcal{M}$ and $\mathcal{N}$ are the mass and angular momentum of a cosmological solution in flat space, which is the corresponding gravity dual to this quantum field theory. Using the same limiting methods as for the $\mathfrak{bms}_3$ case I was also able to determine the thermal entropy of $\mathcal{FW}_3$ invariant quantum field theories which resulted in the following expression

$$S_{\mathcal{FW}_3} = \frac{\pi}{6}\left|c_L\sqrt{\mathcal{M}}\sqrt{1-\frac{3}{4\mathcal{R}}} + c_M\frac{\mathcal{N}\left(4\mathcal{R}-6+3\mathcal{P}\sqrt{\mathcal{R}}\right)}{4\sqrt{\mathcal{M}}(\mathcal{R}-3)\sqrt{1-\frac{3}{4\mathcal{R}}}}\right|, \tag{16.4}$$

where $\mathcal{R}$ and $\mathcal{P}$ parametrize the spin-3 generalizations of $\mathcal{M}$ and $\mathcal{N}$.

Similar to the preceding study of unitarity in non-AdS holography, I then determined unitary representations of various nonlinear $\mathcal{FW}$-algebras, which resulted in a general NO-GO theorem. Under the assumptions I made it is thus not possible to have higher-spins and unitarity at the same time in flat space holography for higher-spin gravity theories whose dual field theories are *nonlinear* $\mathcal{FW}$-algebras. Following this NO-GO theorem I also presented a YES-GO example in the form of the *linear* $\mathcal{FW}_\infty$ algebra.

I then embarked upon the study of flat space higher-spin gravity theories. I reviewed the general construction of higher-spin gravity theories in flat space in terms of a Chern-Simons formulation with an explicit example of spin-3 gravity. I continued by showing how to consistently introduce chemical potentials both in spin-2 and spin-3 gravity for flat space. In order to check this construction I first analyzed flat space Einstein gravity with chemical potentials before showing some applications of my construction. I showed how to determine the entropy of flat space cosmologies including chemical potentials both for spin-2 and spin-3 hair which resulted in (16.3) and (16.4) respectively as well as the corresponding grand canonical free energy. Furthermore, the study of the free energy showed that, similar to AdS, there are also phase transitions from flat space cosmologies to hot flat space and vice versa.
After having investigated flat space (higher-spin) gravity I listed early explicit checks indicating a holographic correspondence in flat space before providing an explicit check by calculating entanglement and thermal entropy holographically. The holographic prescription in flat space followed the same basic principles as in AdS, i.e. using a Wilson line. I showed how to construct a suitable topological probe that yields the correct result for the entanglement entropy of an entangling region which is both timelike ($\Delta u$) and spacelike ($\Delta\phi$) separated

$$S_E^{\text{NO}} = \frac{c_L}{6} \ln\left[\frac{r_0 \Delta\phi}{2}\right] + \frac{c_M}{6}\frac{\Delta u}{\Delta\phi}, \tag{16.5}$$

$$S_E^{\text{GFS}} = \frac{c_L}{6} \ln\left[r_0 \sin\left(\frac{\Delta\phi}{2}\right)\right] + \frac{c_M}{12} \cot\left(\frac{\Delta\phi}{2}\right)\Delta u, \tag{16.6}$$

using the null orbifold (NO) and global flat space (GFS) with a boundary located at $r = r_0$ as gravity duals. Furthermore, I showed explicitly how this topological probe can be used to also determine the thermal entropy of flat space cosmologies, which again yielded (16.3). In addition, I extended this formalism in order to also include higher-spin symmetries and explicitly calculated the thermal entropy of flat space cosmologies with spin-3 hair, which precisely matched (16.4).

# 17 The Bird's Perspective

> *Der Abschied von einer langen und wichtigen Arbeit ist immer mehr traurig als erfreulich.*
> *(Parting with a lengthy and important task is always more sad than joyful.)*
>
> **– Friedrich Schiller**
> Briefe an Goethe, 27. Juni 1796

"*How general is the holographic principle?*"

This was the question I posed right at the beginning of my thesis and which has been the guiding principle during my research so far. Since this is a rather general question and most likely will require a lot of research to give a definite answer to, I will not claim that I was able to answer this question completely. What I can say, however, is that this thesis was able to shed some light on general aspects of holography in $2+1$ dimensions which were not known before. Firstly and maybe most importantly it is indeed possible to establish holographic correspondences that even include higher-spin symmetries, are unitary and which do not rely on spaces with constant negative curvature. Therefore one insight is that holography does not necessarily require AdS spacetimes to work.

From my studies concerning flat space holography I can also conclude that, while it is possible to establish a holographic correspondence for (a part of) flat space, it is not as "straightforward" as in AdS. In hindsight and with a lot of experience with both AdS and flat space holography, it seems that AdS is somehow tailor made for studying holography because of its special structure where many relations between quantum observables and geometry are particularly clear. This can be a blessing and a curse at the same time. Since most of our holographic intuition comes from studying AdS holography, one tends to apply the same or very similar techniques also to other proposed holographic dualities. Flat space holography is a prime example for that. It is close enough to AdS that many things can be treated in an analogous way by simply taking limits from the AdS results or use techniques which are heavily inspired by AdS/CFT. Most of the time this seems to work, however, in many calculations involving flat space holography it so often happens that everything works out, except at certain parts that have to be fixed manually in comparison to AdS, where usually the non-zero cosmological is ready to save the day. For me this is a strong indication that maybe a slightly different framework should be used for flat space holography than for AdS holography.

Probably the strongest hint that something in the description is not quite right is the issue of unitarity for flat space Einstein gravity which is, under the assumption presented in this thesis, i.e. highest-weight representations of the $\mathfrak{bms}_3$ algebra, not unitary. This is at first thought rather surprising. However, there is one argument why it should not be surprising that flat space holography, as I described it in this thesis, is non-unitary. The reason is that one is only looking at one half of the problem, which is either $\mathscr{I}^+$ or $\mathscr{I}^-$. Thus, radiation emitted from $\mathscr{I}^-$ which reaches $\mathscr{I}^+$ would only be seen as radiation coming from "nowhere". Therefore it would not be surprising that one finds a non-unitary theory. There is, however, one flaw in that argument, i.e. that there are no local propagating degrees of freedom in the bulk and thus nothing should be able to propagate from $\mathscr{I}^-$ to $\mathscr{I}^+$. Hence the issue of non-unitarity in flat space is still a deeply puzzling one, and it may require a different approach[1] than the one used up until now.

A similar issue is related to holographic entanglement entropy in flat space. While the topological probe using a Wilson line presented in this thesis works perfectly fine for all practical purposes, its geometric interpretation, however, is not very clear as of yet. One can also perform a similar computation using geodesics in flat space, and at least at first sight it seems that the geometrical object described by this Wilson line is not a geodesic simply attached at the boundary, but rather something more elusive. In $AdS_3$ the geometric picture (geodesics) and the gauge theoretic one (Wilson line) match perfectly, whereas in flat space it seems not to be as straightforward to assign a geometric interpretation to the topological probe used to determine entanglement entropy holographically.

To cut a long story short, let me now give an answer to the central question of this thesis:

"*The holographic principle is* likely *a general principle of nature.*"

The reason why I think that holography is *likely* a general principle of nature lies in the connections found between geometry and quantum field theory in so many different instances. Even though it seems that one cannot apply exactly the same techniques and intuition for each instance of holography, it is still encouraging that one can find holographic correspondences beyond AdS/CFT. But that is exactly what makes holography such an interesting topic for research. With every step you take away from AdS/CFT there are more connections between geometry and quantum field theories to be discovered, opening up a completely novel and interesting way to think about a problem which may have been thought of being unsolvable before.

[1] One possible solution to this issue of non-unitarity might be given by using *induced* representations rather than highest-weight ones for flat space holography as proposed for example in [IX, 73, 175].

# 18 The Wizard's Perspective

> *It's a dangerous business, Frodo, going out your door. You step onto the road, and if you don't keep your feet, there's no knowing where you might be swept off to.*
>
> **– J.R.R. Tolkien**
> The Lord of the Rings

There is this funny phenomenon frequently happening during research, and that is being left with even more questions and projects to do than one originally started out with. Unsurprisingly, this phenomenon also occurred during my time as a PhD student at the TU Wien. Therefore, I would like to list some of the remaining open questions or possible follow-up projects that emerged while doing research on holography during these past years.

In the context of non-AdS spacetimes with higher-spin symmetries and the $\mathcal{W}$-algebras that make their appearance as asymptotic symmetry algebras I only studied but a fraction of the various different possible $\mathcal{W}$-algebras. It would be interesting to extend this line of research towards a systematic classification of other families of $\mathcal{W}$-algebras and study their unitary representations.

Having developed a holographic description of entanglement entropy in Chapter 15, I would also consider it interesting to extend this description so that it incorporates various non-AdS geometries, such as Lobachevsky, Schrödinger, Lifshitz and Warped AdS. Similar to the flat space and AdS case, a lot could be learned about the holographic principle in general by studying holographic entanglement entropy of these spacetimes.

Since flat space holography is a comparatively new field of research there are a lot of things still to be done and understood. In the following I will mention some possible follow-up projects that are related to my thesis or which I think are important to take a look at in order to gain a better understanding of this holographic correspondence without claiming to be exhaustive.

I already mentioned in the conclusions that non-unitarity of Einstein gravity is a puzzling issue. Thus, it would be a natural question to address this issue. In [IX] I proposed together with my collaborators Andrea Campoleoni, Hernan Gonzalez and Blagoje Oblak that the correct unitary representations to consider for flat space holography are not highest-weight representations, but rather so called "rest frame" states. This is an interesting proposal which should be investigated more. It could, however, also be possible that the highest-weight representation is indeed correct but the way one checks for unitary representations has to be altered for $\mathfrak{bms}_3$ invariant quantum field theories.

Even though GCFT techniques have been making huge progress over the last few years, there is still a long way to go until one can say to have the same level of understanding and control as with a CFT. One example of this would be that there is, as of yet, no extension of GCFTs that also allow the treatment of higher-spin fields from a GCFTs side that would provide a check of e.g. (15.61). This would also go hand in hand with a more in-depth understanding of the nonrelativistic nature of GCFTs.

Another remaining problem already mentioned in the concluding chapters would be a geometrical interpretation of the entanglement entropy derived in Chapter 15 using Wilson lines. It would be very interesting to see what kind of geometric object is actually the dual one to entanglement entropy on the boundary. This is also connected to holographic checks of information theoretic properties such as e.g. strong subadditivity and monogamy of mutual information for flat space holography. Of course a proof of the Ryu-Takayanagi proposal along the lines of [114] for flat space holography would also be very interesting.

One crucial aspect that is still missing in the whole discussion revolving around entanglement entropy for $\mathfrak{bms}_3$ invariant quantum field theories is that, as of yet, there are no models known[1] which explicitly realize the full infinite dimensional $\mathfrak{bms}_3$ symmetries. Thus, it would be of vital interest for a better understanding of entanglement entropy of $\mathfrak{bms}_3$ invariant quantum field theories to find models, if they exist at all, that explicitly realize these symmetries.

As a final and at the same time very ambitious point regarding flat space holography I want to mention a *complete* holographic description of asymptotic flat spacetimes. Most of the things that are known about flat space holography up until now are only valid for either $\mathscr{I}^+$ or $\mathscr{I}^-$. Thus, in order to have a complete description one should be able to describe holography on both $\mathscr{I}^+$ and $\mathscr{I}^-$ simultaneously including $i^0$. This is a very ambitious project at the moment but I think there is no way around solving this problem for a better understanding of flat space holography in general.

Since I am ultimately interested in an even better understanding of the generality of the holographic principle, I want to mention in the following some points that are not directly related to this thesis but which I think might be important to study further and might shed more light on the connection between quantum field theory and spacetime geometry.

The ultimate insight one might hope to gain from a fundamental understanding of the holographic principle is how to describe quantum gravity. One of the main conceptual issues that arise when trying to think about a quantum theory of gravity is the question what happens with the usual notion of geometry known from general relativity at scales close to the Planck scale. A very interesting approach to this problem is that gravity at very small scales might actually emerge from underlying quantum information[2]. Gravity and entanglement in $2+1$ dimensions are comparatively simple to study. Therefore, it might be worthwhile to try and check whether or not one can also construct spacetime purely from entanglement information for e.g. flat space. Since it has already been shown that this works in one direction, i.e. take some geometric information and determine the entanglement entropy from that information, it is conceivable that this procedure also works in the other direction. Of course all the work which has been done in this thesis was focused on $2+1$ dimensions and thus is a toy model in comparison to the world we actually live and perform experiments in. Therefore it is also of interest to try and take a step away from the playground which is $2+1$ dimensional gravity, enter the higher-dimensional (real) world and try to implement the insights gained from the lower-dimensional toy model to a higher-dimensional setup. This will hopefully also lead to practical applications and experiments which might be able to explicitly verify that, indeed, the holographic principle is a fundamental property of nature.

[1]Except the flat limit of Liouville theories [176] which, however, results in a highly non-local Lagrangian and is therefore not very easy to work with.

[2] The most promising approach right now is called AdS/MERA correspondence, where MERA stands for "Multi-scale Entanglement Renormalization Ansatz" which has been first proposed in [177].

# Part VI

## Appendix

In these appendices I collect explicit matrix representations of algebras that are used for computations in this thesis. Furthermore I present the details of the canonical analysis for two specific embeddings of $\mathfrak{sl}(2,\mathbb{R}) \hookrightarrow \mathfrak{sl}(4,\mathbb{R})$ and explicit calculations used to determine quantum $\mathcal{W}$-algebras which rely on calculating Jacobi identities. In addition, I include explicit examples of İnönü–Wigner contractions of various $\mathcal{W}$-algebras to $\mathcal{FW}$-algebras as well as explicit expressions of non-constant contributions to spin-2 and spin-3 fields coming from spin-2 and spin-3 chemical potentials.

# A Matrix Representations

## A.1 $\mathfrak{sl}$ Matrix Representations

### A.1.1 $\mathfrak{sl}(2,\mathbb{R})$

**Fundamental representation:**

$$L_1 = \begin{pmatrix} 0 & 0 \\ 1 & 0 \end{pmatrix}, \quad L_0 = \frac{1}{2}\begin{pmatrix} 1 & 0 \\ 0 & -1 \end{pmatrix}, \quad L_{-1} = \begin{pmatrix} 0 & -1 \\ 0 & 0 \end{pmatrix}. \tag{A.1}$$

Invariant Bilinear Form:

$$h_{ab} = \left(\begin{array}{c|ccc} & L_1 & L_0 & L_{-1} \\ \hline L_1 & 0 & 0 & -1 \\ L_0 & 0 & \frac{1}{2} & 0 \\ L_{-1} & -1 & 0 & 0 \end{array}\right). \tag{A.2}$$

**Adjoint representation:**

$$L_1 = \begin{pmatrix} 0 & 0 & 0 \\ 2 & 0 & 0 \\ 0 & 1 & 0 \end{pmatrix}, \quad L_0 = \begin{pmatrix} 1 & 0 & 0 \\ 0 & 0 & 0 \\ 0 & 0 & -1 \end{pmatrix}, \quad L_{-1} = \begin{pmatrix} 0 & -1 & 0 \\ 0 & 0 & -2 \\ 0 & 0 & 0 \end{pmatrix}. \tag{A.3}$$

Invariant Bilinear Form:

$$h_{ab} = 4\left(\begin{array}{c|ccc} & L_1 & L_0 & L_{-1} \\ \hline L_1 & 0 & 0 & -1 \\ L_0 & 0 & \frac{1}{2} & 0 \\ L_{-1} & -1 & 0 & 0 \end{array}\right). \tag{A.4}$$

## A.1.2 $\mathfrak{sl}(3,\mathbb{R})$

**Fundamental representation, principal embedding:**

$$\mathfrak{L}_1 = \begin{pmatrix} 0 & 0 & 0 \\ 1 & 0 & 0 \\ 0 & 1 & 0 \end{pmatrix}, \quad \mathfrak{L}_0 = \begin{pmatrix} 1 & 0 & 0 \\ 0 & 0 & 0 \\ 0 & 0 & -1 \end{pmatrix}, \quad \mathfrak{L}_{-1} = \begin{pmatrix} 0 & -2 & 0 \\ 0 & 0 & -2 \\ 0 & 0 & 0 \end{pmatrix},$$
$$\mathfrak{W}_2 = \begin{pmatrix} 0 & 0 & 0 \\ 0 & 0 & 0 \\ 2 & 0 & 0 \end{pmatrix}, \quad \mathfrak{W}_1 = \begin{pmatrix} 0 & 0 & 0 \\ 1 & 0 & 0 \\ 0 & -1 & 0 \end{pmatrix}, \quad \mathfrak{W}_0 = \begin{pmatrix} \frac{2}{3} & 0 & 0 \\ 0 & -\frac{4}{3} & 0 \\ 0 & 0 & \frac{2}{3} \end{pmatrix},$$
$$\mathfrak{W}_{-2} = \begin{pmatrix} 0 & 0 & 8 \\ 0 & 0 & 0 \\ 0 & 0 & 0 \end{pmatrix}, \quad \mathfrak{W}_{-1} = \begin{pmatrix} 0 & -2 & 0 \\ 0 & 0 & 2 \\ 0 & 0 & 0 \end{pmatrix}. \tag{A.5}$$

## A.1.3 $\mathfrak{sl}(4,\mathbb{R})$

**Fundamental representation, 2-1-1 embedding:**

$\mathfrak{sl}(2,\mathbb{R})$ generators:

$$L_1 = \begin{pmatrix} 0 & 0 & 0 & 0 \\ 0 & 0 & 0 & 0 \\ 0 & 0 & 0 & 0 \\ 1 & 0 & 0 & 0 \end{pmatrix}, \quad L_0 = \begin{pmatrix} \frac{1}{2} & 0 & 0 & 0 \\ 0 & 0 & 0 & 0 \\ 0 & 0 & 0 & 0 \\ 0 & 0 & 0 & -\frac{1}{2} \end{pmatrix}, \quad L_{-1} = \begin{pmatrix} 0 & 0 & 0 & -1 \\ 0 & 0 & 0 & 0 \\ 0 & 0 & 0 & 0 \\ 0 & 0 & 0 & 0 \end{pmatrix}. \tag{A.6}$$

Singlets:

$$J^0 = \begin{pmatrix} 0 & 0 & 0 & 0 \\ 0 & \frac{1}{2} & 0 & 0 \\ 0 & 0 & -\frac{1}{2} & 0 \\ 0 & 0 & 0 & 0 \end{pmatrix}, \quad J^+ = \begin{pmatrix} 0 & 0 & 0 & 0 \\ 0 & 0 & 0 & 0 \\ 0 & 1 & 0 & 0 \\ 0 & 0 & 0 & 0 \end{pmatrix}, \quad J^- = \begin{pmatrix} 0 & 0 & 0 & 0 \\ 0 & 0 & -1 & 0 \\ 0 & 0 & 0 & 0 \\ 0 & 0 & 0 & 0 \end{pmatrix}$$
$$S = \begin{pmatrix} 1 & 0 & 0 & 0 \\ 0 & -1 & 0 & 0 \\ 0 & 0 & -1 & 0 \\ 0 & 0 & 0 & 1 \end{pmatrix}. \tag{A.7}$$

Doublets:

$$
G^{+|+}_{\frac{1}{2}} = \begin{pmatrix} 0 & 0 & 0 & 0 \\ 0 & 0 & 0 & 0 \\ 1 & 0 & 0 & 0 \\ 0 & 0 & 0 & 0 \end{pmatrix}, \quad G^{+|+}_{-\frac{1}{2}} = \begin{pmatrix} 0 & 0 & 0 & 0 \\ 0 & 0 & 0 & 0 \\ 0 & 0 & 0 & -1 \\ 0 & 0 & 0 & 0 \end{pmatrix},
$$

$$
G^{+|-}_{\frac{1}{2}} = \begin{pmatrix} 0 & 0 & 0 & 0 \\ 0 & 0 & 0 & 0 \\ 0 & 0 & 0 & 0 \\ 0 & 0 & 1 & 0 \end{pmatrix}, \quad G^{+|-}_{-\frac{1}{2}} = \begin{pmatrix} 0 & 0 & 1 & 0 \\ 0 & 0 & 0 & 0 \\ 0 & 0 & 0 & 0 \\ 0 & 0 & 0 & 0 \end{pmatrix},
$$

$$
G^{-|+}_{\frac{1}{2}} = \begin{pmatrix} 0 & 0 & 0 & 0 \\ 1 & 0 & 0 & 0 \\ 0 & 0 & 0 & 0 \\ 0 & 0 & 0 & 0 \end{pmatrix}, \quad G^{-|+}_{-\frac{1}{2}} = \begin{pmatrix} 0 & 0 & 0 & 0 \\ 0 & 0 & 0 & -1 \\ 0 & 0 & 0 & 0 \\ 0 & 0 & 0 & 0 \end{pmatrix}
$$

$$
G^{-|-}_{\frac{1}{2}} = \begin{pmatrix} 0 & 0 & 0 & 0 \\ 0 & 0 & 0 & 0 \\ 0 & 0 & 0 & 0 \\ 0 & -1 & 0 & 0 \end{pmatrix}, \quad G^{-|-}_{-\frac{1}{2}} = \begin{pmatrix} 0 & -1 & 0 & 0 \\ 0 & 0 & 0 & 0 \\ 0 & 0 & 0 & 0 \\ 0 & 0 & 0 & 0 \end{pmatrix}. \tag{A.8}
$$

Invariant Bilinear Form:

$$h_{ab} =$$

| | $L_1$ | $L_0$ | $L_{-1}$ | $G^{++}_{+}$ | $G^{++}_{-}$ | $G^{+-}_{+}$ | $G^{+-}_{-}$ | $G^{-+}_{+}$ | $G^{-+}_{-}$ | $G^{--}_{+}$ | $G^{--}_{-}$ | $S^{+}$ | $S^{0}$ | $S^{-}$ | $S$ |
|---|---|---|---|---|---|---|---|---|---|---|---|---|---|---|---|
| $L_1$ | 0 | 0 | −1 | 0 | 0 | 0 | 0 | 0 | 0 | 0 | 0 | 0 | 0 | 0 | 0 |
| $L_0$ | 0 | $\frac{1}{2}$ | 0 | 0 | 0 | 0 | 0 | 0 | 0 | 0 | 0 | 0 | 0 | 0 | 0 |
| $L_{-1}$ | −1 | 0 | 0 | 0 | 0 | 0 | 0 | 0 | 0 | 0 | 0 | 0 | 0 | 0 | 0 |
| $G^{++}_{+}$ | 0 | 0 | 0 | 0 | 0 | 0 | 0 | 0 | 0 | 0 | 1 | 0 | 0 | 0 | 0 |
| $G^{++}_{-}$ | 0 | 0 | 0 | 0 | 0 | 0 | 0 | 0 | 0 | −1 | 0 | 0 | 0 | 0 | 0 |
| $G^{+-}_{+}$ | 0 | 0 | 0 | 0 | 0 | 0 | 0 | 0 | 1 | 0 | 0 | 0 | 0 | 0 | 0 |
| $G^{+-}_{-}$ | 0 | 0 | 0 | 0 | 0 | 0 | 0 | −1 | 0 | 0 | 0 | 0 | 0 | 0 | 0 |
| $G^{-+}_{+}$ | 0 | 0 | 0 | 0 | 0 | 0 | −1 | 0 | 0 | 0 | 0 | 0 | 0 | 0 | 0 |
| $G^{-+}_{-}$ | 0 | 0 | 0 | 0 | 0 | 1 | 0 | 0 | 0 | 0 | 0 | 0 | 0 | 0 | 0 |
| $G^{--}_{+}$ | 0 | 0 | 0 | 0 | −1 | 0 | 0 | 0 | 0 | 0 | 0 | 0 | 0 | 0 | 0 |
| $G^{--}_{-}$ | 0 | 0 | 0 | 1 | 0 | 0 | 0 | 0 | 0 | 0 | 0 | 0 | 0 | 0 | 0 |
| $S^{+}$ | 0 | 0 | 0 | 0 | 0 | 0 | 0 | 0 | 0 | 0 | 0 | 0 | 0 | −1 | 0 |
| $S^{0}$ | 0 | 0 | 0 | 0 | 0 | 0 | 0 | 0 | 0 | 0 | 0 | 0 | $\frac{1}{2}$ | 0 | 0 |
| $S^{-}$ | 0 | 0 | 0 | 0 | 0 | 0 | 0 | 0 | 0 | 0 | 0 | −1 | 0 | 0 | 0 |
| $S$ | 0 | 0 | 0 | 0 | 0 | 0 | 0 | 0 | 0 | 0 | 0 | 0 | 0 | 0 | 4 |

(A.9)

**Fundamental representation, 2-2 embedding:**

$\mathfrak{sl}(2,\mathbb{R})$ generators:

$$
L_1 = \begin{pmatrix} 0 & 0 & 0 & 0 \\ 0 & 0 & 0 & 0 \\ 1 & 0 & 0 & 0 \\ 0 & 1 & 0 & 0 \end{pmatrix}, \quad L_0 = \begin{pmatrix} \frac{1}{2} & 0 & 0 & 0 \\ 0 & \frac{1}{2} & 0 & 0 \\ 0 & 0 & -\frac{1}{2} & 0 \\ 0 & 0 & 0 & -\frac{1}{2} \end{pmatrix}, \quad L_{-1} = \begin{pmatrix} 0 & 0 & -1 & 0 \\ 0 & 0 & 0 & -1 \\ 0 & 0 & 0 & 0 \\ 0 & 0 & 0 & 0 \end{pmatrix}. \tag{A.10}
$$

Triplets:

$$
T_1^+ = \begin{pmatrix} 0 & 0 & 0 & 0 \\ 0 & 0 & 0 & 0 \\ 0 & 0 & 0 & 0 \\ 1 & 0 & 0 & 0 \end{pmatrix}, \quad
T_0^+ = \begin{pmatrix} 0 & 0 & 0 & 0 \\ \frac{1}{2} & 0 & 0 & 0 \\ 0 & 0 & 0 & 0 \\ 0 & 0 & -\frac{1}{2} & 0 \end{pmatrix}, \quad
T_{-1}^+ = \begin{pmatrix} 0 & 0 & 0 & 0 \\ 0 & 0 & -1 & 0 \\ 0 & 0 & 0 & 0 \\ 0 & 0 & 0 & 0 \end{pmatrix},
$$

$$
T_1^0 = \begin{pmatrix} 0 & 0 & 0 & 0 \\ 0 & 0 & 0 & 0 \\ \frac{1}{2} & 0 & 0 & 0 \\ 0 & -\frac{1}{2} & 0 & 0 \end{pmatrix}, \quad
T_0^0 = \begin{pmatrix} \frac{1}{4} & 0 & 0 & 0 \\ 0 & -\frac{1}{4} & 0 & 0 \\ 0 & 0 & -\frac{1}{4} & 0 \\ 0 & 0 & 0 & \frac{1}{4} \end{pmatrix}, \quad
T_{-1}^0 = \begin{pmatrix} 0 & 0 & -\frac{1}{2} & 0 \\ 0 & 0 & 0 & \frac{1}{2} \\ 0 & 0 & 0 & 0 \\ 0 & 0 & 0 & 0 \end{pmatrix},
$$

$$
T_1^- = \begin{pmatrix} 0 & 0 & 0 & 0 \\ 0 & 0 & 0 & 0 \\ 0 & 1 & 0 & 0 \\ 0 & 0 & 0 & 0 \end{pmatrix}, \quad
T_0^- = \begin{pmatrix} 0 & \frac{1}{2} & 0 & 0 \\ 0 & 0 & 0 & 0 \\ 0 & 0 & 0 & -\frac{1}{2} \\ 0 & 0 & 0 & 0 \end{pmatrix}, \quad
T_{-1}^- = \begin{pmatrix} 0 & 0 & 0 & -\frac{1}{2} \\ 0 & 0 & 0 & 0 \\ 0 & 0 & 0 & 0 \\ 0 & 0 & 0 & 0 \end{pmatrix}.
\tag{A.11}
$$

Singlets:

$$
S^+ = \begin{pmatrix} 0 & 0 & 0 & 0 \\ 1 & 0 & 0 & 0 \\ 0 & 0 & 0 & 0 \\ 0 & 0 & 1 & 0 \end{pmatrix}, \quad
S^0 = \frac{1}{2}\begin{pmatrix} \frac{1}{2} & 0 & 0 & 0 \\ 0 & -\frac{1}{2} & 0 & 0 \\ 0 & 0 & \frac{1}{2} & 0 \\ 0 & 0 & 0 & -\frac{1}{2} \end{pmatrix}, \quad
S^- = \begin{pmatrix} 0 & -1 & 0 & 0 \\ 0 & 0 & 0 & 0 \\ 0 & 0 & 0 & -1 \\ 0 & 0 & 0 & 0 \end{pmatrix}.
\tag{A.12}
$$

Invariant Bilinear Form:

$$
h_{ab} = \left(\begin{array}{c|ccccccccccccccc}
 & L_1 & L_0 & L_{-1} & {T_1}^+ & {T_0}^+ & {T_{-1}}^+ & {T_1}^0 & {T_0}^0 & {T_{-1}}^0 & {T_1}^- & {T_0}^- & {T_{-1}}^- & S^+ & S^0 & S^- \\
\hline
L_1 & 0 & 0 & -2 & 0 & 0 & 0 & 0 & 0 & 0 & 0 & 0 & 0 & 0 & 0 & 0 \\
L_0 & 0 & 1 & 0 & 0 & 0 & 0 & 0 & 0 & 0 & 0 & 0 & 0 & 0 & 0 & 0 \\
L_{-1} & -2 & 0 & 0 & 0 & 0 & 0 & 0 & 0 & 0 & 0 & 0 & 0 & 0 & 0 & 0 \\
{T_1}^+ & 0 & 0 & 0 & 0 & 0 & 0 & 0 & 0 & 0 & 0 & 0 & -1 & 0 & 0 & 0 \\
{T_0}^+ & 0 & 0 & 0 & 0 & 0 & 0 & 0 & 0 & 0 & 0 & \frac{1}{2} & 0 & 0 & 0 & 0 \\
{T_{-1}}^+ & 0 & 0 & 0 & 0 & 0 & 0 & 0 & 0 & 0 & -1 & 0 & 0 & 0 & 0 & 0 \\
{T_1}^0 & 0 & 0 & 0 & 0 & 0 & 0 & 0 & 0 & -\frac{1}{2} & 0 & 0 & 0 & 0 & 0 & 0 \\
{T_0}^0 & 0 & 0 & 0 & 0 & 0 & 0 & 0 & \frac{1}{4} & 0 & 0 & 0 & 0 & 0 & 0 & 0 \\
{T_{-1}}^0 & 0 & 0 & 0 & 0 & 0 & 0 & -\frac{1}{2} & 0 & 0 & 0 & 0 & 0 & 0 & 0 & 0 \\
{T_1}^- & 0 & 0 & 0 & 0 & 0 & -1 & 0 & 0 & 0 & 0 & 0 & 0 & 0 & 0 & 0 \\
{T_0}^- & 0 & 0 & 0 & 0 & \frac{1}{2} & 0 & 0 & 0 & 0 & 0 & 0 & 0 & 0 & 0 & 0 \\
{T_{-1}}^- & 0 & 0 & 0 & -1 & 0 & 0 & 0 & 0 & 0 & 0 & 0 & 0 & 0 & 0 & 0 \\
S^+ & 0 & 0 & 0 & 0 & 0 & 0 & 0 & 0 & 0 & 0 & 0 & 0 & 0 & 0 & -2 \\
S^0 & 0 & 0 & 0 & 0 & 0 & 0 & 0 & 0 & 0 & 0 & 0 & 0 & 0 & 1 & 0 \\
S^- & 0 & 0 & 0 & 0 & 0 & 0 & 0 & 0 & 0 & 0 & 0 & 0 & -2 & 0 & 0
\end{array}\right)
\tag{A.13}
$$

**Fundamental representation, 3-1 embedding:**

$\mathfrak{sl}(2,\mathbb{R})$ generators:

$$L_1 = \begin{pmatrix} 0 & 0 & 0 & 0 \\ 0 & 0 & 0 & 0 \\ \sqrt{2} & 0 & 0 & 0 \\ 0 & 0 & \sqrt{2} & 0 \end{pmatrix}, \quad L_0 = \begin{pmatrix} \frac{1}{2} & 0 & 0 & 0 \\ 0 & 0 & 0 & 0 \\ 0 & 0 & 0 & 0 \\ 0 & 0 & 0 & -\frac{1}{2} \end{pmatrix}, \quad L_{-1} = \begin{pmatrix} 0 & 0 & -\sqrt{2} & 0 \\ 0 & 0 & 0 & 0 \\ 0 & 0 & 0 & -\sqrt{2} \\ 0 & 0 & 0 & 0 \end{pmatrix}. \tag{A.14}$$

Triplets:

$$T_1^+ = \begin{pmatrix} 0 & 0 & 0 & 0 \\ 0 & 0 & 0 & 0 \\ 0 & 0 & 0 & 0 \\ 0 & -1 & 0 & 0 \end{pmatrix}, \quad T_0^+ = \begin{pmatrix} 0 & 0 & 0 & 0 \\ 0 & 0 & 0 & 0 \\ 0 & -\frac{1}{\sqrt{2}} & 0 & 0 \\ 0 & 0 & 0 & 0 \end{pmatrix}, \quad T_{-1}^+ = \begin{pmatrix} 0 & -1 & 0 & 0 \\ 0 & 0 & 0 & 0 \\ 0 & 0 & 0 & 0 \\ 0 & 0 & 0 & 0 \end{pmatrix},$$
$$T_1^- = \begin{pmatrix} 0 & 0 & 0 & 0 \\ -1 & 0 & 0 & 0 \\ 0 & 0 & 0 & 0 \\ 0 & 0 & 0 & 0 \end{pmatrix}, \quad T_0^- = \begin{pmatrix} 0 & 0 & 0 & 0 \\ 0 & 0 & \frac{1}{\sqrt{2}} & 0 \\ 0 & 0 & 0 & 0 \\ 0 & 0 & 0 & 0 \end{pmatrix}, \quad T_{-1}^- = \begin{pmatrix} 0 & 0 & 0 & 0 \\ 0 & 0 & 0 & -1 \\ 0 & 0 & 0 & 0 \\ 0 & 0 & 0 & 0 \end{pmatrix}. \tag{A.15}$$

Quintet:

$$W_2 = \begin{pmatrix} 0 & 0 & 0 & 0 \\ 0 & 0 & 0 & 0 \\ 0 & 0 & 0 & 0 \\ -1 & 0 & 0 & 0 \end{pmatrix}, \quad W_1 = \begin{pmatrix} 0 & 0 & 0 & 0 \\ 0 & 0 & 0 & 0 \\ -\frac{1}{2\sqrt{2}} & 0 & 0 & 0 \\ 0 & 0 & -\frac{1}{2\sqrt{2}} & 0 \end{pmatrix}, \quad W_0 = \begin{pmatrix} -\frac{1}{6} & 0 & 0 & 0 \\ 0 & 0 & 0 & 0 \\ 0 & 0 & \frac{1}{3} & 0 \\ 0 & 0 & 0 & -\frac{1}{6} \end{pmatrix},$$
$$W_{-2} = \begin{pmatrix} 0 & 0 & 0 & -1 \\ 0 & 0 & 0 & 0 \\ 0 & 0 & 0 & 0 \\ 0 & 0 & 0 & 0 \end{pmatrix}, \quad W_{-1} = \begin{pmatrix} 0 & 0 & \frac{1}{2\sqrt{2}} & 0 \\ 0 & 0 & 0 & 0 \\ 0 & 0 & 0 & -\frac{1}{2\sqrt{2}} \\ 0 & 0 & 0 & 0 \end{pmatrix}. \tag{A.16}$$

Singlet:

$$S = \begin{pmatrix} \frac{1}{4} & 0 & 0 & 0 \\ 0 & -\frac{3}{4} & 0 & 0 \\ 0 & 0 & \frac{1}{4} & 0 \\ 0 & 0 & 0 & \frac{1}{4} \end{pmatrix}. \tag{A.17}$$

Invariant Bilinear Form:

$$
h_{ab} = \left(\begin{array}{c|ccccccccccccccc}
 & L_1 & L_0 & L_{-1} & {T_1}^+ & {T_0}^+ & {T_{-1}}^+ & {T_1}^- & {T_0}^- & {T_{-1}}^- & W_2 & W_1 & W_0 & W_{1-} & W_{-2} & S \\
\hline
L_1 & 0 & 0 & -4 & 0 & 0 & 0 & 0 & 0 & 0 & 0 & 0 & 0 & 0 & 0 & 0 \\
L_0 & 0 & 2 & 0 & 0 & 0 & 0 & 0 & 0 & 0 & 0 & 0 & 0 & 0 & 0 & 0 \\
L_{-1} & -4 & 0 & 0 & 0 & 0 & 0 & 0 & 0 & 0 & 0 & 0 & 0 & 0 & 0 & 0 \\
{T_1}^+ & 0 & 0 & 0 & 0 & 0 & 0 & 0 & 0 & 1 & 0 & 0 & 0 & 0 & 0 & 0 \\
{T_0}^+ & 0 & 0 & 0 & 0 & 0 & 0 & 0 & -\frac{1}{2} & 0 & 0 & 0 & 0 & 0 & 0 & 0 \\
{T_{-1}}^+ & 0 & 0 & 0 & 0 & 0 & 0 & 1 & 0 & 0 & 0 & 0 & 0 & 0 & 0 & 0 \\
{T_1}^- & 0 & 0 & 0 & 0 & 0 & 1 & 0 & 0 & 0 & 0 & 0 & 0 & 0 & 0 & 0 \\
{T_0}^- & 0 & 0 & 0 & 0 & -\frac{1}{2} & 0 & 0 & 0 & 0 & 0 & 0 & 0 & 0 & 0 & 0 \\
{T_{-1}}^- & 0 & 0 & 0 & 1 & 0 & 0 & 0 & 0 & 0 & 0 & 0 & 0 & 0 & 0 & 0 \\
W_2 & 0 & 0 & 0 & 0 & 0 & 0 & 0 & 0 & 0 & 0 & 0 & 0 & 0 & 1 & 0 \\
W_1 & 0 & 0 & 0 & 0 & 0 & 0 & 0 & 0 & 0 & 0 & 0 & 0 & -\frac{1}{4} & 0 & 0 \\
W_0 & 0 & 0 & 0 & 0 & 0 & 0 & 0 & 0 & 0 & 0 & 0 & \frac{1}{6} & 0 & 0 & 0 \\
W_{-1} & 0 & 0 & 0 & 0 & 0 & 0 & 0 & 0 & 0 & 0 & -\frac{1}{4} & 0 & 0 & 0 & 0 \\
W_{-2} & 0 & 0 & 0 & 0 & 0 & 0 & 0 & 0 & 0 & 1 & 0 & 0 & 0 & 0 & 0 \\
S & 0 & 0 & 0 & 0 & 0 & 0 & 0 & 0 & 0 & 0 & 0 & 0 & 0 & 0 & \frac{3}{4}
\end{array}\right) \tag{A.18}
$$

## A.2 $\mathfrak{isl}$ Matrix Representations

Throughout this thesis I use the following matrix representations of $\mathfrak{isl}(2,\mathbb{R})$ and $\mathfrak{isl}(3,\mathbb{R})$ generators in terms of $4\times 4$ and $6\times 6$ block-diagonal matrices. This block structure is a remnant of the decomposition of the $AdS_3$ symmetry algebra $\mathfrak{so}(2,2)\sim\mathfrak{sl}(2,\mathbb{R})\oplus\mathfrak{sl}(2,\mathbb{R})$ before the İnönü–Wigner contraction. In the following expressions $\epsilon$ denotes the Grassmann parameter ($\epsilon^2=0$) first introduced in [70].

### A.2.1 $\mathfrak{isl}(2,\mathbb{R})$

$$
L_1 = \begin{pmatrix} 0 & 0 & 0 & 0 \\ 1 & 0 & 0 & 0 \\ 0 & 0 & 0 & 0 \\ 0 & 0 & 1 & 0 \end{pmatrix}, \quad
L_0 = \begin{pmatrix} \frac{1}{2} & 0 & 0 & 0 \\ 0 & -\frac{1}{2} & 0 & 0 \\ 0 & 0 & \frac{1}{2} & 0 \\ 0 & 0 & 0 & -\frac{1}{2} \end{pmatrix}, \quad
L_{-1} = \begin{pmatrix} 0 & -1 & 0 & 0 \\ 0 & 0 & 0 & 0 \\ 0 & 0 & 0 & -1 \\ 0 & 0 & 0 & 0 \end{pmatrix},
$$
$$
M_1 = \begin{pmatrix} 0 & 0 & 0 & 0 \\ \epsilon & 0 & 0 & 0 \\ 0 & 0 & 0 & 0 \\ 0 & 0 & -\epsilon & 0 \end{pmatrix}, \quad
M_0 = \begin{pmatrix} \frac{\epsilon}{2} & 0 & 0 & 0 \\ 0 & -\frac{\epsilon}{2} & 0 & 0 \\ 0 & 0 & -\frac{\epsilon}{2} & 0 \\ 0 & 0 & 0 & \frac{\epsilon}{2} \end{pmatrix}, \quad
M_{-1} = \begin{pmatrix} 0 & -\epsilon & 0 & 0 \\ 0 & 0 & 0 & 0 \\ 0 & 0 & 0 & \epsilon \\ 0 & 0 & 0 & 0 \end{pmatrix}. \tag{A.19}
$$

## A.2.2 $\mathfrak{isl}(3,\mathbb{R})$

**Principal embedding:**

All even generators can be obtained as a tensor product of the $\mathfrak{sl}(3,\mathbb{R})$ generators given in (A.5) and $\mathbb{1}_{2\times 2}$ as

$$L_n = \mathfrak{L}_n \otimes \mathbb{1}_{2\times 2}, \quad U_m = \mathfrak{W}_m \otimes \mathbb{1}_{2\times 2}, \tag{A.20}$$

where $n = \pm 1, 0$ and $m = \pm 2, \pm 1, 0$.
All odd generators can be obtained in a similar fashion as a tensor product of the $\mathfrak{sl}(3,\mathbb{R})$ generators given in (A.5) and $\gamma^\star_{(2)}$ as defined in (12.3) i.e.

$$M_n = \epsilon\,\mathfrak{L}_n \otimes \gamma^\star_{(2)}, \quad V_n = \epsilon\,\mathfrak{W}_n \otimes \gamma^\star_{(2)}. \tag{A.21}$$

## A.2.3 $\mathfrak{isl}(3,\mathbb{R})$ in the $8+1$ Representation

For deriving entropy and holonomy conditions I use the following matrix representation of isl$(3)$ generators in terms of $8+1$-dimensional matrices with a "tensor"- and a "vector"-block. Generic generators $G$ are written in the form

$$G = \begin{pmatrix} \mathrm{ad}_{8\times 8} & \mathrm{odd}_{8\times 1} \\ \mathbb{0}_{1\times 8} & 0 \end{pmatrix} \tag{A.22}$$

where $\mathrm{ad}_{8\times 8}$ is an $8\times 8$ matrix that is an element of sl$(3)$ in the adjoint representation and $\mathrm{odd}_{8\times 1}$ is an $8\times 1$ column vector. The even generators $L_n$ and $U_n$ have ad $\neq \mathbb{0}$, odd $= \mathbb{0}$; the odd generators $M_n$ and $V_n$ have ad $= \mathbb{0}$, odd $\neq \mathbb{0}$. In fact, one can use the odd generators as unit basis vectors,

$$\mathrm{odd}_{M_n} = E_{n+2} \qquad \mathrm{odd}_{V_n} = E_{n+6} \tag{A.23}$$

with

$$E_i = (\underbrace{0,\ldots,0}_{i-1}, 1, \underbrace{0,\ldots,0}_{8-i})^T \qquad i = 1..8\,. \tag{A.24}$$

The ad-parts of the even generators compatible with the algebra (12.1) are then given by the following $8 \times 8$ matrices.

$$
\begin{aligned}
\mathrm{ad}_{L_{-1}} &= -\begin{pmatrix}
0&1&0&0&0&0&0&0\\
0&0&2&0&0&0&0&0\\
0&0&0&0&0&0&0&0\\
0&0&0&0&1&0&0&0\\
0&0&0&0&0&2&0&0\\
0&0&0&0&0&0&3&0\\
0&0&0&0&0&0&0&4\\
0&0&0&0&0&0&0&0
\end{pmatrix}, &
\mathrm{ad}_{L_0} &= \begin{pmatrix}
1&0&0&0&0&0&0&0\\
0&0&0&0&0&0&0&0\\
0&0&-1&0&0&0&0&0\\
0&0&0&2&0&0&0&0\\
0&0&0&0&1&0&0&0\\
0&0&0&0&0&0&0&0\\
0&0&0&0&0&0&-1&0\\
0&0&0&0&0&0&0&-2
\end{pmatrix},\\
\mathrm{ad}_{L_1} &= \begin{pmatrix}
0&0&0&0&0&0&0&0\\
2&0&0&0&0&0&0&0\\
0&1&0&0&0&0&0&0\\
0&0&0&0&0&0&0&0\\
0&0&0&4&0&0&0&0\\
0&0&0&0&3&0&0&0\\
0&0&0&0&0&2&0&0\\
0&0&0&0&0&0&1&0
\end{pmatrix}, &
\mathrm{ad}_{U_{-2}} &= \begin{pmatrix}
0&0&0&0&0&0&4&0\\
0&0&0&0&0&0&0&16\\
0&0&0&0&0&0&0&0\\
0&-2&0&0&0&0&0&0\\
0&0&-4&0&0&0&0&0\\
0&0&0&0&0&0&0&0\\
0&0&0&0&0&0&0&0\\
0&0&0&0&0&0&0&0
\end{pmatrix},\\
\mathrm{ad}_{U_{-1}} &= \begin{pmatrix}
0&0&0&0&0&-2&0&0\\
0&0&0&0&0&0&-2&0\\
0&0&0&0&0&0&0&4\\
1&0&0&0&0&0&0&0\\
0&-1&0&0&0&0&0&0\\
0&0&1&0&0&0&0&0\\
0&0&0&0&0&0&0&0\\
0&0&0&0&0&0&0&0
\end{pmatrix}, &
\mathrm{ad}_{U_0} &= \begin{pmatrix}
0&0&0&0&2&0&0&0\\
0&0&0&0&0&0&0&0\\
0&0&0&0&0&0&-2&0\\
0&0&0&0&0&0&0&0\\
2&0&0&0&0&0&0&0\\
0&0&0&0&0&0&0&0\\
0&0&-2&0&0&0&0&0\\
0&0&0&0&0&0&0&0
\end{pmatrix},\\
\mathrm{ad}_{U_1} &= \begin{pmatrix}
0&0&0&-4&0&0&0&0\\
0&0&0&0&2&0&0&0\\
0&0&0&0&0&2&0&0\\
0&0&0&0&0&0&0&0\\
0&0&0&0&0&0&0&0\\
3&0&0&0&0&0&0&0\\
0&1&0&0&0&0&0&0\\
0&0&-1&0&0&0&0&0
\end{pmatrix}, &
\mathrm{ad}_{U_2} &= \begin{pmatrix}
0&0&0&0&0&0&0&0\\
0&0&0&-16&0&0&0&0\\
0&0&0&0&-4&0&0&0\\
0&0&0&0&0&0&0&0\\
0&0&0&0&0&0&0&0\\
0&0&0&0&0&0&0&0\\
4&0&0&0&0&0&0&0\\
0&2&0&0&0&0&0&0
\end{pmatrix}.
\end{aligned}
\tag{A.25}
$$

# Canonical Analysis for $\mathfrak{sl}(4,\mathbb{R})$ Lobachevsky Holography

B

In this appendix I present in more detail the canonical analysis of the two non-principal embeddings of $\mathfrak{sl}(2,\mathbb{R}) \hookrightarrow \mathfrak{sl}(4,\mathbb{R})$, namely the 2-2 and the 3-1 embedding, which were discussed in Section 8.1 of the main text.

## B.1 Canonical Analysis of the 2-2 Embedding

In order to be consistent with the fluctuations as in (8.5) the Chern-Simons gauge connection $A$ has to obey the following boundary conditions

$$a_\rho^{(0)} = L_0, \qquad a_t^{(0)} = 0, \qquad a_\mu^{(1)} = \mathcal{O}(e^{-2\rho}), \tag{B.1a}$$

$$a_\varphi^{(0)} = \sigma L_1 - \frac{2\pi}{k}\left(\frac{\mathcal{L}}{2}L_{-1} + \sum_{a=-1}^{1}\left[(2-a^2)\mathcal{T}^a T_{-1}^a - (1-\frac{3a^2}{2})\mathcal{S}^a S^a\right]\right), \tag{B.1b}$$

where the state dependent functions $\mathcal{L}, \mathcal{T}^a$ and $\mathcal{S}^a$ only depend on the angular coordinate $\varphi$.
The connection $\bar{A}$ on the other hand has to abide the following asymptotic behavior

$$\bar{a}_\rho^{(0)} = -L_0, \qquad \bar{a}_t^{(0)} = 2S^0, \qquad \bar{a}_\mu^{(1)} = \mathcal{O}(e^{-2\rho}), \tag{B.2a}$$

$$\bar{a}_\varphi^{(0)} = -L_{-1} + \frac{2\pi}{k}\bar{\mathcal{S}}S^0. \tag{B.2b}$$

***A*-sector:** The gauge transformations which preserve (B.1) are given by

$$\begin{aligned}\epsilon^{(0)} =& \epsilon^1 L_1 + \epsilon^2 L_0 + \epsilon^3 L_{-1} + \epsilon^4 T_1^+ + \epsilon^5 T_0^+ + \epsilon^6 T_{-1}^+ + \epsilon^7 T_1^0 + \epsilon^8 T_0^0\\ &+ \epsilon^9 T_{-1}^0 + \epsilon^{10} T_1^- + \epsilon^{11} T_0^- + \epsilon^{12} T_{-1}^- + \epsilon^{13} S^+ + \epsilon^{14} S^0 + \epsilon^{15} S^-,\end{aligned} \tag{B.3}$$

with

$$\epsilon^1 = \epsilon, \quad \epsilon^2 = -\sigma\epsilon', \quad \epsilon^3 = \frac{1}{2}\epsilon'' - \frac{\pi\sigma}{k}\left(\mathcal{T}^-\epsilon^+ + \mathcal{T}^+\epsilon^- + \mathcal{T}^0\epsilon^0 + \mathcal{L}\epsilon\right), \tag{B.4a}$$

$$\epsilon^4 = \epsilon^+, \quad \epsilon^5 = -\sigma\epsilon^{+\prime} + \frac{\pi\sigma}{k}\left(2\mathcal{S}^0\epsilon^+ + \mathcal{S}^+\epsilon^0\right), \tag{B.4b}$$

$$\begin{aligned}\epsilon^6 =&\frac{1}{2}\epsilon^{+\prime\prime}+\frac{\pi}{2k^2}\left(-2k\epsilon^+\left(\mathcal{S}^0\right)'-k\epsilon^0\left(\mathcal{S}^+\right)'-4k\mathcal{S}^0\epsilon^{+\prime}\right.\\&-2k\mathcal{S}^+\epsilon^{0\prime}-2k\sigma\mathcal{L}\epsilon^+-4k\sigma\epsilon\mathcal{T}^+-2\pi\mathcal{S}^-\mathcal{S}^+\epsilon^+\\&\left.-2\pi\left(\mathcal{S}^+\right)^2\epsilon^-+4\pi\left(\mathcal{S}^0\right)^2\epsilon^++2\pi\mathcal{S}^0\mathcal{S}^+\epsilon^0\right),\end{aligned}\tag{B.4c}$$

$$\epsilon^7 =\epsilon^0,\quad \epsilon^8=-\sigma\epsilon^{0\prime}-\frac{2\pi\sigma}{k}\left(\mathcal{S}^-\epsilon^++\mathcal{S}^+\epsilon^-\right),\tag{B.4d}$$

$$\begin{aligned}\epsilon^9 =&\frac{1}{2}\epsilon^{0\prime\prime}+\frac{\pi}{k^2}\left(k\epsilon^+\left(\mathcal{S}^-\right)'+k\epsilon^-\left(\mathcal{S}^+\right)'+2k\mathcal{S}^-\epsilon^{+\prime}\right.\\&+2k\mathcal{S}^+\epsilon^{-\prime}-k\sigma\mathcal{L}\epsilon^0-4k\sigma\epsilon\mathcal{T}^0-2\pi\mathcal{S}^-\mathcal{S}^0\epsilon^+\\&\left.-2\pi\mathcal{S}^-\mathcal{S}^+\epsilon^0+2\pi\mathcal{S}^0\mathcal{S}^+\epsilon^-\right),\end{aligned}\tag{B.4e}$$

$$\epsilon^{10} =\epsilon^-,\quad \epsilon^{11}=-\sigma\epsilon^{-\prime}-\frac{\pi\sigma}{k}\left(2\mathcal{S}^0\epsilon^--\mathcal{S}^-\epsilon^0\right),\tag{B.4f}$$

$$\begin{aligned}\epsilon^{12} =&\frac{1}{2}\epsilon^{-\prime\prime}+\frac{\pi}{2k^2}\left(-k\epsilon^0\left(\mathcal{S}^-\right)'+2k\epsilon^-\left(\mathcal{S}^0\right)'-2k\mathcal{S}^-\epsilon^{0\prime}\right.\\&+4k\mathcal{S}^0(\varphi)\epsilon^{-\prime}-2k\sigma\mathcal{L}\epsilon^--4k\sigma\epsilon\mathcal{T}^--2\pi\mathcal{S}^-\mathcal{S}^0\epsilon^0\\&\left.+4\pi\left(\mathcal{S}^0\right)^2\epsilon^--2\pi\left(\mathcal{S}^-\right)^2\epsilon^+-2\pi\mathcal{S}^-\mathcal{S}^+\epsilon^-\right),\end{aligned}\tag{B.4g}$$

$$\epsilon^{13} =\epsilon_0^+,\quad \epsilon^{14}=\epsilon_0^0,\quad \epsilon^{15}=\epsilon_0^-,\tag{B.4h}$$

and

$$\epsilon^{(1)}=\mathcal{O}(e^{-2\rho}),\tag{B.5}$$

where I suppressed again the dependence of the gauge parameters on the angular coordinate $\varphi$.

Having determined the boundary condition preserving gauge transformations the next step is to determine the behavior of the state dependent functions under those gauge transformations. The non-trivial gauge transformations which are relevant for determining the corresponding Dirac brackets are given by

$$\delta_\epsilon\mathcal{L}=\left(\mathcal{L}'\epsilon+2\mathcal{L}\epsilon'\right)\sigma-\frac{k}{2\pi}\epsilon''',\tag{B.6a}$$

$$\delta_\epsilon\mathcal{T}^\pm =\sigma\left(-\frac{2\pi}{k}\left(\mathcal{T}^0\mathcal{S}^\pm+\mathcal{T}^\pm\mathcal{S}^0\right)\epsilon+\left(\mathcal{T}^\pm\right)'\epsilon+2\mathcal{T}^\pm\epsilon'\right),\tag{B.6b}$$

$$\delta_\epsilon\mathcal{T}^0 =\sigma\left(\frac{\pi}{k}\left(\mathcal{T}^+\mathcal{S}^-+\mathcal{T}^-\mathcal{S}^+\right)\epsilon+\left(\mathcal{T}^0\right)'\epsilon+2\mathcal{T}^0\epsilon'\right),\tag{B.6c}$$

$$\begin{aligned}\delta_{\epsilon^\pm}\mathcal{T}^\pm =&\sigma\left(\frac{1}{2}\mathcal{L}'\epsilon^\pm+\mathcal{L}\epsilon^{\pm\prime}\right)-\frac{k}{4\pi}\epsilon^{\pm\prime\prime\prime}\pm\frac{3}{2}\mathcal{S}^{0\prime}\epsilon^{\pm\prime}\pm\frac{1}{2}\mathcal{S}^{0\prime\prime}\epsilon^\pm\pm\frac{3}{2}\mathcal{S}^0\epsilon^{\pm\prime\prime}\\&+\frac{\pi}{k}\left(\frac{1}{2}\mathcal{S}^-\mathcal{S}^{+\prime}+\mathcal{S}^+\mathcal{S}^{-\prime}-3\mathcal{S}^0\mathcal{S}^{0\prime}\mp2\sigma\mathcal{L}\mathcal{S}^0\mp\frac{2\pi}{k}\mathcal{S}^-\mathcal{S}^+\mathcal{S}^0\pm\frac{2\pi}{k}\left(\mathcal{S}^0\right)^3\right)\epsilon^\pm\\&+\frac{3\pi}{k}\left(\frac{1}{2}\mathcal{S}^-\mathcal{S}^+-\left(\mathcal{S}^0\right)^2\right)\epsilon^{\pm\prime},\end{aligned}\tag{B.6d}$$

$$\begin{aligned}\delta_{\epsilon^\pm}\mathcal{T}^0 =&-\frac{3}{4}\mathcal{S}^{\mp\prime}\epsilon^{\pm\prime}-\frac{1}{4}\mathcal{S}^{\mp\prime\prime}\epsilon^\pm-\frac{3}{4}\mathcal{S}^\mp\epsilon^{\pm\prime}\pm\frac{3\pi}{2k}\mathcal{S}^\mp\mathcal{S}^0\epsilon^{\pm\prime}\\&+\frac{\pi}{k}\left(\sigma\mathcal{L}\mathcal{S}^\mp+\frac{\pi}{k}\left(\mathcal{S}^\mp\right)^2\mathcal{S}^\pm-\frac{\pi}{k}\mathcal{S}^\mp\left(\mathcal{S}^0\right)^2\pm\frac{1}{2}\mathcal{S}^0\mathcal{S}^{\mp\prime}\pm\mathcal{S}^\mp\mathcal{S}^{0\prime}\right)\epsilon^+,\end{aligned}\tag{B.6e}$$

$$\delta_{\epsilon^\pm}\mathcal{T}^\mp = \frac{3\pi}{2k}\left(\mathcal{S}^\mp\mathcal{S}^{\mp\prime}\epsilon^\pm + \left(\mathcal{S}^\mp\right)^2\epsilon^{\pm\prime}\right), \tag{B.6f}$$

$$\delta_{\epsilon^0}\mathcal{T}^0 = \frac{\sigma}{2}\left(\frac{1}{2}\mathcal{L}'\epsilon^0 + \mathcal{L}\epsilon^{0\prime}\right) - \frac{k}{8\pi}\epsilon^{0\prime\prime\prime} + \frac{3\pi}{4k}\left(\mathcal{S}^+\mathcal{S}^{-\prime} + \mathcal{S}^-\mathcal{S}^{+\prime} + 2\mathcal{S}^-\mathcal{S}^+\right)\epsilon^0, \tag{B.6g}$$

$$\delta_{\epsilon^a}\mathcal{S}^b = (a+b)(1-b(1+b-2ab))\mathcal{T}^{b-a}\epsilon^a, \tag{B.6h}$$

$$\delta_{\epsilon_0^a}\mathcal{S}^b = (a+b)\mathcal{S}^{b-a}\epsilon_0^a + \frac{k}{2\pi}(1-3a^2)\epsilon_0^{a\prime}\delta_{a,b}. \tag{B.6i}$$

The canonical boundary charge is given by

$$\mathcal{Q}[\varepsilon] = \int \mathrm{d}\varphi\left(\mathcal{L}\epsilon + \sum_{a=-1}^{1}\mathcal{T}^a\epsilon^{-a} + \mathcal{S}^a\epsilon_0^{-a}\right). \tag{B.7}$$

Calculating the Poisson brackets resulting from these BCPGTs one finds again that a Sugawara shift of the form

$$\mathcal{L} \to \hat{\mathcal{L}} = \mathcal{L} + \frac{\pi\sigma}{k}\left(\left(\mathcal{S}^0\right)^2 - \mathcal{S}^+\mathcal{S}^-\right), \tag{B.8}$$

is necessary in order to make all appearing fields proper Virasoro primaries. After having done this shift one obtains the following Dirac brackets

$$\{\hat{\mathcal{L}}(\varphi), \hat{\mathcal{L}}(\bar{\varphi})\} = \sigma\left(2\hat{\mathcal{L}}\delta' - \hat{\mathcal{L}}'\delta\right) - \frac{k}{2\pi}\delta''', \tag{B.9a}$$

$$\{\hat{\mathcal{L}}(\varphi), \mathcal{T}^a(\bar{\varphi})\} = \sigma\left(2\mathcal{T}^a\delta' - \mathcal{T}^{a\prime}\delta\right), \tag{B.9b}$$

$$\{\hat{\mathcal{L}}(\varphi), \mathcal{S}^a(\bar{\varphi})\} = \sigma\mathcal{S}^a\delta', \tag{B.9c}$$

$$\{\mathcal{S}^a(\varphi), \mathcal{S}^b(\bar{\varphi})\} = (a-b)\mathcal{S}^{a+b}\delta + \frac{k}{2\pi}(1-3a^2)\delta'\delta_{a+b,0}, \tag{B.9d}$$

$$\{\mathcal{S}^a(\varphi), \mathcal{T}^b(\bar{\varphi})\} = (a-b)(1-a(1+a+2ab))\mathcal{T}^{a+b}\delta, \tag{B.9e}$$

$$\{\mathcal{T}^\pm(\varphi), \mathcal{T}^\pm(\bar{\varphi})\} = \frac{3\pi}{2k}\left(-\mathcal{S}^\pm\mathcal{S}^{\pm\prime}\delta + \left(\mathcal{S}^\pm\right)^2\delta'\right), \tag{B.9f}$$

$$\begin{aligned}\{\mathcal{T}^0(\varphi), \mathcal{T}^0(\bar{\varphi})\} =& \frac{1}{4}\left(\sigma\left(2\hat{\mathcal{L}}\delta' - \hat{\mathcal{L}}'\delta\right) - \frac{k}{2\pi}\delta'''\right) + \frac{\pi}{k}\left(2\mathcal{S}^-\mathcal{S}^+ - \frac{1}{2}\left(\mathcal{S}^0\right)^2\right)\delta' \\ &- \frac{\pi}{k}\left(\mathcal{S}^+\mathcal{S}^{-\prime} + \mathcal{S}^-\mathcal{S}^{+\prime} - \frac{1}{2}\mathcal{S}^0\mathcal{S}^{0\prime}\right)\delta,\end{aligned} \tag{B.9g}$$

$$\begin{aligned}\{\mathcal{T}^\pm(\varphi), \mathcal{T}^0(\bar{\varphi})\} =& \frac{1}{4}\left(-3\mathcal{S}^{\pm\prime}\delta' + \mathcal{S}^{\pm\prime\prime}\delta + 3\mathcal{S}^\pm\delta''\right) + \frac{2\pi^2}{k^2}\left(\left(\mathcal{S}^0\right)^2\mathcal{S}^\pm - \left(\mathcal{S}^\pm\right)^2\mathcal{S}^\mp\right)\delta \\ &- \frac{\pi\sigma}{k}\mathcal{L}\mathcal{S}^\pm\delta \pm \frac{\pi}{k}\left(\left(\frac{1}{2}\mathcal{S}^0\mathcal{S}^{\pm\prime} + \mathcal{S}^\pm\mathcal{S}^{0\prime}\right)\delta - \frac{3}{2}\mathcal{S}^\pm\mathcal{S}^0\delta'\right),\end{aligned} \tag{B.9h}$$

$$\begin{aligned}\{\mathcal{T}^-(\varphi), \mathcal{T}^+(\bar{\varphi})\} =& \frac{1}{2}\left(\sigma\left(2\hat{\mathcal{L}}\delta' - \hat{\mathcal{L}}'\delta\right) - \frac{k}{2\pi}\delta'''\right) + \frac{1}{2}\left(3\mathcal{S}^{\pm\prime}\delta' - \mathcal{S}^{\pm\prime\prime}\delta - 3\mathcal{S}^\pm\delta''\right) \\ &+ \frac{2\pi\sigma}{k}\hat{\mathcal{L}}\mathcal{S}^0\delta + \frac{4\pi^2}{k^2}\left(\mathcal{S}^-\mathcal{S}^+\mathcal{S}^0 - \left(\mathcal{S}^0\right)^3\right)\delta \\ &+ \frac{\pi}{k}\left(4\mathcal{S}^0\mathcal{S}^{0\prime} - \mathcal{S}^-\mathcal{S}^{+\prime} - \frac{3}{2}\mathcal{S}^+\mathcal{S}^{-\prime}\right) + \frac{\pi}{k}\left(\frac{5}{2}\mathcal{S}^-\mathcal{S}^+ - 4\left(\mathcal{S}^0\right)^2\right)\delta'.\end{aligned} \tag{B.9i}$$

Using the following mode expansions

$$\hat{\mathcal{L}}(\varphi) = \frac{\sigma}{2\pi}\sum_{n\in\mathbb{Z}} L_n e^{-in\varphi}, \qquad \mathcal{T}^a(\varphi) = \frac{1}{2\pi}\sum_{n\in\mathbb{Z}} T^a_n e^{-in\varphi}, \tag{B.10a}$$

$$\mathcal{S}^a(\varphi) = \frac{i}{2\pi}\sum_{n\in\mathbb{Z}} S^a_n e^{-in\varphi}, \quad \delta(\varphi-\bar{\varphi}) = \frac{1}{2\pi}\sum_{n\in\mathbb{Z}} e^{-in(\varphi-\bar{\varphi})}, \tag{B.10b}$$

and replacing $i\{.,.\} \to [.,.]$ one arrives at the classical commutation relations displayed in (8.30).

$\bar{A}$**-sector:** The $\bar{A}$ sector with the boundary conditions (B.2) is considerably simpler as the sector treated before as there is only one state dependent function involved. Thus the gauge transformations which preserve (B.2) are also much simpler and given by

$$\bar{\epsilon}^{(0)} = \bar{\epsilon}_0 S^0, \tag{B.11}$$

and

$$\bar{\epsilon}^{(1)} = \mathcal{O}(e^{-2\rho}). \tag{B.12}$$

The only non-trivial gauge transformations are thus

$$\delta_{\bar{\epsilon}_0}\bar{\mathcal{S}} = \frac{k}{2\pi}\bar{\epsilon}_0'. \tag{B.13}$$

The canonical boundary charge is then given by

$$\bar{\mathcal{Q}}[\bar{\varepsilon}] = \int \mathrm{d}\,\varphi \bar{\mathcal{S}}\bar{\epsilon}_0, \tag{B.14}$$

which immediately implies the following Dirac brackets

$$\{\bar{\mathcal{S}}(\varphi), \bar{\mathcal{S}}(\bar{\varphi})\} = \frac{k}{2\pi}\delta'. \tag{B.15}$$

In terms of Fourier modes

$$\bar{\mathcal{S}}(\varphi) = \frac{1}{2\pi}\sum_{n\in\mathbb{Z}} \bar{S}_n e^{-in\varphi}, \quad \delta(\varphi-\bar{\varphi}) = \frac{1}{2\pi}\sum_{n\in\mathbb{Z}} e^{-in(\varphi-\bar{\varphi})} \tag{B.16}$$

and upon replacing $i\{.,.\} \to [.,.]$ one finally obtains (8.32).

## B.2 Canonical Analysis of the 3-1 Embedding

The connection $A$ whose fluctuations are compatible with (8.5) has to obey the following boundary conditions

$$a_\rho^{(0)} = L_0, \qquad a_t^{(0)} = 0, \qquad a_\mu^{(1)} = \mathcal{O}(e^{-2\rho}), \tag{B.17a}$$

$$a_\varphi^{(0)} = \sigma L_1 + \frac{2\pi}{k}\left(-\frac{1}{4}\mathcal{L}L_{-1} + \mathcal{T}^+T_{-1}^+ + \mathcal{T}^-T_{-1}^- + \mathcal{W}W_{-2} + \frac{4}{3}\mathcal{S}S\right). \tag{B.17b}$$

The connection $\bar{A}$ on the other hand has to satisfy the following boundary conditions

$$\bar{a}_\rho^{(0)} = -L_0, \qquad \bar{a}_\varphi^{(0)} = \frac{8\pi}{3k}\bar{\mathcal{S}}S, \qquad \bar{a}_t^{(0)} = \frac{4\sqrt{2}}{\sqrt{3}}S, \tag{B.18a}$$

$$\begin{aligned}\bar{a}_\rho^{(1)} = \bar{a}_t^{(1)} = \mathcal{O}(e^{-2\rho})&\left(\sum_{i=-1}^{1}\left(L_i + T_i^+ + T_i^-\right) + \sum_{i=0}^{2} W_i + S\right)\\ &+ \mathcal{O}(e^{-3\rho})W_{-1} + \mathcal{O}(e^{-4\rho})W_{-2},\end{aligned} \tag{B.18b}$$

$$\bar{a}_\varphi^{(1)} = \mathcal{O}(e^{-2\rho})\left(\sum_{i=-1}^{1}\left(L_i + T_i^+ + T_i^-\right) + \sum_{i=-1}^{2} W_i + S\right) + \mathcal{O}(e^{-3\rho})W_{-2}. \tag{B.18c}$$

**$A$-sector:** The gauge transformations which preserve (B.17) are given by

$$\begin{aligned}\epsilon^{(0)} =& \epsilon^1 L_1 + \epsilon^2 L_0 + \epsilon^3 L_{-1} + \epsilon^4 T_1^+ + \epsilon^5 T_0^+ + \epsilon^6 T_{-1}^+ + \epsilon^7 T_1^- + \epsilon^8 T_0^-\\ &+ \epsilon^9 T_{-1}^- + \epsilon^{10} W_2 + \epsilon^{11} W_1 + \epsilon^{12} W_0 + \epsilon^{13} W_{-1} + \epsilon^{14} W_{-2} + \epsilon^{15} S,\end{aligned} \tag{B.19}$$

with

$$\epsilon^1 = \epsilon, \quad \epsilon^2 = -\sigma\epsilon', \quad \epsilon^3 = \frac{1}{2}\epsilon'' - \frac{\pi\sigma}{2k}\left(\mathcal{T}^-\epsilon^+ + \mathcal{T}^+\epsilon^- + 2\mathcal{W}\epsilon_\mathcal{W} + \mathcal{L}\epsilon\right), \tag{B.20a}$$

$$\epsilon^4 = \epsilon^+, \quad \epsilon^5 = -\sigma\left(\epsilon^{+\prime} + \frac{2}{3k}\left(4\pi\mathcal{S}\epsilon^+ + 3\pi\mathcal{T}^+\epsilon_\mathcal{W}\right)\right), \tag{B.20b}$$

$$\begin{aligned}\epsilon^6 =& \frac{1}{2}\epsilon^{+\prime\prime} + \frac{\pi}{2k}\left(2\epsilon_\mathcal{W}\left(\mathcal{T}^+\right)' + 3\mathcal{S}'\epsilon^+ + \sigma\left(6\mathcal{T}^+\epsilon - \mathcal{L}\epsilon^+\right)\right.\\ &\left. + \frac{16}{3}\mathcal{S}\epsilon^{+\prime} + 3\mathcal{T}^+\epsilon_\mathcal{W}' + \frac{16\pi}{3k}\left(\mathcal{S}\mathcal{T}^+\epsilon_\mathcal{W} + \frac{4}{3}\mathcal{S}^2\epsilon^+\right)\right),\end{aligned} \tag{B.20c}$$

$$\epsilon^7 = \epsilon^-, \quad \epsilon^8 = -\sigma\left(\epsilon^{-\prime} - \frac{2}{3k}\left(4\pi\mathcal{S}\epsilon^- - 3\pi\mathcal{T}^-\epsilon_\mathcal{W}\right)\right), \tag{B.20d}$$

$$\begin{aligned}\epsilon^9 =& \frac{1}{2}\epsilon^{-\prime\prime} - \frac{\pi}{2k}\left(2\epsilon_\mathcal{W}\left(\mathcal{T}^-\right)' + 3\mathcal{S}'\epsilon^- - \sigma\left(6\mathcal{T}^-\epsilon - \mathcal{L}\epsilon^-\right)\right.\\ &\left. + \frac{16}{3}\mathcal{S}\epsilon^{-\prime} + 3\mathcal{T}^-\epsilon_\mathcal{W}' - \frac{16\pi}{3k}\left(\mathcal{S}\mathcal{T}^-\epsilon_\mathcal{W} + \frac{4}{3}\mathcal{S}^2\epsilon^-\right)\right),\end{aligned} \tag{B.20e}$$

$$\epsilon^{10} = \epsilon_\mathcal{W}, \quad \epsilon^{11} = -\sigma\epsilon_\mathcal{W}' - \frac{\pi}{k}\left(2\mathcal{S}^0\epsilon^- - \mathcal{S}^-\epsilon^0\right), \tag{B.20f}$$

$$\epsilon^{12} = \frac{1}{2}\epsilon_\mathcal{W}'' - \frac{\sigma\pi}{k}\mathcal{L}\epsilon_\mathcal{W}, \tag{B.20g}$$

$$\epsilon^{13} = -\frac{\sigma}{6}\epsilon_{\mathcal{W}}''' + \frac{\pi}{6k}\left(4\sigma\left(\mathcal{T}^{+}\epsilon^{-} - \mathcal{T}^{-}\epsilon^{+}\right) + 2\epsilon_{\mathcal{W}}\mathcal{L}' + 5\mathcal{L}\epsilon_{\mathcal{W}}'\right), \tag{B.20h}$$

$$\begin{aligned}\epsilon^{14} =& \frac{1}{24}\epsilon_{\mathcal{W}}'''' + \frac{\pi}{6k}\left(\epsilon^{+}\left(\mathcal{T}^{-}\right)' - \epsilon^{-}\left(\mathcal{T}^{+}\right)' + 4\left(\mathcal{T}^{-}\epsilon^{+\prime} - \mathcal{T}^{+}\epsilon^{-\prime}\right)\right.\\ &+\sigma\left(12\mathcal{W}\epsilon - \frac{1}{2}\epsilon_{\mathcal{W}}\mathcal{L}'' - \frac{7}{2}\mathcal{L}'\epsilon_{\mathcal{W}}' - 2\mathcal{L}\epsilon_{\mathcal{W}}''\right)\\ &\left.+\frac{8\pi}{k}\mathcal{S}\mathcal{T}^{-}\epsilon^{+} + \frac{8\pi}{k}\mathcal{S}\mathcal{T}^{+}\epsilon^{-} + \frac{12\pi}{k}\mathcal{T}^{-}\mathcal{T}^{+}\epsilon_{\mathcal{W}} + \frac{3\pi}{2k}\mathcal{L}^{2}\epsilon_{\mathcal{W}}\right),\end{aligned} \tag{B.20i}$$

$$\epsilon^{15} = \epsilon^{0}. \tag{B.20j}$$

and

$$\epsilon^{(1)} = \mathcal{O}(e^{-2\rho}). \tag{B.21}$$

This leads to the following non-trivial gauge transformations

$$\delta_{\epsilon}\mathcal{L} = -\frac{k\epsilon'''}{\pi} + \sigma\left(\epsilon\mathcal{L}' + 2\mathcal{L}\epsilon'\right), \tag{B.22a}$$

$$\delta_{\epsilon}\mathcal{T}^{\pm} = \sigma\left(\epsilon\left(\left(\mathcal{T}^{\pm}\right)' \pm \frac{8\pi\mathcal{S}\mathcal{T}^{\pm}}{3k}\right) + 2\mathcal{T}^{\pm}\epsilon'\right), \tag{B.22b}$$

$$\delta_{\epsilon}\mathcal{W} = \sigma\left(\epsilon\mathcal{W}' + 3\mathcal{W}\epsilon'\right), \tag{B.22c}$$

$$\delta_{\epsilon^{\pm}}\mathcal{L} = \sigma\left(\epsilon^{\pm}\left(\left(\mathcal{T}^{\pm}\right)' \pm \frac{8\pi\mathcal{S}\mathcal{T}^{\mp}}{3k}\right) + 2\mathcal{T}^{\mp}\epsilon^{\pm\prime}\right), \tag{B.22d}$$

$$\begin{aligned}\delta_{\epsilon^{\pm}}\mathcal{T}^{\pm} =& \epsilon^{\pm}\left(\pm\frac{128\pi^{2}\mathcal{S}^{3}}{27k^{2}} + \frac{16\pi\mathcal{S}\mathcal{S}'}{3k} \mp \frac{4\pi\sigma\mathcal{S}\mathcal{L}}{3k} \pm \frac{2\mathcal{S}''}{3} \mp \mathcal{W} - \frac{\sigma\mathcal{L}'}{4}\right)\\ &+\left(\frac{16\pi\mathcal{S}^{2}}{3k} \pm 2\mathcal{S}' - \frac{\mathcal{L}}{2}\right)\epsilon^{\pm\prime} + \frac{k\epsilon^{\pm\prime\prime\prime}}{4\pi} \pm 2\mathcal{S}\epsilon^{\pm\prime\prime},\end{aligned} \tag{B.22e}$$

$$\delta_{\epsilon^{\pm}}\mathcal{W} = \epsilon^{\pm}\left(\frac{2\pi}{3k}\left(\mathcal{S}\left(\mathcal{T}^{\pm}\right)' + 3\mathcal{T}^{\mp}\mathcal{S}' \mp \sigma\mathcal{L}\mathcal{T}^{\mp}\right) \pm \frac{\left(\mathcal{T}^{\mp}\right)''}{12} \pm \frac{32\pi^{2}\mathcal{S}^{2}\mathcal{T}^{\mp}}{9k^{2}}\right) \tag{B.22f}$$

$$+\epsilon^{\pm\prime}\left(\pm\frac{5}{12}\left(\mathcal{T}^{\mp}\right)' + \frac{10\pi\mathcal{S}\mathcal{T}^{\mp}}{3k}\right) \pm \frac{5}{6}\mathcal{T}^{\mp}\epsilon^{\pm\prime\prime}, \tag{B.22g}$$

$$\delta_{\epsilon^{\pm}}\mathcal{S} = \mp\mathcal{T}^{\mp}\epsilon^{\pm}, \quad \delta_{\epsilon_{\mathcal{W}}}\mathcal{L} = \sigma\left(2\mathcal{W}'\epsilon_{\mathcal{W}} + 3\mathcal{W}\epsilon_{\mathcal{W}}'\right), \tag{B.22h}$$

$$\begin{aligned}\delta_{\epsilon_{\mathcal{W}}}\mathcal{T}^{\pm} =& \epsilon_{\mathcal{W}}\left(\frac{2\pi}{3k}\left(4\mathcal{S}\left(\mathcal{T}^{\pm} + 2\mathcal{T}^{\pm}\mathcal{S}'\right)' \mp \sigma\mathcal{L}\mathcal{T}^{\pm}\right) \pm \frac{1}{2}\left(\mathcal{T}^{\pm}\right)'' \pm \frac{32\pi^{2}\mathcal{S}^{2}\mathcal{T}^{\pm}}{9k^{2}}\right)\\ &+\epsilon_{\mathcal{W}}'\left(\frac{10\pi\mathcal{S}\mathcal{T}^{\pm}}{3k} \pm \frac{5}{4}\left(\mathcal{T}^{\pm}\right)'\right) \pm \frac{5}{6}\mathcal{T}^{\pm}\epsilon_{\mathcal{W}}'',\end{aligned} \tag{B.22i}$$

$$\begin{aligned}\delta_{\epsilon_{\mathcal{W}}}\mathcal{W} =& \epsilon_{\mathcal{W}}\left(\frac{2\pi}{k}\left(\mathcal{T}^{+}\left(\mathcal{T}^{-}\right)' + \mathcal{T}^{-}\left(\mathcal{T}^{+}\right)'\right) + \sigma\left(\frac{\pi\mathcal{L}\mathcal{L}'}{3k} - \frac{\mathcal{L}'''}{24}\right)\right) + \frac{k\epsilon_{\mathcal{W}}'''''}{48\pi}\\ &+\epsilon_{\mathcal{W}}'\left(\frac{4\pi}{k}\mathcal{T}^{-}\mathcal{T}^{+} + \sigma\left(\frac{\pi\mathcal{L}^{2}}{3k} - \frac{3\mathcal{L}''}{16}\right)\right) - \frac{5\sigma}{16}\left(\mathcal{L}'\epsilon_{\mathcal{W}}'' + \frac{2}{3}\mathcal{L}\epsilon_{\mathcal{W}}'''\right),\end{aligned} \tag{B.22j}$$

$$\delta_{\epsilon^{0}}\mathcal{T}^{\pm} = \mp\mathcal{T}^{\pm}\epsilon^{0}, \quad \delta_{\epsilon^{0}}\mathcal{S} = \frac{3k\epsilon^{0\prime}}{8\pi}. \tag{B.22k}$$

The canonical charge is given by

$$\mathcal{Q}[\varepsilon] = \int \mathrm{d}\varphi \left(\mathcal{L}\epsilon + \mathcal{T}^{+}\epsilon^{-} + \mathcal{T}^{-}\epsilon^{+} + \mathcal{W}\epsilon_{\mathcal{W}} + \mathcal{S}\epsilon^{0}\right). \tag{B.23}$$

As in the previous examples one has to perform a Sugawara shift of the form

$$\mathcal{L} \rightarrow \hat{\mathcal{L}} = \mathcal{L} + \frac{4\pi\sigma}{3k}\mathcal{S}\mathcal{S}, \tag{B.24}$$

in order to bring the resulting algebra of Dirac brackets into a form where all fields are proper Virasoro primaries. Using the gauge transformations and canonical boundary charge obtained before one finds that the state dependent functions obey the following Dirac bracket algebra

$$\{\hat{\mathcal{L}}(\varphi), \hat{\mathcal{L}}(\bar{\varphi})\} = \sigma\left(2\hat{\mathcal{L}}\delta' - \hat{\mathcal{L}}'\delta\right) - \frac{k}{2\pi}\delta''', \tag{B.25a}$$

$$\{\hat{\mathcal{L}}(\varphi), \mathcal{S}(\bar{\varphi})\} = \sigma\mathcal{S}\delta', \tag{B.25b}$$

$$\{\hat{\mathcal{L}}(\varphi), \mathcal{T}^{\pm}(\bar{\varphi})\} = \sigma\left(2\mathcal{T}^{\pm}\delta' - \mathcal{T}^{\pm\prime}\delta\right), \tag{B.25c}$$

$$\{\hat{\mathcal{L}}(\varphi), \mathcal{W}(\bar{\varphi})\} = \sigma\left(3\mathcal{W}\delta' - \mathcal{W}'\delta\right), \tag{B.25d}$$

$$\{\mathcal{S}(\varphi), \mathcal{S}(\bar{\varphi})\} = \frac{3k}{8\pi}\delta', \tag{B.25e}$$

$$\{\mathcal{S}(\varphi), \mathcal{T}^{\pm}(\bar{\varphi})\} = \pm\,\mathcal{T}^{\pm}\delta, \tag{B.25f}$$

$$\begin{aligned}\{\mathcal{T}^{\pm}(\varphi), \mathcal{T}^{\mp}(\bar{\varphi})\} = &\left(-\frac{\sigma}{2}\mathcal{L} + \frac{14\pi}{3k}\mathcal{S}^2 \mp 2\mathcal{S}'\right)\delta' \pm 2\mathcal{S}\delta'' + \frac{k}{4\pi}\delta''' \\ &+ \left(\frac{\sigma}{4}\mathcal{L}' \mp \frac{4\pi\sigma}{3k}\mathcal{L}\mathcal{S} \pm \frac{80\pi^2}{27k^2}\mathcal{S}^2 \mp \mathcal{W} - \frac{7\pi}{3k}\left(\mathcal{S}^2\right)' \pm \frac{2}{3}\mathcal{S}''\right)\delta,\end{aligned} \tag{B.25g}$$

$$\begin{aligned}\{\mathcal{W}(\varphi), \mathcal{T}^{\pm}(\bar{\varphi})\} = &\left(\frac{10\pi}{3k}\mathcal{S}\mathcal{T}^{\pm} \pm \frac{5}{4}\left(\mathcal{T}\pm\right)'\right)\delta' \mp \frac{5}{6}\mathcal{T}^{\pm}\delta'' \\ &- \left(\frac{2\pi}{3k}\left(2\mathcal{T}^{\pm}\mathcal{S}' \mp \sigma\mathcal{L}\mathcal{T}^{\pm} + 4\mathcal{S}\left(\mathcal{T}^{\pm}\right)'\right) \pm \frac{24\pi^2}{9k^2}\mathcal{S}^2\mathcal{T}^{\pm} \pm \frac{\left(\mathcal{T}^{\pm}\right)''}{2}\right)\delta,\end{aligned} \tag{B.25h}$$

$$\begin{aligned}\{\mathcal{W}(\varphi), \mathcal{W}(\bar{\varphi})\} = &\left(\frac{\pi}{k}\left(\frac{\sigma\mathcal{L}^2}{3} + 4\mathcal{T}^{+}\mathcal{T}^{-} - \frac{\left(\mathcal{S}^2\right)'}{4}\right) + \frac{8\pi^2}{9k^2}\sigma\mathcal{L}\mathcal{S}^2 + \frac{16\pi^3}{27k^3}\mathcal{S}^4 - \frac{3\sigma\mathcal{L}''}{16}\right)\delta' \\ &+ \left(\frac{\mathcal{L}'''}{24} - \frac{\pi}{3k}\sigma\mathcal{L}\mathcal{L}' - \frac{4\pi^2}{9k^2}\sigma\left(\mathcal{L}\mathcal{S}^2\right)' - \frac{8\pi^3}{27k^3}\left(\mathcal{S}^4\right)'\right. \\ &\left. - \frac{2\pi}{k}\left(\mathcal{T}^{+}\mathcal{T}^{-}\right)' + \frac{\pi}{18k}\left(\mathcal{S}^2\right)'''\right)\delta + \frac{5}{16}\sigma\mathcal{L}'\delta'' \\ &+ \frac{5\pi}{12k}\left(\mathcal{S}^2\right)'\delta'' - \frac{5}{24}\sigma\mathcal{L}\delta''' - \frac{5\pi}{18k}\mathcal{S}^2\delta''' + \frac{k}{48\pi}\delta'''''.\end{aligned} \tag{B.25i}$$

Defining the Fourier modes as

$$\hat{\mathcal{L}} = \frac{1}{2\pi}\sum_{n\in\mathbb{Z}} L_n e^{-in\varphi}, \qquad \mathcal{T}^{\pm} = \frac{1}{2\pi}\sum_{n\in\mathbb{Z}} T_n^{\pm} e^{-in\varphi}, \tag{B.26a}$$

$$\mathcal{S} = \frac{i}{2\pi}\sum_{n\in\mathbb{Z}} S_n e^{-in\varphi}, \qquad \mathcal{W} = -\frac{i}{2\pi}\sum_{n\in\mathbb{Z}} W_n e^{-in\varphi} \tag{B.26b}$$

$$\delta(\varphi-\bar{\varphi}) = \frac{1}{2\pi}\sum_{n\in\mathbb{Z}} e^{-in(\varphi-\bar{\varphi})}, \tag{B.26c}$$

employing the following zero mode shift

$$L_0 \to L_0 - k\delta_{n,0}, \tag{B.27}$$

and upon replacing $i\{.,.\} \to [.,.]$ one obtains the semi-classical commutation relations (8.36).

$\bar{A}$**-sector:** The $\bar{A}$ sector with the boundary conditions (B.18) is considerably simpler than the sector treated before as there is only one state dependent function involved. Thus, the gauge transformations which preserve (B.18) are also much simpler and given by

$$\bar{\epsilon}^{(0)} = \bar{\epsilon}_0 S^0, \tag{B.28}$$

and

$$\bar{\epsilon}^{(1)} = \mathcal{O}(e^{-2\rho})\left(\sum_{i=-1}^{1}\left(L_i + T_i^{\pm}\right) + \sum_{i=0}^{2} W_i\right) + \mathcal{O}(e^{-3\rho})W_{-1} + \mathcal{O}(e^{-4\rho})W_{-2}. \tag{B.29}$$

The only non-trivial gauge transformations are thus

$$\delta_{\bar{\epsilon}_0}\bar{\mathcal{S}} = \frac{3k}{8\pi}\bar{\epsilon}_0'. \tag{B.30}$$

The canonical boundary charge is then given by

$$\bar{\mathcal{Q}}[\bar{\varepsilon}] = \int \mathrm{d}\varphi\, \bar{\mathcal{S}}\bar{\epsilon}_0, \tag{B.31}$$

which immediately implies the following Dirac brackets

$$\{\bar{\mathcal{S}}(\varphi), \bar{\mathcal{S}}(\bar{\varphi})\} = \frac{3k}{8\pi}\delta'. \tag{B.32}$$

In terms of Fourier modes

$$\bar{\mathcal{S}}(\varphi) = \frac{i}{2\pi}\sum_{n\in\mathbb{Z}} \bar{S}_n e^{-in\varphi}, \quad \delta(\varphi-\bar{\varphi}) = \frac{1}{2\pi}\sum_{n\in\mathbb{Z}} e^{-in(\varphi-\bar{\varphi})} \tag{B.33}$$

and upon replacing $i\{.,.\} \to [.,.]$ one finally obtains (8.37).

# Quantum $\mathcal{W}$-Algebras and Jacobi Identities

# C

In this appendix I show in detail how one can obtain the quantum version of the $\mathcal{W}$-algebras considered in chapter 8. The easiest way to obtain the quantum $\mathcal{W}$-algebra from the semiclassical one is by first introducing an appropriate normal ordering description for the nonlinear terms appearing in the algebra. Then one has to take into account possible additional terms on the right-hand side of the commutation relations which might be suppressed for large values of the central charges but will be present for finite values. In the same manner the structure constants of the $\mathcal{W}$-algebra will be different for a finite value of the central charges. Thus, the final step in this procedure will be to assume modified structure constants whose precise form will be fixed by the requirement that the algebra has to satisfy the Jacobi identities.

## C.1 Normal Ordering Definitions and Shorthands

In order to find the quantized version of the classical $\mathcal{W}$-algebras used in this thesis I define normal ordering for bilinear terms in the following way

$$:AB:_n = \sum_{p\geq -s+1} A_{n-p}B_p + \sum_{p<-s+1} B_p A_{n-p}, \tag{C.1}$$

where $s$ is the spin of the operator $B$. Consequently this leads to the following expression for the normal ordered trilinear terms

$$\begin{aligned} :ABC:_n = ::AB:C:_n &= \sum_{q\geq -c+1} :AB:_{n-q} C_q + \sum_{q<-c+1} C_q :AB:_{n-q} \\ &= \sum_{\substack{p\geq -b+1\\ q\geq -c+1}} A_{n-p-q}B_pC_q + \sum_{\substack{p<-b+1\\ q\geq -c+1}} B_pA_{n-p-q}C_q \\ &\quad + \sum_{\substack{p\geq -b+1\\ q<-c+1}} C_qA_{n-p-q}B_p + \sum_{\substack{p<-b+1\\ q<-c+1}} C_qB_pA_{n-p-q}, \end{aligned} \tag{C.2}$$

where $b, c$ denote the spin of the operators $B$ and $C$ respectively.

## C.2 Convenient Identities

In this section I will collect some identities of nonlinear operators which have proven to be very useful when calculating the Jacobi identities.

**Differences of nonlinear operators:**

$$\sum_{i\in\mathbb{Z}} i\,(:A_{-i+n}B_{i+p}: - :A_{-i+n+p}B_i:) = -p\sum_{i\in\mathbb{Z}} :A_{-i+n+p}B_i: + \sum_{i=0}^{p-1}(i-p)[B_i, A_{-i+m+n+p}], \tag{C.3a}$$

$$\sum_{i\in\mathbb{Z}} (:A_{-i+m+n}B_{i+p}: - :A_{-i+n+p}B_{i+m}:) = \sum_{i=0}^{p-m-1} [B_{i+m}, A_{-i+n+p}]. \tag{C.3b}$$

**Other useful identities:**

$$[A_n, \sum_p :B_{m-p}C_p:] = \sum_p (:[A_n, B_{m-p}]C_p: + :B_{n+m-p}[A_n, C_{p-n}]:) + \sum_{p=-[a,c]+1}^{n-c} [[A_n, C_{p-n}], B_{n+m-p}], \tag{C.4a}$$

$$[A_n, \sum_p p:B_{m-p}C_p:] = \sum_p (p:[A_n, B_{m-p}]C_p: + (p-n):B_{n+m-p}[A_n, C_{p-n}]:) + \sum_{p=-[a,c]+1}^{n-c} (p-n)[[A_n, C_{p-n}], B_{n+m-p}], \tag{C.4b}$$

$$\begin{aligned}[A_n, \sum_{p,q} :B_{m-p-q}C_pD_q:] = &\sum_q :\sum_p (:[A_n, B_{m-p-q}]C_p: + :B_{n+m-p-q}[A_n, C_{p-n}]:)\, D_q: \\ &+ \sum_q :\sum_{p=-[a,c]+1}^{n-c} [[A_n, C_{p-n}], B_{n+m-p-q}]D_q: \\ &+ \sum_q :\sum_p :B_{n+m-p-q}C_p: [A_n, D_{q-n}]: \\ &+ \sum_{q=-[a,d]+1}^{n-d} [[A_n, D_{q-n}], \sum_p :B_{n+m-p-q}C_p:],\end{aligned} \tag{C.4c}$$

$$\sum_p :A_{n-p}B_p: = \sum_p :B_{n-p}A_p: + \sum_{p=-a+1}^{n+b-1} [A_p, B_{n-p}], \tag{C.4d}$$

$$\sum_p p:A_{n-p}B_p: = \sum_p (n-p):B_{n-p}A_p: + \sum_{p=-a+1}^{n+b-1} (n-p)[A_p, B_{n-p}]. \tag{C.4e}$$

In the expressions above $a, b, c, d$ represent the spin of the operators $A, B, C, D$ respectively and $[a, c]$ and $[a, d]$ are the spin of the operator on the right hand side of the commutators $[A, C]$ and $[A, D]$.

## C.3 Quantum $\mathcal{W}$-Algebras

### C.3.1 Quantum $\mathcal{W}_4^{(2-1-1)}$

Starting from (8.17) and after introducing normal ordering for the nonlinear operators as in (C.3) and (C.4) I assume that the quantum $\mathcal{W}_4^{(2-1-1)}$ algebra has the following form

$$\begin{aligned}
[\hat{L}_n, \hat{L}_m] =& (n-m)\hat{L}_{n+m} + \frac{\hat{\mathfrak{c}}}{12} n(n^2-1)\delta_{m+n,0}, && \text{(C.5a)}\\
[\hat{L}_n, J^a_m] =& -mJ^a_{n+m}, && \text{(C.5b)}\\
[\hat{L}_n, S_m] =& -mS_{n+m}, && \text{(C.5c)}\\
[\hat{L}_n, G^{a|b}_m] =& \left(\frac{n}{2}-m\right) G^{a|b}_{n+m}, && \text{(C.5d)}\\
[J^a_n, J^b_m] =& (a-b)J^{a+b}_{m+n} + C_1\left(1-3a^2\right) n\delta_{a+b,0}\delta_{n+m,0}, && \text{(C.5e)}\\
[J^a_n, G^{b|c}_m] =& \frac{a-b}{2} G^{2a+b|c}_{n+m}, && \text{(C.5f)}\\
[S_n, G^{a|b}_m] =& 2bG^{a|b}_{n+m}, && \text{(C.5g)}\\
[S_n, S_m] =& C_2 n\delta_{n+m}, && \text{(C.5h)}\\
[G^{a|b}_n, G^{c|d}_m] =& \delta_{b+d,0}\left(\delta_{a,c}\left(C_3(m-n)J^c_{n+m} + dC_4\left\{:SJ:\right\}^c_{n+m}\right)\right.\\
& +\delta_{a+c,0}\left(cC_5\hat{L}_{n+m} - \frac{cd}{2}C_6(m-n)S_{n+m} + C_7(m-n)J^0_{n+m}\right.\\
& +cC_8\left(n^2-\frac{1}{4}\right)\delta_{n+m,0} + c\left(C_9 :JJ:^{\{+|-\}}_{n+m} + C_{10} :SS:_{n+m}\right.\\
& \left.\left.\left. +cdC_{11}\left\{:SJ:\right\}^0_{n+m} + C_{12} :JJ:^{0|0}_{n+m}\right)\right)\right), && \text{(C.5i)}
\end{aligned}$$

for some structure constants $\hat{\mathfrak{c}}, C_1, \ldots, C_{12}$ where in addition I used again the shorthand notations (8.18) but with the normal ordered expressions for the nonlinear operators instead of the semiclassical ones.
After calculating the Jacobi identities one finds that the following equations have to be satisfied

$$\begin{gathered}
2C_3 = C_2C_4, \quad 4C_8 = -C_2C_6, \quad C_5 + 2C_6 = -2C_{10}C_2, \quad 2C_7 = C_{11}C_2,\\
C_3 = C_7, \quad C_6 = 4C_4C_1, \quad C_4 = C_{11}, \quad C_5 + C_3 + 2C_{12} = 4C_9C_1,\\
2C_9 = -C_{12}, \quad \hat{\mathfrak{c}}C_5 - 6C_8 + 2C_{10}C_2 + 2C_{12}C_1 - 8C_9C_1 = 0,\\
3C_5 - 2C_6 - C_7 - 4C_{11} - 2C_9 + \frac{1}{2}C_{12} + 8C_{10} = 0, \quad C_{12} = -4C_{11},\\
C_9 = 2C4, \quad 4C_{10} = -C_{11} - 2C_4, \quad C_5 - C_3 = 4C_4,\\
-2C_6 + C_7 + C_5 = 8C_4, \quad 2C_3 = 2C_6 - C_7 + C_5, \quad C_8 = -2C_1C_3,
\end{gathered} \quad \text{(C.6)}$$

where the highlighted parts are the quantum corrections which modify the semiclassical relations among the structure constants. Now imposing $C_2 = -4k$ and $C_5 = 1$[1] one can solve the system of equations (C.6), which leads to the following set of coefficients

$$
\begin{gathered}
C_1 = -\frac{k+1}{2}, \quad C_2 = -4k, \quad C_3 = \frac{k}{k-2}, \quad C_4 = -\frac{1}{2(k-2)}, \\
C_5 = 1, \quad C_6 = \frac{k+1}{k-2}, \quad C_7 = \frac{k}{k-2}, \quad C_8 = \frac{k(k+1)}{k-2}, \\
C_9 = -\frac{1}{k-2}, \quad C_{10} = \frac{3}{8(k-2)}, \quad C_{11} = -\frac{1}{2(k-2)}, \quad C_{12} = \frac{2}{k-2}, \\
\hat{\mathfrak{c}} = \frac{3(k+2)(2k+1)}{k-2}.
\end{gathered} \tag{C.7}
$$

After applying the following rescaling of the $G$-modes

$$G_n^{a|b}\sqrt{-(k-2)} \to G_n^{a|b}, \tag{C.8}$$

one obtains the quantum $\mathcal{W}_4^{2-1-1}$-algebra displayed in (9.10).

## C.3.2 Quantum $\mathcal{W}_4^{(2-2)}$

In order to determine the quantum $\mathcal{W}_4^{(2-2)}$-algebra I assume the following deformations of the original classical algebra (8.30)

$$[L_n, L_m] = (n-m)L_{n+m} + \frac{C_0}{12}n(n^2-1)\delta_{n+m,0}, \tag{C.9a}$$

$$[L_n, T_m^a] = (n-m)T_{n+m}^a, \tag{C.9b}$$

$$[L_n, S_m^a] = -mS_{n+m}^a, \tag{C.9c}$$

$$[S_n^a, S_m^b] = (a-b)S_{n+m}^{a+b} + C_1(1-3a^2)n\delta_{a+b,0}\delta_{n+m,0}, \tag{C.9d}$$

$$[S_n^a, T_m^b] = f(a,b)T_{n+m}^{a+b}, \tag{C.9e}$$

$$[T_n^\pm, T_m^\pm] = C_2(n-m) :SS:_{n+m}^{\pm|\pm}, \tag{C.9f}$$

$$
\begin{aligned}
[T_n^0, T_m^0] = &C_3(n-m)L_{n+m} + C_5(n-m) :SS:_{n+m}^{0|0} \\
&+ C_6(n-m) :SS:_{n+m}^{\{+|-\}} + C_7 n(n^2-1)\delta_{n+m,0},
\end{aligned} \tag{C.9g}
$$

$$
\begin{aligned}
[T_n^\pm, T_m^0] = &\hat{g}(n,m)S_{n+m}^\pm \pm \Big((C_{91}n + C_{92}m) :SS:_{n+m}^{0|\pm} + C_{93} :\Omega SS:_{n+m}^{[0|\pm]} \\
&-(C_{92}n + C_{91}m) :SS:_{n+m}^{\pm|0}\Big) + C_{10}\{:SL:\}_{n+m}^\pm \\
&+ C_{11}\left(:SSS:_{n+m}^{\{\pm|\pm|\mp\}} - :SSS:_{n+m}^{\{0|0|\pm\}}\right),
\end{aligned} \tag{C.9h}
$$

[1]This choice can be motivated by the requirement that the quantum algebra has to agree with the semiclassical one in the limit of large central charges, or in this case equivalently for large $k$.

$$
\begin{aligned}
[T^-_n, T^+_m] =& C_{18}(n-m)L_{n+m} + \hat{g}(n,m)S^0_{n+m} + C_{20}n(n^2-1)\delta_{n+m,0} \\
&+ C_{21}(n-m) :SS:^{0|0}_{n+m} + \Big((C_{221}n + C_{222}m) :SS:^{+|-}_{n+m} \\
&+ C_{223} :\Omega SS:^{[+|-]}_{n+m} - (C_{222}n + C_{221}m) :SS:^{-|+}_{n+m}\Big) \\
&+ C_{23}\{:SL:\}^0_{n+m} + C_{24} :SSS:^{\{-|+|0\}}_{n+m} + C_{25} :SSS:^{0|0|0}_{n+m},
\end{aligned}
\tag{C.9i}
$$

with

$$
\hat{g}(n,m) = C_{81} + C_{82}n^2 + C_{83}m^2 + C_{84}nm + C_{85}n + C_{86}m. \tag{C.10}
$$

With the help of (C.4) and by denoting the difference between the quantum and semiclassical commutation relations as $\Delta[A, :B:] = [A, :B:] - [A, B]$, for some linear generator $A$ and nonlinear generator $B$, one obtains the following quantum corrections

$$
\begin{aligned}
\Delta[L_n, :\Omega SS:^{a|b}_m] =& (a-b)\frac{n}{6}(1+3n+2n^2)S^{a+b}_{n+m} \\
&- C_1(1-3a^2)\frac{n^2}{12}(n^2-1)\delta_{a+b,0}\delta_{n+m,0},
\end{aligned}
\tag{C.11a}
$$

$$
\begin{aligned}
\Delta[S^a_n, :\Omega SS:^{b|c}_m] =& -(a-c)\left((a+c-b)\frac{n}{2}(n+1)S^{a+b+c}_{n+m}\right. \\
&\left.+ C_1(1-3b^2)\frac{n}{6}(n^2-1)\delta_{a+b+c,0}\delta_{n+m,0}\right),
\end{aligned}
\tag{C.11b}
$$

$$
\Delta[T^a_n, :\Omega SS:^{b|c}_m] = -f(c,a)f(b,a+c)\frac{1}{2}(n+1)(n+2)T^{a+b+c}_{n+m}, \tag{C.11c}
$$

$$
\begin{aligned}
\Delta[L_n, :SS:^{a|b}_m] =& -(a-b)\frac{n}{2}(n+1)S^{a+b}_{n+m} \\
&+ C_1\frac{n}{6}(n^2-1)(1-3a^2)\delta_{a+b,0}\delta_{n+m,0},
\end{aligned}
\tag{C.11d}
$$

$$
\begin{aligned}
\Delta[S^a_n, :SS:^{b|c}_m] =& (a-c)\left((a+c-b)nS^{a+b+c}_{n+m}\right. \\
&\left.+ C_1(1-3b^2)\frac{n}{2}(n-1)\delta_{a+b+c,o}\delta_{n+m,0}\right),
\end{aligned}
\tag{C.11e}
$$

$$
\Delta[T^a_n, :SS:^{b|c}_m] = f(c,a)f(b,a+c)(n+1)T^{a+b+c}_{n+m}, \tag{C.11f}
$$

$$
\Delta[L_n, :LS:^a_m] = \frac{n}{6}(n^2-1)S^a_{n+m}, \tag{C.11g}
$$

$$
\Delta[S^a_n, :LS:^b_m] = (a-b)\frac{n}{2}(n-1)S^{a+b}_{n+m}, \tag{C.11h}
$$

$$
\Delta[T^a_n, :LS:^b_m] = f(b,a)(m+2)(n+1)T^{a+b}_{n+m}, \tag{C.11i}
$$

$$
\Delta[L_n, :SL:^a_m] = -\frac{n}{6}(n+1)(13+9m+5n)S^a_{n+m}, \tag{C.11j}
$$

$$
\begin{aligned}
\Delta[S^a_n, :SL:^b_m] =& (a-b)n(n-1)S^{a+b}_{n+m} \\
&+ C_1(1-3a^2)\frac{n}{2}(n-1)(n-2)\delta_{a+b,0}\delta_{n+m,0},
\end{aligned}
\tag{C.11k}
$$

$$
\Delta[T^a_n, :SL:^b_m] = -f(b,a)\frac{3}{2}n(n+1)T^{a+b}_{n+m}, \tag{C.11l}
$$

$$\begin{aligned}\Delta[L_n, :SSS:^{a|b|c}_m] =& \frac{n}{2}(n+1)\left((b-a) :SS:^{a+b|c}_{n+m} +(c-a) :SS:^{a+c|b}_{n+m}\right.\\ &+(c-b) :SS:^{a|b+c}_{n+m}\Big) + \frac{n}{6}(n^2-1)\Big((c-b)(c+b-a)S^{a+b+c}_{n+m}\\ &+C_1(1-3c^2)\left(S^b_{n+m}\delta_{a+c,0} + S^a_{n+m}\delta_{b+c,0}\right)\\ &+C_1(1-3b^2)\delta_{a+b,0}S^c_{n+m}\Big)\\ &+ (c-b)C_1(1-3a^2)\frac{n}{24}(n^2-1)(n-2)\delta_{a+b+c,0}\delta_{n+m,0},\end{aligned} \tag{C.11m}$$

$$\begin{aligned}\Delta[S^a_n, :SSS:^{b|c|d}_m] =& n\Big((a-c)(a+c-b) :SS:^{a+b+c|d}_{n+m}\\ &+ (a-d)\left((a+d-b) :SS:^{a+b+d|c}_{n+m} +(a+d-c) :SS:^{b|a+c+d}_{n+m}\right)\Big)\\ &+ C_1\frac{n}{2}(n-1)\Big((a-c)(1-3b^2)\delta_{a+b+c,0}S^d_{n+m}\\ &+(a-d)\left((1-3b^2)\delta_{a+b+d,0}S^c_{n+m} + (1-3c^2)\delta_{a+c+d,0}S^b_{n+m}\right)\Big)\\ &+ (a-d)(a+d-c)(a+d+c-b)\frac{n}{2}(n-1)S^{a+b+c+d}_{n+m}\\ &+ (a-d)(a+d-c)C_1(1-3b^2)\frac{n}{6}(n-1)(n-2)\delta_{a+b+c+d,0}\delta_{n+m,0},\end{aligned} \tag{C.11n}$$

$$\begin{aligned}\Delta[T^a_n, :SSS:^{b|c|d}_m] =& (n+1)\Big(f(c,a)f(b,a+c) :TS:^{a+b+c|d}_{n+m}\\ &+f(d,a)f(b,a+d) :TS:^{a+b+d|c}_{n+m}\\ &+f(d,a)f(c,a+d) :ST:^{b|a+c+d}_{n+m}\Big)\\ &- f(d,a)f(c,a+d)f(b,a+c+d)\frac{n}{2}(n+1)T^{a+b+c+d}_{n+m}.\end{aligned} \tag{C.11o}$$

Calculating the Jacoby identities with these deformations and quantum corrections leads to the following set of equations for the deformed structure constants

$$\begin{gathered}C_0 = \frac{-12C_1^2+7C_1+2}{C_1+2},\quad C_2 = \frac{(3C_1+2)C_3}{2C_1(C_1+2)},\quad C_5 = -\frac{(C_1-2)C_3}{2C_1(C_1+2)},\quad C_6 = \frac{C_3}{C_1+2},\\ C_7 = -\frac{\left(6C_1^2+C_1-2\right)C_3}{6(C_1+2)},\quad C_{82} = \frac{\left(-12C_1^2+3C_1+2\right)C_3+12C_1(C_1+2)C_{92}}{12C_1(C_1+2)},\\ C_{81} = \frac{\left(3C_1^2-3C_1-8\right)C_3}{3C_1(C_1+2)},\quad C_{83} = \frac{\left(3C_1^2+3C_1-2\right)C_3+6C_1(C_1+2)C_{92}}{3C_1(C_1+2)},\\ C_{84} = \frac{\left(-12C_1^2+3C_1+2\right)C_3+12C_1(C_1+2)C_{92}}{12C_1(C_1+2)},\quad C_{11} = \frac{2C_3}{3C_1^2+6C_1},\quad C_{10} = \frac{C_3}{C_1},\\ C_{85} = \frac{4C_1(C_1+2)C_{92}-(3C_1+10)C_3}{4C_1(C_1+2)},\quad C_{86} = \frac{4C_1(C_1+2)C_{92}-(3C_1+10)C_3}{4C_1(C_1+2)},\\ C_{91} = \frac{(3C_1+2)C_3+2C_1(C_1+2)C_{92}}{2C_1(C_1+2)},\quad C_{93} = \frac{C_3}{2C_1},\quad C_{20} = -\frac{\left(6C_1^2+C_1-2\right)C_3}{3(C_1+2)},\\ C_{18} = 2C_3,\quad C_{223} = -\frac{C_3}{2C_1},\quad C_{23} = \frac{2C_3}{C_1},\quad C_{24} = \frac{2C_3}{3C_1^2+6C_1},\quad C_{25} = -\frac{4C_3}{C_1(C_1+2)},\\ C_{222} = -\frac{(C_1+2)C_{92}+2C_3}{C_1+2},\quad C_{21} = -\frac{4C_3}{C_1+2},\quad C_{221} = \frac{(C_1-2)C_3-2C_1(C_1+2)C_{92}}{2C_1(C_1+2)}.\end{gathered} \tag{C.12}$$

If one chooses

$$C_3 = \frac{1}{4} \quad \text{and} \quad C_{92} = -\frac{1}{4C_1}, \tag{C.13}$$

and solves

$$C_0 = \frac{-12C_1^2 + 7C_1 + 2}{C_1 + 2}, \tag{C.14}$$

in terms of $C_0$, while keeping in mind that for large values of $C_0$ the semi-classical expression $C_0 = -12C_1$ has to be satisfied, one obtains the following quantum $\mathcal{W}_4^{(2-2)}$-algebra

$$[L_n, L_m] =(n-m)L_{n+m} + \frac{c}{12}n(n^2-1)\delta_{n+m,0}, \tag{C.15a}$$

$$[L_n, T^a_m] =(n-m)T^a_{n+m}, \tag{C.15b}$$

$$[L_n, S^a_m] =-\, mS^a_{n+m}, \tag{C.15c}$$

$$[S^a_n, S^b_m] =(a-b)S^{a+b}_{n+m} + \kappa(1-3a^2)n\delta_{a+b,0}\delta_{n+m,0}, \tag{C.15d}$$

$$[S^a_n, T^b_m] =f(a,b)T^{a+b}_{n+m}, \tag{C.15e}$$

$$[T^\pm_n, T^\pm_m] =\frac{3\kappa+2}{8\kappa(\kappa+2)}(n-m) :SS:^{\pm|\pm}_{n+m}, \tag{C.15f}$$

$$\begin{aligned}[T^0_n, T^0_m] =&\frac{1}{4}(n-m)L_{n+m} - \frac{\kappa-2}{8\kappa(\kappa+2)}(n-m) :SS:^{0|0}_{n+m} \\ &+ \frac{1}{4(\kappa+2)}(n-m) :SS:^{\{+|-\}}_{n+m} - \frac{(2\kappa-1)(3\kappa+2)}{24(\kappa+2)}n(n^2-1)\delta_{n+m,0},\end{aligned} \tag{C.15g}$$

$$\begin{aligned}[T^\pm_n, T^0_m] =&q(n,m)S^\pm_{n+m} \pm \left(\frac{\kappa-2}{8\kappa(\kappa+2)}n - \frac{1}{4\kappa}m\right) :SS:^{0|\pm}_{n+m} \pm \frac{1}{8\kappa} :\Omega SS:^{[0|\pm]}_{n+m} \\ &\pm \left(\frac{1}{4\kappa}n - \frac{\kappa-2}{8\kappa(\kappa+2)}m\right) :SS:^{\pm|0}_{n+m} + \frac{1}{4\kappa}\{:LS:\}^\pm_{n+m} \\ &+ \frac{1}{6\kappa(\kappa+2)}\left(:SSS:^{\{\pm|\pm|\mp\}}_{n+m} - :SSS:^{\{0|0|\pm\}}_{n+m}\right),\end{aligned} \tag{C.15h}$$

$$\begin{aligned}[T^-_n, T^+_m] =&\frac{1}{2}(n-m)L_{n+m} + 2q(n,m)S^0_{n+m} \\ &- \frac{(2\kappa-1)(3\kappa+2)}{12(\kappa+2)}n(n^2-1)\delta_{n+m,0} - \frac{1}{\kappa+2}(n-m) :SS:^{0|0}_{n+m} \\ &+ \left(\frac{3\kappa+2}{8\kappa(\kappa+2)}n - \frac{\kappa-2}{4\kappa(\kappa+2)}m\right) :SS:^{+|-}_{n+m} - \frac{1}{8\kappa} :\Omega SS:^{[+|-]}_{n+m} \\ &+ \left(\frac{\kappa-2}{4\kappa(\kappa+2)}n - \frac{3\kappa+2}{8\kappa(\kappa+2)}m\right) :SS:^{-|+}_{n+m} + \frac{1}{2\kappa}\{:LS:\}^0_{n+m} \\ &+ \frac{1}{6\kappa(\kappa+2)} :SSS:^{\{-|+|0\}}_{n+m} - \frac{1}{\kappa(\kappa+2)} :SSS:^{0|0|0}_{n+m},\end{aligned} \tag{C.15i}$$

with

$$
\begin{aligned}
q(n,m) =& \frac{3\kappa^2-3\kappa-8}{12\kappa(\kappa+2)} - \frac{12\kappa^2+9\kappa+22}{48\kappa(\kappa+2)}(n^2+m^2) \\
&+ \frac{3\kappa^2-3\kappa-14}{12\kappa(\kappa+2)}nm - \frac{7\kappa+18}{16\kappa(\kappa+2)}(n+m), && \text{(C.16a)} \\
\kappa =& \frac{1}{24}\left(7-c-\sqrt{c^2-110c+145}\right). && \text{(C.16b)}
\end{aligned}
$$

and

$$
c = 12k. \tag{C.17}
$$

### C.3.3 Quantum $\mathcal{W}_4^{(2)}$

In contrast to the other embeddings treated in this thesis determining the quantum $\mathcal{W}_4^{(2)}$ algebra in the same way as before is not a very efficient course of action. The reason for this is that this algebra is a special case of a Feigin-Semikhatov algebra which are usually denoted $\mathcal{W}_N^{(2)}$ whose OPEs have been first described in [136]. Thus, one can simply convert the OPEs found in [136] into commutation relations in order to obtain the quantum $\mathcal{W}_4^{(2)}$ algebra. This procedure yields

$$
\begin{aligned}
[L_n, L_m] =& (n-m)L_{n+m} + \frac{c}{12}n(n^2-1)\delta_{n+m,0}, && \text{(C.18a)} \\
[L_n, J_n] =& -mJ_{n+m}, && \text{(C.18b)} \\
[L_n, \hat{T}^{\pm}_m] =& (n-m)\hat{T}^{\pm}_{n+m}, && \text{(C.18c)} \\
[L_n, \hat{W}_m] =& (2n-m)\hat{W}_{n+m}, && \text{(C.18d)} \\
[J_n, J_m] =& -\frac{3k}{4}n\delta_{n+m}, && \text{(C.18e)} \\
[J_n, \hat{T}^{\pm}_m] =& \pm\hat{T}^{\pm}_{n+m}, && \text{(C.18f)} \\
[\hat{T}^{\pm}_n, \hat{T}^{\mp}_m] =& -\frac{1}{4}(n-m)L_{n+m} \mp \hat{W}_{n+m} + \frac{k(6k+1)}{4(4-3k)}n(n^2-1)\delta_{n+m,0} \\
&\pm \left(\frac{2(3\hat{k^2}-k+2)}{3k(3k-4)}(n^2+m^2) - \frac{6k^2+7k-8}{3k(3k-4)}(1+mn) - \frac{1}{k}(n+m)\right) J_{n+m} \\
&+ \frac{9}{4(4-3k)}(n-m) :JJ:_{n+m} \pm \frac{1}{3k}\{:LJ:\}_{n+m} \\
&\pm \frac{4(33k-8)}{27k^2(3k-4)} :JJJ:_{n+m}, && \text{(C.18g)}
\end{aligned}
$$

$$
\begin{aligned}
[\hat{W}_n, \hat{T}^{\pm}_m] = &\Bigg(\pm\frac{16-102k+72k^2+27k^3}{27k^2(3k+2)} \mp \frac{-32+60k+18k^2+27k^3}{108k^2(3k+2)}n^2 \\
&\pm\frac{(3k-2)(3k+8)}{12k(3k+2)}nm \mp \frac{3k-1}{2(3k+2)}m^2 \pm \frac{(3k-8)(3k-1)}{9k^2(3k+2)}n \\
&\pm\frac{21k-16}{6k(3k+2)}m\Bigg)\hat{T}^{\pm}_{n+m} \mp \frac{3k-4}{6k(3k+2)}\{:L\hat{T}^{\pm}:\}_{n+m} \\
&-\left(\frac{15k-8}{9k^2(3k+2)} + \frac{3}{2(3k+2)}n - \frac{9k^2+3k-8}{9k^2(3k+2)}m\right):J\hat{T}^{\pm}:_{n+m} \\
&-\left(\frac{15k-8}{9k^2(3k+2)} + \frac{(3k-2)(3k+8)}{18k^2(3k+2)}n - \frac{2(3k-1)}{3k(3k+2)}m\right):\hat{T}^{\pm}J:_{n+m} \\
&+\frac{9k^2-9k+8}{9k^2(3k+2)}\left(:\Omega J\hat{T}^{\pm}:_{n+m} - :\Omega\hat{T}^{\pm}J:_{n+m}\right) \\
&\mp\frac{4}{3k(3k+2)}\{:JJ\hat{T}^{\pm}:\}_{n+m},
\end{aligned}
\tag{C.18h}
$$

$$
\begin{aligned}
[\hat{W}_n, \hat{W}_m] = &\hat{g}(n,m)\left(L_{n+m} + \frac{2}{3k}:JJ:_{n+m}\right) - \frac{3k+8}{27k^2(3k+2)}(n-m)\{:LJJ:\}_{n+m} \\
&-\frac{(3k+1)(6k+1)(15k-8)}{2160k(3k-4)}(n^2-4)(n^2-1)n\delta_{n+m,0} \\
&+\frac{3}{2(3k+2)}(n-m)\left((n+m+3)\hat{W}_{n+m} - :\hat{T}\hat{T}:^{\{+|-\}}_{n+m}\right) \\
&-\frac{6k+1}{4(3k+2)(3k-4)}(n-m)(n+m+3)(n+m+2)(n+m+1)J_{n+m} \\
&+\left(\frac{6k+1}{3k(3k+2)(3k-4)}(n-m)(11+9m+2m^2+9n+4mn+2n^2)\right):JJ:_{n+m} \\
&+\frac{6k+1}{3k(3k+2)(3k-4)}\left((m^2-n^2):\Omega JJ:_{n+m} + :\Omega^2 JJ:_{n+m}\right) \\
&+\frac{2}{k(3k+2)}(n-m)\{:\hat{W}L:\}_{n+m} - \frac{3k-4}{12k(3k+2)}(n-m):LL:_{n+m} \\
&-\frac{1}{2k(3k+2)}(n-m)(n+m+3)\{:LJ:\}_{n+m} \\
&+\frac{27k^2+12k+8}{6k^2(3k-4)^2(3k+2)}(n-m)(n+m+3):JJJ:_{n+m} \\
&-\frac{9k^2+144k-64}{27k^3(3k-4)(3k+2)}:JJJJ:_{n+m},
\end{aligned}
\tag{C.18i}
$$

with

$$
\begin{aligned}
\hat{g}(n,m) = &-\frac{3}{4k(180k^2+123k-34)}\left(-\frac{16}{9}(3k-4)(15k-8)\right.\\
&+\frac{2}{9}(180k^3-687k^2+407k-198)m-5(18k^2-3k+2)m^2\\
&+\frac{1}{18}(-180k^3-303k^2+64k-20)m^3\\
&-\frac{2}{27}(540k^3+279k^2-3147k+1070)n-\frac{4}{3}(3k-4)(15k-8)nm\\
&+\frac{1}{12}(3k-4)(60k^2+k+10)m^2n+\frac{1}{3}(90k^2+291k-98)n^2\\
&-\frac{1}{36}(3k-4)(180k^2+243k-98)mn^2\\
&\left.-\frac{1}{54}(-540k^3-549k^2-480k+196)\right)\,,
\end{aligned}
\tag{C.19}
$$

and the Virasoro central charge

$$
c=\frac{3k(24k+13)}{3k-4}. \tag{C.20}
$$

Please note that the Chern-Simons level $k$ which appears in all these expressions is *not* the same parameter $k$ as used in [136]. Denoting the level used in [136] as $k_{\text{FS}}$ one can obtain the results presented in this Appendix by a shift (and an additional rescaling of the $T^a_n$ generators)

$$
k_{\text{FS}}=-(k+\frac{8}{3}). \tag{C.21}
$$

# Contracting $\mathcal{W} \to \mathcal{FW}$

# D

This appendix is a collection of non-relativistic contractions of various $\mathcal{W}$-algebras which will be of interest for answering questions about unitarity of flat space higher-spin theories.

## D.1 $\mathcal{W}_4^{(2-1-1)} \to \mathcal{FW}_4^{(2-1-1)}$

I start with two copies of the $\mathcal{W}_4^{(2-1-1)}$ algebra, generated by $\mathcal{L}_n, J_n, S_n^\pm, S_n^0, G_n^{a|b}$, with $a, b = \pm$,

$$[\mathcal{L}_n,\, \mathcal{L}_m] = (n-m)\mathcal{L}_{n+m} + \frac{c}{12}\, n(n^2-1)\, \delta_{m+n,\,0}, \tag{D.1a}$$

$$[\mathcal{L}_n,\, S_m^a] = -\, m S_{n+m}^a, \tag{D.1b}$$

$$[\mathcal{L}_n,\, J_m] = -\, m J_{n+m}, \tag{D.1c}$$

$$[\mathcal{L}_n,\, G_m^{a|b}] = (\frac{n}{2} - m)\, G_{n+m}^{a|b}, \tag{D.1d}$$

$$[S_n^a,\, S_m^b] = (a-b) S_{m+n}^{a+b} - \frac{k+1}{2}\,(1-3a^2)\, n\, \delta_{a+b,\,0}\, \delta_{n+m,\,0}, \tag{D.1e}$$

$$[S_n^a,\, G_m^{b|d}] = \frac{a-b}{2}\, G_{n+m}^{2a+b|d}, \tag{D.1f}$$

$$[J_n,\, G_m^{a|b}] = 2b G_{n+m}^{a|b}, \tag{D.1g}$$

$$[J_n,\, J_m] = -\, 4kn\, \delta_{n+m,\,0}, \tag{D.1h}$$

$$[G_n^{\pm|\pm},\, G_m^{\pm|\mp}] = \frac{k}{\alpha k+\beta}\,(n-m) S_{m+n}^{\pm} \mp \frac{1}{\alpha k+\beta}\; :JS:_{n+m}^{\pm}, \tag{D.1i}$$

$$\begin{aligned}
[G_n^{a|\pm},\, G_m^{-a|\mp}] = {} & a\frac{(k-2)}{\alpha k+\beta}\, \mathcal{L}_{n+m} \mp \frac{a}{2}\, \frac{(k+1)}{\alpha k+\beta}\,(n-m) J_{m+n} + \frac{k}{\alpha k+\beta}\,(n-m) S_{m+n}^0 \\
& a\frac{k(k+1)}{\alpha k+\beta}\,(n^2 - \frac{1}{4})\, \delta_{m+n,\,0} - \frac{a}{\alpha k+\beta}\,\Big(\, :SS:_{n+m}^{\{-|+\}} \\
& - \frac{3}{8}\, :JJ:_{n+m} \mp b\; :JS:_{n+m}^{0} - 2\; :SS:_{n+m}^{0|0} \Big),
\end{aligned} \tag{D.1j}$$

with the central charge

$$c = \frac{3(k+2)(2k+1)}{k-2}\,, \tag{D.2}$$

and $\alpha, \beta \in \mathbb{R}$. I define the following linear combinations

$$L_n := \mathcal{L}_n + \bar{\mathcal{L}}_n, \qquad M_n := -\epsilon\left(\mathcal{L}_n - \bar{\mathcal{L}}_n\right), \tag{D.3a}$$

$$O_n := J_n + \bar{J}_n, \qquad P_n := -\epsilon\left(J_n - \bar{J}_n\right), \tag{D.3b}$$

$$Q^a_n := S^a_n + \bar{S}^a_n, \qquad R^a_n := -\epsilon(S^a_n - \bar{S}^a_n), \tag{D.3c}$$

$$U^{a|b}_n := G^{a|b}_n + \bar{G}^{a|b}_n, \qquad V^{a|b}_n := -\epsilon(G^{a|b}_n - \bar{G}^{a|b}_n). \tag{D.3d}$$

In the limit $\epsilon \to 0$ one obtains the following non-vanishing linear commutation relations

$$[L_n, L_m] = (n-m)L_{m+n} + \frac{c_L}{12} n(n^2-1)\,\delta_{n+m,0}, \tag{D.4a}$$

$$[L_n, M_m] = (n-m)M_{m+n} + \frac{c_M}{12} n(n^2-1)\,\delta_{n+m,0}, \tag{D.4b}$$

$$[L_n, O_m] = -\,mO_{n+m}, \tag{D.4c}$$

$$[L_n, P_m] = -\,mP_{n+m}, \tag{D.4d}$$

$$[L_n, Q^a_m] = -\,mQ^a_{n+m}, \tag{D.4e}$$

$$[L_n, R^a_m] = -\,mR^a_{n+m}, \tag{D.4f}$$

$$[L_n, U^{a|b}_m] = (\frac{n}{2} - m)U^{a|b}_{n+m}, \tag{D.4g}$$

$$[L_n, V^{a|b}_m] = (\frac{n}{2} - m)V^{a|b}_{n+m}, \tag{D.4h}$$

$$[M_n, O_m] = -\,mP_{n+m}, \tag{D.4i}$$

$$[M_n, Q^a_m] = -\,mR^a_{n+m}, \tag{D.4j}$$

$$[M_n, U^{a|b}_m] = (\frac{n}{2} - m)V^{a|b}_{n+m}, \tag{D.4k}$$

$$[O_n, O_m] = \frac{2(54-c_L)}{3}\, n\,\delta_{n+m,0}, \tag{D.4l}$$

$$[O_n, P_m] = -\,\frac{2c_M}{3}\, n\,\delta_{n+m,0}, \tag{D.4m}$$

$$[O_n, U^{a|b}_m] = 2bU^{a|b}_{n+m}, \tag{D.4n}$$

$$[O_n, V^{a|b}_m] = 2bV^{a|b}_{n+m}, \tag{D.4o}$$

$$[P_n, U^{a|b}_m] = 2bV^{a|b}_{n+m}, \tag{D.4p}$$

$$[Q^a_n, Q^b_m] = (a-b)Q^{a+b}_{n+m} + \frac{42-c_L}{12}(1-3a^2)n\,\delta_{a+b,0}\,\delta_{n+m,0}, \tag{D.4q}$$

$$[Q^a_n, R^b_m] = (a-b)R^{a+b}_{n+m} - \frac{c_M}{12}(1-3a^2)n\,\delta_{a+b,0}\,\delta_{n+m,0}, \tag{D.4r}$$

$$[Q^a_n, U^{b|d}_m] = \frac{a-b}{2}U^{2a+b|d}_{n+m}, \tag{D.4s}$$

$$[Q^a_n, V^{b|d}_m] = \frac{a-b}{2}V^{2a+b|d}_{n+m}, \tag{D.4t}$$

$$[R^a_n, U^{b|d}_m] = \frac{a-b}{2}V^{2a+b|d}_{n+m}, \tag{D.4u}$$

and the following nonlinear relations

$$[U_n^{\pm|\pm},\, U_m^{\pm|\mp}] = (n-m)(\frac{1}{\alpha}Q^{\pm}_{n+m} - \frac{12\beta}{\alpha^2 c_M}R^{\pm}_{n+m}) \mp \frac{6}{\alpha c_M}(\,:OR:^{\pm}_{n+m} + :PQ:^{\pm}_{n+m}\,)$$
$$\pm \frac{6(\alpha c_L + 12\beta - 54\alpha)}{\alpha^2 c_M^2}\;:PR:^{\pm}_{n+m}, \tag{D.5a}$$
$$[U_n^{a|\pm},\, U_m^{-a|\mp}] = a(\frac{1}{\alpha}L_{n+m} - \frac{12(2\alpha+\beta)}{\alpha^2 c_M}M_{n+m}) \mp \frac{a}{2}(n-m)(\frac{1}{\alpha}O_{n+m} + \frac{12(\alpha-\beta)}{\alpha^2 c_M}P_{n+m})$$
$$+ (n-m)(\frac{1}{\alpha}Q^0_{n+m} - \frac{12\beta}{\alpha^2 c_M}R^0_{n+m}) + a\frac{\alpha c_L - 6(7\alpha+2\beta)}{6\alpha^2}\,(n^2-\frac{1}{4})\,\delta_{n+m,0}$$
$$- \frac{6}{c_M}\,(\{:QR:\}^{\{+|-\}}_{n+m} - \tfrac{3}{8}\{:OP:\}_{n+m} \pm a(:OR:^0_{n+m} + :PQ:^0_{n+m})$$
$$- 2\{:QR:\}^{0|0}_{n+m}) + \frac{6(ac_L + 12b - 54a)}{ac_M^2}\,(2\;:RR:^{\{-|+\}}_{n+m}$$
$$- \frac{3}{8}\;:PP:_{n+m} \pm a\;:PR:^0_{n+m} - 2\;:RR:^{0|0}_{n+m}\,), \tag{D.5b}$$
$$[U_n^{\pm|\pm},\, V_m^{\pm|\mp}] = (n-m)\frac{1}{\alpha}R^{\pm}_{n+m} \mp \frac{6}{\alpha c_M}\;:PR^{\pm}:_{n+m}$$
$$[U_n^{a|\pm]},\, V_m^{-a|\mp}] = \frac{a}{\alpha}M_{n+m} \mp \frac{a}{2\alpha}(n-m)P_{n+m} + (n-m)\frac{1}{\alpha}R^0_{m+n} + a\frac{c_M}{6\alpha}(n^2-\frac{1}{4})\,\delta_{n+m,0}$$
$$- \frac{6a}{\alpha c_M}\,(2\;:RR:^{\{+|-\}}_{n+m} - \frac{3}{8}\;:PP:_{n+m} \pm a\;:PR:^0_{n+m} - 2\;:RR:^{0|0}_{n+m}\,)\,. \tag{D.5c}$$

## D.2 $\mathcal{W}_4^{(2-2)} \to \mathcal{FW}_4^{(2-2)}$

I contract now two copies of the $\mathcal{W}_4^{(2-2)}$-algebra (C.15) with generators $\mathcal{L}_n, T_n^a, S_n^a$ and $\bar{\mathcal{L}}_n, \bar{T}_n^a, \bar{S}_n^a$ respectively and define the usual linear combinations.

$$L_n := \mathcal{L}_n + \bar{\mathcal{L}}_n, \qquad M_n := -\epsilon\left(\mathcal{L}_n - \bar{\mathcal{L}}_n\right), \tag{D.6a}$$
$$O_n^a := S_n^a + \bar{S}_n^a, \qquad P_n^a := -\epsilon\left(S_n^a - \bar{S}_n^a\right), \tag{D.6b}$$
$$U^a := T_n^a + \bar{T}_n^a, \qquad V^a := -\epsilon\left(T_n^a - \bar{T}_n^a\right). \tag{D.6c}$$

The non-relativistic contraction $\epsilon \to 0$ yields the following algebra.

$$[L_n,\, L_m] = (n-m)L_{m+n} + \frac{c_L}{12}n(n^2-1)\,\delta_{n+m,0}, \tag{D.7a}$$
$$[L_n,\, M_m] = (n-m)M_{m+n} + \frac{c_M}{12}n(n^2-1)\,\delta_{n+m,0}, \tag{D.7b}$$
$$[L_n,\, O_m^a] = -\,mO^a_{n+m}, \tag{D.7c}$$
$$[L_n,\, P_m^a] = -\,mP^a_{n+m}, \tag{D.7d}$$
$$[L_n,\, U_m^a] = (n-m)U^a_{n+m}, \tag{D.7e}$$
$$[L_n,\, V_m^a] = (n-m)V^a_{n+m}, \tag{D.7f}$$
$$[M_n,\, O_m^a] = -\,mP^a_{n+m}, \tag{D.7g}$$

$$[M_n,\, U^a_m] =(n-m)V^a_{n+m}, \tag{D.7h}$$

$$[O^a_n,\, O^b_m] =(a-b)O^{a+b}_{n+m} + \frac{62-c_L}{12}(1-3a^2)n\,\delta_{a+b,\,0}\delta_{n+m,\,0}, \tag{D.7i}$$

$$[O^a_n,\, P^b_m] =(a-b)P^{a+b}_{n+m} - \frac{c_M}{12}(1-3a^2)\,\delta_{a+b,\,0}\delta_{n+m,\,0}, \tag{D.7j}$$

$$[O^a_n,\, U^b_m] =f(a,b)U^{a+b}_{n+m}, \tag{D.7k}$$

$$[O^a_n,\, V^b_m] =f(a,b)V^{a+b}_{n+m}, \tag{D.7l}$$

$$[P^a_n,\, U^b_m] =f(a,b)V^{a+b}_{n+m}, \tag{D.7m}$$

$$[U^\pm_n, V^\pm_m] =\frac{9}{2c_M}(n-m):PP:^{\pm|\pm}_{n+m}, \tag{D.7n}$$

$$\begin{aligned}[U^0_n, V^0_m] =&\frac{1}{4}(n-m)M_{n+m} + \frac{3}{2c_M}(n-m):PP:^{0|0}_{n+m} - \frac{3}{c_M}(n-m):PP:^{\{+|-\}}_{n+m}\\ &+ \frac{c_M}{48}n(n^2-1)\delta_{n+m,\,0},\end{aligned} \tag{D.7o}$$

$$\begin{aligned}[U^\pm_n, V^0_m] =&-\frac{1}{4}g(n,m)P^\pm_{n+m} \mp \frac{3}{2c_M}n:PP:^{0|\pm}_{n+m} \pm \frac{3}{c_M}m:PP:^{0|\pm}_{n+m}\\ &\mp \frac{3}{2c_M}:\Omega PP:^{[0|\pm]}_{n+m} \mp \frac{3}{c_M}n:PP:^{\pm|0}_{n+m} \pm \frac{3}{2c_M}m:PP:^{\pm|0}_{n+m}\\ &- \frac{3}{2c_M}\{:MP:\}^\pm_{n+m} + \frac{24}{c_M^2}\left(:PPP:^{\{\pm|\pm|\mp\}}_{n+m} - :PPP:^{\{0|0|\pm\}}_{n+m}\right),\end{aligned} \tag{D.7p}$$

$$\begin{aligned}[U^-_n, V^+_m] =&\frac{1}{2}(n-m)M_{n+m} - \frac{1}{2}g(n,m)P^0_{n+m} + \frac{c_M}{24}n(n^2-1)\delta_{n+m,0}\\ &(n-m)\frac{12}{c_M}:PP:^{0|0}_{n+m} - \frac{9}{2c_M}n:PP:^{+|-}_{n+m} + \frac{3}{c_M}m:PP:^{+|-}_{n+m}\\ &- \frac{3}{2c_M}:\Omega PP:^{[+|-]}_{n+m} - \frac{3}{c_M}n:PP:^{-|+}_{n+m} + \frac{9}{2c_M}n:PP:^{-|+}_{n+m}\\ &- \frac{3}{c_M}\{:MP:\}^0_{n+m} + \frac{24}{c_M^2}\left(:PPP:^{\{-|+|0\}}_{n+m} - 6:PPP:^{0|0|0}_{n+m}\right),\end{aligned} \tag{D.7q}$$

$$\begin{aligned}[U^\pm_n, U^0_m] =&-\frac{1}{4}g(n,m)O^\pm_{n+m} + p(n,m)\frac{3}{c_M}P^\pm_{m+n}\\ &\pm n\left(\frac{3(c_L-158)}{2c_M^2}:PP:^{0|\pm}_{n+m} - \frac{3}{2c_M}\{:OP:\}^{0|\pm}_{n+m}\right)\\ &\mp m\left(\frac{3(c_L-62)}{c_M^2}:PP:^{0|\pm}_{n+m} - \frac{3}{c_M}\{:OP:\}^{0|\pm}_{n+m}\right)\\ &\pm\left(\frac{3(c_L-62)}{2c_M^2}:\Omega PP:^{[0|\pm]}_{n+m} - \frac{3}{2c_M}\{:\Omega OP:\}^{[0|\pm]}_{n+m}\right)\\ &\pm n\left(\frac{3(c_L-62)}{c_M^2}:PP:^{\pm|0}_{n+m} - \frac{3}{c_M}\{:OP:\}^{\pm|0}_{n+m}\right)\\ &\mp m\left(\frac{3(c_L-158)}{2c_M^2}:PP:^{\pm|0}_{n+m} - \frac{3}{2c_M}\{:OP:\}^{\pm|0}_{n+m}\right)\\ &+ \frac{3(c_L-62)}{2c_M^2}\{:MP:\}^\pm_{n+m} - \frac{3}{2c_M}\left(\{:MO:\}^\pm_{n+m} + \{:LP:\}^\pm_{n+m}\right)\\ &+ \frac{24}{c_M^2}\left(\{:PPO:\}^{\{\pm|\pm|\mp\}}_{n+m} - \{:PPO:\}^{\{0|0|\pm\}}_{n+m}\right)\\ &- \frac{48(c_L-86)}{c_M^3}\left(:PPP:^{\{\pm|\pm|\mp\}}_{n+m} - :PPP:^{\{0|0|\pm\}}_{n+m}\right),\end{aligned} \tag{D.7r}$$

$$
\begin{aligned}
[U^-_n, U^+_m] =& \frac{1}{2}(n-m)L_{n+m} - \frac{1}{2}g(n,m)O^0_{n+m} + p(n,m)\frac{6}{c_M}P^0_{m+n} + \frac{c_L-18}{24}n(n^2-1)\delta_{n+m,0} \\
&+ (n-m)\left(-\frac{12(c_L-110)}{c_M^2}:PP:^{0|0}_{n+m} + \frac{12}{c_M}\{:OP:\}^{0|0}_{n+m}\right) \\
&+ n\left(\frac{9(c_L-94)}{2c_M^2}:PP:^{+|-}_{n+m} - \frac{9}{2c_M}\{:OP:\}^{+|-}_{n+m}\right) \\
&+ m\left(\frac{(474-3c_L)}{c_M^2}:PP:^{+|-}_{n+m} + \frac{3}{c_M}\{:OP:\}^{+|-}_{n+m}\right) \\
&- \left(\frac{3(c_L-62)}{2c_M^2}:\Omega PP:^{[+|-]}_{n+m} - \frac{3}{2c_M}\{:\Omega OP:\}^{[+|-]}_{n+m}\right) \\
&- n\left(\frac{(474-3c_L)}{c_M^2}:PP:^{-|+}_{n+m} + \frac{3}{c_M}\{:OP:\}^{-|+}_{n+m}\right) \\
&- m\left(\frac{9(c_L-94)}{2c_M^2}:PP:^{-|+}_{n+m} - \frac{9}{2c_M}\{:OP:\}^{-|+}_{n+m}\right) \\
&+ \frac{3(c_L-62)}{c_M^2}\{:MP:\}^0_{n+m} - \frac{3}{c_M}\left(\{:MO:\}^0_{n+m} + \{:LP:\}^0_{n+m}\right) \\
&+ \frac{24}{c_M^2}\left(\{:PPO:\}^{\{-|+|0\}}_{n+m} - 6\{:PPO:\}^{0|0|0}_{n+m}\right) \\
&- \frac{48(c_L-86)}{c_M^3}\left(:PPP:^{\{-|+|0\}}_{n+m} - 6:PPP:^{0|0|0}_{n+m}\right), \qquad \text{(D.7s)}
\end{aligned}
$$

$$
[U^\pm_n, U^\pm_m] = \frac{9(c_L-94)}{2c_M^2}(n-m):PP:^{\pm|\pm}_{n+m} + \frac{9}{2c_M}(n-m)\{:OP:\}^{\pm|\pm}_{n+m}, \qquad \text{(D.7t)}
$$

$$
\begin{aligned}
[U^0_n, U^0_m] =& \frac{1}{4}(n-m)L_{n+m} + \frac{474-3c_L}{2c_M^2}(n-m):PP:^{0|0}_{n+m} + \frac{3}{2c_M}\{:OP:\}^{0|0}_{n+m} \\
&+ \frac{3(c_L-110)}{c_M^2}(n-m):PP:^{\{+|-\}}_{n+m} - \frac{3}{c_M}\{:OP:\}^{\{+|-\}}_{n+m} \\
&+ \frac{c_L-18}{48}n(n^2-1)\,\delta_{n+m,0}, \qquad \text{(D.7u)}
\end{aligned}
$$

(D.7v)

with

$$
p(n,m) = -5(n^2+m^2) + 6(1+nm) - 7(n+m). \qquad \text{(D.8)}
$$

## D.3 $\mathcal{W}_N^{(2)} \to \mathcal{FW}_N^{(2)}$

I construct now the non-relativistic contraction of two copies of the Feigin–Semikhatov algebra, which provides an infinite family of examples. Since I will be mainly interested in unitary representations of the resulting algebras I will only take a detailed look at the resulting central charges. Starting with two copies of the following $\mathcal{W}_N^{(2)}$ algebra

$$
\begin{aligned}
[J_n,\, J_n] =& \kappa n \delta_{n+m,\,0}, && \text{(D.9a)}\\
[J_n,\, \mathcal{L}_m] =& n J_{n+m}, && \text{(D.9b)}\\
[J_n,\, G^{\pm}_m] =& \pm\, G^{\pm}_{n+m}, && \text{(D.9c)}\\
[\mathcal{L}_n,\, \mathcal{L}_m] =& (n-m)\mathcal{L}_{n+m} + \frac{c}{12} n(n^2-1)\delta_{n+m,\,0}, && \text{(D.9d)}\\
[\mathcal{L}_n,\, G^{\pm}_m] =& (n(\tfrac{N}{2}-1)-m)\, G^{\pm}_{n+m}, && \text{(D.9e)}\\
[G^{+}_n,\, G^{-}_m] =& \frac{\lambda_{N-1}(N,k)}{(N-1)!}\, f(n)\, \delta_{n+m,\,0} + g(n,m)\lambda_{N-2}(N,k) J_{n+m}\ldots, && \text{(D.9f)}\\
[\hat{W}^s_n,\, \text{anything}] =& \ldots, && \text{(D.9g)}
\end{aligned}
$$

with

$$
\begin{aligned}
\kappa =& \frac{N-1}{N} k + N - 2, && \text{(D.10a)}\\
c =& -\frac{((N+k)(N-1)-N)((N+k)(N-2)N-N^2+1)}{N+k}, && \text{(D.10b)}\\
\lambda_n(N,k) =& \prod_{i=1}^{n} (i(k+N-1)-1), && \text{(D.10c)}
\end{aligned}
$$

and $f(n)$ and $g(n,m)$ being some functions of their respective arguments whose explicit form does not matter for the discussion of unitary representations. In order to have a well defined contraction in the limit $\epsilon \to 0$ some of the generators have to be rescaled first in an appropriate way. The main hurdle for having a well defined contraction is the $k$-behavior of some of the structure constants. If one parametrizes the central charges of the two copies in the following way

$$
c_i = \frac{1}{2\epsilon} \left(\alpha c_M + \beta \epsilon c_L\right), \tag{D.11}
$$

with $\alpha = \beta = 1$ for $c_1 \equiv c$ and $\alpha = -\beta = -1$ for $c_2 \equiv \bar{c}$ then $k_1 \equiv k$ ($k_2 \equiv \bar{k}$) is for $\epsilon \to 0$ approximated by the following expression

$$
k_i \sim -\frac{\alpha c_M}{2N(2-3N+N^2)\epsilon} - \frac{\beta c_L + 2(N(N((N-5)N+5)+1)-1)}{2N(N-2)(N-1)} + \mathcal{O}(\epsilon), \tag{D.12}
$$

which means that every power of $k$ ($\bar{k}$) is proportional to $\frac{c_M}{\epsilon}$. This in turn shows that the central terms can at most be polynomials of degree one in terms of $k$ ($\bar{k}$) and all other structure constants can at most be of degree zero in $k$ ($\bar{k}$). Thus, this rescaling only applies to the higher spin generators $\hat{W}_n^s$ and the generators $G_n^\pm$. After rescaling one can define the following linear combinations

$$
\begin{aligned}
L_n :=& \mathcal{L}_n + \bar{\mathcal{L}}_n, & M_n &:= -\epsilon(\mathcal{L}_n - \bar{\mathcal{L}}_n), & \text{(D.13a)}\\
O_n :=& J_n + \bar{J}_n, & K_n &:= -\epsilon(J_n - \bar{J}_n), & \text{(D.13b)}\\
U_n^\pm :=& G_n^\pm + \bar{G}_n^\pm, & V_n^\pm &:= -\epsilon(G_n^\pm - \bar{G}_n^\pm), & \text{(D.13c)}\\
W_n^s :=& \hat{W}_n^s + \bar{\hat{W}}_n^s, & X_n^s &:= -\epsilon(\hat{W}_n^s - \bar{\hat{W}}_n^s), & \text{(D.13d)}
\end{aligned}
$$

and calculate the resulting algebra in the limit $\epsilon \to 0$. This leads to the following algebra

$$
\begin{aligned}
[L_n, L_m] =& (n-m)L_{m+n} + \frac{c_L}{12}n(n^2-1)\delta_{n+m,0}, & \text{(D.14a)}\\
[L_n, M_m] =& (n-m)M_{m+n} + \frac{c_M}{12}n(n^2-1)\delta_{n+m,0}, & \text{(D.14b)}\\
[L_n, O_m] =& -mO_{n+m}, & \text{(D.14c)}\\
[L_n, K_m] =& -mK_{n+m}, & \text{(D.14d)}\\
[L_n, U_m^\pm] =& \left(\frac{n}{2} - m\right) U_{n+m}^\pm, & \text{(D.14e)}\\
[L_n, V_m^\pm] =& \left(\frac{n}{2} - m\right) V_{n+m}^\pm, & \text{(D.14f)}\\
[M_n, O_m] =& -mK_{n+m}, & \text{(D.14g)}\\
[M_n, U_m^\pm] =& \left(\frac{n}{2} - m\right) V_{n+m}^\pm, & \text{(D.14h)}\\
[O_n, O_m] =& -\frac{2(N-1)^2(N+1) - c_L}{(N-2)N^2}n\delta_{n+m,0}, & \text{(D.14i)}\\
[O_n, K_m] =& -\frac{c_M}{(N-2)N^2}n\delta_{n+m,0}, & \text{(D.14j)}\\
[O_n, U_m^\pm] =& \pm U_{n+m}^\pm, & \text{(D.14k)}\\
[O_n, V_m^\pm] =& \pm V_{n+m}^\pm, & \text{(D.14l)}\\
[K_n, U_m^\pm] =& \pm V_{n+m}^\pm, & \text{(D.14m)}\\
[U_n^+, U_m^-] =& \ldots, & \text{(D.14n)}\\
[U_n^\pm, V_m^\mp] =& \ldots, & \text{(D.14o)}\\
[W_n^s, \text{anything}] =& \ldots, & \text{(D.14p)}\\
[X_n^s, \text{anything}] =& \ldots, & \text{(D.14q)}
\end{aligned}
$$

with the central charges (10.25).

## D.4 $\mathfrak{hs}[\lambda] \to \mathfrak{ihs}[\lambda]$

In this section I show how to contract two copies of $\mathfrak{hs}[\lambda]$ by making use of an İnönü–Wigner contraction. I assume that the two copies are generated by generators $\mathcal{V}^s_n$ and $\bar{\mathcal{V}}^s_n$ (following the definition in [178]), with $s \geq 2$ and $|n| < s$ which obey the following commutations relations

$$[\mathcal{V}^s_n, \mathcal{V}^t_m] = \sum_{\substack{u=2\\ \text{even}}}^{s+t-1} g^{st}_u(m,n;\lambda)\mathcal{V}^{s+t-u}_{n+m}, \quad [\bar{\mathcal{V}}^s_n, \bar{\mathcal{V}}^t_m] = \sum_{\substack{u=2\\ \text{even}}}^{s+t-1} g^{st}_u(m,n;\lambda)\bar{\mathcal{V}}^{s+t-u}_{n+m}, \quad \text{(D.15)}$$

with

$$\begin{aligned} g^{st}_u(m,n;\lambda) &= \frac{2q^{u-2}}{(u-1)!}\phi^{st}_u(\lambda)N^{st}_u(m,n), \\ N^{st}_u(m,n) &= \sum_{k=0}^{u-1}(-1)^k \binom{u-1}{k}[s-1+m]_{u-1-k}[s-1-m]_k \times \\ &\quad \times [t-1+n]_k[t-1-n]_{u-1-k}, \\ \phi^{st}_u(\lambda) &= {}_4F_3\left[\begin{matrix} \frac{1}{2}+\lambda, & \frac{1}{2}-\lambda, & \frac{2-u}{2}, & \frac{1-u}{2} \\ \frac{3}{2}-s, & \frac{1}{2}-t, & \frac{1}{2}+s+t-u & \end{matrix} \middle| 1\right], \end{aligned} \quad \text{(D.16)}$$

where I make use of the descending Pochhammer symbol

$$[a]_n = a(a-1)\dots(a-n+1), \quad \text{(D.17)}$$

and ${}_4F_3$ denotes a generalized hypergeometric series.
A non-relativistic contraction of the algebra $\mathfrak{hs}[\lambda] \oplus \mathfrak{hs}[\lambda]$ can be achieved in the following way. First one defines the following generators in terms of the old $\mathfrak{hs}[\lambda]$ generators and the contraction parameter $\epsilon$

$$V^s_n := \mathcal{V}^s_n + \bar{\mathcal{V}}^s_n, \quad U^s_n := -\epsilon\left(\mathcal{V}^s_n - \bar{\mathcal{V}}^s_n\right). \quad \text{(D.18)}$$

One can now easily determine the commutation relations between these new generators in the limit $\epsilon \to 0$ by using (D.15). These commutation relations which I will call $\mathfrak{ihs}[\lambda]$ are given by

$$[V^s_n, V^t_m] = \sum_{\substack{u=2\\ \text{even}}}^{s+t-1} g^{st}_u(m,n;\lambda)V^{s+t-u}_{n+m}, \quad \text{(D.19a)}$$

$$[V^s_n, U^t_m] = \sum_{\substack{u=2\\ \text{even}}}^{s+t-1} g^{st}_u(m,n;\lambda)U^{s+t-u}_{n+m}, \quad \text{(D.19b)}$$

$$[U^s_n, U^t_m] = 0. \quad \text{(D.19c)}$$

## D.5 $\mathcal{W}_\infty \to \mathcal{FW}_\infty$

The $\mathcal{W}_\infty$-algebra can be seen as a Virasoro-like extension of the $\mathfrak{hs}[1]$ algebra and appears as the asymptotic symmetry algebra of Chern-Simons theories based on the $\mathfrak{hs}[1] \oplus \mathfrak{hs}[1]$ algebra which describes an infinite number of higher-spin fields in AdS$_3$ with integer spins $s \geq 2$. Thus the commutation relations of $\mathcal{W}_\infty$ [155, 156] are given by

$$[\mathcal{V}^s_n, \mathcal{V}^t_m] = \sum_{\substack{u=2\\ \text{even}}}^{s+t-1} g^{st}_u(m,n;1)\mathcal{V}^{s+t-u}_{n+m} + c^s(m)\delta^{st}\delta_{n+m,0}, \tag{D.20a}$$

$$[\bar{\mathcal{V}}^s_n, \bar{\mathcal{V}}^t_m] = \sum_{\substack{u=2\\ \text{even}}}^{s+t-1} g^{st}_u(m,n;1)\bar{\mathcal{V}}^{s+t-u}_{n+m} + \bar{c}^s(m)\delta^{st}\delta_{n+m,0}, \tag{D.20b}$$

where the structure constants are given by (D.16) and the central terms by

$$c^s(m) = f^s(m)c, \quad \bar{c}^s(m) = f^s(m)\bar{c} \tag{D.21}$$

with

$$f^s(m) = \frac{2^{2s-3}s!(s+2)!}{(2s+1)!!(2s+3)!!}\prod_{j=-(s+1)}^{s+1}(m+j). \tag{D.22}$$

Now defining the contracted generators as in (D.18) one obtains immediately $\mathcal{FW}_\infty$ in the $\epsilon \to 0$ limit as

$$[V^s_n, V^t_m] = \sum_{\substack{u=2\\ \text{even}}}^{s+t-1} g^{st}_u(m,n;\lambda)V^{s+t-u}_{n+m} + c^s_L(m)\delta^{st}\delta_{n+m,0}, \tag{D.23a}$$

$$[V^s_n, U^t_m] = \sum_{\substack{u=2\\ \text{even}}}^{s+t-1} g^{st}_u(m,n;\lambda)U^{s+t-u}_{n+m} + c^s_M(m)\delta^{st}\delta_{n+m,0}, \tag{D.23b}$$

$$[U^s_n, U^t_m] = 0. \tag{D.23c}$$

where $c^s_L(m)$, $c^s_M(m)$ are defined in the same way as in (D.21) but with

$$c_L = c + \bar{c}, \quad c_M = -\epsilon(c - \bar{c}). \tag{D.24}$$

# E Higher-Spin Chemical Potentials in Flat Space

In this appendix[1] I collect contributions to the metric $g_{\mu\nu}$ (spin-2 field) and the spin-3 field $\Phi_{\mu\nu\lambda}$ that vanish identically for zero mode solutions with constant chemical potentials, $\mathcal{M}' = \mathcal{N}' = \mu_{\mathrm{M}}' = \mu_{\mathrm{L}}' = \mu_{\mathrm{V}}' = \mu_{\mathrm{U}}' = 0$. The expressions for the metric appearing in (13.12a) are given by

$$\begin{aligned} g_{uu}^{(r)} &= \tfrac{16}{3}\,\mathcal{M}\,(\mu_{\mathrm{U}}'\mu_{\mathrm{V}} - \mu_{\mathrm{U}}\mu_{\mathrm{V}}') - \tfrac{8}{3}\mathcal{N}\mu_{\mathrm{U}}'\mu_{\mathrm{U}} - \tfrac{8}{3}\mathcal{N}'\mu_{\mathrm{U}}^2 \\ &\quad + 2\,(\mu_{\mathrm{L}}'(1+\mu_{\mathrm{M}}) - \mu_{\mathrm{L}}\mu_{\mathrm{M}}') - \tfrac{4}{3}\,(\mu_{\mathrm{U}}'''\mu_{\mathrm{V}} - \mu_{\mathrm{U}}\mu_{\mathrm{V}}''') + 2\,(\mu_{\mathrm{U}}''\mu_{\mathrm{V}}' - \mu_{\mathrm{U}}'\mu_{\mathrm{V}}''), \end{aligned} \tag{E.1a}$$

$$\begin{aligned} g_{uu}^{(0')} &= -\tfrac{2}{3}\mathcal{M}''\mu_{\mathrm{V}}^2 - \tfrac{4}{3}\mathcal{N}''\mu_{\mathrm{U}}\mu_{\mathrm{V}} - \tfrac{5}{3}\mathcal{M}'\mu_{\mathrm{V}}'\mu_{\mathrm{V}} + \tfrac{4}{3}\mathcal{N}'(\mu_{\mathrm{U}}\mu_{\mathrm{V}}' - \tfrac{7}{2}\mu_{\mathrm{U}}'\mu_{\mathrm{V}}) - \tfrac{5}{3}\mathcal{M}(2\mu_{\mathrm{V}}\mu_{\mathrm{V}}'' - \mu_{\mathrm{V}}'^2) \\ &\quad - \tfrac{4}{3}\mathcal{N}(\mu_{\mathrm{U}}\mu_{\mathrm{V}}'' - \tfrac{5}{2}\mu_{\mathrm{U}}'\mu_{\mathrm{V}}' + 4\mu_{\mathrm{U}}''\mu_{\mathrm{V}}) - 2(1+\mu_{\mathrm{M}})\mu_{\mathrm{M}}'' + \mu_{\mathrm{M}}'^2 + \tfrac{2}{3}(\mu_{\mathrm{V}}\mu_{\mathrm{V}}'''' - \mu_{\mathrm{V}}'\mu_{\mathrm{V}}''') + \tfrac{1}{3}\mu_{\mathrm{V}}''^2\,. \end{aligned} \tag{E.1b}$$

The four coefficient-functions of the spin-3 field in $\Phi_{uuu}$ contained in (13.14a) read explicitly

$$\Phi_{uuu}^{(r^3)} = -\tfrac{4}{3}\mathcal{M}'\mu_{\mathrm{U}}^3 - \tfrac{4}{3}\mathcal{M}\mu_{\mathrm{U}}^2\mu_{\mathrm{U}}' - \mu_{\mathrm{L}}^2\mu_{\mathrm{U}}' + 2\mu_{\mathrm{L}}\mu_{\mathrm{L}}'\mu_{\mathrm{U}} + \tfrac{4}{3}\mu_{\mathrm{U}}^2\mu_{\mathrm{U}}''' - 2\mu_{\mathrm{U}}\mu_{\mathrm{U}}'\mu_{\mathrm{U}}'' + \mu_{\mathrm{U}}'^3, \tag{E.2a}$$

$$\Phi_{uuu}^{(r^2)} = \Phi_{uuu}^{(r^2,Q'')} + \Phi_{uuu}^{(r^2,Q')} + \Phi_{uuu}^{(r^2,Q)} + \Phi_{uuu}^{(r^2,\mathrm{rest})}, \tag{E.2b}$$

$$\begin{aligned} \Phi_{uuu}^{(r)} &= \Phi_{uuu}^{(r,Q^2)} + \Phi_{uuu}^{(r,Q\cdot Q')} + \Phi_{uuu}^{(r,Q'')} + \Phi_{uuu}^{(r,Q')} + \mathcal{M}\Phi_{uuu}^{(r,\mathcal{M})} + \mathcal{N}\Phi_{uuu}^{(r,\mathcal{N})} \\ &\quad + \mathcal{V}\Phi_{uuu}^{(r,\mathcal{V})} + \mathcal{Z}\Phi_{uuu}^{(r,\mathcal{Z})} + \Phi_{uuu}^{(r,\mathrm{rest})}, \end{aligned} \tag{E.2c}$$

$$\begin{aligned} \Phi_{uuu}^{(0)} &= \Phi_{uuu}^{(Q^2)} + \Phi_{uuu}^{(Q\cdot Q)} + \mathcal{M}''\Phi_{uuu}^{(\mathcal{M}'')} + \mathcal{N}''\Phi_{uuu}^{(\mathcal{N}'')} + \mathcal{M}'\Phi_{uuu}^{(\mathcal{M}')} + \mathcal{N}'\Phi_{uuu}^{(\mathcal{N}')} \\ &\quad + \mathcal{M}\Phi_{uuu}^{(\mathcal{M})} + \mathcal{N}\Phi_{uuu}^{(\mathcal{N})} + \mathcal{V}\Phi_{uuu}^{(\mathcal{V})} + \mathcal{Z}\Phi_{uuu}^{(\mathcal{Z})} + \Phi_{uuu}^{(\mathrm{rest})}, \end{aligned} \tag{E.2d}$$

with the quadratic part

$$\Phi_{uuu}^{(r^2,Q'')} = \tfrac{2}{3}\mathcal{M}''\mu_{\mathrm{U}}^2\mu_{\mathrm{V}} + \tfrac{4}{3}\mathcal{N}''\mu_{\mathrm{U}}^3, \tag{E.3a}$$

$$\Phi_{uuu}^{(r^2,Q')} = \mathcal{M}'\mu_{\mathrm{U}}(\tfrac{11}{3}\mu_{\mathrm{U}}\mu_{\mathrm{V}}' - 2\mu_{\mathrm{U}}'\mu_{\mathrm{V}}) + \tfrac{10}{3}\mathcal{N}'\mu_{\mathrm{U}}^2\mu_{\mathrm{U}}', \tag{E.3b}$$

$$\Phi_{uuu}^{(r^2,Q)} = \mathcal{M}(\mu_{\mathrm{U}}^2\mu_{\mathrm{V}}'' + \tfrac{4}{3}\mu_{\mathrm{U}}\mu_{\mathrm{U}}''\mu_{\mathrm{V}} + \tfrac{2}{3}\mu_{\mathrm{U}}\mu_{\mathrm{U}}'\mu_{\mathrm{V}}' - \tfrac{7}{3}\mu_{\mathrm{U}}'^2\mu_{\mathrm{V}}) + 4\mathcal{N}(\mu_{\mathrm{U}}^2\mu_{\mathrm{U}}'' - \tfrac{1}{3}\mu_{\mathrm{U}}\mu_{\mathrm{U}}'^2), \tag{E.3c}$$

$$\begin{aligned} \Phi_{uuu}^{(r^2,\mathrm{rest})} &= (1+\mu_{\mathrm{M}})(\mu_{\mathrm{L}}'\mu_{\mathrm{U}}' - \mu_{\mathrm{L}}\mu_{\mathrm{U}}'') + 2\mu_{\mathrm{M}}'(\mu_{\mathrm{L}}\mu_{\mathrm{U}}' - \mu_{\mathrm{L}}'\mu_{\mathrm{U}}) - 2\mu_{\mathrm{M}}''\mu_{\mathrm{L}}\mu_{\mathrm{U}} \\ &\quad + \mu_{\mathrm{L}}'(\mu_{\mathrm{L}}'\mu_{\mathrm{V}} - \mu_{\mathrm{L}}\mu_{\mathrm{V}}') + \tfrac{1}{3}\mu_{\mathrm{L}}^2\mu_{\mathrm{V}}'' - \tfrac{2}{3}\mu_{\mathrm{U}}^2\mu_{\mathrm{V}}'''' - \tfrac{4}{3}\mu_{\mathrm{U}}\mu_{\mathrm{U}}'''\mu_{\mathrm{V}}' + \tfrac{2}{3}\mu_{\mathrm{U}}\mu_{\mathrm{U}}''\mu_{\mathrm{V}}'' \\ &\quad + \tfrac{2}{3}\mu_{\mathrm{U}}\mu_{\mathrm{U}}'\mu_{\mathrm{V}}'^3 - \mu_{\mathrm{U}}'^2\mu_{\mathrm{V}}'' - \mu_{\mathrm{U}}''^2\mu_{\mathrm{V}} + \mu_{\mathrm{U}}''\mu_{\mathrm{U}}'\mu_{\mathrm{V}}' + \tfrac{4}{3}\mu_{\mathrm{U}}'''\mu_{\mathrm{U}}'\mu_{\mathrm{V}}, \end{aligned} \tag{E.3d}$$

[1] Please note that this appendix is almost identical to the one that was already published in [VII]. I decided to include this appendix nevertheless for the sake of completeness.

the linear part

$$\Phi^{(r,Q^2)}_{uuu} = \tfrac{16}{9}\mathcal{M}^2\mu_{\mathrm{V}}(\mu_{\mathrm{U}}\mu'_{\mathrm{V}} - \mu'_{\mathrm{U}}\mu_{\mathrm{V}}) + \tfrac{20}{9}\mathcal{M}\mathcal{N}\mu_{\mathrm{U}}^2\mu'_{\mathrm{V}} - \tfrac{8}{3}\mathcal{M}\mathcal{N}\mu_{\mathrm{U}}\mu'_{\mathrm{U}}\mu_{\mathrm{V}} - \tfrac{8}{9}\mathcal{N}^2\mu_{\mathrm{U}}^2\mu'_{\mathrm{U}}, \tag{E.4a}$$

$$\Phi^{(r,Q\cdot Q')}_{uuu} = -\tfrac{4}{3}\mathcal{M}'\mathcal{N}\mu_{\mathrm{U}}^2\mu_{\mathrm{V}} + \tfrac{8}{9}\mathcal{M}\mathcal{N}'\mu_{\mathrm{U}}^2\mu_{\mathrm{V}} - \tfrac{8}{9}\mathcal{N}\mathcal{N}'\mu_{\mathrm{U}}^3, \tag{E.4b}$$

$$\Phi^{(r,Q'')}_{uuu} = \tfrac{2}{3}\mathcal{M}''\mu_{\mathrm{V}}(\mu'_{\mathrm{U}}\mu_{\mathrm{V}} - \mu_{\mathrm{U}}\mu'_{\mathrm{V}}) + \tfrac{4}{3}\mathcal{N}''\mu_{\mathrm{U}}(\mu'_{\mathrm{U}}\mu_{\mathrm{V}} - \mu_{\mathrm{U}}\mu'_{\mathrm{V}}), \tag{E.4c}$$

$$\begin{aligned}\Phi^{(r,Q')}_{uuu} &= \tfrac{1}{3}\mathcal{M}'\big((1+\mu_{\mathrm{M}})^2\mu_{\mathrm{U}} - (1+\mu_{\mathrm{M}})\mu_{\mathrm{L}}\mu_{\mathrm{V}} + 2\mu_{\mathrm{V}}(\mu_{\mathrm{U}}\mu''_{\mathrm{V}} - \mu''_{\mathrm{U}}\mu_{\mathrm{V}}) + 8\mu'_{\mathrm{V}}(\mu'_{\mathrm{U}}\mu_{\mathrm{V}} - \mu_{\mathrm{U}}\mu'_{\mathrm{V}})\big)\\ &\quad - \tfrac{2}{3}\mathcal{N}'\big((1+\mu_{\mathrm{M}})\mu_{\mathrm{L}}\mu_{\mathrm{U}} - 7\mu'^2_{\mathrm{U}}\mu_{\mathrm{V}} + 2\mu_{\mathrm{U}}\mu''_{\mathrm{U}}\mu_{\mathrm{V}} - \tfrac{2}{3}\mu_{\mathrm{U}}^2\mu''_{\mathrm{V}} + 6\mu_{\mathrm{U}}\mu'_{\mathrm{U}}\mu'_{\mathrm{V}}\big),\end{aligned} \tag{E.4d}$$

$$\begin{aligned}\Phi^{(r,\mathcal{M})}_{uuu} &= \tfrac{4}{3}(1+\mu_{\mathrm{M}})^2\mu'_{\mathrm{U}} - 2(1+\mu_{\mathrm{M}})\big(\mu'_{\mathrm{M}}\mu_{\mathrm{U}} + \tfrac{2}{3}(\mu_{\mathrm{L}}\mu'_{\mathrm{V}} - \mu'_{\mathrm{L}}\mu_{\mathrm{V}})\big) + \tfrac{2}{3}\mu_{\mathrm{L}}\mu'_{\mathrm{M}}\mu_{\mathrm{V}}\\ &\quad + \tfrac{28}{9}\mu'_{\mathrm{U}}\mu_{\mathrm{V}}\mu''_{\mathrm{V}} - \tfrac{16}{9}\mu_{\mathrm{U}}\mu''_{\mathrm{V}}\mu'_{\mathrm{V}} - \tfrac{4}{3}\mu''_{\mathrm{U}}\mu_{\mathrm{V}}\mu'_{\mathrm{V}} - \tfrac{4}{9}\mu_{\mathrm{U}}\mu_{\mathrm{V}}\mu'''_{\mathrm{V}} + \tfrac{4}{9}\mu'''_{\mathrm{U}}\mu_{\mathrm{V}}^2,\end{aligned} \tag{E.4e}$$

$$\begin{aligned}\Phi^{(r,\mathcal{N})}_{uuu} &= \tfrac{2}{3}(1+\mu_{\mathrm{M}})\big(\mu'_{\mathrm{L}}\mu_{\mathrm{U}} - \mu_{\mathrm{L}}\mu'_{\mathrm{U}}\big) - \tfrac{2}{3}\mu_{\mathrm{L}}\mu'_{\mathrm{M}}\mu_{\mathrm{U}} + 2\mu_{\mathrm{L}}\mu'_{\mathrm{L}}\mu_{\mathrm{V}} - \mu_{\mathrm{L}}^2\mu'_{\mathrm{V}} + \tfrac{8}{9}\mu_{\mathrm{U}}\mu'''_{\mathrm{U}}\mu_{\mathrm{V}}\\ &\quad - \tfrac{14}{3}\mu_{\mathrm{U}}\mu''_{\mathrm{U}}\mu'_{\mathrm{V}} + 2\mu''_{\mathrm{U}}\mu'_{\mathrm{U}}\mu_{\mathrm{V}} + \tfrac{4}{9}\mu_{\mathrm{U}}^2\mu'''_{\mathrm{V}} + \tfrac{5}{3}\mu'^2_{\mathrm{U}}\mu'_{\mathrm{V}} - \tfrac{2}{9}\mu_{\mathrm{U}}\mu'_{\mathrm{U}}\mu''_{\mathrm{V}},\end{aligned} \tag{E.4f}$$

$$\Phi^{(r,\mathcal{V})}_{uuu} = 8(1+\mu_{\mathrm{M}})\mu_{\mathrm{U}}\mu'_{\mathrm{V}} - 8\mu_{\mathrm{L}}\mu_{\mathrm{V}}\mu'_{\mathrm{V}} - 16\mu'_{\mathrm{M}}\mu_{\mathrm{U}}\mu_{\mathrm{V}} + 16\mu'_{\mathrm{L}}\mu_{\mathrm{V}}^2 \tag{E.4g}$$

$$\Phi^{(r,\mathcal{Z})}_{uuu} = 16(1+\mu_{\mathrm{M}})\mu_{\mathrm{U}}\mu'_{\mathrm{U}} - 16\mu_{\mathrm{L}}\mu_{\mathrm{V}}\mu'_{\mathrm{U}} - 32\mu'_{\mathrm{M}}\mu_{\mathrm{U}}^2 + 32\mu'_{\mathrm{L}}\mu_{\mathrm{U}}\mu_{\mathrm{V}}, \tag{E.4h}$$

$$\begin{aligned}\Phi^{(r,\mathrm{rest})}_{uuu} &= -\tfrac{1}{3}(1+\mu_{\mathrm{M}})^2\mu'''_{\mathrm{U}} + (1+\mu_{\mathrm{M}})\big(\mu'_{\mathrm{M}}\mu''_{\mathrm{U}} - \mu''_{\mathrm{M}}\mu'_{\mathrm{U}}\big) + \tfrac{1}{3}(1+\mu_{\mathrm{M}})\big(\mu_{\mathrm{L}}\mu'''_{\mathrm{V}} - \mu'_{\mathrm{L}}\mu''_{\mathrm{V}}\big)\\ &\quad - \mu'^2_{\mathrm{M}}\mu'_{\mathrm{U}} + 2\mu''_{\mathrm{M}}\mu'_{\mathrm{M}}\mu_{\mathrm{U}} - 2\mu'_{\mathrm{L}}\mu''_{\mathrm{M}}\mu_{\mathrm{V}} - \tfrac{2}{3}\mu_{\mathrm{L}}\mu'_{\mathrm{M}}\mu''_{\mathrm{V}} + \mu_{\mathrm{L}}\mu''_{\mathrm{M}}\mu'_{\mathrm{V}} + \mu'_{\mathrm{L}}\mu'_{\mathrm{M}}\mu'_{\mathrm{V}}\\ &\quad + \tfrac{2}{3}\mu''_{\mathrm{U}}\mu_{\mathrm{V}}\mu'''_{\mathrm{V}} - \tfrac{2}{3}\mu'_{\mathrm{U}}\mu_{\mathrm{V}}\mu''''_{\mathrm{V}} - \tfrac{4}{9}\mu'''_{\mathrm{U}}\mu_{\mathrm{V}}\mu''_{\mathrm{V}} - \tfrac{2}{9}\mu_{\mathrm{U}}\mu'''_{\mathrm{V}}\mu''_{\mathrm{V}} + \tfrac{1}{3}\mu'_{\mathrm{U}}\mu''^2_{\mathrm{V}}\\ &\quad - \tfrac{1}{3}\mu'_{\mathrm{U}}\mu'''_{\mathrm{V}}\mu'_{\mathrm{V}} + \tfrac{2}{3}\mu_{\mathrm{U}}\mu''''_{\mathrm{V}}\mu'_{\mathrm{V}} - \tfrac{1}{3}\mu''_{\mathrm{U}}\mu''_{\mathrm{V}}\mu'_{\mathrm{V}} + \tfrac{1}{3}\mu'''_{\mathrm{U}}\mu'^2_{\mathrm{V}},\end{aligned} \tag{E.4i}$$

and the constant part

$$\Phi^{(Q^2)}_{uuu} = \tfrac{2}{3}\mathcal{M}^2\big(\tfrac{5}{3}\mu_{\mathrm{V}}^2\mu''_{\mathrm{V}} - \mu_{\mathrm{V}}\mu'^2_{\mathrm{V}}\big) - \tfrac{4}{9}\mathcal{N}^2\big(\mu_{\mathrm{U}}^2\mu''_{\mathrm{V}} - \tfrac{5}{2}\mu_{\mathrm{U}}\mu'_{\mathrm{U}}\mu'_{\mathrm{V}} - 8\mu_{\mathrm{U}}\mu''_{\mathrm{U}}\mu_{\mathrm{V}} + \tfrac{25}{4}\mu'^2_{\mathrm{U}}\mu_{\mathrm{V}}\big), \tag{E.5a}$$

$$\Phi^{(Q\cdot Q)}_{uuu} = \tfrac{4}{9}\mathcal{M}\mathcal{N}\big(6\mu_{\mathrm{U}}\mu_{\mathrm{V}}\mu''_{\mathrm{V}} - \mu_{\mathrm{U}}\mu'^2_{\mathrm{V}} - 5\mu'_{\mathrm{U}}\mu_{\mathrm{V}}\mu'_{\mathrm{V}} + 4\mu''_{\mathrm{U}}\mu_{\mathrm{V}}^2\big), \tag{E.5b}$$

$$\Phi^{(\mathcal{M}'')}_{uuu} = \tfrac{2}{9}\mathcal{M}\mu_{\mathrm{V}}^3 + \tfrac{4}{9}\mathcal{N}\mu_{\mathrm{U}}\mu_{\mathrm{V}}^2 - \tfrac{1}{6}(1+\mu_{\mathrm{M}})^2\mu_{\mathrm{V}} - \tfrac{2}{9}\mu_{\mathrm{V}}^2\mu''_{\mathrm{V}} + \tfrac{1}{6}\mu_{\mathrm{V}}\mu'^2_{\mathrm{V}}, \tag{E.5c}$$

$$\Phi^{(\mathcal{N}'')}_{uuu} = \tfrac{4}{9}\mathcal{M}\mu_{\mathrm{U}}\mu_{\mathrm{V}}^2 + \tfrac{8}{9}\mathcal{N}\mu_{\mathrm{U}}^2\mu_{\mathrm{V}} - \tfrac{1}{3}(1+\mu_{\mathrm{M}})^2\mu_{\mathrm{U}} - \tfrac{4}{9}\mu_{\mathrm{U}}\mu_{\mathrm{V}}\mu''_{\mathrm{V}} + \tfrac{1}{3}\mu_{\mathrm{U}}\mu'^2_{\mathrm{V}}, \tag{E.5d}$$

$$\begin{aligned}\Phi^{(\mathcal{M}')}_{uuu} &= -\tfrac{1}{9}\mathcal{M}'\mu_{\mathrm{V}}^3 - \tfrac{4}{9}\mathcal{N}'\mu_{\mathrm{U}}\mu_{\mathrm{V}}^2 + \tfrac{1}{3}\mathcal{M}\mu_{\mathrm{V}}^2\mu'_{\mathrm{V}} + \tfrac{16}{9}\mathcal{N}\mu_{\mathrm{U}}\mu_{\mathrm{V}}\mu'_{\mathrm{V}} - \tfrac{10}{9}\mathcal{N}\mu'_{\mathrm{U}}\mu_{\mathrm{V}}^2\\ &\quad - \tfrac{7}{12}(1+\mu_{\mathrm{M}})^2\mu'_{\mathrm{V}} + \tfrac{1}{3}(1+\mu_{\mathrm{M}})\mu'_{\mathrm{M}}\mu_{\mathrm{V}} + \tfrac{2}{9}\mu_{\mathrm{V}}^2\mu'''_{\mathrm{V}} - \tfrac{8}{9}\mu_{\mathrm{V}}\mu''_{\mathrm{V}}\mu'_{\mathrm{V}} + \tfrac{7}{12}\mu'^3_{\mathrm{V}},\end{aligned} \tag{E.5e}$$

$$\begin{aligned}\Phi^{(\mathcal{N}')}_{uuu} &= -\tfrac{4}{9}\mathcal{N}'\mu_{\mathrm{U}}^2\mu_{\mathrm{V}} + \tfrac{8}{9}\mathcal{M}\mu_{\mathrm{V}}\big(\tfrac{7}{4}\mu'_{\mathrm{U}}\mu_{\mathrm{V}} - \mu_{\mathrm{U}}\mu'_{\mathrm{V}}\big) + \tfrac{8}{9}\mathcal{N}\mu_{\mathrm{U}}\big(2\mu'_{\mathrm{U}}\mu_{\mathrm{V}} + \mu_{\mathrm{U}}\mu'_{\mathrm{V}}\big) - \tfrac{7}{6}(1+\mu_{\mathrm{M}})^2\mu'_{\mathrm{U}}\\ &\quad + \tfrac{2}{3}(1+\mu_{\mathrm{M}})\mu'_{\mathrm{M}}\mu_{\mathrm{U}} + \tfrac{4}{9}\mu_{\mathrm{U}}\mu_{\mathrm{V}}\mu'''_{\mathrm{V}} - \tfrac{2}{9}\mu_{\mathrm{U}}\mu'_{\mathrm{V}}\mu''_{\mathrm{V}} - \tfrac{14}{9}\mu'_{\mathrm{U}}\mu_{\mathrm{V}}\mu''_{\mathrm{V}} + \tfrac{7}{6}\mu'_{\mathrm{U}}\mu'^2_{\mathrm{V}},\end{aligned} \tag{E.5f}$$

$$\begin{aligned}\Phi^{(\mathcal{M})}_{uuu} &= -\tfrac{5}{6}(1+\mu_{\mathrm{M}})^2\mu''_{\mathrm{V}} + \tfrac{4}{3}(1+\mu_{\mathrm{M}})\mu'_{\mathrm{M}}\mu'_{\mathrm{V}} - \tfrac{4}{3}(1+\mu_{\mathrm{M}})\mu''_{\mathrm{M}}\mu_{\mathrm{V}} - \tfrac{1}{3}\mu'^2_{\mathrm{M}}\mu_{\mathrm{V}}\\ &\quad - \tfrac{2}{9}\mu_{\mathrm{V}}^2\mu''''_{\mathrm{V}} + \tfrac{4}{9}\mu_{\mathrm{V}}\mu'_{\mathrm{V}}\mu'''_{\mathrm{V}} - \tfrac{7}{9}\mu_{\mathrm{V}}\mu''^2_{\mathrm{V}} + \tfrac{7}{18}\mu'^2_{\mathrm{V}}\mu''_{\mathrm{V}},\end{aligned} \tag{E.5g}$$

$$\begin{aligned}\Phi^{(\mathcal{N})}_{uuu} &= -\tfrac{4}{3}(1+\mu_{\mathrm{M}})^2\mu''_{\mathrm{U}} + \tfrac{5}{3}(1+\mu_{\mathrm{M}})\mu'_{\mathrm{M}}\mu'_{\mathrm{U}} - \tfrac{2}{3}(1+\mu_{\mathrm{M}})\mu''_{\mathrm{M}}\mu_{\mathrm{U}} - \tfrac{1}{3}(1+\mu_{\mathrm{M}})\mu_{\mathrm{L}}\mu''_{\mathrm{V}}\\ &\quad - \tfrac{2}{3}\mu'^2_{\mathrm{M}}\mu_{\mathrm{U}} + \mu'_{\mathrm{M}}\mu_{\mathrm{L}}\mu'_{\mathrm{V}} - 2\mu''_{\mathrm{M}}\mu_{\mathrm{L}}\mu_{\mathrm{V}} - \tfrac{2}{9}\mu_{\mathrm{U}}\big(2\mu_{\mathrm{V}}\mu''''_{\mathrm{V}} + \mu'_{\mathrm{V}}\mu'''_{\mathrm{V}} - \mu''^2_{\mathrm{V}}\big)\\ &\quad + \tfrac{10}{9}\mu'_{\mathrm{U}}\mu_{\mathrm{V}}\mu'''_{\mathrm{V}} - \tfrac{5}{9}\mu'_{\mathrm{U}}\mu'_{\mathrm{V}}\mu''_{\mathrm{V}} - \tfrac{16}{9}\mu''_{\mathrm{U}}\mu_{\mathrm{V}}\mu''_{\mathrm{V}} + \tfrac{4}{3}\mu''_{\mathrm{U}}\mu'^2_{\mathrm{V}},\end{aligned} \tag{E.5h}$$

$$\Phi^{(\mathcal{V})}_{uuu} = -2(1+\mu_{\rm M})\mu_{\rm V}'^{\,2} + 8\mu_{\rm M}'\mu_{\rm V}\mu_{\rm V}' - 16\mu_{\rm M}''\mu_{\rm V}^2, \tag{E.5i}$$

$$\Phi^{(\mathcal{Z})}_{uuu} = -\tfrac{16}{3}(1+\mu_{\rm M})\mu_{\rm U}\mu_{\rm V}'' + 16\mu_{\rm M}'\mu_{\rm U}\mu_{\rm V}' - 32\mu_{\rm M}''\mu_{\rm U}\mu_{\rm V} + \tfrac{16}{3}\mu_{\rm L}\mu_{\rm V}\mu_{\rm V}'' - 4\mu_{\rm L}\mu_{\rm V}'^{\,2}, \tag{E.5j}$$

$$\begin{aligned}\Phi^{(\text{rest})}_{uuu} = {} & \tfrac{1}{6}(1+\mu_{\rm M})^2\mu_{\rm V}'''' - \tfrac{1}{3}(1+\mu_{\rm M})\mu_{\rm M}'\mu_{\rm V}''' + \tfrac{1}{3}(1+\mu_{\rm M})\mu_{\rm M}''\mu_{\rm V}'' + \tfrac{1}{3}\mu_{\rm M}'^{\,2}\mu_{\rm V}'' - \mu_{\rm M}'\mu_{\rm M}''\mu_{\rm V}' \\ & + \mu_{\rm M}''^{\,2}\mu_{\rm V} + \tfrac{2}{9}\mu_{\rm V}\mu_{\rm V}''\mu_{\rm V}'''' - \tfrac{1}{9}\mu_{\rm V}\mu_{\rm V}'''^{\,2} + \tfrac{1}{9}\mu_{\rm V}'\mu_{\rm V}''\mu_{\rm V}''' - \tfrac{1}{6}\mu_{\rm V}'^{\,2}\mu_{\rm V}'''' - \tfrac{1}{27}\mu_{\rm V}''^{\,3}\,. \end{aligned} \tag{E.5k}$$

The remaining non-constant contributions appearing in (13.14) are given by

$$\Phi^{(r)}_{ruu} = -(1+\mu_{\rm M})\mu_{\rm U}' + 2\mu_{\rm M}'\mu_{\rm U} + \mu_{\rm L}\mu_{\rm V}' - 2\mu_{\rm L}'\mu_{\rm V}, \tag{E.6a}$$

$$\Phi^{(0)}_{ruu} = \tfrac{1}{3}(1+\mu_{\rm M})\mu_{\rm V}'' - \mu_{\rm M}'\mu_{\rm V}' + 2\mu_{\rm M}''\mu_{\rm V}, \tag{E.6b}$$

and

$$\Phi^{(r^3)}_{uu\varphi} = -2(\mu_{\rm L}\mu_{\rm U}' - \mu_{\rm L}'\mu_{\rm U}), \tag{E.7a}$$

$$\Phi^{(r^2)}_{uu\varphi} = -(1+\mu_{\rm M})\mu_{\rm U}'' + 2\mu_{\rm M}'\mu_{\rm U}' - 2\mu_{\rm M}''\mu_{\rm U} + \tfrac{2}{3}\mu_{\rm L}\mu_{\rm V}'' - \mu_{\rm L}'\mu_{\rm V}', \tag{E.7b}$$

$$\begin{aligned}\Phi^{(r)}_{uu\varphi} = {} & -\tfrac{4}{3}\mathcal{M}(1+\mu_{\rm M})\mu_{\rm V}' - \tfrac{1}{3}\mathcal{M}'(1+\mu_{\rm M})\mu_{\rm V} + \tfrac{2}{3}\mathcal{M}\mu_{\rm M}'\mu_{\rm V} - 2\mathcal{N}(\mu_{\rm L}\mu_{\rm V}' - \mu_{\rm L}'\mu_{\rm V}) \\ & - \tfrac{2}{3}\big(\mathcal{N}(1+\mu_{\rm M})\mu_{\rm U}\big)' - 8\mathcal{V}\mu_{\rm V}\mu_{\rm V}' - 16\mathcal{Z}\mu_{\rm U}'\mu_{\rm V} + \tfrac{1}{3}(1+\mu_{\rm M})\mu_{\rm V}''' - \tfrac{2}{3}\mu_{\rm M}'\mu_{\rm V}'' + \mu_{\rm M}''\mu_{\rm V}', \end{aligned} \tag{E.7c}$$

$$\Phi^{(0)}_{uu\varphi} = \mathcal{N}\big(-\tfrac{1}{3}(1+\mu_{\rm M})\mu_{\rm V}'' + \mu_{\rm M}'\mu_{\rm V}' - 2\mu_{\rm M}''\mu_{\rm V}\big) + 8\mathcal{Z}\big(\tfrac{2}{3}\mu_{\rm V}\mu_{\rm V}'' - \tfrac{1}{2}\mu_{\rm V}'^{\,2}\big)\,. \tag{E.7d}$$